\documentclass[a4paper,11pt]{article}
\pdfoutput=1 

\usepackage{jheppub} 

\usepackage[T1]{fontenc} 
\usepackage{amsmath}
\usepackage{amssymb}
\usepackage{amsfonts}
\usepackage{mathtools}
\usepackage{microtype}
\usepackage{tikz}
\usepackage{bm}
\usepackage[utf8]{inputenc}
\usepackage{blkarray}
\usepackage{bigstrut}
\usepackage{nccmath}
\usepackage{enumitem}
\usepackage{dsfont}
\usepackage{physics}
\usepackage{float}
\usepackage{subfig}

\newcommand{\masters}{\nu}
\newcommand{\lbl}{Loop-by-Loop~}

\def \be {\begin{equation}}
\def \ee {\end{equation}}

\title{\boldmath Decomposition of Feynman Integrals on the Maximal Cut by Intersection Numbers}

\author[a,b]{Hjalte Frellesvig,}
\author[a,b]{\ Federico Gasparotto,}
\author[a,b]{\ Stefano Laporta,}
\author[a,b]{\ Manoj K. Mandal,} 
\author[a,b]{\ Pierpaolo Mastrolia,}
\author[b,a]{\ Luca Mattiazzi,}
\author[c,d]{\ Sebastian Mizera}

\affiliation[a]{Dipartimento di Fisica e Astronomia, Universit\`a di Padova, Via Marzolo 8, 35131 Padova, Italy}

\affiliation[b]{INFN, Sezione di Padova, Via Marzolo 8, 35131 Padova, Italy}

\affiliation[c]{Perimeter Institute for Theoretical Physics, Waterloo, ON N2L 2Y5, Canada}
\affiliation[d]{Department of Physics \& Astronomy, University of Waterloo, Waterloo, ON N2L 3G1, Canada}

\emailAdd{hjalte.frellesvig@pd.infn.it}
\emailAdd{federico.gasparotto@pd.infn.it} 
\emailAdd{stefano.laporta@pd.infn.it}
\emailAdd{manojkumar.mandal@pd.infn.it}
\emailAdd{pierpaolo.mastrolia@pd.infn.it}
\emailAdd{luca.mattiazzi@pd.infn.it}
\emailAdd{smizera@pitp.ca}

\abstract{
We elaborate on the recent idea of a direct decomposition of Feynman integrals onto a basis of master integrals on maximal cuts using intersection numbers.
We begin by showing an application of the method to the derivation of contiguity relations for special functions, such as the Euler beta function, the Gauss ${}_2F_1$ hypergeometric function, and the Appell $F_1$ function. Then, we apply the new method to decompose Feynman integrals whose maximal cuts admit 1-form integral representations, including examples that have from two to an arbitrary number of loops, and/or from zero to an arbitrary number of legs. 
Direct constructions of differential equations and dimensional recurrence relations for Feynman integrals are also discussed.
We present two novel approaches to decomposition-by-intersections 
in cases where the maximal cuts admit a 2-form integral representation, 
with a view towards the extension of the formalism to $n$-form representations.
The decomposition formulae computed through the use of intersection numbers are directly verified to agree with the ones obtained using integration-by-parts identities.
}

\begin{document} 
\maketitle
\flushbottom

\newpage
\section{Introduction}

The highly interactive nature of elementary-particle dynamics is an extremely complex problem to describe. 
Feynman diagrams play a fundamental role by encoding how, within the perturbative approach to quantum field theory, complex scattering reactions may arise from the multiple interactions among simpler network components, representing either external or internal particles. 
For each scattering process, the sum of all diagrams gives the so called {\it scattering amplitude}, 
a complex function of external kinematics (and internal quantum numbers) whose absolute squared value determines the quantum-mechanical density of probability for that reaction to happen. 

The shape of the diagrams informs directly about the complexity level of the process they describe, which increases with the number of interacting external particles, 
the number of interaction vertices,
and with the number of loops. 
In general, Feynman diagrams represent functions of kinematic invariants formed by the external momenta and by the masses of the particles.
While scattering amplitudes associated to tree-level graphs can be written in terms of rational functions, those coming from multi-loop graphs are usually decomposed into special functions admitting multifold integral representations.

Feynman integrals in dimensional regularization are known to be not mutually independent. 
A novel way to access their algebraic structure has been recently unveiled by some of us, who showed that relations for multi-loop Feynman integrals are controlled by {\it intersection numbers} \cite{Mastrolia:2018uzb}.
 
The by-now standard evaluation techniques of Feynman integrals exploit the loop-momentum shift invariance to establish  \emph{integration-by-parts} (IBP) relations  \cite{Chetyrkin:1981qh} among integrals whose integrands are built out of products of the same set of denominators (and scalar products), but raised to different powers. 
IBP identities have been playing a crucial role in the calculation of multi-loop integrals, because they yield the identification of 
a minimal set of elements, dubbed
\emph{master integrals} (MIs), which can be used as a basis for the decomposition of multi-loop amplitudes.
At the same time, IBP-decomposition algorithms can be applied to special integrands, 
built by acting on the master integrand with differential operators (w.r.t. kinematic invariants), or by multiplying their numerators by polynomials  
which modify their dimensions, or by considering
arbitrary denominator powers, respectively turning the decomposition formulas into 
 differential equations \cite{Barucchi:1973zm,KOTIKOV1991158,KOTIKOV1991123,Bern:1993kr,Remiddi:1997ny,Gehrmann:1999as,Henn:2013pwa,Henn:2014qga}, dimensional recurrence relations \cite{Tarasov:1996br,Lee:2009dh}, and finite difference equations \cite{Laporta:2001dd,Laporta:2003jz} obeyed by MIs.
 Solving them amounts to the actual determination of the MIs themselves, as an alternative to the use of  direct integration techniques.

The derivation of the IBP-decomposition formulas requires the solution of a large system of linear relations, generated by imposing that integrals of total derivatives vanish on the integration boundary \cite{Laporta:2001dd}, see also  \cite{smirnov2005evaluating,Grozin:2011mt,Zhang:2016kfo,Kotikov:2018wxe}. 
For multi-loop multi-scale scattering amplitudes, solving the system of IBP relations may however represent a formidable task, whose accomplishment has been motivating  
important refinements of the system-solving strategies \cite{vonManteuffel:2014ixa,Peraro:2016wsq,Boehm:2018fpv,Kosower:2018obg,Liu:2018dmc,Kardos:2018uzy,Chawdhry:2018awn,vonManteuffel:2012np,Lee:2012cn,Smirnov:2008iw,Maierhoefer:2017hyi,Georgoudis:2016wff,Maierhofer:2018gpa,Smirnov:2019qkx}. 
Together with novel algorithms for simplifying the solution of systems of differential equations \cite{Henn:2013pwa}, which triggered further studies \cite{Argeri:2014qva,Lee:2014ioa,Gehrmann:2014bfa,Gituliar:2017vzm,Lee:2017oca,Adams:2018yfj}, the calculations of several multi-scale, multi-loop, multi-particle amplitudes became feasible  
\cite{Laporta:2017okg,Anastasiou:2015ema, Bonciani:2015eua,Gehrmann:2015dua, Bonetti:2017ovy,Borowka:2016ypz,Baglio:2018lrj,Lindert:2017pky,Jones:2018hbb,Maltoni:2018zvp,Gehrmann:2015bfy,Badger:2017jhb,Abreu:2018jgq,Abreu:2018zmy,Chicherin:2018yne,Abreu:2018aqd,Badger:2018enw,Chicherin:2019xeg,Abreu:2019rpt}.

The most recent developments in the research of mathematical methods for the evaluation of 
Feynman integrals have been benefiting from a special representation known as \emph{Baikov representation} \cite{Baikov:1996iu}, where
instead of the components of the loop momenta, the propagators themselves supplemented by independent scalar products between external and internal momenta, are the integration variables. 
This change of variables introduces a Jacobian equal to the Gram determinant of the scalar products formed by both types of momenta, referred to as the \emph{Baikov polynomial}.
The Baikov polynomial fully characterizes the space on which the integrals are defined, and in particular
the number of MIs can be inferred from the number of its critical points \cite{Lee:2013hzt}, see also \cite{Marcolli:2008vr,marcolli2010feynman,Bitoun:2017nre}.
IBP identities may relate integrals corresponding to a given graph to integrals that correspond to its sub-graphs. We may consider these two sets of integrals, respectively, as to the homogeneous and the non-homogenous terms of the IBP relations. 
The homogeneous terms of IBP identies can be detected by {\it maximal cuts}, since the multiple cut-conditions annihilate the terms corresponding to subdiagrams \cite{Larsen:2015ped,Bosma:2017hrk,Frellesvig:2017aai,Zeng:2017ipr,Bosma:2017ens,Harley:2017qut,Boehm:2018fpv}. By the same arguments, maximal cuts of MIs correspond to the homogeneous solutions of dimensional recurrence relations and of the differential equations, which, in general, are non-homogeneous equations \cite{Lee:2012te,Remiddi:2016gno,Primo:2016ebd,Primo:2017ipr}. Similar ideas were introduced \cite{Anastasiou:2002yz} and lead to relation between multi-loop integrals and phase-space integrals, known as {\it reverse unitarity}.
The homogeneous solutions play an important role in the construction of canonical systems of differential equation for MIs \cite{Henn:2013pwa}, as it was observed in \cite{Remiddi:2016gno,Primo:2016ebd,Primo:2017ipr}, generalizing the role of Magnus exponential matrix \cite{Argeri:2014qva} to the case of elliptic equations. 

The number of MIs for a given integral family, the order of the differential equations they obey, and the classification of the homogeneous solutions according to the independent components of the integration domain, revealed 
a natural correspondence between the MIs and the geometric properties of the integration domain
 \cite{Laporta:2004rb,Remiddi:2016gno,Primo:2016ebd,Primo:2017ipr,Bosma:2017ens,vonManteuffel:2017hms,Frellesvig:2017aai,Harley:2017qut,Adams:2018kez}, easily accessed within Baikov representation. \\

Let us imagine, for a moment, that the objective of a calculation is simply the decomposition in terms of master integrals of just {\it one} multi-loop Feynman integral.
The IBP reduction algorithm can be seen as a {\it collective} integral decomposition. The computational machinery does not act on individual integrals, one at a time, but it is based on the solution of systems of equations where {\it the wanted} integral appears related, within linear relations, to many additional integrals. Its decomposition is then achieved together with the decomposition of other integrals -- even if the latter might not be of interests, for instance.
The IBP decomposition, although very effective, is computationally expensive.

The new computational strategy proposed in ref.~\cite{Mastrolia:2018uzb} offers a change of perspective: it targets the direct decomposition of individual integrals in terms of master integrals,  bypassing the system solving procedure characterizing the integration-by-parts reduction.

This task can be achieved by applying to Feynman integrals concepts and computational tools borrowed from  
the {\it intersection theory} of differential forms 
\cite{cho1995,matsumoto1998,Mizera:2017rqa}. 
It is a recent branch of algebraic geometry and topology, which was developed to study the Aomoto--Gel'fand hypergeometric functions \cite{aomoto2011theory}. 
This class of functions has two important properties: their integrands are multivalued, and vanish at the boundaries of the integration domain - exactly like Feynman integrals in dimensional regularization. 
Baikov representation makes these properties manifest, and allows to establish an explicit correspondence between Feynman integrals and the functions which can be studied {\it via} intersection theory. 
In particular, any Feynman integral is cast as 
a $n$-form integral, characterized by three basic elements: the integration contour, part of its integrand given by a multivalued function (associated to the Baikov polynomial), and a differential form (corresponding to the genuine product of denominators times the integration measure). 

In general, two integrals can give the same result if:
they have the same integration domain, but
their integrands differ by a term whose primitive vanishes on the integration boundaries; and/or they 
have the same integrand, but their integration 
domains differ by a contour on which the primitive vanish anyhow.
Therefore each integral is actually a pairing of representatives of two equivalence classes: characterized by the integration variety (homology class) or 
by the integrand (cohomology class).
Intersection theory allows for a derivation of relations among integrals belonging to those equivalence classes, which in the case of hypergeometric functions correspond to Gauss' contiguity relations.
In ref.~\cite{Mastrolia:2018uzb}, 
the concepts of intersection theory were applied to Feynman integrals in order to show that it is possible: 
{\it i)} to identify a basis of master integrals; 
{\it ii)} to decompose any individual integral in the chosen basis simply by a projection technique; 
{\it iii)} to derive differential equations for master integrals. 
The so-called {\it intersection number} of differential forms \cite{cho1995,matsumoto1998} constitutes the crucial novel operation that allows to implement the notion of {\it scalar products} between differential forms, which ultimately determine the coefficients of the integral decomposition.
In this way, the problem of reducing a given Feynman integral in terms of master integrals can be solved by {\it projections}: any integral can be decomposed just like an arbitrary vector can be projected onto a chosen basis of a vector space. This analysis was performed by considering integrals on maximal cuts admitting 1-form representations \cite{Mastrolia:2018uzb}. \\

In this work, we elaborate on the decomposition-by-intersection of Feynman integrals onto a basis of master integrals, and we systematically apply it to an extensive list of cases, in order to show its advantages.

We begin by recalling basics of intersection theory for hypergeometric functions,
and show their correspondence to the Baikov representation, both in the standard formulation \cite{Baikov:1996iu} and in the \lbl version \cite{Frellesvig:2017aai}. 
We then define how intersection theory allows to determine the dimension of the integral space, 
and discuss different options for the choice of the integral bases. Afterwards, we introduce the intersections numbers and give the 
{\it master decomposition formula} for the direct evaluation of the coefficients of the reduction in terms of basis integrals. This formula can be also applied to derive differential equations and dimensional recurrence relations for generic basis integrals.

Before addressing Feynman calculus,
we consider the derivation of contiguity relations for special functions, such as the Euler $\beta$ function, the Gauss ${}_2F_1$ hypergeometric function, and the Appell $F_1$ function, which 
belong to the more general class of Lauricella functions.
Then, we apply the new method to decompose Feynman integrals whose maximal cuts admit 1-form integral representations, including examples that have from two to an arbitrary number of loops, and/or from zero to an arbitrary number of legs. 
The 1-form integral representations accounts for multi-loop integrals (on maximal cuts) which have either one irreducible scalar product (ISP), or that have multiple ISPs but can be expressed as a one-fold integral using the \lbl approach,

In a few instructive cases, we show the direct constructions of differential equations and dimensional recurrence relations for master integrals, and discuss how the different choice of the basis may impact of the form of the result. Special emphasis is given to basis of {\it monomial forms} and to basis of 
{\it dlog forms}, 
in particular showing how, the latter obey a {\it canonical} systems of differential equations.

As stressed, the main part of this work deals with the application of intersection theory to 1-forms. The complete decomposition of multi-loop Feynman integrals in terms of master integrals (not just the ones belonging to maximal-cut diagrams, but to the complete chain of sub-diagrams, which would correspond to a smaller number of cuts) requires the application of the intersection theory for $n$-forms.
In the literature, the case of intersection numbers of dlog $n$-forms has been understood \cite{matsumoto1998,Mizera:2017rqa}, but Feynman integrals belong to the wider class of generic rational $n$-forms.

As additional main results of this manuscript, 
we present two novel approaches to decomposition-by-intersections 
in cases where the maximal cuts admit a 2-form integral representation. 
They are important new development both for physical and mathematical research areas, as they represent 
the first step towards the extension of the formalism to generic $n$-form representations. Owing to the results of the research presented in this work, we are confident that this objective is within reach. 

The decomposition formulae computed through the use of intersection numbers for 1- and 2-forms are directly verified to agree with the ones obtained using integration-by-parts identities on the maximal cut. 
As reported in the examples discussed later, we employed several codes for checking our results, and when available, we compared them with the literature. 
Let us finally observe that, in a few cases, the number of master integrals (on the maximal cuts) found by means of intersection theory is smaller than the one found by applying the IBP-decomposition to the interested integral families: the mismatch has been mitigated by finding the additional, missing relations at the cost of  applying the IBP-reduction to integrals families with a larger number of denominators. \\

The results presented in this paper demonstrate that a fascinating property of Feynman integrals has been found, which opens a completely new path to access their algebraic structures. Together with the idea of dimensional regularization, its main application to streamlining reductions unto master integrals, is expected to yield significant computational advantages for the evaluation of high-multiplicity scattering amplitudes at higher orders in perturbation theory.\\

The paper is organized as follows.
In Sec. \ref{sec:basics} we introduce the basics of hypergeometric integrals, and their description in terms of (co)homology and intersection theory. Significantly this section introduces the {\it master decomposition formula} eq. \eqref{basis-projection}. Then follows Sec. \ref{sec:feynmanintegrals} in which we discuss Feynman integrals, the Baikov representations, unitarity cuts, and the connection to intersection theory. The section also discusses relations between Feynman integrals such as reductions unto master integrals, differential equations, and dimensional recurrence. 

In Sec. \ref{sec:specialfunction} we discuss certain specific mathematical functions to which our theory is applicable, which serve as our first examples. These are the $\beta$-function, the ${}_2 F_1$, the Appell $F_1$, and the Lauricella $F_D$. The next two sections, Secs. \ref{sec:fourloopvacuum} and \ref{sec:threeloop}, contain our first examples of the use of the theory to Feynman integrals: a four-loop vacuum integral, and a three-loop two-point function respectively. These two integrals are put in a one-form representation by the standard Baikov parametrization. In Sec. \ref{sec:sunrise} we discuss the sunrise integral in a massless and a massive version. These integrals can be put into a one-form representation by the \lbl version of Baikov parametrization, where the standard version of Baikov representation would have left them as a two-fold integrals. In Sec. \ref{sec:nonplanartriangle} we return to the non-planar triangle of ref. \cite{Mastrolia:2018uzb}, and show how to use our method to deal with doubled propagators. Then in Sec. \ref{sec:planar-triangle} follows an example of a planar triangle diagram with only one master integral, showing the intersection approach alongside a traditional cut-based extraction. Sec. \ref{sec:planardoubletriangle} discusses a certain planar diagram, that is of interest due to peculiar properties of the Baikov polynomial, something that in the past has led to ambiguities in the counting of master integrals in that sector. Secs. \ref{sec:massless-double-box} and \ref{sec:intmass} discuss two double-box integrals, with and without an internal mass. The massive case is of interest as this is our first example in which the intersection theory on the maximal cut, detects a relation that is not usually found by IBP identities. The next two sections applies the theory to some cases of physical interest, namely Bhabha scattering in Sec. \ref{sec:bhabha}, and associated Higgs production ($H{+}j$ and $HH$) in Sec. \ref{sec:higgsproduction}. These two cases are at the edge of what is possible to fully reduce with traditional IBP methods. Then follows Sec. \ref{sec:pentabox} on the celebrated penta-box in a planar and a non-planar version and Sec. \ref{sec:multileg} about its generalizations to cases with massive legs, and significantly to cases with more legs, including an $n$-leg example. In Sec. \ref{sec:arbitraryloop} we look some high-loop integrals (planar and non-planar) that contribute to $H{+}j$ production, and we show several $n$-loop generalizations thereof (that we denote ``rocket diagrams''), including their reductions unto master integrals, and their differential equations.

The following two sections contain discussions on the extension of the one-form algorithm described in this paper, to higher forms. Sec. \ref{sec:iteratedoneforms} discusses one approach that consists of iterating the one-form algorithm combined with direct integration. Sec. \ref{sec:twoforms} discusses another genuinely multivariate approach in which the intersection numbers are computed using intersecting hyperplanes. Finally Sec. \ref{sec:conclusions} contains our conclusions and discussion. The paper ends with two appendices. First App.~\ref{app:criticalpoints} in which we discuss the relation between the number of critical points and the number of master integrals. In particular we calculate the number of critical points with both the standard and the \lbl approaches to Baikov parametrization on the maximal cut, and we discuss in detail the cases in which the two numbers thereby obtained are in disagreement. Then follows App.~\ref{app:LBLiterative} in which we we use the iterated one-form algorithm to derive expressions for the integrand in the \lbl approach when the ISPs that get integrated out are present.

\section{Basics of Hypergeometric Integrals}
\label{sec:basics}

In this section we review a few concepts from the theory of hypergeometric functions and Feynman integrals that serve as a basis for the remainder of the paper.

Consider an integral $I$ over the variables ${\bf z} = (z_1,z_2,\ldots,z_m)$ of the general form:
\be\label{hypergeometric-integral}
I = \int_{\mathcal{C}} u(\mathbf{z})\, \varphi(\mathbf{z}),
\ee
where $u(\mathbf{z})$ is a multi-valued function and $\varphi(\mathbf{z}) = \hat{\varphi}(\mathbf{z}) d^m \mathbf{z}$ is a differential $m$-form. We assume that $u(\mathbf{z})$ {\it vanishes on the boundaries} of $\mathcal{C}$, $u(\partial {\cal C})=0$, so that, upon integration no surface-term is leftover. For example, choosing
\be
u(z) = z^a (z-1)^b, \qquad \varphi(z) = \frac{dz}{z(z-1)}, \qquad \mathcal{C} = [0,1]
\ee
gives the Euler beta function $B(a,b)$ for $\text{Re}(a),\text{Re}(b)>0$. More generally, integrals of the type \eqref{hypergeometric-integral} are called Aomoto--Gel'fand hypergeometric functions \cite{aomoto1977structure,Gelfand}, or simply hypergeometric functions.

As with any integral, there could exist many forms $\varphi$ that integrate to give the same result $I$. Let us consider the total derivative of $u$ times any $(m{-}1)$-differential form $\xi$:
\be
\int_{\mathcal{C}} d \left( u\, \xi \right) = 0.
\ee
By Stokes' theorem, the result is zero due to our choice of the integration domain $\mathcal{C}$. Let us manipulate the above integral so that it is of the form \eqref{hypergeometric-integral}:
\be
0 = \int_{\mathcal{C}} d \left( u\, \xi \right) = \int_{\mathcal{C}} \left( du\wedge \xi  + u\, d\xi\right) = \int_{\mathcal{C}} u \left( \frac{du}{u}\wedge  +\, d \right) \xi \equiv \int_{\mathcal{C}} u\, \nabla_{\omega} \xi.
\ee
In the final equality we defined a connection $\nabla_{\omega}$, which differs from the usual derivative by the one-form $\omega$:
\be\label{nabla}
\nabla_{\omega} \equiv d + \omega\wedge, \qquad \text{where} \qquad \omega \equiv d\log u.
\ee
Since the above expression integrates to zero, we have
\be
\int_{\mathcal{C}} u\, \varphi = \int_{\mathcal{C}} u \left(\varphi + \nabla_{\omega} \xi\right).
\ee
Hence $\varphi$ and $\varphi + \nabla_{\omega} \xi$ carry the same information and we can talk about equivalence (cohomology) classes ${_\omega}\langle \varphi |$ of forms that integrate to the same result:
\be\label{twisted-cocycle}
{_\omega}\langle \varphi | : \; \varphi \sim \varphi + \nabla_{\omega} \xi.
\ee
In other words, whenever two forms are equal to each other up to integration-by-parts identities, they belong to the same equivalence class. This class is called a \emph{twisted cocycle}. The word \emph{twisted} refers to the fact that the usual derivative operator $d$ is replaced by the covariant derivative $\nabla_{\omega}$ given in \eqref{nabla}, as a consequence of the presence of the multi-valued function $u$ in the hypergeometric integral. We often refer to any representative of the class \eqref{twisted-cocycle} as twisted cocycle, as well as drop the subscript ${}_\omega$ when it is clear from the context.\footnote{For completeness, let us mention that, similarly, there are equivalence (homology) classes of integration domains $\mathcal{C}$ that give the same result for the integral \eqref{hypergeometric-integral}, called \emph{twisted cycles} $|\mathcal{C}]_{\omega}$, 
though we do not make use of this fact in the current manuscript.} A remarkable observation is that we can pair up $\langle \varphi|$ and $| \mathcal{C} ]$ to obtain the integral from \eqref{hypergeometric-integral}, which we denote by
\be
\langle \varphi| \mathcal{C} ] \equiv \int_{\mathcal{C}} u\, \varphi.
\ee
This integral representation, as a bilinear in $\langle \varphi|$ and $| \mathcal{C} ]$,
is suitable for establishing linear relations between hypergeometric functions. In fact,
let us assume that the number of linearly-independent twisted cocycles is $\masters$, and
indicate an arbitrary basis of forms,
\be
\langle e_1 |, \quad \langle e_2 |, \quad\cdots,\quad \langle e_{\masters} |.
\ee
A basis decomposition is achieved by expressing an arbitrary twisted cocycle, say $\langle \varphi |$, as a linear combination of the above ones. This goal be achieved as follows. Introduce a \emph{dual} (and auxiliary) space of twisted cocycles, whose basis is denoted by $| h_i \rangle$ for $i=1,2,\ldots,\masters$, and consider the matrix ${\bf C}$, whose entries are the pairing 
$\langle e_i | h_j \rangle$, 
\begin{eqnarray}
\mathbf{C}_{ij} = \langle e_i | h_j \rangle \qquad \text{for}\qquad i,j=1,2,\ldots, \masters \ . 
\end{eqnarray}
This pairing is called the {\it intersection number} of $\langle e_i |$ and $| h_j \rangle$.
We then construct the $(\masters{+}1)\times (\masters{+}1)$ matrix ${\bf M}$, defined as,
\be
\mathbf{M} = \left(
\begin{array}{ccccc}
\langle \varphi | \psi \rangle & \langle \varphi |  h_1 \rangle & 
\langle \varphi |  h_2 \rangle & \ldots &
\langle \varphi |  h_{\masters} \rangle \\
\langle  e_1 | \psi \rangle & \langle  e_1 |  h_1 \rangle & 
\langle  e_1 |  h_2 \rangle & \ldots &
\langle  e_1 |  h_{\masters} \rangle \\
\langle  e_2 | \psi \rangle &\langle  e_2 |  h_1 \rangle & 
\langle  e_2 |  h_2 \rangle & \ldots & \langle  e_2 |  h_{\masters} \rangle \\
\vdots & \vdots &  \vdots & \ddots & \vdots \\
\langle  e_{\masters} | \psi \rangle & \langle  e_{\masters} |  h_1 \rangle & 
\langle  e_{\masters} |  h_2 \rangle & \ldots & \langle  e_{\masters} |  h_{\masters} \rangle 
\end{array}
\right)
\equiv
\left(\begin{array}{cc}
	\langle \varphi | \psi \rangle & \mathbf{A}^\intercal\\
	\mathbf{B} & \mathbf{C} 
\end{array}\right).
\ee
The columns of the matrix $\mathbf{M}$ are labelled by $|\psi\rangle, |h_1\rangle, |h_2\rangle, \ldots, |h_{\masters} \rangle$ for an arbitrary $|\psi\rangle$, while the rows are labelled by $\langle \varphi|, \langle e_1 |, \langle e_2 |, \ldots, \langle e_{\masters} |$. Each entry is given by a pairing (bilinear) of the corresponding row and column. In the second equality, we expose the structure of $\mathbf{M}$ as a $\masters \times \masters$ submatrix $\mathbf{C}$, 
a column vector $\mathbf{B}$ and a row vector $\mathbf{A}^\intercal$,
respectively with elements $\mathbf{B}_i = \langle e_i | \psi \rangle$ and 
$\mathbf{A}_i = \langle \varphi | h_i \rangle$ (for $i = 1,2, \ldots, \masters$).

The fact that the $\masters{+}1$ cocycles labelling the rows and columns are necessarily linearly dependent (since the basis is $\masters$-dimensional) and that each entry of $\mathbf{M}$ is a bilinear, implies that the determinant of this matrix vanishes. Using the well-known identity for the determinant of a block matrix, we find:
\be
\det \mathbf{M} = \det \mathbf{C}\, \bigg( \langle \varphi| \psi \rangle - \mathbf{A}^\intercal\, \mathbf{C}^{-1}\, \mathbf{B} \bigg) = 0.
\ee
In addition, $\det \mathbf{C}$ cannot be zero (by definition), since it is formed from bilinears between two bases. Therefore we conclude that:
\begin{align}
\langle \varphi | \psi \rangle &= \mathbf{A}^\intercal\, \mathbf{C}^{-1}\, \mathbf{B}\nonumber\\
&= \sum_{i,j=1}^{\masters} \langle \varphi | h_j \rangle \, (\mathbf{C}^{-1})_{ji}\, \langle e_i | \psi \rangle . 
\end{align}
Given the arbitrariness of $|\psi \rangle$, we obtain the {\it master decomposition formula}
\be\label{basis-projection}
\langle \varphi | = \sum_{i,j=1}^{\masters} \langle \varphi | h_j \rangle \, \left( \mathbf{C}^{-1} \right)_{ji}\, \langle e_i |,
\ee
which provides an explicit way of projecting $\langle \varphi|$ onto a basis of $\langle e_i|$. Following \cite{Mastrolia:2018uzb}, in this paper we use \eqref{basis-projection} to perform the decomposition of Feynman integrals in terms of master integrals, on the maximal cut. For example, by contracting both sides with the twisted cycle $|\mathcal{C}]$ (which boils down to multiplying by $u$ and integrating over $\mathcal{C}$), we have a linear identity between integrals:
\be
\int_{\mathcal{C}} u\, \varphi = \sum_{i,j=1}^{\masters} \langle \varphi | h_j \rangle \, \left(\mathbf{C}^{-1} \right)_{ji} \int_{\mathcal{C}} u\, e_i.
\ee

Similarly, the same idea can be used to derive linear system of differential equations satisfied by the basis integrals $\langle e_i | \mathcal{C}]$ in some external variable $x$. It is enough to notice that
\be
\partial_x \,  \langle e_i | {\cal C} ]
=
\partial_x 
\int_{\cal C} u\, e_i  =
\int_{\cal C} u\, (\partial_x + \sigma \wedge) e_i  
=
 \langle (\partial_x + \sigma \wedge) e_i | \mathcal{C}],
\ee
where $\sigma \equiv \partial_x \log(u)$.
Let us remark that even if ${\cal C}$ depends on $x$, the differential operator $\partial_x$ 
commutes with the integral sign, due to 
the vanishing of $u$ on the boundary of $\mathcal{C}$. Therefore, the problem reduces to projecting $\langle (\partial_x + \sigma \wedge) e_i |$ on the right-hand side back onto a basis using \eqref{basis-projection}.

One should think of $\langle e_i |$ and $|h_j \rangle$ as parameterizing a vector space of inequivalent integrands of a hypergeometric function. In this sense $\mathbf{C}$ provides a metric on this space. Naturally, the prescription \eqref{basis-projection} is only useful if computing invariants of the type $\langle \varphi_L | \varphi_R \rangle$ is efficient. We argue that this is the case. It turns out that the dual space of twisted cocycles has a straightforward interpretation as the equivalence classes:
\be
| \varphi \rangle_{\omega} : \; \varphi \sim \varphi + \nabla_{-\omega} \xi,
\ee
where the only difference to \eqref{twisted-cocycle} is the use of the connection $\nabla_{-\omega} \equiv d - \omega \wedge$ instead of $\nabla_{\omega}$. The resulting bilinear:
\be
\langle \varphi_L | \varphi_R \rangle_\omega
\ee
is called the \emph{intersection number} of $\langle\varphi_L|$ and $|\varphi_R\rangle$. This term is conventionally used in the literature on hypergeometric functions, but it does not mean that $\langle \varphi_L | \varphi_R \rangle_\omega$ is an integer. In general, it can be a rational function of external parameters. The characteristic property of the intersection number is that it is a bilinear in the two equivalence classes. We give multiple ways of computing it throughout the text.

In this brief review, we only scratched the surface of the fascinating theory of hypergeometric functions. We refer the interested reader to \cite{aomoto2011theory,yoshida2013hypergeometric} for review of twisted (co)homologies and their intersection theory, as well as \cite{Mizera:2016jhj,Mizera:2017rqa,Mizera:2017cqs,Mastrolia:2018uzb} and \cite{delaCruz:2017zqr,Tourkine:2019ukp} 
for some recent applications of these ideas to physics.

In the following, we focus on {\it Feynman integrals}. 
In order to translate them into the form \eqref{hypergeometric-integral} we make use of the Baikov representation in the standard form~\cite{Baikov:1996iu} and the \lbl approach developed in \cite{Frellesvig:2017aai}.

\section{Feynman Integral Decomposition}
\label{sec:feynmanintegrals}
		
Consider scalar Feynman integrals with $L$ loops, $E{+}1$ external momenta, and $N = LE + \frac{1}{2} L(L{+}1)$ (generalised) denominators\footnote{$N$ amounts to the total number of scalar products which can be built with the loop momenta $k_i$ and the independent external momenta $p_j$, and corresponds to the sum of the so called {\it reducible} and {\it irreducible} scalar products. The former can be expressed in terms of the denominators of graph propagators, while the latter are independent of them. Nevertheless, they also 
can be interpreted as auxiliary denominators, not related to any internal line of the graph.}
in a generic dimension $d$:
\begin{eqnarray}
	\label{eq:Feynman-integral}
	I_{a_1, a_2, \ldots, a_N} \equiv \int
	\prod_{i=1}^{L} \frac{d^d k_i}{\pi^{d/2}}  
	\prod_{j=1}^{N} \frac{1}{D_j^{a_j}}.
\end{eqnarray}
where $D_j$ stands for either a genuine denominator or an irreducible scalar product (ISP). 

In Baikov representation, one changes the integration variables, from the loop momenta $k_i$ to the denominators $D_j$, at the cost of introducing a Jacobian, see, e.g., \cite{Lee:2009dh,Lee:2010wea} or Appendix~A of \cite{Mastrolia:2018uzb}. Here we summarize the final forms of the standard and \lbl Baikov representations.
		
\begin{enumerate}[leftmargin=1.3em]
	\item {\bf Standard Baikov Representation}. 
	In this case,~\cite{Baikov:1996iu}, after the change of variables,
	the Feynman integral may be written as,
	\begin{equation}
		I_{a_1,a_2,\ldots, a_N} 
		\equiv
		K \int_{\cal C} u \, \varphi 
	\end{equation}
	where
	\begin{equation}
		u = B^\gamma\,, \qquad \gamma \equiv (d {-} E {-} L {-}1)/2
	\end{equation}
	and
	\begin{equation}
		\varphi \equiv {\hat \varphi} \, d^N{\bf z}\,, \qquad
		{\hat \varphi} \equiv \frac{1}{z_1^{a_1} z_2^{a_2} \cdots z_N^{a_N}}\,, \qquad
		d^N{\bf z} \equiv dz_1 \wedge dz_2 \wedge \cdots \wedge dz_N\,,
	\end{equation}
	and where $B$ is the Baikov polynomial computed as a determinant of the Gram matrix of scalar products, depending on loop momenta, and $K$ is a constant pre-factor (independent of the integration variables), which may depend on the external kinematic invariants and on the dimensional regulator $d$. The integration contour $\mathcal{C}$ is defined such that $B$ vanishes on its boundaries.
			
	We can re-express it, in the language of intersection theory, as a bilinear pairing,
    \begin{equation}
		I_{a_1,a_2,\ldots, a_N} 
		\equiv
		K \, \langle \varphi | {\cal C} ]_\omega \, ,
	\end{equation}
	with
	\begin{equation}
		\omega \equiv d \log(u) = \gamma d \log(B).
	\end{equation}
			
	\item {\bf \lbl (LBL) Baikov Representation}. 
	In this case \cite{Frellesvig:2017aai}, after the change of variables, 
			the number of integration variables $M$ can be smaller than the $N$ (because $N-M$ ISPs have been integrated out).
    For this case, the integral have the form	
	\begin{equation}
		I_{a_1, a_2, \ldots, a_M, a_{M+1}, \ldots, a_N}
		\equiv K \, \int_{\cal C} u \, \varphi
		=
		K \, 
		 \langle \varphi | {\cal C} ]_{\omega}
	\end{equation}
	with
	\begin{equation}
		u = B_1^{\gamma_1} B_2^{\gamma_2} \cdots B_n^{\gamma_n},  \qquad
		\omega \equiv d\log(u) 
		= \sum_{i=1}^{n} \gamma_i \, d\log(B_i) \ , \qquad (n \le 2L-1) \ ,
	\end{equation}
	and where
	\begin{equation}
		\varphi \equiv {\hat \varphi} \, d^M{\bf z}, \qquad  
		{\hat \varphi}  \equiv
		\frac{f(z_1,\ldots,z_M)}{z_1^{a_1} z_2^{a_2} \cdots z_M^{a_M}}, \qquad
		d^M{\bf z} \equiv dz_1 \wedge dz_2 \wedge \cdots \wedge dz_M
		\label{eq:phifromloopbyloop}
	\end{equation}
	where $f$ is a rational function of the $z_i$ (that is $1$ if all $a_i$ with $i>M$ are 0). 
	
	More explicitly the set of $B_i$ in the \lbl approach generally consists of $L$ Baikov polynomials for the individual loops, and $L-1$ additional Gram determinants, in accordance with the prescription in ref. \cite{Frellesvig:2017aai}. How small $M$ can be made depends in general on the underlying Feynman graph\footnote{For two-loop diagrams $M = 2+E+E_2$ where $E$ is the number of independent external momenta, and $E_2$ is the number of independent momenta external to the loop that is integrated out first.}.

	\item{\bf Cut Integrals.} Within the Baikov representation, the on-shell cut-conditions $D_i=0$ are most naturally expressed as a contour integration. Any multiple $m$-cut integral,
	with $D_1 = D_2 = \cdots = D_m = 0$, becomes 
	\begin{align}
		I_{a_1,a_2,\ldots, a_N} \Big|_{m\text{-cut}} \, \equiv \, K \int_{{\cal C}_{m\text{-cut}}} \!\!\!\!\!\!\! u \, \varphi
	\end{align}
	where the deformed contour is defined as
	\begin{align}\label{eq:Kcut}
		{\cal C}_{m\text{-cut}} &=\; \circlearrowleft_1 \wedge \circlearrowleft_2 \wedge \ldots \wedge \circlearrowleft_m \wedge\; \mathcal{C}' 
	\end{align}
	with the $\circlearrowleft_i$-contours denoting a small loop in the complex plane around the pole at $z_i=0$. 
	Accordingly, the integration domain of the cut-integral is given by the geometric intersection of ${\cal C}$ with the planes $z_i = 0, (i=1,2,\ldots,m)$ identifying the on-shell conditions,
	\begin{eqnarray}
	{\cal C}' \equiv \bigcap_{i=1}^m \{ z_i = 0 \}  \cap {\cal C}.
	\end{eqnarray}
	In general, the domain ${\cal C}'$ may admit a decomposition into subregions, 
	\begin{eqnarray}
	{\cal C}' = \bigcup_{j} {{\cal C}_j}' \ ,
	\label{eq:subdomains:deco}
	\end{eqnarray}
	though only $\masters$ of them can be independent.
	After integrating over the cut variables, the left over (phase-space) integral reads as,
	\begin{eqnarray}
		\label{eq:def:cut-integral}
		I_{a_1,a_2,\ldots, a_N} \Big|_{m\text{-cut}}
		&=& K' \int_{{\cal C}'} u' \, \varphi' \ ,
	\end{eqnarray}
	with
	\begin{gather}
	\label{eq:cut:genpow}
		K' u' = (K u) \Big|_{z_1 = \ldots  = z_m = 0} \ , \qquad \varphi' \equiv {\hat \varphi}' \, d^{N-m}{\bf z}' \ , \\  
		{\hat \varphi}' \equiv
		\frac{f(z_{m+1}, \ldots, z_N)}{z_{m+1}^{a_{m+1}} \cdots z_N^{a_N}} 
		\left( \frac{{\cal D}_m ( u )}{u} \right)
	    \Bigg|_{z_1 = \ldots  = z_m = 0} \ , \\
     {\cal D}_m \equiv 
     \prod_{i=1}^m \frac{ \partial_{z_i}^{(a_i-1)} }{ (a_i - 1) !} \ , \\
		d^{N-m}{\bf z}' \equiv dz_{m+1} \wedge \cdots \wedge dz_N \ , 
	\end{gather}
	where $u'$ vanishes on the boundary ${\cal C}'$, and $f$ is a rational function (see eqs. \eqref{eq:phifromloopbyloop}).
	Therefore, also the $m$-cut integral keeps admitting a bilinear pairing representation, 
	\begin{eqnarray}
	I_{a_1,a_2,\ldots, a_N} \Big|_{m\text{-cut}} = 
		I_{a_{m+1},\ldots, a_N} = K' \, {}_{\omega'} \! \langle \varphi' | {\cal C}' ] \  
		\qquad {\rm with } \qquad 
		\omega' \equiv d \log(u') \ .
	\end{eqnarray}

     \paragraph{Notation.}		
		In the following examples, for ease of notation, we 
		drop the prime symbol ${}^\prime$, and use directly $K$, $u$, $\omega$, $\varphi$ and $z$ to express the various quantities on the cut. Moreover, in the univariate case where after the maximal cut the integrals are characterized by a single ISP, we use the notation $I_{a_1,a_2,\ldots, a_N} \big|_{m\text{-cut}} \equiv
			I_{a_{1},\ldots, a_m;a_{m+1}}$, where $a_{m+1}$ is the power of the remaining irreducible scalar product. 
\end{enumerate}

\subsection{Intersection Numbers of One-Forms} 
	
In this section we specialize to the case when $\varphi$ are $1$-forms. Consider,
\begin{eqnarray}
	\masters = \{ {\rm the\ number\ of\ solutions\ of} \ \omega=0  \} \ ,
\end{eqnarray}
and define ${\cal P}$ as the set of {\it poles} of $\omega$ , 
\begin{eqnarray}
	{\cal P} \equiv \{\, z \ | \ z\ {\rm is\ a\ pole\ of\ }\omega \, \} \, .
\end{eqnarray}
Note that $\mathcal{P}$ can also include the pole at infinity if $\Res_{z=\infty} (\omega) \neq 0$.\footnote{
The number $\masters$ of master integrals is equal, up to a sign, to the Euler characteristic $\chi = -\masters$ of the space $\mathbb{CP}^1 \setminus \mathcal{P}$, on which the forms are defined, where the number of poles in $\mathcal{P}$ is exactly $\masters{+}2$, provided that all $\Res_{z=p}(\omega)$ are not non-negeative integers. See also \cite{Lee:2013hzt,Bitoun:2017nre} for discussion of Euler characteristic in the context of Feynman integrals.
Earlier considerations on possible relations between the number of MIs and geometric properties of differential manifolds can be found in \cite{Kosower:2011ty,CaronHuot:2012ab}.}

Given two (univariate) 1-forms $\varphi_L$ and $\varphi_R$,
we define the {\it intersection number} as 
\cite{cho1995,matsumoto1998}
\begin{eqnarray}\label{one-form-intersection}
	\langle \varphi_L | \varphi_R \rangle_\omega = 
	\sum_{p \in {\cal P}} {\rm Res}_{z=p} \Big( \psi_p \, \varphi_R \Big),
\end{eqnarray}
where, $\psi_p$ is a function (0-form), solution to the differential equation
$\nabla_{\omega} \psi = \varphi_{L}$, around $p$, {\it i.e.},
\begin{equation}
	\label{eq:diffeqaroundp}
	\nabla_{\omega_p} \psi_p = \varphi_{L,p} \ , 
\end{equation}
where $\nabla_{\omega}$ was defined in eq.~\eqref{nabla} (the notation $f_p$ indicates the Laurent expansion of $f$ around $z=p$). The above equation can be also solved globally, however only a handful of terms in the Laurent expansion around $z=p$ are needed to evaluate the residue in \eqref{one-form-intersection}.
In particular, after defining 
$\tau \equiv z-p$, and the {\it ansatz}, 
\begin{gather}
	\psi_p =
	\sum_{j={\rm min}}^{{\rm max}} 
	\psi_p^{(j)}
	\tau^{j} 
	+ {\cal O}\left( \tau^{{\rm max}+1} \right) \ , \\
	{\rm min} = {\rm ord}_p(\varphi_L)+1 \ ,
	\qquad 
	{\rm max} = -{\rm ord}_p(\varphi_R)-1 \ ,
\end{gather}
the differential equation in eq. (\ref{eq:diffeqaroundp})
freezes all unknown coefficients $\psi_p^{(j)}$. 
In other words, the Laurent expansion of  $\psi_p$ around each $p$, is determined by the Laurent expansion of $\varphi_{L,R}$
and of $\omega$. A given point $p$ contributes only if the condition $\text{min} \leq \text{max}$ is satisfied, and the above expansion exists only if $\Res_{z=p}(\omega)$ is not a non-positive integer.

\paragraph{\bf Symmetry Properties.} Intersection numbers of one-forms have the following symmetry property under the exchange of $\varphi_L$ and $\varphi_R$,
\begin{equation}\label{intersection-symmetry}
    \langle \varphi_L |
    \varphi_R \rangle_{\omega} = -
    \langle \varphi_R |
    \varphi_L \rangle {}_{-\omega} \ ,
\end{equation}
Notice that on the {\it r.h.s.} the intersection number is evaluated with respect to the form $-\omega$ (instead of $\omega$).

\paragraph{\bf Logarithmic Forms.} 
When both $\varphi_L$ and $\varphi_R$ are logarithmic, meaning that $\text{ord}_p (\varphi_{L/R}) \geq -1$ for all points $p \in {\cal P}$, then the formula \eqref{one-form-intersection} simplifies to
\be
\langle \varphi_L | \varphi_R \rangle_{\omega} = \sum_{p\in {\cal P}} \frac{\Res_{z=p} (\varphi_L) \, \Res_{z=p} (\varphi_R)}{\Res_{z=p} (\omega)}.
\ee
Note that in this case the intersection number becomes symmetric in $\varphi_L$ and $\varphi_R$, {\it i.e.},
\be
\langle \varphi_L |
        \varphi_R \rangle_{\omega} = 
       \langle \varphi_R |
         \varphi_L \rangle {}_{\omega} \ ,
\ee
while \eqref{intersection-symmetry} still holds.

	\paragraph{\bf Vector Space Metric, Integral Decomposition and Master Integrals.}
Following the discussion in Sec.~\ref{sec:basics}, consider an $\masters$-dimensional vector space, 
and its dual space, whose basis are respectively represented as, $\langle e_i|$ and $| h_i \rangle$ with $i=1,2,\ldots,\masters$. We use intersection numbers to define a {\it metric} on this space
	\begin{eqnarray}
		{\bf C}_{ij} \equiv \langle e_i | h_j \rangle \ ,
	\end{eqnarray}
	which gives rise to $\masters \times \masters$ matrix ${\bf C}$.
	According to the {\it master decomposition formula} eq.~\eqref{basis-projection},
    any element $\langle \varphi |$ of the space can be decomposed in terms of $\langle e_i|$, as
	\begin{equation}
		\langle \varphi |  
		=
		\sum_{i,j=1}^{\masters}  
		\langle \varphi | h_j \rangle\,
		\left( {\bf C}^{-1}\right)_{ji} \,
		\langle e_i | \ .
		\label{eq:masterdeco:}
	\end{equation} 
	Therefore,
	the pairing of $\langle \varphi |$ on the {\it l.h.s.} and $\langle e_i |$ on the {\it r.h.s.} with the integration cycle $|{\cal C}]$, univocally gives rise to the decomposition (on the cut) of the Feynman integral $I$ in terms of {\it master integrals} $J_i$,
	by means of projections built with intersection numbers, {\it i.e.}
	\begin{eqnarray}
	I &=& K \langle \varphi | {\cal C} ] = 
	\sum_{i=1}^{\nu} c_i \, J_i \ ,
	\end{eqnarray}
	where
	\begin{eqnarray}
    J_i &\equiv& K \, E_i \ , 
    \quad
{\rm with} \quad 
E_i \equiv \langle e_i | {\cal C} ] \ ,
\end{eqnarray}
and
\begin{eqnarray}
c_i &\equiv& \sum_{j=1}^{\masters}  
		\langle \varphi | h_j \rangle\,
		\left( {\bf C}^{-1}\right)_{ji} \ . 
\end{eqnarray} 
The main goal of this work is to show that 
the decomposition formulas for Feynman integrals obtained by intersection numbers are equivalent to the one derived by the standard integration-by-parts identities (IBPs). Very interestingly, using intersection numbers, the system-solving strategy inherent to the IBP-decomposition is completely bypassed \cite{Mastrolia:2018uzb}.

\paragraph{\bf Reducible Integrals and Maximal Cuts.} As shown, the number of independent basis forms, and hence MIs, is given by $\nu$. Therefore, for any given integral family the existence of MIs is due to the existence of the solutions for $\omega=0$. 
It is possible to identify a few special cases: 
\begin{itemize}
    \item {\bf Reducibility}. Absence of master integrals, amounting to $\nu=0$, can happen
    either when Baikov polynomial on the maximal cut is vanishing, $B=0$, or when $B$ is linear in the integration variable, $B=z$: in the former case, $\omega$ does not exist; in the latter case, $u=B^\gamma$, therefore $\omega=\gamma\, dz/z$, and $\omega=0$ has no solutions. In these cases, the integral family is {\it reducible}, namely the corresponding integrals can be expressed as a combination of the master integrals of the subtopologies.
    \item {\bf Maximal Cuts}. Baikov polynomial $B$ is a non-zero {\it constant} on the maximal cut. This means that no ISP is left over to parametrize the cut integral. In other words, the integral is fully localized by the cut-conditions.
    In this case, the condition $\omega=0$ is {\it always} satisfied, and there is $\nu=1$ master integral. \\
    This situation may occur, for instance, at one-loop, where maximal cuts are indeed maximum cuts.
\end{itemize}

	\paragraph{\bf Choices of Bases.} 
	The bases $|h_i \rangle$ and $|e_i \rangle$ can be different from each other, but 
	$|h_i \rangle = |e_i \rangle$ is a possible choice too. 
	We decompose 1-form employing either a {\it monomial basis} 
	\begin{eqnarray}
		\langle  e_i | = 
		\langle  \phi_i | \equiv z^{i-1} dz \ ,
	\end{eqnarray}
	or a {\it dlog-basis}, of the type,
	\begin{eqnarray}
		\langle  e_i | = 
		\langle  \varphi_i | \equiv \frac{dz}{z - z_i} \ ,
	\end{eqnarray}
	where $z_i$ are {\it poles} of $\omega$.

	Alternatively, \emph{orthonormal bases} for twisted cocycles can be chosen as follows. Out of the set of poles $\mathcal{P} = \{ z_1, z_2, \ldots, z_{\masters+1}, z_{\masters+2}\}$ pick two special ones, say $z_{\masters+1}$ and $z_{\masters+2}$. Then construct bases of $\masters$ one-forms using:
	\be
	\langle e_i | \equiv d \log \frac{z - z_i}{z - z_{\masters+1}}, \qquad | h_i \rangle \equiv \Res_{z=z_i} (\omega)\; d\log \frac{z - z_i}{z - z_{\masters+2}}
	\ee
	for $i=1,2,\ldots,\masters$. With this choice, the intersection matrix $\mathbf{C}$ becomes the identity matrix,
	\be
	\mathbf{C}_{ij} = \delta_{ij}
	\ee
	as can be shown directly using the residue prescription \eqref{one-form-intersection}, and therefore the basis decomposition formula simplifies to
	\be
	\langle \varphi | = \sum_{i=1}^{\masters} \langle \varphi | h_i \rangle \langle e_i | \ .
	\ee

\subsection{\label{sec:DE}System of Differential Equations}

Let us give more details about deriving systems of differential equations using intersection numbers.

Consider the system of differential equations in $x$ for the basis $\langle e_i |$,
\begin{eqnarray}
\label{eq:sysdiffeq}
\partial_x \langle e_i | = {\bf \Omega }_{ij} \,
\langle e_j | \ ,
\qquad {\bf \Omega } = {\bf \Omega }(d,x) ,
\end{eqnarray}
in general depending on the space-time dimension $d$ and external variables $x$.
Let us consider the  {\it l.h.s.} of eq. (\ref{eq:sysdiffeq}),
after taking the derivative in $x$,

\begin{eqnarray}
\partial_x \langle e_i |  = 
\langle (\partial_x + \sigma \wedge) e_i |  
\equiv \langle \Phi_i | \ ,
\end{eqnarray}
where $\sigma = \partial_x \log u$. Here $\langle \Phi_i |$ can be decomposed 
in terms of $\langle e_i |$, by means of intersection numbers,
\begin{eqnarray}
\langle \Phi_i |
&=&
\langle \Phi_i | h_k \rangle \,
\left( {\bf C}^{-1} \right)_{kj} \,
\langle e_j | \\ 
&=&
{\bf F}_{ik} 
\left( {\bf C}^{-1} \right)_{kj} 
\langle e_j | \\
&=&
{\bf \Omega}_{ij} 
\langle e_j | \ ,
\end{eqnarray}
where summation over indices $j,k$ is implied and we introduced the intersection matrix 
\begin{eqnarray}
{\bf F}_{ik} \equiv 
\langle \Phi_i | h_k \rangle
\end{eqnarray}
as well as defined the matrix ${\bf \Omega}$ as,
\begin{eqnarray}
{\bf \Omega}
&\equiv&
{\bf F}
{\bf C}^{-1}
\label{eq:Omega:FInvC}
\end{eqnarray}
appearing in the r.h.s. of eq.~(\ref{eq:sysdiffeq}).

In \cite{Mastrolia:2018uzb}, it was observed that 
in the case of dlog-basis defined for integrals within the standard Baikov representation (for which $u=B^\gamma$), 
the matrix ${\bf C}^{-1}$ is $\gamma$-factorized, 
and so it is the ${\bf \Omega}$ matrix. Therefore the system of differential equations for the dlog-basis is {\it canonical} \cite{Henn:2013pwa} 
by construction, around the critical dimension $\gamma=0$.

Master Integrals in $d$ dimensions  correspond to integrals of the form
\begin{eqnarray}
J_i \equiv K \, E_i \ , 
\quad
{\rm with} \quad 
E_i \equiv \langle e_i | {\cal C} ],
\end{eqnarray}
where $K$ may depend on $x$ as well.
Therefore, if,
\begin{eqnarray}
\partial_x \langle e_i | = {\bf \Omega }_{ij} \,
\langle e_j | \ ,
\label{eq:hdeqsystem:ei}
\end{eqnarray}
then
the system of differential equations for $J_i$ reads,
\begin{eqnarray}
\partial_x J_i &=& {\bf A}_{ij} \, J_j \ , \\ 
{\rm where \ \ } {\bf A} \equiv  {\bf \Omega} + {\bf K} \ , && \quad
{\rm with \ \ }
\label{eq:A:diffeq}
{\bf K} = \partial_x \log(K) \, \mathbb{I} \ .
\end{eqnarray}

\paragraph{Solutions.}
The system of differential equations in eq.~\eqref{eq:hdeqsystem:ei} can be used to deduce a single homogeneous differential equation of order $\nu$ for each $\langle e_i |$ separately ($i=1,2,\ldots,\masters$).
For each $i$, the $\masters$ independent solutions of such an equation can be found by building the pairing
\begin{eqnarray}
{\bf P}_{ij} = \langle e_i | {\cal C}_j ] = \int_{{\cal C}_j} u \, e_i\ , \qquad i,j=1,2,\ldots,\nu \ ,
\end{eqnarray}
where ${\cal C}_j$ are the independent sub-regions considered in eq. \eqref{eq:subdomains:deco}, see, {\it e.g.,} \cite{Primo:2016ebd,Bosma:2017ens,Primo:2017ipr}.
The $\masters \times \masters$ matrix ${\bf P}$ is the {\it resolvent} matrix of the system of differential equations.
For instance, by choosing a $\masters$-dimensional basis formed by $\langle e_i |$ and its derivatives up the $(\masters-1)^{\rm th}$-order, 
${\bf P}$ becomes the Wronski matrix, whose determinant is the Wronskian of the differential equation obeyed by 
$\langle e_i |$.

The matrix ${\bf P}$ plays an important role in the construction of canonical systems of differential equation \cite{Henn:2013pwa}, as it was observed in \cite{Remiddi:2016gno,Primo:2016ebd,Primo:2017ipr}, generalizing the role of Magnus exponential matrix \cite{Argeri:2014qva} to the case of elliptic equations. 
More generally, in the theory of hypergeometric functions, ${\bf P}$ is known as {\it twisted period matrix}. It can be used, for instance, to build the so called {\it twisted Riemann period relations} \cite{cho1995}, a fundamental identity giving {\it quadratic} relations between hypergeometric functions. A proper study of twisted Riemann period relations to Feynman integrals goes beyond the scope of the current manuscript, and it is left to future investigations.

\subsection{Dimensional Recurrence Relation}
Within the standard Baikov representation,
the $d$ dependence of Feynman integrals 
is carried solely by the prefactor $K$ and 
by the exponent $\gamma$ of the Baikov polynomial $B$.
Let us write the MIs in $d+2n$ dimensions as,
\begin{eqnarray}
J_i^{(d+2n)} \equiv K(d+2n) \, E_i^{(d+2n)}, \end{eqnarray}
with 
\begin{eqnarray}
E_i^{(d+2n)} \equiv 
\langle B^n e_i | {\cal C} ]
= 
\int_{{\cal C}} u \, (B^n \, e_i)  \ , 
\qquad 
i = 1,2,\ldots,\masters \ , 
\end{eqnarray}
and consider the decomposition of the $\langle B^n e_i |$ in terms of the basis $\langle e_j |$, 
\begin{eqnarray}
\langle B^n e_i | = 
({\bf R}_n)_{ij} \, \langle e_j | \ , 
\qquad n=0,1,\ldots, \masters -1 \ .
\label{eq:ric_rel_d+2n_MI}
\end{eqnarray}
This equation can be interpreted as a change of basis,
from $\langle e_i |$ with $(i=1,2,\ldots, \masters)$ to $\langle B^n e_i |$ with $(n=0,1,\ldots, \masters-1)$. 
We can, therefore, decompose 
$\langle B^{\masters} e_i |$ in terms 
of the new basis $\langle B^n e_i |$, as 
\begin{eqnarray}
\langle B^{\masters} e_i |
= \sum_{n=0}^{\masters-1} c_n \,
\langle B^n e_i | \ ,
\end{eqnarray}
which can be written in the suggestive fashion, 
\begin{eqnarray}
\sum_{n=0}^{\masters} c_n \,
\langle B^n e_i |
= 0 \ ,
\end{eqnarray} 
with $c_{\masters} \equiv -1 $.
Upon the pairing with $|\cal C]$, it yields the recursion formula  
for the integral $E_i$,
\begin{eqnarray}
\sum_{n=0}^{\masters} c_n \, E_i^{(d+2n)}
= 0 \ ,
\end{eqnarray}
where the coefficients $c_n$, computed by means of the master decomposition formula eq.~(\ref{eq:masterdeco:}), may depend on $d$ and on the kinematics. 
Finally, 
by a simple redefinition of the coefficients,
the dimensional recurrence relation for the MIs $J_i$ arises, 
\begin{eqnarray}
\sum_{n=0}^{\masters} \alpha_n \, J_i^{(d+2n)}
= 0 \ ,
\end{eqnarray}
with $\alpha_n \equiv c_n/K(d+2n) \, .$

\section{Special Functions}
\label{sec:specialfunction}

One-variable integrals of the hypergeometric type considered in this paper, may always\footnote{If the integrand is just a product of linear terms $\prod_i (z - a_i)^{\gamma_i}$ with the integration path being between two of the $a_i$, a M{\"o}bius transform can bring it into the form discussed in the text.} be expressed in the form 
\begin{align}
    I^{(\alpha)} &\propto \int_0^1 z^{\gamma_1} \, (1 - z)^{\gamma_2} \prod_{i=3}^{\alpha} (1 - x_i z)^{\gamma_i} \, d z \,.
\end{align}
For $\alpha = 2, 3, 4$, this integral (up to pre-factors) corresponds to the Euler beta-function, the Gauss hypergeometric function ${}_2F_1$, and the Appell $F_1$ function repectively, and the general case is known as the Lauricella $F_D$ functions.

In this section, we apply the ideas of intersection theory to these paradigmatic cases with their increasing level of complexity, in order to derive contiguity relations, which for hypergeometric functions play the same role that IBP identities play for Feynman integrals\footnote{Recent applications of the theory of hypergeometric functions to the coaction of one-loop (cut)Feynman integrals can be found in \cite{Abreu:2017enx,Abreu:2018nzy}.}. 

\subsection{Euler Beta Integrals}

We start by discussing integral relations associated to a simple class of integrals such as the Euler {\it beta function}, defined as
\begin{eqnarray}
\label{eq:def:Eulerbeta}
\beta(a,b) \equiv \int_0^1 dz \, z^{a-1} \, (1-z)^{b-1} = \frac{\Gamma(a) \Gamma(b)}{\Gamma(a+b)} \ . 
\end{eqnarray}

\subsubsection{Direct Integration}

Let us consider integrals of the type
\begin{eqnarray}
I_n \equiv \int_{\cal C}  u \ z^n dz \ , 
\qquad u \equiv B^\gamma \ , 
\qquad B \equiv z (1-z) \ , 
\qquad {\cal C} \equiv [0,1] \ .
\end{eqnarray}
These integrals admit a closed-form expression in terms of $\Gamma$ functions,
\begin{eqnarray}
I_n = \frac{\Gamma(1+\gamma)\Gamma(1+\gamma+n) }{\Gamma(2+2\gamma+n)} \ , 
\end{eqnarray}
from which it is possible to derive a relation between $I_n$ and $I_0$,
\begin{eqnarray}
I_{n} = \frac{\Gamma(1+\gamma+n)\Gamma(2+2\gamma)}{\Gamma(1+\gamma)\Gamma(2+2\gamma+n)} \, I_0 \ . 
\end{eqnarray}
For instance, when $n=1$, it reads
\begin{eqnarray}
I_{1} = \frac{1}{2} I_0 \ . 
\label{eq:betarel}
\end{eqnarray}

\subsubsection{Integration-by-Parts Identities}

Let us recover the same relation from integration by parts identities.
With the choice of ${\cal C}$ as above, the following 
integration-by-parts identity holds

\begin{eqnarray}
\int_{\cal C} d( B^{\gamma+1} z^{n-1}) = 0 \ .
\end{eqnarray}
The action of the differential operator under the integral sign yields the following
equation,
\begin{eqnarray}
(\gamma+n) I_{n-1} - (1+2\gamma+n) I_{n} = 0 \ .
\end{eqnarray}
Therefore we obtain the recurrence relation
\begin{eqnarray}
I_{n} = \frac{(\gamma+n)}{(1+2\gamma+n)} \, I_{n-1} \ ,
\end{eqnarray}
which, for $n=1$, gives 
\begin{eqnarray}
I_{1} = \frac{1}{2} I_{0} \ .
\end{eqnarray}

\subsubsection{Intersections}

We are going to (re)derive, once more, the relations between Euler beta integrals using intersection numbers. We consider integrals defined as,
\begin{eqnarray}
I_n &\equiv& 
\int_{\cal C} \, u \, \phi_{n+1} 
\equiv {}_{\omega}\langle \phi_{n+1} | {\cal C}] \ , 
\qquad \phi_{n+1} \equiv z^n dz \ ,
\end{eqnarray}
with
\begin{gather}
u = B^\gamma \, 
\qquad B = z(1-z) \ , \qquad \omega = d\log u = \gamma \left(\frac{1}{z} + \frac{1}{z-1}\right) dz \ , \\
\masters=1 \ , \qquad {\cal P} = \{0,1,\infty\}.
\end{gather}

\paragraph{Monomial Basis.}
$\masters=1$ implies the existence of 1 master integral, which we choose as 
$I_0 = {}_\omega\!\langle \phi_{1}|\mathcal{C}]$.
The goal of this calculation is to derive the relation between $I_1$ and $I_0$,
\begin{eqnarray}
I_1 = c_{1} \ I_0 \qquad \Longleftrightarrow \qquad 
{}_\omega\!\langle \phi_{2} | {\cal C}]
= c_1 \ 
{}_\omega\!\langle \phi_{1} | {\cal C}]
\end{eqnarray}
which can be derived by decomposing $\langle \phi_2 |$ in terms of $\langle \phi_1 |$,
\begin{eqnarray}
\langle \phi_2 | &=& c_ 1\langle \phi_1 | \ , \qquad 
c_1 = 
\langle \phi_2 | \phi_1 \rangle
\langle \phi_1 | \phi_1 \rangle^{-1} 
\end{eqnarray}
Notice that since $\masters=1$, the intersection matrix ${\bf C}_{ij}$ has just one element ${\bf C}_{11} = \langle \phi_1 | \phi_1 \rangle$. 

We need to evaluate the intersection numbers 
$\langle \phi_1 | \phi_1 \rangle$, and 
$\langle \phi_2 | \phi_1 \rangle$.

For each pole $p \in {\cal P}$,
we identify $\phi_{i,p}$ (the series expansion of $\phi_i$ around $z=p$), 
and determine the associated function $\psi_{i,p}$ (the series expansion of $\psi_i$ around $z=p$),
by solving the following differential equation,
\begin{eqnarray}
\nabla_\omega \, \psi_{i,p} = \phi_{i,p} \ .
\end{eqnarray}

After inserting the series expansion of $\phi_{i,p}$ and an ansatz for $\psi_{i,p}$ in the above equation, we get an equation at each order on $p$, which together determines the coefficients in the ansatz for $\psi_{i,p}$.
In practice, we introduce a local coordinate $\tau$, 
defined as $\tau = z - p$, for finite poles, or $\tau = 1/z$ for the pole at infinity, 
and consider the Laurent expansions around $\tau \to 0$ of,
\begin{eqnarray}
\phi_{i,p} &=& \sum_{k=\text{min}-1} \phi_{i,p}^{(k)} \tau^k \ , \qquad 
\omega_p = \sum_{k=-1} \omega_{p}^{(k)} \tau^k \ , 
\qquad ({\rm known}) 
\end{eqnarray}
and the ansatz,
\begin{eqnarray}
\psi_{p} &=& \sum_{k={\rm min}}^{\rm max} 
\alpha_k \, \tau^k \ ,
\qquad ( \alpha_k \ 
{\rm unknown})
\end{eqnarray}
to solve the following differential equation,
\begin{eqnarray}
\frac{d}{d \tau} \psi_{p} + \omega_p \, \psi_{p} - \phi_{i,p} = 0 \ . 
\end{eqnarray}
In our case we have, 
\begin{itemize}[leftmargin=1.3em]

\item For $\varphi_L = \phi_1 = dz$, $\;\varphi_R = \phi_1 = dz$:

\begin{center}
\begin{tabular}{|c||c|c|c|c|}
\hline 
$p$ & min & max & $\varphi_{L,p}$ & $\psi_p$ \\ 
\hline 
\hline
$0$ & $1$ & $-1$ & $d \tau$ & $-$ \\
\hline
$1$ & $1$ & $-1$ & $d \tau$ & $-$ \\
\hline
$\infty$ & $-1$ & $1$ & $- d \tau/ \tau^2$ & $\sum_{i=-1}^1 \alpha_i \, \tau^i $ \\
\hline
\end{tabular}
\end{center}
with
\begin{eqnarray}
\alpha _{-1} = 
\frac{1}{2 \gamma+1} \ , \qquad 
\alpha _0 =  
-\frac{1}{2 (2 \gamma+1)} \ , \qquad 
\alpha _1 = 
   -\frac{\gamma}{2 (2 \gamma-1) (2 \gamma+1)} \ . 
\end{eqnarray}
Around $p=0,1,$ the solution $\psi_p$ does not exist (owing to the values of min and max), 
therefore
\begin{eqnarray}
\langle \phi_1 | \phi_1 \rangle &=& 
{\rm Res}_{z = \infty}( \psi_\infty \phi_1 )
=
\frac{\gamma }{2 (2 \gamma -1) (2 \gamma +1)} \ . 
\end{eqnarray}

\item For $\varphi_L = \phi_2 = z \, dz$, $\;\varphi_R = \phi_1 = dz$: 
\begin{center}
\begin{tabular}{|c||c|c|c|c|}
\hline 
$p$ & min & max & $\varphi_{L,p}$ & $\psi_p$ \\ 
\hline 
\hline
$0$ & $2$ & $-1$ & $\tau \, d \tau$ & $-$ \\
\hline
$1$ & $1$ & $-1$ & $d \tau$ & $-$ \\
\hline
$\infty$ & $-2$ & $1$ & $- d \tau/ \tau^3$ & $\sum_{i=-2}^1 \alpha_i \, \tau^i $ \\
\hline
\end{tabular}
\end{center}
with
\begin{gather}
\alpha _{-2} = 
\frac{1}{2 (\gamma+1)} \ , \quad 
\alpha _{-1} =  
-\frac{\gamma}{2 (\gamma+1) (2\gamma+1)}, \\ 
\alpha _0 = 
   -\frac{1}{4 (2 \gamma+1)} \ , \quad 
   \alpha _1 =  
   -\frac{\gamma}{4 (2 \gamma-1) (2 \gamma+1)} \ .
\end{gather}
Around $p=0,1,$ the solution $\psi_p$ does not exist, 
therefore
\begin{eqnarray}
\langle \phi_2 | \phi_1 \rangle &=&
{\rm Res}_{z = \infty}( \psi_\infty \phi_1 )
=
\frac{\gamma }{4 (2 \gamma -1) (2 \gamma +1)} \ .
\end{eqnarray}

\end{itemize}

Notice that in the above formulas only the $p=\infty$ gave a non-trivial contribution.
In general, the situation depends on the form of the integrands,
and in particular on on the values of min and max, which are dictated by the Laurent series expansions
around $p$ of $\varphi_L$ and $\varphi_R$ paired in the intersection number $\langle \varphi_L | \varphi_R \rangle$ \ .

\noindent
Finally, we get the decomposition of $I_1$ in terms of $I_0$,
\begin{eqnarray}
I_1 &=& c_1 \, I_0 \ ,\\
c_1 &=& 
\langle \phi_2 | \phi_1 \rangle
\langle \phi_1 | \phi_1 \rangle^{-1} 
= \frac{1}{2},
\end{eqnarray}
in agreement with eq. \eqref{eq:betarel}. \\

\paragraph{dlog-basis.}

Consider the master integral associated to the form
\begin{eqnarray}
 \varphi_1  = d \log \frac{z}{z-1} = \left(\frac{1}{z} - \frac{1}{z-1} \right){dz}\ , 
\end{eqnarray}
and let us decompose both $ \langle \phi_1 | $
and $\langle \phi_2 |$ in the basis of $\langle \varphi_1|$,

\begin{eqnarray}
\langle \phi_1 | = 
\langle \phi_1 | \varphi_1 \rangle
\langle \varphi_1 | \varphi_1 \rangle^{-1} 
\langle \varphi_1 |,
\end{eqnarray}

\begin{eqnarray}
\langle \phi_2 | = 
\langle \phi_2 | \varphi_1 \rangle
\langle \varphi_1 | \varphi_1 \rangle^{-1} 
\langle \varphi_1 |.
\end{eqnarray}
We need the intersection numbers,
\begin{equation}
\langle \varphi_1 | \varphi_1 \rangle
=
\frac{2}{\gamma }, \qquad
\langle \phi_1 | \varphi_1 \rangle 
=
\frac{1}{2 \gamma +1}, \qquad
\langle \phi_2 | \varphi_1 \rangle 
= 
\frac{1}{2 (2 \gamma +1)}.
\end{equation}
Therefore
\begin{eqnarray}
\langle \phi_1 | = 
\frac{\gamma }{2 (2 \gamma +1)}
\langle \varphi_1 |, \qquad
\langle \phi_2 | = 
\frac{\gamma }{4 (2 \gamma +1)}
\langle \varphi_1 |
\end{eqnarray}
from which one can also deduce 
$\langle \phi_2 |  = 1/2 \langle \phi_1 | $.

Please note, that in this basis the metric term 
$\langle \varphi_1 | \varphi_1 \rangle$ is very simple,
and that $\langle \varphi_1 | \varphi_1 \rangle^{-1} = \gamma/2$, has $\gamma$ factorizing out. \\

This simple example contains all the relevant ingredients for the decomposition of Feynman integrals
in terms of master integrals. It corresponds to a case with 1 master integral. 
We now consider two other cases, with respectively 2 and 3 master integrals,
in order to show the  algorithmic procedure of the decomposition
by intersection numbers.

\subsection{Gauss ${}_2F_1$ Hypergeometric Function}
Gauss ${}_2F_1$ Hypergeomeric function is defined as
\begin{align}
{\beta(b,c{-}b)} \ {}_2 F_1 (a,b,c;x) = &\!\int_{0}^{1} z^{b-1} (1-z)^{c-b-1} (1{-}xz)^{-a}\, dz
\end{align}
The integration contour $\mathcal{C}$ is $[0,1]$, which is the twisted cycle. $\beta(b,c{-}b)$ is the Euler beta function defined in
eq.~\eqref{eq:def:Eulerbeta}.
In order to use intersection theory, we re-express this integral in terms of the pairing of the twisted cycle and the twisted cocycle:
\begin{eqnarray}
&&{\beta(b,c{-}b)} \ {}_2 F_1 (a,b,c;x) = \int_{\cal C} u \, \varphi 
= \ {}_{\omega} \langle \varphi | {\cal C} ]
\ , 
\end{eqnarray}
where
\begin{eqnarray}
u &=& z^{b-1} (1-x z)^{-a} (1-z)^{-b+c-1} \ , \\
\omega &=& d \log u= \frac{x z^2 (c-a-2)+z (a x-c+x+2)-b x z+b-1}{(z-1) z (x z-1)} \, dz \,, \\
\varphi &=& dz \, .
\end{eqnarray}
In this case, we have
\begin{eqnarray}
\masters = 2\,, \qquad 
\mathcal{P} = \lbrace 0,\, 1,\, \tfrac{1}{x},\, \infty \rbrace
\end{eqnarray}
indicating the existence of 2 independent integrals.
Contiguity relations for Gauss Hypergeometric functions can be obtained through intersection theory, {\it 
via} the master decomposition formula
in eq. (\ref{eq:masterdeco:}), 
requiring the knowledge of the (inverse of the) matrix ${\bf C}$.
We build this matrix for various different choices of the integral basis.

\paragraph{Monomial Basis.}
We choose the basis as $\lbrace \langle \phi_i | \rbrace_{i=1,2}$, we build the metric matrix \textbf{C}, 
\begin{align}
\mathbf{C} = \left(
\begin{array}{cc}
 \langle \phi_1 | \phi_1 \rangle  
&\langle \phi_1 | \phi_2 \rangle  \\
 \langle \phi_2 | \phi_1 \rangle 
 & \langle \phi_1 | \phi_2 \rangle \\
\end{array}
\right)
\end{align}
whose entries are
\allowdisplaybreaks{
\begin{align}
\langle \phi_1 | \phi_1 \rangle &= \Big( x^2 (-(a-b+1)) (b-c+1)-2 a x (-b+c-1)+a (c-2) \Big) \, / \, 
\Big( x^2 (a \nonumber \\
& -c+1) (a-c+2) (a-c+3) \Big), \\
\langle \phi_1 | \phi_2 \rangle &= \Big(x^3 (-(a-b+1)) (a-b+2) (b-c+1)-a x^2 (-b+c-1) (2 a-3 b \nonumber \\
&+ {} c+2)+a x (a+2 c-5) (-b+c-1)-a (c-3) (c-2)\Big)
\, / \, \Big(x^3 (a-c+1) \nonumber \\
& (a-c+2) (a-c+3) (a-c+4) \Big),
\\
\langle \phi_2 | \phi_1 \rangle &= \Big(x^3 (-(a-b)) (a-b+1) (b-c+1)-a x^2 (-b+c-1) (2 a-3 b+c) \nonumber \\
&+a x (a+2 c-3) (-b+c-1)-a (c-2) (c-1)\Big) 
\, / \, \Big(x^3 (a-c) (a-c+1) \nonumber \\
& (a-c+2) (a-c+3) \Big),
\\
\langle \phi_2 | \phi_2 \rangle &= \Big(
-a x^2 (a^2 b-a^2 c 
+a^2-3 a b^2+7 a b c-8 a b-4 a c^2+9 a c-5 a-3 b^2 c \nonumber \\
&+ {} 6 b^2+4 b c^2-10 b c +6 b-c^3+2 c^2-c)+x^4 (-(a^3-3
   a^2 b+3 a^2+3 a b^2 \nonumber \\
&- {} 6 a b+2 a-b^3+3 b^2 -2 b))(b-c+1)+2 a x^3 (a-b+1) (a b-a c+a \nonumber \\
&- {} 2 b^2+3 b c-2 b-c^2+c)+2 a (c-2) x
   (a+c-2) (b-c+1)+a (c^3-6 c^2 \nonumber \\
&+ {} 11 c-6)
\Big)
\, / \, {\Big(x^4 (a-c) (a-c+1) (a-c+2) (a-c+3) (a-c+4)\Big)}.
\end{align}
}
Now, we can derive any functional relation using the following decomposition.
\begin{align}
\langle \phi_n | = \sum_{i,j=1}^{2} \langle \phi_n |\phi_{j} \rangle\, \left(\textbf{C}^{-1}\right)_{ji} \, \langle \phi_{i}|.
\label{eq:2F1:GD}
\end{align}

Let us consider the decomposition of $\beta(b+2,c-b) {}_2 F_1 (a,b+2,c+2;x) \equiv \langle \phi_3|{\cal C}]$ in terms of $\beta(b,c-b) {}_2 F_1 (a,b,c;x)$ and $\beta(b+1,c-b) {}_2 F_1 (a,b+1,c+1;x)$. Using the eq.~(\ref{eq:2F1:GD}) we obtain
\begin{align}
& \beta(b+2,c-b) {}_2 F_1 (a,b+2,c+2;x) = \left(\frac{b}{x (a-c-1)}\right) \beta(b,c-b) {}_2 F_1 (a,b,c;x) \nonumber \\
& \qquad\qquad\qquad +\left( \frac{(b-a+1)x + c}{x (c-a+1)}\right) \beta(b+1,c-b) {}_2 F_1 (a,b+1,c+1;x)
\end{align}
or correspondingly
\begin{align}
& {}_2 F_1 (a,b+2,c+2;x) = \frac{(c+1)}{x (b+1) (c-a+1)} \times \nonumber \\
& \quad\quad \Big( \big( (b-a+1)x + c \big) \; {}_2 F_1 (a,b+1,c+1;x) - c \; {}_2 F_1 (a,b,c;x)\Big) \ ,
\label{eq:hygoidentity}
\end{align}
as verified using \textsc{Mathematica}.

\paragraph{dlog-basis.}
Let us consider the following dlog-basis.
\begin{align}
\varphi_1 &= \left( \frac{1}{z} - \frac{1}{z-1} \right) dz \\
\varphi_2 &= \left( \frac{1}{z-1} - \frac{x}{x z-1} \right) dz.
\end{align}
The \textbf{C} matrix with entries $\mathbf{C}_{ij} = \langle \varphi_i | \varphi_j \rangle$ for this case is as follows
\begin{align}
\mathbf{C} = \frac{1}{c-b-1} \left(
\begin{array}{cc}
 \frac{c-2}{b-1} & -1 \\
 -1 & \frac{a+b-c+1}{a} \\
\end{array}
\right).
\end{align}
The above relations between hypergeometric functions can be obtained using the dlog-basis as well. The \textbf{C} matrix in this case takes a very simple form and it is factorized. If we consider the powers of all the factors to be equal, for example $a=-\gamma,\, b = \gamma{+}1,\, c= 2(\gamma{+}1)$, then $\gamma$ factorizes out, and as a result the system of differential equations for $\varphi_i$ is canonical, according to eq.~\eqref{eq:Omega:FInvC}. 

In particular, let us introduce the prefactor 
\begin{align}
    K &= \frac{(c - b - 1) \, (b - 1)}{(c - 1) \, (c - 2) \, \beta(b,c-b)}\,,
\end{align}
and consider the two integrals,
\begin{align}
I_1 &= \langle \varphi_1 | \mathcal{C} ] = {}_2 F_1 (a, b-1, c-2; x) \, , \\ I_2 &= \langle \varphi_2 | \mathcal{C} ] = \frac{(b-1) (x-1)}{c-2} {}_2 F_1 (a+1, b, c-1; x) \, ,
\end{align}
which, for $a=-\gamma,\, b = \gamma{+}1,\, c= 2(\gamma{+}1)$, read,
\begin{align}
I_1 = {}_2 F_1 (- \gamma, \gamma, 2 \gamma; x) \,, \qquad I_2 = \frac{x-1}{2} {}_2 F_1 (1 - \gamma, 1 + \gamma, 1 + 2 \gamma; x)\,.
\end{align}
Following the method of Sec. \ref{sec:DE}, we derive the system of differential equations with respect to $x$, 
\begin{align}
\partial_x I_i = \mathbf{A}_{ij} I_j \,, \qquad \text{with} \; \qquad \mathbf{A} = \gamma \left( \begin{array}{cc} 0 & \; \frac{-1}{x-1} \\[1mm] \frac{-1}{x} & \;\; \frac{2}{x-1} - \frac{2}{x} \end{array} \right) \ , 
\end{align}
which is {\it canonical}, namely it is fuchsian and $\gamma$-factorised.
It is easily seen that the system can be integrated up order-by-order in $\gamma$, yielding a result where the coefficient at order $\gamma^n$ can be expressed in terms of harmonic polylogarithms (HPLs)~\cite{Remiddi:1999ew} of weight $n$, therefore making explicit the relation between HPLs and the series expansion of ${}_2F_1$ around $\gamma=0$.

\paragraph{Mixed bases.} 
By using mixed bases, namely a monomial-basis  
$\langle e_i | = \langle \phi_i |,$ and a dlog-basis   
$| h_j \rangle = | \varphi_j \rangle$ , 
we can decompose our integrals in terms of a monomial basis, which can be directly mapped onto 
eq.~\eqref{eq:hygoidentity},
without loosing the advantages of simpler expressions due to the dlog-basis algebra. In this case, the intersection matrix becomes
\begin{align}
\textbf{C} = \langle \phi_i | \varphi_j \rangle = \left( \begin{array}{cc}
\frac{1}{c-a-1} & \frac{x-1}{(1 + a - c) x} \\[1mm]
\frac{a - a x + b x}{(a - c) (1 + a - c) x} & \; \frac{(x-1) (1 - c + a x - b x)}{(a - c) (1 + a - c) x^2} \end{array} \right)
\end{align} 
whose entries look slightly more involved than in the dlog case, but much simpler than in the monomial case. To reproduce eq.~\eqref{eq:hygoidentity}, we also need the intersections
\begin{align}
    \langle \phi_3 | \varphi_1 \rangle &= \frac{a (x-1) (c + (2b-a+1) x) - b (1 + b) x^2}{(a-c-1) (a - c) (1 + a - c) x^2} \\
   \langle \phi_3 | \varphi_2 \rangle &= \frac{(x-1) \big( b x (a + c + (b - 2a + 1) x - 1) + (c - a x -1) (c + x - a x) \big)}{(a-c-1) (a - c) (1 + a - c) x^3}
\end{align}
both of which are much simpler than in the monomial basis. As expected, using them in eq.~\eqref{eq:masterdeco:} yields eq.~\eqref{eq:hygoidentity}.

\subsection{Appell $F_1$ Function}

Let us consider the Appell $F_1$ function:

\begin{eqnarray}
\beta(a,c-a) \ F_1 (a,b_1,b_2,c;x,y) = \int_{\cal C} z^{a-1} (1-z)^{-a+c-1} (1-x z)^{-b_1} (1-y z)^{-b_2} \, dz 
\ , 
\end{eqnarray}
the integration contour $\mathcal{C}$ is $[0,1]$, which is the twisted cycle. $\beta(a,c{-}a)$ is the Euler beta function.
In order to use intersection theory, we re-express this integral in terms of the pairing of the twisted cycle and the twisted cocycle:
\begin{eqnarray}
&&{\beta(a,c{-}a)} \ F_1 (a,b_1,b_2,c;x,y) = \int_{\cal C} u \, \varphi 
= \ {}_{\omega} \langle \varphi | {\cal C} ]
\ , 
\end{eqnarray}
where,
\begin{gather}
u = z^{a-1} (1-z)^{-a+c-1} (1-x z)^{-b_1} (1-y z)^{-b_2}, \\
\omega = 
\left(
\frac{-a+c-1}{z-1}+\frac{a-1}{z}-\frac{b_1 x}{x z-1}-\frac{b_2 y}{y z-1}
\right) dz , \\
\varphi = dz , 
\end{gather}
In this case, we have
\begin{gather}
\masters = 3\,, \qquad \mathcal{P} = \lbrace 0,\, 1,\, \tfrac{1}{x},\, \tfrac{1}{y},\, \infty \rbrace.
\end{gather}
indicating the existence of 3 independent integrals.
Contiguity relations for Appell $F_1$ functions \cite{goto2015} can be obtained through intersection theory, {\it 
via} the master decomposition formula
in eq. (\ref{eq:masterdeco:}), 
requiring the knowledge of the (inverse of the) $3 \times 3$ matrix ${\bf C}$. For this purpose, we can choose any basis as per convenience.


\paragraph{dlog-basis.}
Let us consider the following dlog-basis.
\begin{align}
\varphi_1 &= \left( \frac{1}{z}-\frac{1}{z-1} \right) dz, \\
\varphi_2 &= \left( \frac{1}{z-1}-\frac{x}{x z-1} \right) dz, \\
\varphi_3 &= \left( \frac{x}{x z-1}-\frac{y}{y z-1} \right) dz.
\end{align}
The \textbf{C} matrix for this case reads as,
\begin{align}
\mathbf{C} = \frac{1}{c-a-1}\left(
\begin{array}{ccc}
 \frac{c-2}{a-1} & -1 & 0 \\
 -1 & \frac{a-c+b_1+1}{b_1} & \frac{-a+c-1}{b_1} \\
 0 & \frac{-a+c-1}{b_1} & \frac{(a-c+1) \left(b_1+b_2\right)}{b_1 b_2} \\
\end{array}
\right).
\end{align}
Here as well, we observe that the \textbf{C} matrix takes a very simple form and also that it can be factorized. For example, when $b_1=-\gamma,\, b_2 = -\gamma,\, a = \gamma{+}1,\, c= 2(\gamma{+}1)$ the overall power $\gamma$ factors out.

\paragraph{Mixed Bases}
Let us consider projections unto a monomial basis $\phi_1 = 1 \, dz$, $\phi_2 = z \, dz$, $\phi_3 = z^2 \, dz$. Picking as right basis the dlog-basis considered above, we get the entries of the \textbf{C}-matrix $\langle \phi_i | \varphi_j \rangle$ to be

\begin{align}
\langle \phi_1 | \varphi_1 \rangle &= \frac{-1}{1 + b_1 + b_2 - c}\,, \qquad\quad \langle \phi_1 | \varphi_2 \rangle = \frac{x-1}{(1 + b_1 + b_2 - c) x}\,, \\
\langle \phi_1 | \varphi_3 \rangle &= \frac{y-x}{(1 + b_1 + b_2 - c) x y}\,, \\
\langle \phi_2 | \varphi_1 \rangle &= \frac{b_1 y + x (b_2 - (b_1 + b_2 - a) y)}{(b_1 + b_2 - c) (1 + b_1 + b_2 - c) x y}\,, \\
\langle \phi_2 | \varphi_2 \rangle &= \frac{(x-1) ((1 + b_2 - c + (b_1 + b_2 - a) x) y - b_2 x)}{(b_1 + b_2 - c) (1 + b_1 + b_2 - c) x^2 y}\,, \\
\langle \phi_2 | \varphi_3 \rangle &= \frac{(y - x) ((1 + b_2 - c) y + x (1 + b_1 + c (y-1) - (1 + a) y))}{(b_1 + b_2 - c) (1 + b_1 + b_2 - c) x^2 y^2)}\,,
\end{align}
\begin{align}
\langle \phi_3 | \varphi_1 \rangle &= \Big( x^2 \big( b_2 (b_1 - c) + b_2 (b_1 + b_2 + c - 1 - 2 a) y - (a - b_1 - b_2) (1 + a - b_1 - b_2) y^2 \big) \nonumber \\ 
& \qquad + b_1 (b_2 - c) y^2 + x (-2 b_1 b_2 y + b_1 (b_1 + b_2 + c - 2 a - 1) y^2)\Big) \Big/ \nonumber \\
& \quad\;\; \Big( (b_1 + b_2 - c - 1) (b_1 + b_2 - c) (1 + b_1 + b_2 - c) x^2 y^2 \Big)
\end{align}

\begin{align}
\langle \phi_3 | \varphi_2 \rangle &= \Big( (x-1) (b_2 (c - b_1) x^2 + b_2 x (b_1 - b_2 + c + x + 2 a x - 1 - (b_1 + b_2 + c) x) y \nonumber \\
& \qquad + (b_2 + b_2^2 - c - 2 b_2 c + c^2 + (b_1 + b_2 (b_1 + b_2) + c - 1 - (2 b_1 + b_2) c \nonumber \\
& \qquad + a (b_1 - b_2 + c - 1)) x + (b_1 + b_2 - a - 1) (b_1 + b_2 - a) x^2) y^2) \Big) \Big/ \nonumber \\
& \quad\;\; \Big( (b_1 + b_2 - c - 1) (b_1 + b_2 - c) (1 + b_1 + b_2 - c) x^3 y^2 \Big)
\end{align}

\begin{align}
\langle \phi_3 | \varphi_3 \rangle &= \Big( x^2 (b_1 + b_1^2 - 2 b_1 b_2 - b_1 c + b_2 c - (1 + a - c) (c - b_1 + b_2 - 1) x) y \times \nonumber \\
& \qquad + (c - b_1) (1 + b_1 - c) x^3 - x (b_2 - 2 b_1 b_2 + b_2^2 + b_1 c - b_2 c + 2 (b_1 - b_2) (1 + a - c) x \nonumber \\
& \qquad + (b_1 + b_2 - a - 1) (c - a - 1) x^2) y^2 + (b_2 + b_2^2 - c - 2 b_2 c + c^2 \nonumber \\
& \qquad + (1 + a - c) (b_1 - b_2 + c - 1) x + (b_1 + b_2 - a - 1) (c - a - 1) x^2) y^3 \Big) \Big/ \nonumber \\
& \quad\;\; \Big( (b_1 + b_2 - c - 1) (b_1 + b_2 - c) (1 + b_1 + b_2 - c) x^3 y^3 \Big)
\end{align}

Let us, as an example, derive the reduction of the function corresponding to $\phi_4 = z^3 \, dz$. This results in the reduction
\begin{align}
& F_1(a{+}3,b_1,b_2,c{+}3;x,y) = \frac{(c+2)}{(a+1) (a+2) x y (c+2-b_1-b_2)} \times \nonumber \\
& \quad \bigg( (1{+}a) \Big( (1{-}b_2{+}c) y + x (1{+}c + (2{+}a{-}b_2) y - b_1 (1{+}y)) \Big) F_1(a{+}2,b_1,b_2,c{+}2;x,y) \nonumber \\
& \qquad - (c+1) \big( c + (1 + a - b_1) x + (1 + a - b_2) y \big) F_1(a{+}1,b_1,b_2,c{+}1;x,y) \nonumber \\
& \qquad + c (c+1) F_1(a,b_1,b_2,c;x,y) \bigg)\,,
\end{align}
an example of a contiguity relations for Appell $F_1$. The relation has been checked numerically using \textsc{Mathematica}.

\subsection{Lauricella $F_D$ Function}
Finally, let us comment on the Lauricella $F_D$ function \cite{Keiji-MATSUMOTO2013367,goto2015b,goto2017}, which in general depends on $2m{+}2$ external variables and admits the following integral representation
\begin{eqnarray}
\beta(a,c-a) \,
F_D (a,b_1,b_2,\ldots,b_m,c;x_1,\ldots,x_m) = \int_{\cal C} u \, \varphi 
= \ {}_{\omega} \langle \varphi | {\cal C} ]
\ , 
\end{eqnarray}
where
\begin{gather} 
u = z^{a-1} \, (1-z)^{-a+c-1} \, \prod_{i=1}^{m} (1-x_i z)^{-b_i} \ , \\
{\cal C} = [0,1], \qquad \varphi = dz \ , \qquad \omega = d\log(u),\\
\masters = m{+}1, \qquad \mathcal{P} = \{ 0, \frac{1}{x_1}, \frac{1}{x_2}, \ldots, \frac{1}{x_m}, 1,\infty\}
\end{gather}
Contiguity relations for $F_D$ can be found using intersection numbers along the lines of the algorithm discussed in the previous sections \cite{matsumoto2018relative}.

\vspace*{1cm}
\paragraph{}
We now apply the decomposition by intersection numbers to those Feynman integrals, which admit a 1-form integral representation on the maximal cut.
In particular, we show how to build integral relations analogous to the integration-by-parts identities, 
directly generated by projections, 
using the master decomposition formula in eq.~\eqref{eq:masterdeco:}. 
For some cases, we build also the dimensional recurrence relation and systems of differential equations for the master integrals. \\ 
When possible, our results have been successfully checked with the automatic tools 
{\sc SYS} \cite{Laporta:2001dd},
{\sc Reduze2} \cite{vonManteuffel:2012np}, 
{\sc FIRE5} \cite{Smirnov:2014hma}, 
{\sc LiteRed} \cite{Lee:2012cn},
{\sc Kira} \cite{Maierhoefer:2017hyi},
and compared with the available literature.

\paragraph{}
In what follows all propagators are taken to be on-shell, or in other words the internal propagators are cut. Hence, we generally do not indicate the cuts explicitly in the figures, unless when it is specifically required.

\paragraph{}
We begin with two cases where the standard Baikov representation generates 1-form integrals.

\section{Four-Loop Vacuum Diagram}
\label{sec:fourloopvacuum}

\begin{figure}[H]
    \centering
    \includegraphics[width=0.18\textwidth]{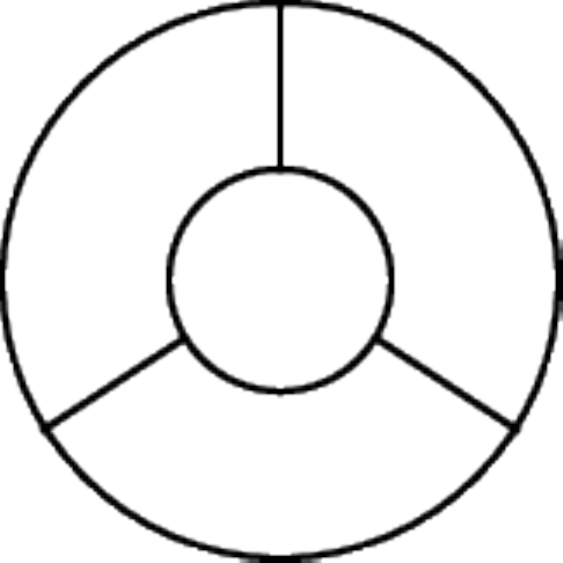}
    \caption{Four-loop vacuum diagram.}
    \label{fig:four_loop_vacuum}
\end{figure}
Let us consider the four-loop vacuum diagram from Fig.~\ref{fig:four_loop_vacuum}, first derived in ref. \cite{Laporta:2002pg}.
The denominators read (the internal mass that is present on all the propagators, is set to unity):
\begin{align}
&D_{1}=k_1^2-1 \, , \qquad D_{2}=k_2^2-1 \, ,\qquad D_{3}=k_3^2-1 \, , \nonumber\\
&D_{4}=\left(k_1-k_2\right){}^2-1 \, , \qquad D_{5}= \left(k_1-k_3\right){}^2-1 \, , \qquad D_{6}=\left(k_2-k_3\right){}^2-1 \, , \\
&D_{7}=\left(k_1-k_4\right){}^2-1 \, , \qquad D_{8}=\left(k_2-k_4\right){}^2-1 \, , \qquad D_{9}=  \left(k_3-k_4\right){}^2-1,\nonumber
\end{align}
while the ISP is
\begin{equation}
z=D_{10}=k_4^2 \, .
\end{equation}
After applying standard Baikov representation, the corresponding integral family is characterized by:
\begin{align}
& u=\left(\frac{z}{2}-\frac{3 z^2}{16}\right)^{\frac{d-5}{2}} \, , \quad \omega=\frac{(d-5) (3 z-4)}{z (3 z-8)}dz \, .
\end{align}
The equation $\omega=0$ a has $1$ solution, indicating 1 master integral. Then, we define,
\begin{align}
& \masters=1 \, , \hspace{2.5cm} \mathcal{P}=\{0, \tfrac{8}{3} , \infty \}.
\end{align}
 Using the master decomposition formula eq.~(\ref{eq:masterdeco:}), we can express any integral in terms of the chosen master integral employing either monomial or dlog-basis.
\paragraph{Monomial Basis.} Let us consider the decomposition of $I_{1,1,1,1,1,1,1,1,1;-2}= \langle \phi_{3} | \mathcal{C}]$ in terms of $J_{1}=I_{1,1,1,1,1,1,1,1,1;0}= \langle \phi_{1} | \mathcal{C}]$.
We obtain the following decomposition in this case
\begin{align}
    \langle \phi_n | = \langle \phi_n |\phi_1 \rangle \,\textbf{C}_{11}^{-1}\, \langle \phi_1|.
    \label{eq:fourlooptad:GD}
\end{align}
We build the metric matrix \textbf{C}, containing a single element
\begin{equation}
\mathbf{C} = \bra{\phi_1}\ket{\phi_1}=\frac{16 (d-5)}{9 (d-6) (d-4)} \, ,
\end{equation}
and the other necessary intersection number
\begin{equation}
\bra{\phi_{3}}\ket{\phi_{1}}=\frac{256 (d-5) (d-1)}{81 (d-6) (d-4) (d-2)}.
\end{equation}
Here, the $\textbf{C}^{-1}$ is trivial to compute as the \textbf{C} contains only one element.
Using these in eq.~(\ref{eq:fourlooptad:GD}), we obtain
\begin{equation}
I_{1,1,1,1,1,1,1,1,1;-2}=\frac{16 (d-1)}{9 (d-2)} \, J_{1} ,
\end{equation}
in agreement with \textsc{SYS}. \\
 
\paragraph{dlog-Basis.} On the other hand, we can compute the decomposition of $\langle{ \phi_{3}} | \mathcal{C}]$ in terms of $\langle{\varphi_1 } | \mathcal{C}]$, with:
\begin{equation}
\hat{\varphi}_1=\frac{1}{z}-\frac{3}{3 z-8}.
\end{equation}
We then compute the intersections:
\begin{equation}
\bra{\varphi_1}\ket{\varphi_1}=\frac{4}{d-5} \, ,
\end{equation}
and
\begin{equation}
\bra{\phi_{3}}\ket{\varphi_1}=\frac{128 (d-1)}{27 (d-4) (d-2)} \, .
\end{equation}
This gives us the following basis decomposition:
\begin{equation}
\langle{\phi_{3}}|= \frac{32 (d-5) (d-1)}{27 (d-4) (d-2)} \, \langle{\varphi_1}|.
\end{equation}
in agreement with \textsc{SYS}. \\

\section{Three-Loop Triple-Cross}
\label{sec:threeloop}

\begin{figure}[H]
    \centering
    \includegraphics[width=0.38\textwidth]{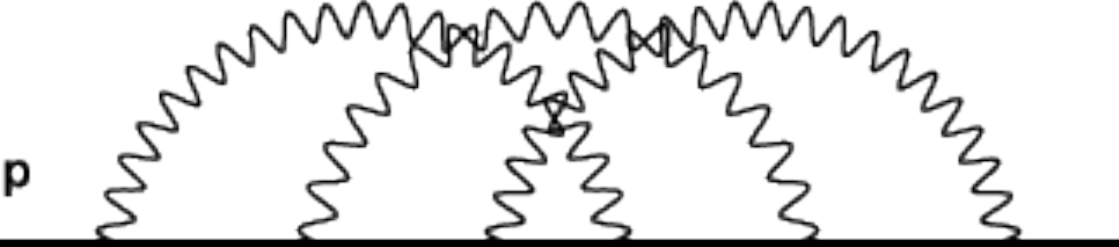}
    \caption{Triple-cross two-point function.}
    \label{fig:triplecross}
\end{figure}
Let us consider the triple-cross two-point function in Fig.~\ref{fig:triplecross}, first derived in \cite{Laporta:1996mq, Laporta:2001rc}. The incoming momentum is labelled by $p$, with $p^2=s$.
The denominators are (the internal mass is set to unity):
\begin{equation}
\begin{split}
& D_{1}=k_1^2 \, , \quad D_{2}=k_2^2 \, , \quad D_{3}=k_3^2 \, , \quad D_{4}=\left(p-k_1\right){}^2-1 \, ,\\
& D_{5}=\left(-k_1-k_2+p\right){}^2-1 \, , \quad D_{6}=\left(-k_1-k_2-k_3+p\right){}^2-1 \, , \\
& D_{7}=\left(-k_2-k_3+p\right){}^2-1 \, , \quad D_{8}=\left(p-k_3\right){}^2-1 \, ;
\end{split}
\end{equation}
We choose the ISP as:
\begin{equation}
z=D_{9}=k_2 \cdot p.
\end{equation}
Within the standard Baikov representation, the corresponding integral family is characterized by,
\begin{eqnarray}
u &=& \left(\frac{1}{4} z^2 (s-2 z-1) (s-2 z+3)\right)^{\frac{d-5}{2}} \, , \\
 K &=& -\frac{s^{\frac{2-d}{2}}}{32 \pi ^3 \Gamma \left(\frac{d-3}{2}\right) \Gamma
   \left(\frac{d-2}{2}\right) \Gamma \left(\frac{d-1}{2}\right)} \, , \\
  \omega&=&(d-5) \left(\frac{1}{-s+2 z-3}+\frac{1}{-s+2 z+1}+\frac{1}{z}\right)dz \, .
\end{eqnarray}
 The 2 solutions of the equation $\omega=0$, imply the existence of 2 master integrals. 
We define
 \begin{eqnarray}
\masters=2 \, , \qquad \mathcal{P}= \{ 0 \, , \frac{s-1}{2} \, , \frac{s+3}{2} \, ,\infty \} ,
\end{eqnarray}
and use the master decomposition formula eq.~(\ref{eq:masterdeco:}) to express any integral in terms of the chosen master integrals, employing either monomial or dlog-basis.

\paragraph{Monomial Basis.} 
Here we choose the basis as $\lbrace \langle \phi_i | \rbrace_{i=1,2}$ corresponding to the following 2 master integrals
\begin{equation}
J_{1}=I_{1,1,1,1,1,1,1,1;0} = \langle \phi_1 | {\cal C} ] \, , 
\quad J_{2}=I_{1,1,1,1,1,1,1,1;-1} = \langle \phi_2 | {\cal C} ] \,.
\label{MIs_monomial_basis_triple_cross}
\end{equation}
Let us consider the decomposition of $I_{1,1,1,1,1,1,1,1;-2}= \langle{\phi_{3}}|\mathcal{C}]$ in terms of ($\ref{MIs_monomial_basis_triple_cross}$). The decomposition formula reads
\begin{align}
\langle \phi_n | = \sum_{i,j = 1}^{2} \langle \phi_n |\phi_j \rangle 
\left( \textbf{C}^{-1}\right)_{ji} \langle \phi_i|.
\label{eq:threelooptriplecross:GD}
\end{align}
We compute the $\mathbf{C}$ matrix:
\begin{equation}
\mathbf{C}_{ij}=\bra{\phi_{i}}\ket{\phi_{j}} \, , \quad i,j=1,2;
\end{equation}
with:
\begin{align}
\bra{\phi_{1}}\ket{\phi_{1}}= & \frac{(d-5) (s (s+2)+9)}{8 (4 (d-10) d+99)} \, , \\
\bra{\phi_{1}}\ket{\phi_{2}}= & \frac{(d-5) (s+1) (d (s (s+2)+33)-6 (s (s+2)+31))}{32 (d-6) (2 d-11) (2 d-9)} \, , \\
\bra{\phi_{2}}\ket{\phi_{1}}= & \frac{(d-5) (s+1) (d (s (s+2)+33)-4 (s (s+2)+36))}{32 (d-4) (2 d-11) (2 d-9)} \, , \\
\bra{\phi_{2}}\ket{\phi_{2}}=& (d-5) \Big(d^2 (s (s+2) (s (s+2)+90)+153)-10 d (s (s{+}2) (s(s{+}2){+}90){+}153) \nonumber \\
& +24 (s (s{+}2) (s (s{+}2){+}92){+}157) \Big) / \Big(   128 (d{-}6) (d{-}4) (2 d{-}11)(2 d{-}9)  \Big) \, .
\end{align}
The other necessary intersection numbers read: 
\begin{align}
\bra{\phi_{3}}\ket{\phi_{1}}= & (d{-}5) \Big(d^2 (s (s+2) (s (s+2)+90)+153)-d (s (s{+}2) (7 s(s{+}2){+}722){+}1227)  \nonumber \\
& + 4 (s (s{+}2) (3 s (s{+}2){+}359){+}612)\Big) /   \Big( 64 (d{-}4) (2 d{-}11) (2d{-}9) (2 d{-}7) \Big), \, \\
\bra{\phi_{3}}\ket{\phi_{2}}= & (d-5) (s+1) \Big(d^3 (s (s+2) (s (s+2)+210)+657)- \nonumber \\
& d^2 (s (s{+}2) (13 s
   (s{+}2){+}2874){+}8973){+}
    6 d (s (s{+}2) (9 s (s{+}2){+}2146){+}6705)- \nonumber \\
   &   24 (s (s{+}2) (3 s
   (s{+}2){+}788){+}2471) \Big) / \Big( 256 (d{-}6) (d{-}4) (2 d{-}11) (2 d{-}9) (2 d{-}7)  \Big)  \, .
\end{align}
Using these in eq.~(\ref{eq:threelooptriplecross:GD}) we obtain the following final reduction
\begin{equation}
I_{1,1,1,1,1,1,1,1;-2}= c_{1} \, J_{1} + c_{2} \, J_{2} \, ,
\end{equation}
where,
\begin{equation}
\begin{split}
& c_{1}=-\frac{(d-4) (s-1) (s+3)}{4 (2 d-7)} \, , \\
& c_{2}=\frac{(3 d-11) (s+1)}{2 (2 d-7)} \, ,
\end{split}
\end{equation}
in agreement with {\sc SYS}. 

\paragraph{dlog-basis.}
On the other hand, let us consider the dlog-basis. Here, we can choose the basis as  
\begin{equation}
\hat{\varphi}_{1}=\frac{1}{z}-\frac{2}{-s+2 z+1} \, , \quad
\hat{\varphi}_{2}=\frac{2}{-s+2 z+1}-\frac{2}{-s+2 z-3} \, .
\end{equation}
The decomposition formula reads:
\begin{equation}
\bra{\phi_{3}}=\sum_{i,j=1}^{2}=\bra{\phi_{3}}\ket{\varphi_{j}} \left(\mathbf{C}^{-1}\right)_{ji}\bra{\varphi_{i}}.
\end{equation}
Then, we compute the \textbf{C} matrix:
\begin{equation}
{\bf C}=
\left(
\begin{array}{cc}
\bra{\varphi_{1}}\ket{\varphi_{1}} &  \bra{\varphi_{1}}\ket{\varphi_{2}} \\
\bra{\varphi_{2}}\ket{\varphi_{1}} &  \bra{\varphi_{2}}\ket{\varphi_{2}} \\
\end{array}
\right)
=
\frac{1}{(d-5)}
\left(
\begin{array}{cc}
 3 & -2 \\
 -2 & 4 \\
\end{array}
\right).
\end{equation}
Moreover we can compute the intersection numbers:
\begin{align}
\bra{\phi_{3}}\ket{\varphi_{1}} & =\frac{(s-1) \left(d^2 \left(s^2-10 s-3\right)+d \left(-7 s^2+84 s+27\right)+2
   \left(6 s^2-87 s-29\right)\right)}{16 (d-4) (2 d-9) (2 d-7)} \, , \\
\bra{\phi_{3}}\ket{\varphi_{2}} & = \frac{d^2 \left(7 s^2+14 s+15\right)-8 d \left(7 s^2+14 s+15\right)+111 s^2+222
   s+239}{4 (d-4) (2 d-9) (2 d-7)}.
\end{align}
The final reduction reads:
\begin{equation}
\bra{\phi_{3}}=c_{1}\bra{\varphi_{1}} + c_{2} \bra{\varphi_{2}} \, ,    
\end{equation}
with:
\begin{align}
c_{1} &= \frac{(d-5) (s+1) \left(d^2 \left(s^2+2 s+33\right)-d \left(7 s^2+14
   s+267\right)+4 \left(3 s^2+6 s+134\right)\right)}{32 (d-4) (2 d-9) (2 d-7)} \, , \\
c_{2} &= \left((d-5) \left(d^2 \left(s^3+31 s^2+91 s+93\right)-d \left(7 s^3+245 s^2+729
   s+747\right) \right. \right. \nonumber \\
   & \left. \left. +4 \left(3 s^3+120 s^2+362 s+373\right)\right) \right)/ \left(64 (d-4) (2 d-9)
   (2 d-7) \right) \, ,
\end{align}
which is verified with {\sc SYS}.
\paragraph{Dimensional Recurrence relation.}
For this case, we show how to build $2^{\rm nd}$-order dimensional recurrence relations for the master integrals $J_1$ and $J_2$ . 
Following eq.~($\ref{eq:ric_rel_d+2n_MI}$), we have
\begin{equation}
\bra{B e_{1}} \equiv \bra{B \phi_{1}}=\sum_{j=1}^{2} \mathbf{R}_{1j} \bra{\phi_{j}}.
\end{equation}
with:
\begin{align}
\mathbf{R}_{11} & =\frac{(d-4) (d-3) \left(s^2+2 s-3\right) \left(s^2+2 s+9\right)}{32 (2 d-7) (2
   d-5)}, \\
\mathbf{R}_{12} & =-\frac{(d-3) (s+1) \left(d \left(s^2+2 s+33\right)-4 \left(s^2+2
   s+30\right)\right)}{16 (2 d-7) (2 d-5)}.
\end{align}
Then, restoring the \emph{proper d-dependent} $K$ factor, we have:
\begin{equation}
\begin{split}
J_{1}^{(d+2)}= & \left(\frac{(d-4) (s-1) (s+3) (s (s+2)+9)}{4 (d-2) (d-1) (2 d-7) (2 d-5) s} \right) J_{1}^{(d)} \\
+ & \left( -\frac{(s+1) (d (s (s+2)+33)-4 (s (s+2)+30))}{2 (d-2) (d-1) (2 d-7) (2 d-5) s} \right) J_{2}^{(d)}.
\end{split}
\end{equation}
The result is in agreement with \textsc{LiteRed}.\\
In a similar manner, we can consider:
\begin{equation}
\begin{split}
\bra{B \phi_{2}} = \sum_{j=1}^{2} \mathbf{R}_{2j} \bra{\phi_{j}},
\end{split}
\end{equation}
with:
\begin{align}
\mathbf{R}_{21}= & \frac{(d-4) (s+1) \left(s^2+2 s-3\right) \left(d^2 \left(s^2+2 s+33\right)-d
   \left(5 s^2+10 s+177\right)+6 \left(s^2+2 s+39\right)\right)}{128 (d-2) (2 d-7)
   (2 d-5)}, \\
\mathbf{R}_{22}= &
-\left( (d-3) \left(d^2 (s (s+2) (s (s+2)+90)+153)-6 d (s (s+2) (s (s+2)+90)+153) \right. \right. \nonumber \\
& \left. \left. +8(s (s+2) (s (s+2)+96)+165)\right) \right) / \left(64 (d-2) (2 d-7) (2 d-5) \right).
\end{align}
Then, reintroducing the \emph{proper d-dependent} $K$ factor, we have:
\begin{equation}
\begin{split}
J_{2}^{(d+2)}= & \left( \frac{(d-4) (s-1) (s+1) (s+3) (d s (s+2)+33 d-2 s (s+2)-78)}{16 (d-2)^2 (d-1) (2
   d-7) (2 d-5) s} \right) J_{1}^{(d)} \\
& - \left( \left( d^2 (s (s+2) (s (s+2)+90)+153)-6 d (s (s+2) (s (s+2)+90)+153) \right. \right. \nonumber \\
& \left. \left. +8 (s (s+2) (s(s+2)+96)+165) \right)/ \left(8 (d-2)^2 (d-1) (2 d-7) (2 d-5) s \right) \right) J_{2}^{(d)}.
\end{split}
\end{equation}
The result is in agreement with \textsc{LiteRed}.

\vspace*{1em}
\paragraph{}
In the following, we consider maximal cuts of Feynman integrals where 1-form representations are obtained within the \lbl Baikov approach.

\section{\label{sec:sunrise}Two-Loop Sunrise}

In this section we will discuss two instances of the two-loop sunrise integral. First the case where all the internal masses are zero, and then the case where all the internal masses are the same but non-zero.

\subsection{\label{sec:massless-sunrise}Massless Sunrise}
\begin{figure}[H]
    \centering
    \includegraphics[width=0.2\textwidth]{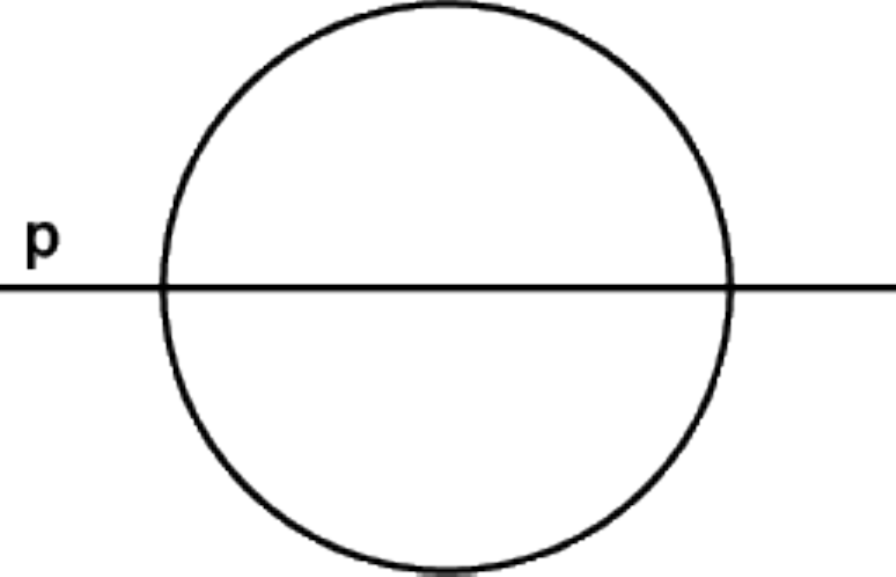}
    \caption{Massless sunrise.}
    \label{fig:Massles_Sunrise}
\end{figure}

Let us consider the massless sunrise diagram at two loops, Fig.~\ref{fig:Massles_Sunrise}. The incoming momentum is $p$, where $p^2=s$. The denominators are defined as 
\begin{align}\label{sunrise-propagators}
D_1=k_1^2,\qquad D_2=(k_1-k_2)^2,\qquad D_3=(k_2-p)^2.
\end{align}
Introducing the ISP as $z = D_4 = k_2^2$,
we obtain:
\begin{align}
u &=  z^{\frac{d}{2}-2} (z-s)^{d-3}, \hspace{1cm} 
\omega = \left(\frac{d-3}{z-s}+\frac{d-4}{2 z}\right) dz \, , \\
\masters &= 1, \hspace{2.3cm} \mathcal{P} = \lbrace 0,s, \infty \rbrace.
\end{align}
Now,
any integral can be expressed in terms of the chosen master integral either by using a monomial or a dlog-basis. Here, for illustration, we choose the monomial basis only.

\paragraph{Monomial Basis.}
We choose the basis as $\lbrace \langle \phi_i | \rbrace_{i=1}$ and the chosen master integral becomes
\begin{align}
    J_1= I_{1,1,1;0} = \langle \phi_1 | \cal{C} ] \ .
\end{align}
Let us consider the decomposition of $I_{1,1,1;-1} = \langle \phi_2 | \cal{C} ]$ in terms of $J_1$. The decomposition formula in this case reads
\begin{align}
\langle \phi_n | = \langle \phi_n |\phi_1 \rangle \,\textbf{C}_{11}^{-1}\, \langle \phi_1|.
\label{eq:2LSunrise:GD}
\end{align}
We build the metric matrix \textbf{C}, which has a single entry in this case.
\begin{align}
\textbf{C} =  \langle \phi_1 | \phi_1 \rangle  = \frac{4 (d-3) s^2}{3 (3 d-10) (3 d-8)}.
\end{align}
In this case the $\textbf{C}^{-1}$ is trivial to compute as the \textbf{C} contains only one element.
The other necessary intersection number is the following.
\begin{align}
\langle \phi_2 | \phi_1 \rangle &= \frac{4 (d-3) s^3}{9 (3 d-10) (3 d-8)} \ .
\end{align}
Using these in eq.~(\ref{eq:2LSunrise:GD}) we obtain the following reduction formula on the maximal cut
\begin{align}\label{massless-sunrise-result}
I_{1,1,1;-1} = \frac{s}{3} I_{1,1,1;0},
\end{align}
which agrees with the \textsc{LiteRed}. \\

\subsection{\label{sec:massive-sunrise}Massive Sunrise}
\begin{figure}[H]
    \centering
    \includegraphics[width=0.2\textwidth]{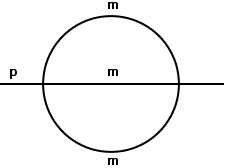}
    \caption{Massive sunrise.}
    \label{fig:Massive_Sunrise}
\end{figure}
We consider here the massive sunrise, Fig.~\ref{fig:Massive_Sunrise}~\cite{Broadhurst:1993mw, Laporta:2004rb}. The incoming momentum is denoted by $p$, with $p^2=s$ and $m^2 = 1$. 
The denominators are:
\begin{equation}
D_{1}=k_1^2-1 \ , \quad 
D_{2}=(k_1-k_2)^2-1\ , \quad 
D_{3}=(k_2-p)^2-1 \ ,
\end{equation}
while the ISP is chosen as: $z=D_{4}=k_2^2$.\\
Therefore, we obtain:
\begin{align}
u= & 
z^{-{1 \over 2}} \, 
(z-4)^{\frac{d-3}{2}} \left(z-s-2 \sqrt{s}-1\right)^{\frac{d-3}{2}} \left(z-s+2 \sqrt{s}-1\right)^{\frac{d-3}{2}}
\ , \\
\omega= & \frac{1}{2} \left(\frac{d-3}{z-s-2 \sqrt{s}-1}+\frac{d-3}{z-s+2 \sqrt{s}-1}+\frac{d-3}{z-4}-\frac{1}{z}\right)
dz \ , \\
& \nonumber \\
& {\rm with \ } \masters=3, \qquad \mathcal{P}=\{0,\, 4,\, s-2 \sqrt{s}+1,\, s+2 \sqrt{s}+1,\, \infty    \}.
\end{align}
\paragraph{dlog Basis.} Let us consider the decomposition of $I_{1,1,1;0}=\langle{\phi_{1}}| \mathcal{C}]$ in terms of $J_{1}=\langle{\varphi_{1}}|\mathcal{C}]$, $J_{2}=\langle{\varphi_{2}}|\mathcal{C}]$ and $J_{3}=\langle{\varphi_{3}}|\mathcal{C}]$, with:
\begin{align}
\varphi_{1}=& \left( \frac{1}{z-s+2 \sqrt{s}-1} \right) dz, \\
\varphi_{2}=& \left( \frac{1}{z-s-2 \sqrt{s}-1} \right) dz, \\
\varphi_{3}=& \left(\frac{1}{z-4} \right) dz.
\end{align}
We compute the $\mathbf{C}$ matrix:
\begin{equation}\label{MasssunC}
\mathbf{C}=
\left(
\begin{array}{ccc}
\bra{\varphi_1}\ket{\varphi_1} & \bra{\varphi_1}\ket{\varphi_2} & \bra{\varphi_1}\ket{\varphi_3} \\
\bra{\varphi_2}\ket{\varphi_1} & \bra{\varphi_2}\ket{\varphi_2} & \bra{\varphi_2}\ket{\varphi_3} \\
\bra{\varphi_3}\ket{\varphi_1} & \bra{\varphi_3}\ket{\varphi_2} & \bra{\varphi_3}\ket{\varphi_3}
\end{array}
\right)
=
-\frac{2}{3 d-10}
\left(
\begin{array}{ccc}
 -\frac{2 d-7}{d-3} & 1 & 1 \\
 1 & -\frac{2 d-7}{d-3} & 1 \\
 1 & 1 & -\frac{2 d-7}{d-3} \\
\end{array}
\right)
\end{equation}
and the intersection numbers:
\begin{align}
\bra{\phi_{1}}\ket{\varphi_{1}}= & \frac{-2 (d-4) s+4 (3 d-10) \sqrt{s}+6 d-16}{9 (d-6) d+80},\\
\bra{\phi_{1}}\ket{\varphi_{2}}= & \frac{8 \left(s+5 \sqrt{s}-2\right)-2 d \left(s+6 \sqrt{s}-3\right)}{9 (d-6) d+80}, \\
\bra{\phi_{1}}\ket{\varphi_{3}}= & \frac{4 (d (s-3)-3 s+11)}{9 (d-6) d+80}.
\end{align}
Finally, eq.~(\ref{eq:masterdeco:}) gives:
\begin{equation}
I_{1,1,1;0}=c_1 \, J_1 + c_2 \, J_2 + c_3 \, J_3,
\end{equation}
with:
\begin{align}
c_1= & -\frac{(d-3) \left(s-2 \sqrt{s}+1\right)}{3 d-8}, \\
c_2= & -\frac{(d-3) \left(s+2 \sqrt{s}+1\right)}{3 d-8}, \\
c_3= & -\frac{4 (d-3)}{3 d-8}.
\end{align}
\paragraph{Symmetry relation.} Public codes show that the number of MIs, without the contribution of symmetry relations, is \emph{four}, while taking into account such relations, the number of MIs is reduced to \emph{two}.\\
The intersection method, within the {\rm LBL} approach, provides \emph{three} MIs, and we show how, thanks to the contribution of symmetries, we can obtain the minimal number of MIs, namely \emph{two}.\\
It is sufficient to consider two, a priori, different integrals, which are known to be equal thanks to symmetry relations, and decompose them in the original basis. Doing so, we obtain an extra-relation among the original MIs, and therefore one of them can be expressed in terms of the others.\\  
In the case at hand, we can consider the symmetry relations 
between integrals with one denominator raised to a squared power:
\begin{equation}
I_{2,1,1;0}=I_{1,2,1;0}=I_{1,1,2;0} \ .
\label{eq:massive_sunrise_symmetry_relation_dots_all}
\end{equation}
On the maximal-cut, within the \lbl approach
$I_{2,1,1;0}$ and $I_{1,2,1;0}$ admit the same (univariate) integrand representation:
\begin{align}
& I_{2,1,1;0}=\langle{\widetilde{\phi}_{1}}|\mathcal{C}]=I_{1,2,1;0}=\langle{\widetilde{\phi}_{2}}|\mathcal{C}]\\
& \qquad \widetilde{\phi}_{1}=\widetilde{\phi}_{2}=\left( \frac{3-d}{z-4} \right) dz \ .
\end{align}
By means of the master decomposition formula eq.~(\ref{eq:masterdeco:})  $\langle{\widetilde{\phi}_{1,2}}|$ are decomposed in the dlog basis 
$\{ \langle \varphi_i| \}_{i=1,2,3} $. 
In particular one finds that
\begin{eqnarray}
\langle \widetilde{\phi}_{1} |
= 
\langle \widetilde{\phi}_{2} |
=
(3-d) \langle \varphi_{3} | \ ,
\end{eqnarray}
implying:
\begin{equation}
I_{2,1,1;0}=I_{1,2,1;0}=(3-d) \,  J_{3} \ .
\label{eq:massive_sunrise_dots_1_2}
\end{equation}
On the other hand, we can consider the decomposition by means of intersection numbers of $I_{1,1,2;0}=\langle{\widetilde{\phi}_{3}}|\mathcal{C}]$, with:
\begin{equation}
\widetilde{\phi}_{3}=\left(\frac{(d-3) (s+z-1)}{s^2-2 s (z+1)+(z-1)^2} \right) dz \ ,
\end{equation}
in terms of: $\{ J_{i} \}_{i=1,2,3}$.\\
In this case, we obtain:
\begin{equation}
I_{1,1,2;0}=\frac{1}{2} (d-3) \left[ \left(\sqrt{s}-1\right) \,  J_{1}- \left(\sqrt{s}+1\right) \, J_{2} \right] \ .
\label{eq:massive_sunrise_dots_3}
\end{equation}
Finally, by equating  eqs.~(\ref{eq:massive_sunrise_symmetry_relation_dots_all}, \ref{eq:massive_sunrise_dots_1_2}, \ref{eq:massive_sunrise_dots_3}), $J_3$ can be expressed in terms of $J_1$ and $J_2$:
\begin{equation}\label{eq:massive_sunrise_symmetry_relation}
J_{3}=-\frac{1}{2}\left[ \left(\sqrt{s}-1\right) \,  J_{1}- \left(\sqrt{s}+1\right) \, J_{2} \right] \ ,
\end{equation}
hence bringing down to {\it two} the number of independent MIs.
Needless to say, the relations found by means of intersection numbers are in agreement with the IBP identities obtained through public codes.

\paragraph{Differential Equation.} We now derive the differential equation for the d-log basis with respect to $x=\sqrt{s}$.\\
We therefore obtain
\begin{equation}
    \sigma (x) = \frac{2 (d-3) x \left(x^2-z-1\right)}{x^4-2
   x^2 (z+1)+(z-1)^2}
\end{equation}
The derivative of the dlog-basis elements $\langle \Phi_i(x) | \equiv \langle (\partial_x + \sigma(x))\varphi_i |$ is given by
\begin{eqnarray}
\langle \Phi_1(x) | &=&
\frac{2 z ((d-4) x+1)-2 \left(x^2-1\right) ((d-4) \
x-1)}{\left((x-1)^2-z\right)^2 \left((x+1)^2-z\right)} dz,
   \\
\langle \Phi_2(x) | &=&
 \frac{2 z ((d-4) x-1)-2 \left(x^2-1\right) ((d-4)x+1)}{\left((x-1)^2-z\right) \left((x+1)^2-z\right)^2} dz,
   \\
\langle \Phi_3(x) | &=&
   \frac{2 (d-3) x \left(x^2-z-1\right)}{(z-4) \left(x^4-2 x^2 \
(z+1)+(z-1)^2\right)}] dz.
\end{eqnarray}

The intersection matrix ${\bf F}$ with entries ${\bf F}_{ij} = \langle \Phi_i | \varphi_j \rangle$ reads,
\begin{eqnarray}
{\bf F} &=&\left(
\begin{array}{ccc}
 -\frac{2}{(d-3)
   (x-1)}+\frac{1}{x}+\frac{1}{x+1}+\frac{1}{x
   -3} & -\frac{1}{x} & -\frac{2 (x-1)}{(x-3)
   (x+1)} \\
 -\frac{1}{x} & \frac{1}{x}-\frac{2}{(d-3)
   (x+1)}+\frac{1}{x+3}+\frac{1}{x-1} &
   -\frac{2 (x+1)}{x^2+2 x-3} \\
 -\frac{2 (x-1)}{(x-3) (x+1)} & -\frac{2
   (x+1)}{x^2+2 x-3} & \frac{4 x
   (x^2-5)}{x^4-10 x^2+9} \\
\end{array}
\right)
\end{eqnarray}
and using the inverse of ${\bf C}$ computed in eq.~(\ref{MasssunC})
we finally obtain
\begin{eqnarray*}
{\bf \Omega} = {\bf F}{\bf C}^{-1} &=& \left(
\begin{array}{ccc}
 \frac{1}{2} \left(\frac{2
   (d-4)}{x-1}+\frac{d-3}{x-3}+\frac{d-3}{x}+\frac{d-3}{x+1}\right) & \frac{(d-3)
   (x+1)}{2 (x-1) x} & -\frac{4 (d-3)}{(x-3)
   (x^2-1)} \\
 \frac{(d-3) (x-1)}{2 x (x+1)} & \frac{1}{2}
   \left(\frac{2
   (d-4)}{x+1}+\frac{d-3}{x-1}+\frac{d-3}{x}+\frac{d-3}{x+3}\right) & -\frac{4
   (d-3)}{(x+3) (x^2-1)} \\
 \frac{(d-3) (x-1)}{-x^2+2 x+3} & -\frac{(d-3)
   (x+1)}{x^2+2 x-3} & \frac{2 (d-3) x
   (x^2-5)}{x^4-10 x^2+9} \\
\end{array}
\right).
\end{eqnarray*}
Using now the symmetry relation in eq.~(\ref{eq:massive_sunrise_symmetry_relation}) the independent functions in the system of differential equation becomes {\it two}, each obeying a $2^{nd}$ order differential equation. In particular, $\varphi_1$ is found to obey, 
\begin{equation}
    P_2 (x) \, \varphi_1 ''(x) + P_1 (x) \, \varphi_1 '(x) + P_0 (x) \, \varphi_1(x) = 0 \ ,
\end{equation}
with
\begin{align}
    P_2 (x) = & x \left(-x^6+6 x^5+13 x^4-60 x^3-39 x^2+54 x+27\right) \, , \\
    P_1 (x) = & d \left(5 x^6-30 x^5-45 x^4+180 x^3+99 x^2-54 x-27\right) + \\
    & - 2 \left(9 x^6-55 x^5-96
   x^4+342 x^3+237 x^2-63 x-54\right) \nonumber \, , \\
    P_0 (x) = & (10-3 d) \left(x \left(2 d ((x-6) x-3) \left(x^2-3\right)+x (x ((45-7 x)
   x+66)-138)-99\right)-27\right) \, .
\end{align}
At $d=2$ the solutions are
\begin{align}
    \varphi_1^{(1)} & = \frac{(x-1) (x+3) 
    E\left( \chi(x) \right)
    -(x-3) (x+1) K\left(\chi(x)\right)}{
    2 (x-3) x (x+1) (x-1) \sqrt{(x-1) (x+3)}} \ , \\
   \varphi_1^{(2)} & = 
   \frac{4 \, x \, 
   K\left(1-\chi(x)\right)
   +(x-1)^2 E\left(1-\chi(x)\right)}{
   2 x (x+1)^2 (x-1) \sqrt{(x-1) (x+3)}} \ ,
\end{align}
with
\begin{equation}
\chi(x) = \frac{16 x}{(x-1)^3(x+3)} \ ,
\end{equation}
We notice the presence of the complete elliptic integrals of the first and second kind $K$ and $E$, consistently with the results in the literature.

\section{Two-Loop Non-Planar Triangle}
\label{sec:nonplanartriangle}

\begin{figure}[H]
    \centering
    \includegraphics[width=0.38\textwidth]{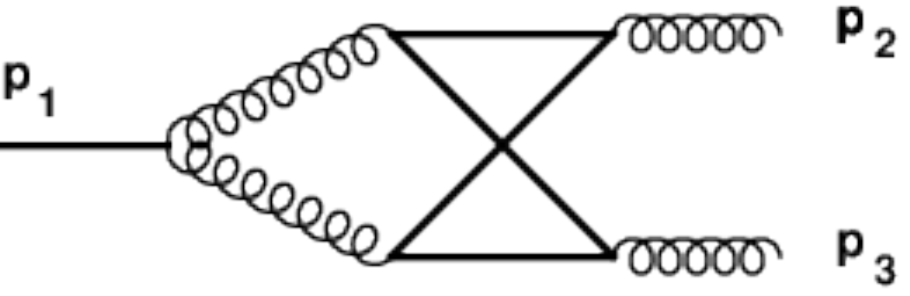}
    \caption{Non-planar triangle.}
    \label{fig:Non_planar_triangle}
\end{figure}

 In this section we discuss the two-loop non-planar triangle~\cite{vonManteuffel:2017hms} in Fig.~\ref{fig:Non_planar_triangle}, also presented in \cite{Mastrolia:2018uzb}, with 

\begin{eqnarray}
		&&D_1 = k_1^2 \ , \ 
		D_2 = k_2^2 - m^2 \ , \ 
		D_3 = (p_1 - k_1)^2 \ , \ 
		D_4 = (p_3 - k_1 + k_2)^2 - m^2,\\
		&&D_5 = (k_1 - k_2)^2 - m^2 \ , \ 
		D_6 = (p_2 - k_2)^2 - m^2 \ . 
		\nonumber 
\end{eqnarray}
We choose the ISP $z = D_{7} = 2(p_2 + k_1)^2 - p_1^2$ \, . \\
Consider a generic case where $B$ is a factorized quartic
polynomial (paradigmatic of elliptic cases), of the type, 

\begin{align}
u = B^\gamma \ , \qquad 
B = \left(z^2-\tau _1^2\right) \left(z^2-\tau _2^2\right) \, ,
\end{align}
where
\begin{align}
\tau_1=s\sqrt{1+(4m)^2/s}\,, \qquad \tau_2 = s \,, \qquad \gamma = \frac{d-5}{2} \, .
\end{align}
From this we may compute
\begin{align}
\omega = \frac{2 \gamma  z \left(2 z^2 - \tau_1^2 - \tau_2^2\right)}{\left(z^2-\tau _1^2\right)
   \left(z^2-\tau _2^2\right)} {dz} \,, \qquad \masters=3 \,, \qquad
{\cal P} = \left\{ -\tau _1, -\tau _2, \tau _2, \tau _1, \infty \right\} \, .
\end{align}

\paragraph{dLog-Basis.}
Let us consider the following dlog-basis, 
\begin{eqnarray}
\varphi_1 &=& \bigg(\frac{1}{\tau _1+z}-\frac{1}{\tau _2+z}\bigg)dz \, , \\
\varphi_2 &=& \bigg(\frac{1}{\tau _2+z}-\frac{1}{z-\tau _2}\bigg)dz \, , \\
\varphi_3 &=& \bigg(\frac{1}{z-\tau _2}-\frac{1}{z-\tau _1}\bigg)dz \, , 
\end{eqnarray}
which gives
\begin{eqnarray}
{\bf C} &=&
\left(
\begin{array}{ccc}
\langle \varphi_1 | \varphi_1 \rangle & 
\langle \varphi_1 | \varphi_2 \rangle & 
\langle \varphi_1 | \varphi_3 \rangle \\
\langle \varphi_2 | \varphi_1 \rangle & 
\langle \varphi_2 | \varphi_2 \rangle & 
\langle \varphi_2 | \varphi_3 \rangle \\
\langle \varphi_3 | \varphi_1 \rangle & 
\langle \varphi_3 | \varphi_2 \rangle & 
\langle \varphi_3 | \varphi_3 \rangle 
\end{array}
\right) 
=
\frac{1}{\gamma}\left(
\begin{array}{ccc}
 2 & -1 & 0 \\
 -1 & 2 & -1 \\
 0 & -1 & 2 \\
\end{array}
\right)
\end{eqnarray}
with inverse matrix,
\begin{eqnarray}
{\bf C}^{-1} &=&
\gamma
\left(
\begin{array}{ccc}
 \frac{3  }{4} & \frac{1 }{2} & \frac{1 }{4} \\
 \frac{1 }{2} & 1  & \frac{1 }{2} \\
 \frac{1 }{4} & \frac{1 }{2} & \frac{3  }{4} \\
\end{array}
\right).
\label{eq:invC:dlog}
\end{eqnarray}
For instance, the projection of $\phi_1 = dz $ is
\begin{eqnarray}
\langle \phi_1 | &=& 
\frac{\gamma  \tau _1}{4 \gamma +1} \langle \varphi_1|
+\frac{\gamma  \left(\tau _1+\tau
   _2\right)}{4 \gamma +1} \langle \varphi_2|
+   \frac{\gamma  \tau _1}{4 \gamma +1} \langle \varphi_3|
\end{eqnarray}
which can be verified with {\sc Reduze}.

\subsection{Denominator Powers Bigger Than One}

Following eq.~(\ref{eq:cut:genpow}), we consider 
the maximal cut $z_1= \ldots = z_6 = 0$
($z_7 = z$) of,
\begin{eqnarray}
I_{1,1,1,2,1,1;0}\bigg|_{z_1=\ldots=z_6=0} 
&=& 
K
\int d z \,
u \, 
{\hat \varphi} \ ,
\end{eqnarray}
with 
\begin{eqnarray}
{\hat \varphi} &=& -\frac{4 \gamma  \left(\tau _2+z\right)}{\left(z-\tau _1\right) \left(\tau
   _1+z\right)} \ , 
\end{eqnarray}
where the expression for $K$ is not needed.
Its decomposition in terms of the dlog-basis reads,
\begin{eqnarray}
\langle \varphi | &=&
-\frac{\gamma  \left(\tau _1-2 \tau _2\right)}{\tau _1} 
\langle \varphi_1 |
+\frac{2 \gamma  \tau
   _2}{\tau _1} \langle \varphi_2 |
   +
   \frac{\gamma  \left(\tau _1+2 \tau _2\right)}{\tau _1}
   \langle \varphi_3 |
\end{eqnarray}
which can be verified with {\sc Reduze}.

\subsection{System of Differential Equations for the dLog-Basis}

In the case of the the Feynman integral considered here, we define the variable 

\begin{eqnarray}
x \equiv \frac{\tau_1}{\tau_2} \ , 
\qquad \Leftrightarrow \qquad
\tau_1 = x \, \tau_2
\end{eqnarray}

To build the system of differential equations, consider
\begin{eqnarray}
u(z) &=& B(z)^\gamma, \\
B(z,x) &=& B(z) \bigg|_{\tau_1 = x \, \tau_2} 
\\
\hat{\omega}(x) &=& \hat{\omega} \bigg|_{\tau_1 = x \, \tau_2} = 
\partial_z \log\Big( B(z,x)^\gamma \Big) 
\\
\sigma(x) &=& 
\partial_x \log\Big( B(z,x)^\gamma \Big) = 
-\frac{2 \gamma  \tau _2^2 x}{z^2-\tau _2^2 x^2}.
\end{eqnarray}
The derivative of the dlog-basis elements $\langle \Phi_i(x) | \equiv \langle (\partial_x + \sigma(x))\varphi_i |$ is given by
\begin{eqnarray}
\langle \Phi_1(x) | &=&
-\frac{\tau _2 \left(2 \gamma  \tau _2^2 x^2-2 \gamma  \tau _2^2
   x+\tau _2^2 x+\tau _2 x z-z^2-\tau _2 z\right)}{\left(\tau _2+z\right)
   \left(\tau _2 x-z\right) \left(\tau _2 x+z\right){}^2} dz,
   \\
\langle \Phi_2(x) | &=&
   \frac{4 \gamma 
   \tau _2^3 x}{\left(\tau _2-z\right) \left(\tau _2+z\right) \left(\tau
   _2 x-z\right) \left(\tau _2 x+z\right)} dz,
   \\
\langle \Phi_3(x) | &=&
   -\frac{\tau _2 \left(2 \gamma 
   \tau _2^2 x^2-2 \gamma  \tau _2^2 x+\tau _2^2 x-\tau _2 x z-z^2+\tau _2
   z\right)}{\left(\tau _2-z\right) \left(\tau _2 x-z\right){}^2
   \left(\tau _2 x+z\right)} dz.
\end{eqnarray}

The intersection matrix ${\bf F}$ with entries ${\bf F}_{ij} = \langle \Phi_i | \varphi_j \rangle$ reads,
\begin{eqnarray}
{\bf F} &=&\left(
\begin{array}{ccc}
 \frac{7 x^2+2 x-1}{(x-1) x (x+1)} & -\frac{2}{x-1} & -\frac{x-1}{x (x+1)}
   \\
 -\frac{2}{x-1} & \frac{4 x}{(x-1) (x+1)} & -\frac{2}{x-1} \\
 -\frac{x-1}{x (x+1)} & -\frac{2}{x-1} & \frac{7 x^2+2 x-1}{(x-1) x (x+1)}
   \\
\end{array}
\right)
\end{eqnarray}
and using ${\bf C}^{-1}$ computed in eq.~(\ref{eq:invC:dlog})
we finally obtain
\begin{eqnarray}
{\bf \Omega} = {\bf F}{\bf C}^{-1} &=& \gamma \, 
\left(
\begin{array}{ccc}
 \frac{4 x^2+x-1}{(x-1) x (x+1)} & \frac{1}{x} & \frac{1}{x (x+1)} \\
 -\frac{2}{(x-1) (x+1)} & \frac{2}{x+1} & -\frac{2}{(x-1) (x+1)} \\
 \frac{1}{x (x+1)} & \frac{1}{x} & \frac{4 x^2+x-1}{(x-1) x (x+1)} \\
\end{array}
\right).\end{eqnarray}

We observe that since ${\bf C}^{-1}$ is a constant matrix with $\gamma$ factored out,
the system of differential equations for the dlog-basis is canonical, being 
$\gamma$ factorized as well as Fuchsian.

\section{Two-Loop Planar Triangle}
\label{sec:planar-triangle}

\begin{figure}[H]
    \centering
    \includegraphics[width=0.38\textwidth]{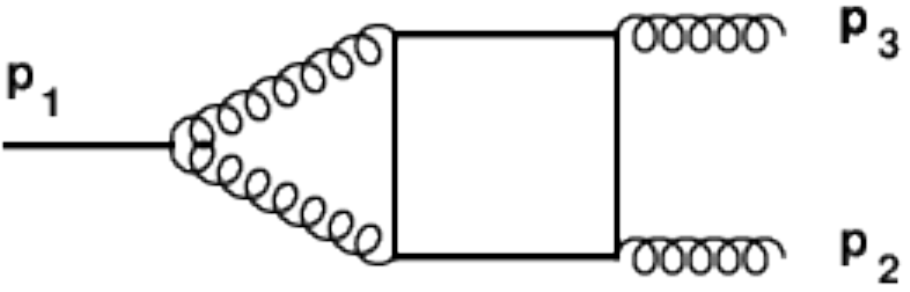}
    \caption{Planar 2-loop triangle.}
    \label{fig:planar-triangle}
\end{figure}

Let us consider the 2-loop planar triangle in Fig.~\ref{fig:planar-triangle}. The independent (incoming) external momenta are chosen to be $p_{1}$ and $p_{2}$, with $p_{1}^{2}=s$, $p_{2}^{2}=0$ and $(p_1+p_2)^2=0$.\\
The denominators are given by:
\begin{align}
& D_{1}=k_1^2 \, , \quad D_{2}=\left(k_1 + p_1\right){}^2 \, , \quad D_{3}=\left(k_2 + p_1\right){}^2 - m{}^2 \, , \nonumber\\ 
& D_{4}=\left(k_2 + p_1 + p_2\right){}^2 - m{}^2 \, , \quad
D_{5}=k_2{}^2 - m{}^2 \, , \quad 
D_{6}=\left(k_1 - k_2\right){}^2 - m{}^2 \, .
\end{align}
We choose the ISP $z=D_{7}=\left(k_1 + p_2 \right){}^2$ \, .
\subsection{Standard Baikov Representation}
The standard Baikov representation, on the maximal cut, gives us:
\begin{equation}
\begin{split}
u & =\left((s-z) \left(4 m^2 z+s^2-s z\right)\right)^{\frac{d-5}{2}}, \\
\omega & =  \left( \frac{(d-5) \left(2 m^2 (s-2 z)+s (z-s)\right)}{(s-z) \left(4 m^2 z+s^2-s z\right)} \right) dz.
\end{split}
\end{equation}
Thus, we infer:
\begin{equation}
\masters=1, \quad \mathcal{P}=\{ s, \frac{s^2}{s-4 m^2}, \infty \}.
\end{equation}
\paragraph{Denominator Power Bigger Than One.}
Let us consider the decomposition of $I_{2,1,1,1,1,1;0}= \langle{\varphi}|\mathcal{C}]$, with:
\begin{equation}
\hat{\varphi}=\frac{(d-5) \left(-4 m^2 s+6 m^2 z+s^2-s z\right)}{(s-z) \left(4 m^2 z+s^2-s z\right)},
\end{equation}
in terms of $J_{1}=I_{1,1,1,1,1,1;0}=\langle{\phi_{1}}|\mathcal{C}]$.\\
We build the $\mathbf{C}$ matrix, containing just a single element:
\begin{equation}
\mathbf{C}_{11}=\bra{\phi_{1}}\ket{\phi_{1}}=\frac{4 (d-5) m^4 s^2}{(d-6) (d-4) \left(s-4 m^2\right)^2},
\end{equation}
and the intersection number:
\begin{equation}
\bra{\varphi}\ket{\phi_{1}}=-\frac{4 (d-5) m^4 s}{(d-6) \left(s-4 m^2\right)^2}.
\end{equation}
Finally, eq.~($\ref{eq:masterdeco:}$) leads to:
\begin{equation}
I_{2,1,1,1,1,1;0}=\frac{4-d}{s} J_{1}.
\end{equation}
The result is in agreement with \textsc{Reduze}.

\subsection{Loop-by-Loop Approach}
Let us consider again the decomposition of $I_{2,1,1,1,1,1;0}=\langle{\varphi}|\mathcal{C}]$ in terms of the integral $I_{1,1,1,1,1,1;0}=\langle{\phi_{1}}|\mathcal{C}]$. \\
In the loop by loop approach the Baikov Polynomial of such integrals do not depend on the ISP. Therefore the maximal cut can be computed by means of residues:
\begin{equation}
I_{a_1, \ldots, a_6;0}= \prod_{i=1}^{6} \frac{1}{(a_{i}-1)!} \left( \frac{\partial^{a_i-1}}{\partial z_{i}^{a_i-1}}u \right)  \Bigg|_{z_{1}= \ldots = z_{6}=0.}.  
\end{equation}
Thus, we have:
\begin{align}
      I_{2,1,1,1,1,1;0}= (4-d)2^{\frac{3 d}{2}-5}  \left(-m^2\right)^{d-4} s^{\frac{3 d - 15}{2}} \left(4
   m^2-s\right)^{\frac{3 - d}{2}} \, , \\
      I_{1,1,1,1,1,1;0}= 2^{\frac{3 d}{2}-5} \left(-m^2\right)^{d-4} s^{\frac{3 d - 13}{2}} \left(4
   m^2-s\right)^{\frac{3-d}{2}} \, .
\end{align}
The ratio of their maximal cut directly gives the coefficient of the IBP decomposition, resulting in
\begin{equation}
     I_{2,1,1,1,1,1;0}=c_1 \, I_{1,1,1,1,1,1;0}
\end{equation}
with
\begin{equation}
    c_1= \frac{\langle{\varphi}|\mathcal{C}]}{\langle{\phi_{1}}|\mathcal{C}]}=\frac{4-d}{s}
\end{equation}
in agreement with the result given by intersection theory and by \textsc{Reduze}.

\section{Planar Double Triangle}
\label{sec:planardoubletriangle}

\begin{figure}[H]
    \centering
    \includegraphics[width=0.38\textwidth]{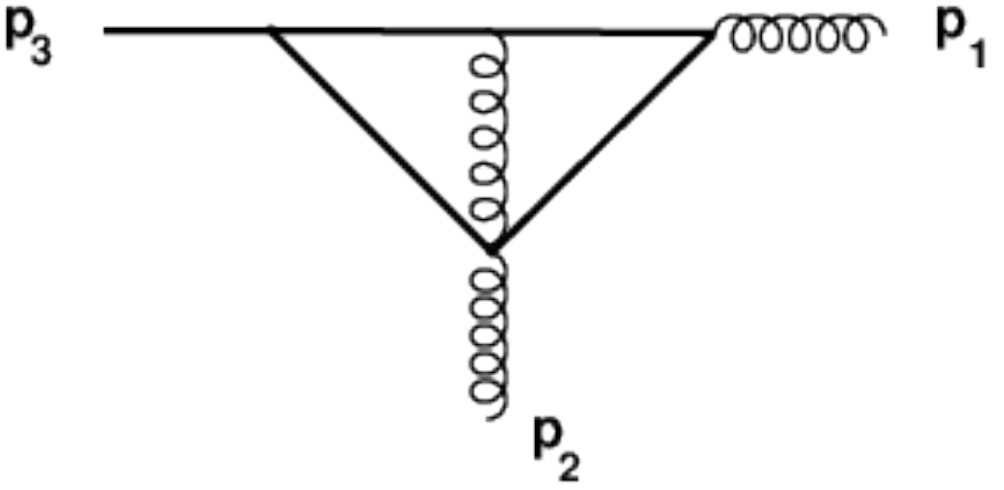}
    \caption{The planar double triangle.}
    \label{fig:Yang_planar_triangle}
\end{figure}

We consider the planar two-Loop triangle in Fig.~\ref{fig:Yang_planar_triangle}, where the independent (incoming) momenta are chosen to be $p_1$ and $p_2$ with $p_1^2=p_2^2=0$. The kinematic invariant is $s=(p_1+p_2)^2$ and the denominators are defined as 
\begin{align}
D_1&=k_1^2 - m^2,\qquad D_2=k_2 - m^2,\qquad D_3=(k_1-k_2)^2, \nonumber\\ 
D_4&=(k_2-p_1)^2 - m^2,\qquad D_5 = (k_1-p_1-p_2)^2-m^2.
\end{align}
We choose the ISP $z=D_6=(k_1 - p_1)^2 - m^2$.

Using the \lbl form of Baikov representation and performing the maximal cut as defined in eq.~(\ref{eq:def:cut-integral}) we find
\begin{gather}
u = \frac{(z (s+z) + s m^2)^{\frac{d-4}{2}}}{z}, \qquad
\omega = \frac{z ((d-6) s + 2 (d-5) z) - 2 m^2 s }{2 z \, (z (s + z) + s m^2)} \, dz
\end{gather}
with the $\omega$ corresponding to $\masters = 2$ in agreement with the literature \cite{Aglietti:2006tp,Anastasiou:2006hc}.

Let us notice that $u$ has a factor of $z$ risen to an {\it integer} power, violating one of the assumption for the applicability of intersection theory \cite{aomoto2011theory}. We solve this issue, by introducing a regulating exponent $\rho$, $z^{-1} \rightarrow z^{\rho-1}$, which we put to zero at the end of the calculation. Additionally, we factorize the polynomial appearing in $u$, so that,
\begin{gather}
u = ((z-r_1)(z-r_2))^{\frac{d-4}{2}} z^{\rho - 1}\,, \\
\omega = \frac{2 r_1 r_2 (\rho-1) - (r_1 + r_2) (d-6 + 2 \rho) z + 2 (d-5 + \rho) z^2}{2 z (z - r_1) (z - r_2)} \, dz \,, \\
\masters = 2\,, \qquad \mathcal{P} = \{ 0,\, r_1,\, r_2,\, \infty \}\,.
\end{gather}
with
\begin{eqnarray}
r_1=\frac{1}{2} \left(-\sqrt{s^2-4 m^2 s}-s\right) , \quad r_2=\frac{1}{2} \left(\sqrt{s^2-4 m^2 s}-s\right).
\end{eqnarray}
We observe that introducing the regulator changed neither the number $\masters$ of master integrals, nor introduce any spurious singularity.

\paragraph{Mixed Bases.}

We choose a monomial basis of master integrals
\begin{align}
J_1 = I_{1,1,1,1,1;0} \,, \qquad J_2 = I_{1,1,1,1,1;-1}\,,    
\end{align}
corresponding to $\phi_1 = 1 \, dz$ and $\phi_2 = z \, dz$. Additionally we pick the right basis of
\begin{align}
\hat{\varphi}_1 = \frac{1}{z} - \frac{1}{z-r_1} \,, \qquad \hat{\varphi}_2 = \frac{1}{z-r_1} - \frac{1}{z-r_2} \,.
\end{align}

This gives the $\textbf{C}$-matrix to be
\begin{align}
\textbf{C} = \langle \phi_i | \varphi_j \rangle = 
\left(
\begin{array}{cc}
\frac{r_1}{d-4}  & \frac{r_2-r_1}{d-4} \\[1mm]
 \frac{r_1 (r_1-r_2)}{2(d-3)} & \frac{(r_2-r_1)(r_1+r_2)}{2(d-3)} \\
\end{array}
\right) \,,
\end{align}
where we have inserted $\rho \rightarrow 0$ as in the following.

Let us perform the reduction of $I_{1,1,1,1,1;-2}$ in the basis of $J_1$ and $J_2$. The twisted cocycle corresponding to $I_{1,1,1,1,1;-2}$ is $\phi_3 = z^2 dz$. Therefore, to obtain the decomposition, we need also
\begin{align}
\langle \phi_3 | \varphi_1 \rangle =  \frac{r_1 (r_1 - r_2) (r_1 + r_2)}{4 (d-3)} \,, \qquad
\langle \phi_3 | \varphi_2 \rangle =  \frac{(r_2 - r_1) (r_1 + r_2)^2}{4 (d-3)}\,.
\end{align}
Using eq.~(\ref{eq:masterdeco:}), we obtain the following reduction formula on the maximal cut
\begin{align}
I_{1,1,1,1,1;-2} = - \frac{s}{2} \, I_{1,1,1,1,1;-1},
\end{align}
which agrees with the reduction obtained from {\sc LiteRed}.

\paragraph{Further Considerations.}

In this case, we have observed that integrals with positive powers of $z$ (in the sense that $z^n$ appears in the numerator) always have zero coefficient of $J_0$ (and non-zero of $J_1$).

\begin{figure}[H]
    \centering
    \includegraphics[width=0.38\textwidth]{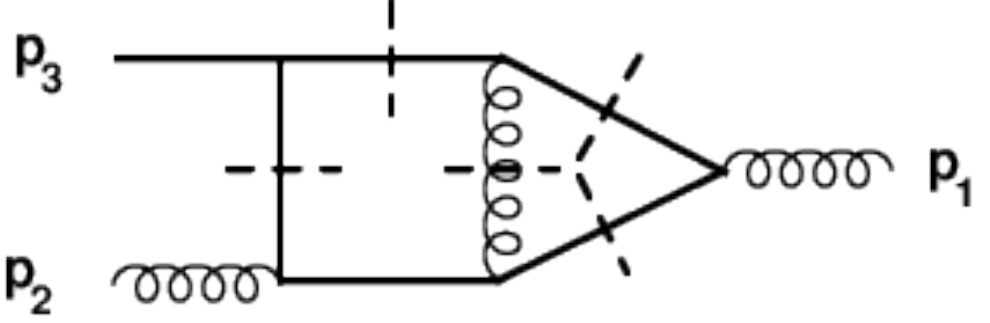}
    \caption{A six-denominator two loop sector.}
    \label{fig:sixdeno:triangle:Yang}
\end{figure}

Integrals with negative powers of $z$ (meaning that $z$ appears in the denominator as an actual propagator) correspond to a penta-cut, of the six denominator triangle graph shown in Fig.~\ref{fig:sixdeno:triangle:Yang}.
This six-propagator sector has no master integrals, but for the integrals associated with the indicated penta-cut, the coefficients of both $J_1$ and $J_2$ are different from zero. As an example, let us consider,
$I_{1,1,1,1,1;1}$, for which the relevant cocycle is $\phi_0 = \tfrac{1}{z} dz$. By computing
\begin{align}
\langle \phi_0 | \varphi_1 \rangle = -1 \,, \qquad
\langle \phi_0 | \varphi_2 \rangle = 0 \,,
\end{align}
the complete reduction eq.~(\ref{eq:masterdeco:}) reads,
\begin{align}
I_{1,1,1,1,1;1} = \frac{d-4}{2 m^2} I_{1,1,1,1,1;0} + \frac{d-3}{s m^2} I_{1,1,1,1,1;-1},
\end{align}
in agreement with {\sc{LiteRed}}.

For more discussion of this sector, see Appendix \ref{app:doubletriangle}.

\section{Massless Double-Box}
\label{sec:massless-double-box}

\begin{figure}[H]
    \centering
    \includegraphics[width=0.38\textwidth]{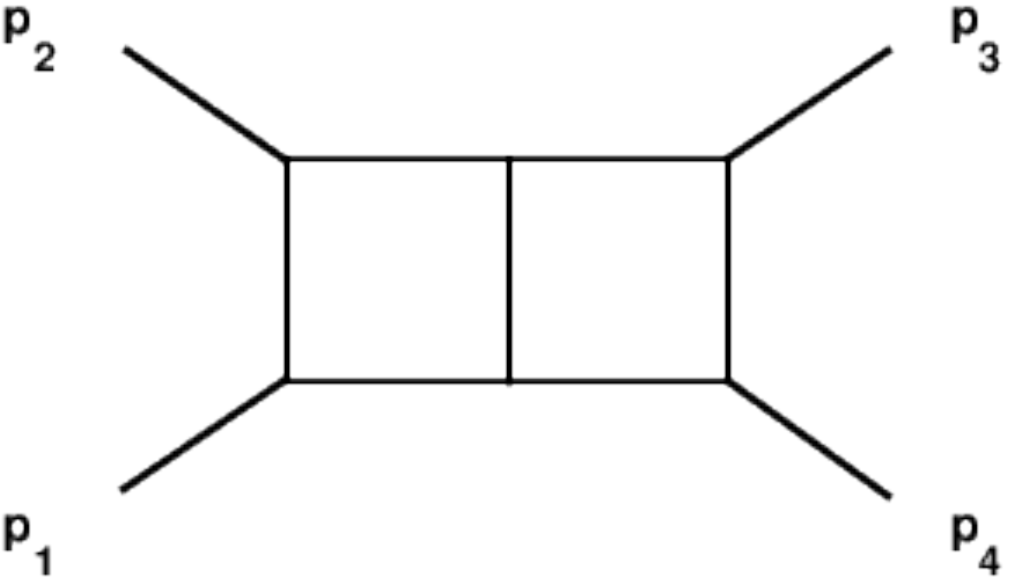}
    \caption{Massless double-box.}
    \label{fig:MasslessDoubleBox}
\end{figure}

Let us consider the massless double-box~\cite{Smirnov:1999gc, Smirnov:1999wz} in Fig.~\ref{fig:MasslessDoubleBox}. The external (outgoing) momenta are denoted $p_{i}$ with $p_{i}^{2}=0$ for $i=1,2,3,4$. We define the kinematic invariants to be $s=(p_{1}+p_{2})^2$, and $t=(p_{2}+p_{3})^2$.

The denominators are given by:
\begin{align}
& D_{1}=k_1^2 \, , \quad D_{2}=\left(k_1-p_1\right){}^2 \, , \quad D_{3}=\left(k_1-p_1-p_2\right){}^2 \, , \quad D_{4}=\left(k_1-k_2\right){}^2 \, ,\nonumber\\
& D_{5}=\left(k_2-p_1-p_2\right){}^2 \, , \quad D_{6}=\left(k_2-p_1-p_2-p_3\right){}^2 \, , \quad D_{7}=k_2^2 \, .
\label{eq:double-box-propagators}
\end{align}
The leftover ISP is:
\begin{equation}
z = D_{8} = (k_2-p_1)^2.
\label{eq:dbisp1}
\end{equation}
The \lbl Baikov representation, after a hepta-cut, gives,
\begin{align}
& u = z^{\frac{d}{2}-3} (s+z)^{2-\frac{d}{2}} (t-z)^{d-5} \, ,\quad 
\omega =
\bigg(\frac{4-d}{2 (s+z)}+\frac{d-5}{z-t}+\frac{d-6}{2 z} 
\bigg) dz 
\label{eq:u_LBL_Massless_DB}
\, , \\
& \masters=2 \, , \hspace{3.25cm} \mathcal{P}=\{ 0 \,,\; -s\,,\; t\,,\; \infty \} .
\end{align}

\paragraph{Mixed Bases.}

We pick the two master integrals
\begin{equation}
J_{1} = I_{1,1,1,1,1,1,1;0}\,, \qquad J_{2} = I_{1,1,1,1,1,1,1;-1}\,,
\end{equation}
corresponding to $\phi_1 = 1 \, dz$ and $\phi_2 = z \, dz$.

Additionally we pick the right basis as
\begin{equation}
\hat{\varphi}_1 = \frac{1}{z} - \frac{1}{z+s} \,, \qquad \hat{\varphi}_2 = \frac{1}{z+s} - \frac{1}{z-t}\,,
\label{eq:right_dlog_basis_Massless_DB}
\end{equation}

This gives the intersection matrix $\mathbf{C}$ to be
\begin{align}
    \mathbf{C} = \langle \phi_i | \varphi_j \rangle = \left( \begin{array}{cc}
    \frac{-s}{d-5} & \frac{s + t}{d-5} \\[1mm]
 \frac{s ((3 d - 14) s + 2 (d-5) t)}{2 (d-5) (d-4)} & \;\frac{-(3d-14) s (s + t)}{2 (d-5) (d-4)}
    \end{array} \right)
\label{eq:C_matrix_Mixed_basis_Massless_DB}
\end{align}

If we want to reduce $I_{1,1,1,1,1,1,1;-2}$ corresponding to $\phi_3 = z^2 \, dz$, we also need the intersections
\begin{align}
    \langle \phi_3 | \varphi_1 \rangle &= \frac{s (4 (d-5) t^2 - 3 (d-4) (3d-14) s^2 - 4 (d-5) (2d-9) s t)}{4 (d-5) (d-4) (d-3)} \,,\\
    \langle \phi_3 | \varphi_2 \rangle &= \frac{s (s + t) (3 (d-4) (3d-14) s + 2 (d-6) (d-5) t)}{4 (d-5) (d-4) (d-3)} \,.
\end{align}

Putting this together, the final reduction reads:
\begin{equation}\label{double-box-result}
I_{1,1,1,1,1,1,1;-2}=c_{1} \, J_{1}+c_{2} \, J_{2} \ ,
\end{equation}
with
\begin{equation}
c_{1}=\frac{(d-4) s t}{2 (d-3)} \, , \quad
c_{2}=\frac{2 t-3 (d-4) s}{2 (d-3)} \ ,
\end{equation}
in agreement with \textsc{Reduze}.

\subsection{A Second Example: The Other ISP}
\label{sec:theotherISP}

For most of the examples in this paper, 
the \lbl version of Baikov parametrization
has been employed.
As this approach generally integrates out some degrees of freedom, one might fear that some information is irrevocably lost: but this is not the case, as we shall see. Specifically, we consider 
$I_{1,1,1,1,1,1,1;-1,-1}$, belonging to the the same double-box family as above, but for which the other potential ISP
\begin{equation}
D_{9} = (k_1-p_1-p_2-p_3)^2
\label{eq:dbisp2}
\end{equation}
is present as well. Let us apply the \lbl procedure by integrating out $k_1$ first.
 Expanding $D_9$ gives a number of scalar products, all of which can be written in terms of the  propagators, or Baikov variables, considered in the former case, with the exception of $k_1 \cdot p_3$. The vector $p_3$ can be decomposed in a basis of vectors formed by $p_1, p_2, k_2$ and a complementary, perpendicular vector $\eta$, as
\begin{align}
p_3 &= \kappa_1 p_1 + \kappa_2 p_2 + \kappa_3 k_2 + \kappa_4 \eta
\label{eq:p3expansion}
\end{align}
with
\begin{align}
\eta^{\mu} &\equiv \varepsilon_{\;\; \nu_1 \nu_2 \nu_3}^{\mu} \, p_1^{\nu_1} p_2^{\nu_2} k_2^{\nu_3} \,.
\end{align}
Contracting each side of eq. \eqref{eq:p3expansion} with $p_1$, $p_2$, $k_2$, and $p_3$ gives four equations that allows us to identify all four $\kappa$s, and inserting this expression in $k_1 \cdot p_3$ give four terms, three of which can be re-expressed in terms of the other propagators. The remaining term $k_1 \cdot \eta$ can easily be seen to integrate to zero. Putting all of this together yields a rather lengthy expression, but on the maximal cut it simplifies, with the result
\begin{equation}
D_{9} \rightarrow f(z) = \frac{s (t - z)}{2 (s + z)}
\label{eq:massless_DB_D9_pow1}
\end{equation}
where the $f(z)$ refers to the function introduced in eq. \eqref{eq:phifromloopbyloop}.
See App.~\ref{app:decompISP} for a different way of obtaining eq. \ref{eq:massless_DB_D9_pow1}, valid also when $D_9$ appears in the numerator at a higher power.
According to eq. \eqref{eq:phifromloopbyloop}, the maximal cut of $I_{1,1,1,1,1,1,1;-1,-1}$  corresponds to 
\begin{align}
\hat{\phi} &= \frac{s \, (t - z) \, z}{2 (s + z)}\,.
\end{align}
To apply \eqref{eq:masterdeco:},
the necessary intersection numbers are,
\begin{align}
\langle \phi | \varphi_1 \rangle &= \frac{s^2 ((42 - 9 d) s + (38 - 8 d) t)}{4 (d-5) (d-4)} \, , \\
\langle \phi | \varphi_2 \rangle &= \frac{s (s + t) ((9d-42) s + 2 (d-4) t)}{4 (d-5) (d-4)} \, ,
\end{align}
and then, the decomposition formula reads,
\begin{align}
I_{1,1,1,1,1,1,1;-1,-1} &= \frac{st}{2} \, J_{1} + \frac{-3s}{2} \, J_{2} \, ,
\end{align}
in agreement with {\sc FIRE}.

For a different approach to the other ISP based on intersection theory, see appendix \ref{app:decompISP}.

\section{Internally Massive Double-box}
\label{sec:intmass}

\begin{figure}[H]
    \centering
    \includegraphics[width=0.38\textwidth]{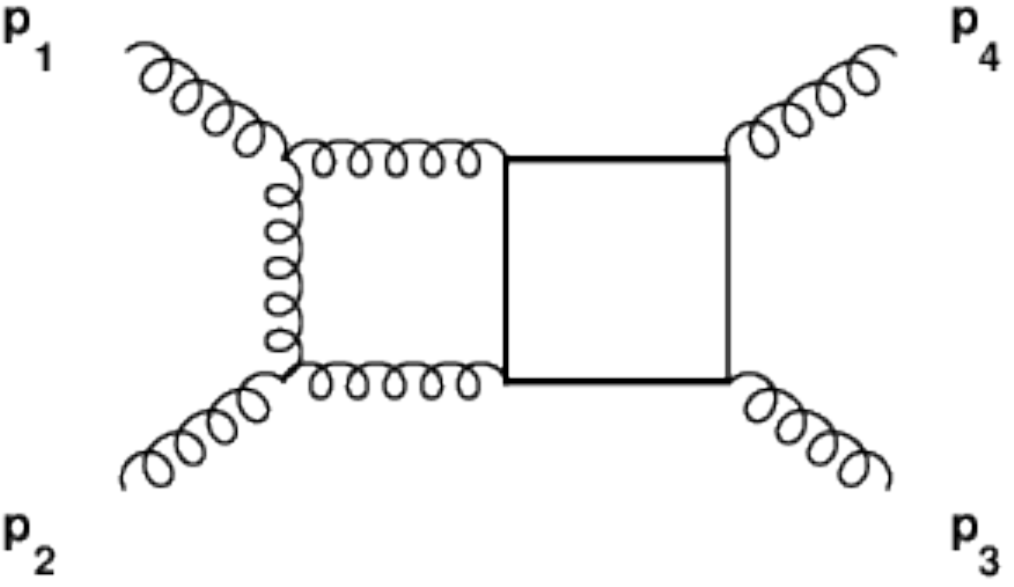}
    \caption{Four points two-loop diagram}
    \label{fig:Bonciani_diagram}
\end{figure}
Let us consider the four point two-loop integral depicted on Fig.~\ref{fig:Bonciani_diagram}. This diagram has recently been computed in ref. \cite{Becchetti:2017abb} in the context of top-mass corrections to $2j$ or $2 \gamma$ production. \\
The external kinematics is as in the previous section.
The set of propagators reads:
\begin{align}
& D_{1}=k_1^2,\;\;\; D_{2}=(k_1 + p_1)^2,\;\;\; D_{3}=(k_1 - p_3 - p_4)^2, \;\;\; D_{4}=(k_2 - p_3 - p_4)^2 - m^2, \nonumber \\
& D_{5}=(k_2 - p_4)^2 - m^2,\;\;\; D_{6}= k_2^2 - m^2,\;\;\; D_{7} =(k_1 - k_2)^2 - m^2 \, ,
\label{eq:intmasprops}
\end{align}
while the additional ISP is :
\begin{align}
z = D_8 = (k_1 - p_4)^2. 
\end{align}
With \lbl Baikov parametrization we have two options, we can either integrate the massive or the mass-less loop out first. Starting with the massless loop gives four master integrals in agreement with ref. \cite{Becchetti:2017abb} and most public IBP codes. However, starting with the massive one, the number of master integrals is 3 : 
In this case, we get,
\begin{align}
u &= (z - t)^{d-5} \, (z + s)^{(4 - d)/2} \, (4 m^2 s + (4 m^2 - s) z)^{(d - 5)/2} \, z^{-1/2} \ ,
\end{align}
corresponding to
\begin{align}
\omega &= \frac{ q_0 + q_1 z + q_2 z^2 + q_3 z^3}{2 z (s + z) (z-t) (4 m^2 s + (4 m^2 - s) z)} dz \ ,
\label{eq:omegaintmass}
\end{align}
with
\begin{align}
q_0 &= 4 m^2 s^2 t \ , \nonumber \\
q_1 &= s ((d-6) s t + 4 m^2 ((2 d - 11) s + 3 t)) \ , \nonumber \\
q_2 &= (4 (4d-23) m^2 s + (16 - 3 d) s^2 + 8 m^2 t - 2 s t) \ , \\
q_3 &= 2 (d-6) (4 m^2 - s) \ , \nonumber
\end{align}
and here it is easy to see that $\omega=0$ has three solutions, corresponding to $\masters=3$ master integrals for this sector.

We can get 3 master integrals out of the public code {\sc Kira} 1.1 \cite{Maierhoefer:2017hyi} if we also search for relations in sectors in which the ISP $z$ is allowed to appear as a propagator. What is happening is the following:
Picking as master integrals in the seven-propagator sector defined by eqs. \eqref{eq:intmasprops}, the four integrals
\begin{align}
I_{1111111;0} \,, \;\; I_{1211111;0} \,, \;\; I_{1111211;0} \,, \;\; I_{1111112;0} \,,
\end{align}
{\sc Kira} finds the relation
\begin{align}
I_{0111121;1} &= \tfrac{1}{2} I_{1111112;0} - \tfrac{1}{2} I_{1111211;0} - \tfrac{d-4}{4 m^2} I_{1111011;1} + \tfrac{d-4}{4 m^2} I_{1111110;1} + \text{subtopologies},
\label{eq:themagicrelation}
\end{align}
where ``subtopologies'' refers to integrals in sectors with less than seven propagators. This relation is an example of an IBP relation that relate different sectors with the same number of propagators (what is referred to as a ``magic relation'' in ref. \cite{Maierhofer:2018gpa}). On the cut of the first seven propagators, this identity reduces to
\begin{align}
I_{1111112;0} &= I_{1111211;0} \, ,
\end{align}
which is the relation that reduces the number of master integrals in this sector to 3 on the hepta-cut. 
We have verified this number with a numerical evaluation of the integrals combined with the high-precision arithmetic PSLQ algorithm
\cite{Bailey09pslq:an} (80 digits accuracy).

Proceeding with the reduction-by-intersections starting from the $\omega$ of eq. \eqref{eq:omegaintmass}, the decomposition formulas are in agreement with {\sc Kira}.

For more discussion of this sector, see appendix \ref{app:intmass}.

\section{Two-Loop Bhabha Scattering}
\label{sec:bhabha}

In this section, we discuss the seven-propagator sectors that contribute to Bhabha scattering (i.e. $e^+ e^- \rightarrow e^+ e^-$ in QED) at two loops. We take the external electrons to be on-shell so the integrals are functions of three variables $s = (p_1+p_2)^2$, $t = (p_2+p_3)^2$, and $m^2 = p_i^2$. There are three such seven-propagator sectors, two planar and one non-planar.

\subsection{First Planar Sector}
\label{sec:bhabha1}

\begin{figure}[H]
    \centering
    \includegraphics[width=0.38\textwidth]{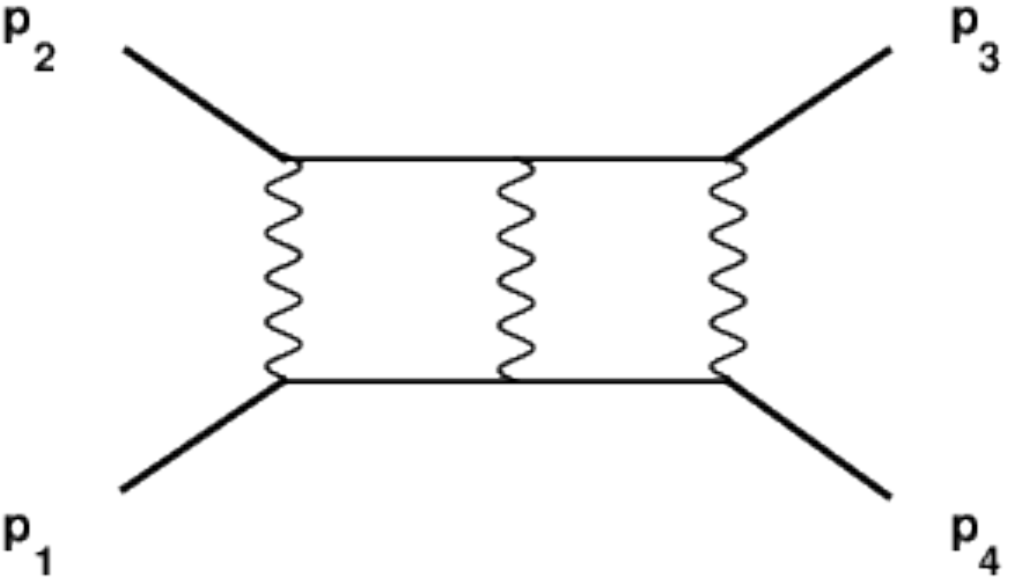}
    \caption{Bhabha - First planar sector.}
    \label{fig:Bhabha_planar_1}
\end{figure}

This planar sector is depicted on Fig. \ref{fig:Bhabha_planar_1}, and was first computed in ref. \cite{Henn:2013woa}. It may be expressed in terms of the seven propagators
\begin{align}
& D_1 = k_2^2 - m^2, \quad D_2 = (k_2 {+} p_1)^2, \quad D_3 = (k_2 + p_1 + p_2)^2 - m^2, \quad D_4 = (k_1 {+} p_1 {+} p_2)^2 {-} m^2, \nonumber \\
& D_5 = (k_1 + p_1 + p_2 + p_3)^2, \quad D_6 = k_1^2 - m^2, \quad D_7 = (k_1 - k_2)^2.
\end{align}
Additionally we need the variable $z = D_8 = (k_1+p_1)^2$, which play the roles of ISP and eighth integration variable.

Using the \lbl form of Baikov representation, by first integrating out $k_2$, we obtain:
\begin{align}
u &= (z + s - 4 m^2)^{(4 - d)/2} (z - t)^{d-5} z^{(d-6)/2} \, , \\
\omega & = \frac{(d-6) (4 m^2 - s) t + (2 t - (3 d - 16) (4 m^2 - s)) z + 2 (d-6) z^2}{2 z (z + s - 4 m^2) (z - t)} \, dz \, , \\
\masters & =2 \, , \quad \quad \quad \quad \quad \quad \mathcal{P}  =\{ 0 \, , 4m^2-s \, , t \, , \infty \} \, .
\end{align}

\paragraph{Mixed Bases.} Let us consider the decomposition of $I_{1,1,1,1,1,1,1;-1}=\langle{\phi_{2}}|\mathcal{C}]$.\\
The MIs can be chosen as:
\begin{equation}
J_{1}=I_{1,1,1,1,1,1,1;0}=\langle{e_{1}}| \mathcal{C}] \, , \quad J_{2}=I_{1,1,1,1,1,1,2;0}=\langle{e_{2}}|\mathcal{C}], \, 
\end{equation}
with:
\begin{align}
\hat{e}_{1}  = 1 \, , \quad
\hat{e}_{2}  = \frac{5-d}{z} \, .
\end{align}
We introduce the dlog differential-stripped cocycles:
\begin{align}
\hat{\varphi}_{1} = \frac{1}{z}-\frac{1}{z-t} \, ,    
\quad \hat{\varphi}_{2}=\frac{1}{z-t}-\frac{1}{z+s-4m^2} \, .
\end{align}
We can compute the $\mathbf{C}$ matrix:
\begin{equation}
\mathbf{C}_{ij}= \bra{e_{i}}\ket{\varphi_{j}} \, , \quad i,j=1,2 \, ,    
\end{equation}
with:
\begin{align}
\bra{e_{1}}\ket{\varphi_{1}} & = \frac{t}{d-5} \, , \quad \hspace{1cm} \bra{e_{1}}\ket{\varphi_{2}}=\frac{4m^2-s-t}{d-5} \, , \\
\bra{e_{2}}\ket{\varphi_{1}} & =\frac{-2(d-5)}{d-6} \, , \quad\;\, \bra{e_{2}}\ket{\varphi_{2}}=0 \, .
\end{align}
Moreover we can compute the intersection numbers:
\begin{align}
\bra{\phi_{2}}\ket{\varphi_{1}}=\frac{(4m^2-s)t}{2(d-5)} \, , \quad 
\bra{\phi_{2}}\ket{\varphi_{2}}=\frac{(3 d - 14) (4 m^2 - s) (4 m^2 - s - t)}{2 (d-5) (d-4)} \, .
\end{align}
The final reduction, given by eq.~($\ref{eq:masterdeco:}$) is:
\begin{equation}
I_{1,1,1,1,1,1,1;-1}= c_{1} \, J_{1}+ c_{2} \, J_{2}
\end{equation}
with:
\begin{equation}
c_{1}=\frac{(3 d - 14) (4 m^2 - s)}{2 (d-4)} \, , \quad c_{2}= \frac{(d-6) (4 m^2 - s) t}{2 (d - 5) (d - 4)} \, ,
\end{equation}
 in agreement with \textsc{FIRE}.

\subsection{Second Planar Sector}
\label{sec:bhabha2}

\begin{figure}[H]
    \centering
    \includegraphics[width=0.38\textwidth]{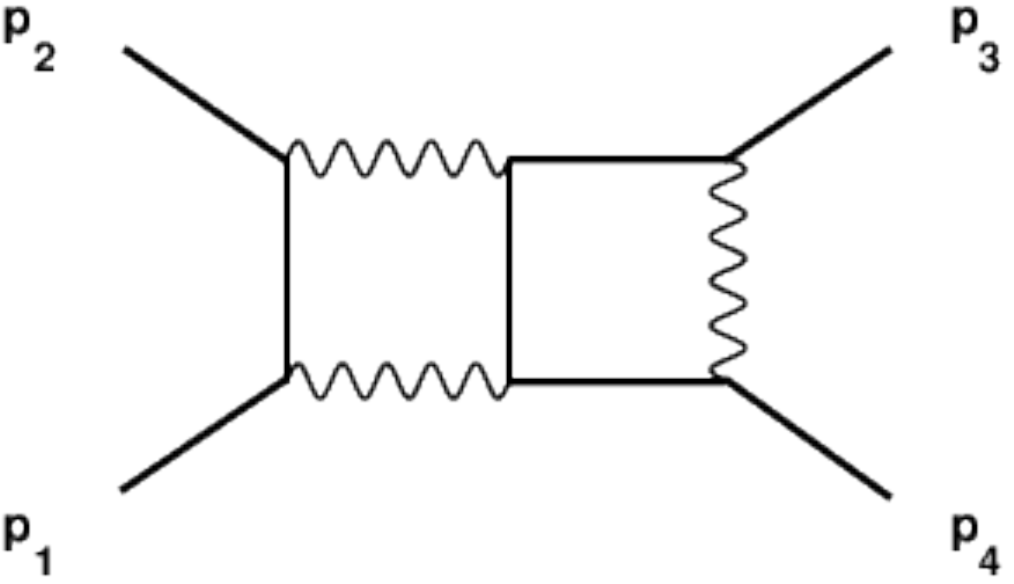}
    \caption{Bhabha - second planar sector.}
    \label{fig:Bhabha_planar_family_2}
\end{figure}
The second planar seven-propagator sector contributing to Bhabha scattering, shown in Fig.~\ref{fig:Bhabha_planar_family_2}, can be expressed in terms of the propagators
\begin{align}
& D_1 =  k_2^2, \;\; D_2 = (k_2 + p_1)^2 - m^2, \;\; D_3 = (k_2 + p_1 + p_2)^2, \;\; D_4 = (k_1 + p_1 + p_2)^2 - m^2, \nonumber \\
& D_5 = (k_1 + p_1 + p_2 + p_3)^2, \;\; D_6 = k_1^2 - m^2, \;\; D_7 = (k_1 - k_2)^2 - m^2.
\end{align}
We also need the ISP variable $z = D_8 = (k_1 + p_1)^2$.

After using the \lbl Baikov representation and the hepta-cut, we get 
\begin{align}
u =& \;(4 m^2 - z)^{(d-5)/2} (4 m^2 - s - z)^{(4 - d)/2} (t - z)^{d-5} z^{-1/2} \, , \\
\omega =& \Big( \frac{4 m^2 (4 m^2 - s) t + (4 (2d-11) m^2 (4 m^2 - s) - (12 m^2 + (d-6) s) t) z}{2 z (4 m^2 - z) (4 m^2 - s - z) (z-t)} \nonumber \\ 
& + \frac{((92-16 d) m^2 + (3d-16) s + 2 t) z^2 + 2 (d-6) z^3}{2 z (4 m^2 - z) (4 m^2 - s - z) (z-t)}\Big) dz \, , \\
\masters  &=3 \, , \quad \quad \quad \quad \quad \quad \mathcal{P}  =\{ 0 \, , 4m^2 \, , 4m^2-s \, , t \, , \infty \} \, .
\end{align}

\paragraph{Mixed Bases.} Let us consider the decomposition of $I_{1,1,1,1,1,1,1;-1}$ in terms of the three master integrals
\begin{align}
J_1 = I_{1,1,1,1,1,1,1;0}=\langle{e_{1}}| \mathcal{C}] \, , \quad J_2 = I_{1,1,1,1,2,1,1;0}=\langle{e_{2}}| \mathcal{C}] \, , \quad J_3 = I_{1,1,1,1,1,2,1;0}=\langle{e_{3}}| \mathcal{C}] \, ,
\end{align}
with:
\begin{gather}
\hat{e}_{1}  = 1 \, , \qquad
\hat{e}_{2}  = \frac{(5-d) ((4 m^2 - s) t + (4 m^2 - s - 2t) z)}{(4 m^2 - s) (t-z)^2} \, , \\
\quad \hat{e}_{3}  = \frac{(d-6) (4 m^2 - s) t + ((d-4) (4 m^2 - s) - 2 (d-5) t) z}{2 (4 m^2 - s) (4 m^2 - s - z) (t-z)} \, .
\end{gather}
The dlog differential-stripped cocycles reads:
\begin{align}
\hat{\varphi}_1 = \frac{1}{z} - \frac{1}{z {-} 4 m^2} ,\;\;\; \hat{\varphi}_2 =  \frac{1}{z {-} 4 m^2} - \frac{1}{z {+} s {-} 4 m^2} ,\;\;\; \hat{\varphi}_3 = \frac{1}{z {+} s {-} 4 m^2} - \frac{1}{z {-} t} \, .
\end{align}
We can compute the $\mathbf{C}$ matrix:
\begin{equation}
\mathbf{C}_{ij}= \bra{e_{i}}\ket{\varphi_{j}} \, , \quad i,j=1,2 \, ,    
\end{equation}
with:
\begin{equation}
\begin{split}
\bra{e_{1}}\ket{\varphi_{1}} & = \frac{4 m^2}{d-5} \, , \quad \hspace{1cm} \bra{e_{1}}\ket{\varphi_{2}}=\frac{-s}{d-5} \, , \quad \hspace{1cm} \bra{e_{1}}\ket{\varphi_{3}}=\frac{s + t - 4 m^2}{d-5} \,, \\
\bra{e_{2}}\ket{\varphi_{1}} & =\frac{8 (d-5) m^2 (s + t - 4 m^2)}{(d-6) (4 m^2 - s) (4 m^2 - t)} \, , \quad \bra{e_{2}}\ket{\varphi_{2}}=\frac{2 (5-d) s t}{(d-6) (4 m^2 - s) (4 m^2 - t)} \,, \\ 
& \qquad \qquad \qquad  \bra{e_{2}}\ket{\varphi_{3}}=\frac{4 (d-5) m^2 (4 m^2 - s - t)}{(d-6) (4 m^2 - s) (4 m^2 - t)} \,, \\
\bra{e_{3}}\ket{\varphi_{1}} & =0 \, , \quad \hspace{1cm} \bra{e_{3}}\ket{\varphi_{2}}=1 \, , \quad \hspace{1cm} \bra{e_{3}}\ket{\varphi_{3}}=-\frac{4 m^2 - s - t}{4 m^2 - s} \, .
\end{split}
\end{equation}
The additional intersection numbers are 
\begin{align}
\langle \phi_2 | \varphi_1 \rangle &= \frac{2 m^2 \, (4 (2d-9) - m^2 (d-4) s - 2 (d-5) t)}{(d-5) (d-4)} \,, \nonumber \\
\langle \phi_2 | \varphi_2 \rangle &= \frac{s \, ((76 - 16 d) m^2 + (3d-14) s + 2 (d-5) t)}{2 (d-5) (d-4)} \,, \\
\langle \phi_2 | \varphi_3 \rangle &= \frac{(4 (9-2d) m^2 + (3d-14) s) \, (4 m^2 - s - t)}{2 (d-5) (d-4)} \,, \nonumber
\end{align}
The final reduction, given by eq.~($\ref{eq:masterdeco:}$) is:
\begin{equation}
I_{1,1,1,1,1,1,1;-1}= c_{1} \, J_{1}+ c_{2} \, J_{2} + c_{3} \, J_{3}
\end{equation}
with:
\begin{gather}
    c_{1}=-\frac{8 (9 - 2 d) m^4 + (d-6) s t + 2 (d-4) m^2 (s + 2 t)}{2 (d-4) (2 m^2 - t)} \, , \\
    \quad c_{2}= \frac{(d-6) \, (4 m^2 - s) \, (4 m^2 - t) \, t}{2 (d-5) (d-4) (2 m^2 - t)} \, \quad c_{3}= -\frac{2 \, (4 m^2 - s) \, s \, (m^2 - t)}{(d-4) (2 m^2 - t)} ,
\end{gather}
 in agreement with \textsc{Kira}.

\subsection{Non-Planar Sector}
\label{sec:bhabha3}

\begin{figure}[H]
    \centering
    \includegraphics[width=0.38\textwidth]{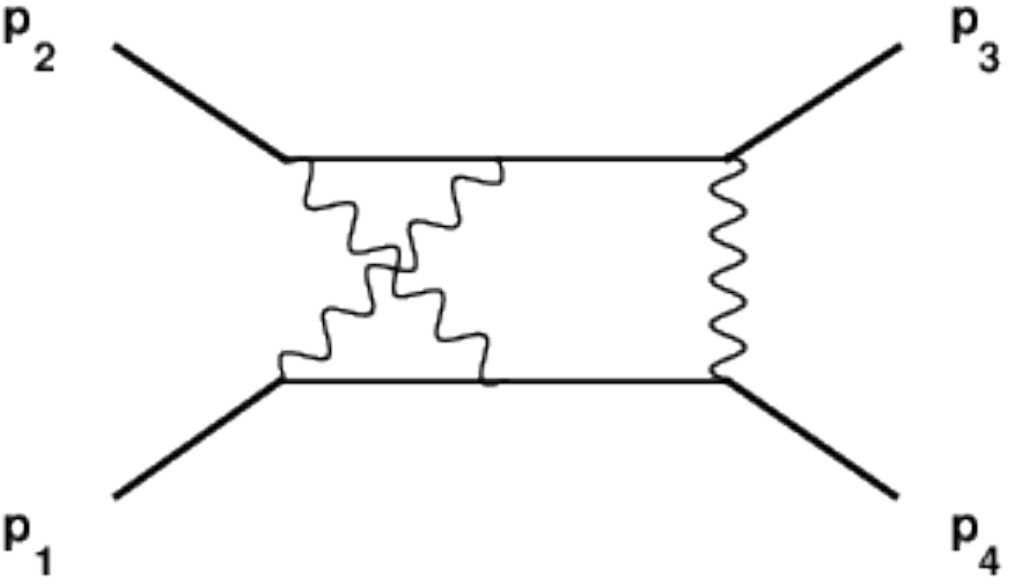}
    \caption{Bhabha non-planar sector.}
    \label{fig:Bhabha_non_planar_family}
\end{figure}
We consider the non-planar contribution to Bhabha scattering in Fig.~\ref{fig:Bhabha_non_planar_family}.
The seven denominators can be parametrized as:
\begin{align}
& D_1 = k_2^2 - m^2, \;\; D_2 = (k_2 + p_1)^2, \;\; D_3 = (k_1 + p_1 + p_2)^2 - m^2, \;\; D_4 = (k_1 + p_1 + p_2 + p_3)^2, \nonumber \\
& D_5 = k_1^2 - m^2, \;\; D_6 = (k_1 - k_2)^2, \;\; D_7 = (k_1 - k_2 + p_2)^2 - m^2 \, ,
\end{align}
and the leftover ISP is:
\begin{equation}
z = D_8 = (k_1+p_1)^2.    
\end{equation}
We find:
\begin{align}
  u & = (z - t)^{d-5} \, z^{(d-6)/2} \, (z + s)^{(d-5)/2} \, (z + s - 4 m^2)^{-1/2} \, \\
\omega & = \frac{c_0 + c_1 z + c_2 z^2 + c_3 z^3}{2 z (z+s) (s + z - 4 m^2) (z-t)} \, dz 
\label{eq:omegat_BhaBha_Non_Planar}
\end{align}
where in eq.~($\ref{eq:omegat_BhaBha_Non_Planar}$) we defined:
\begin{align}
c_0 &= (d-6) (4 m^2 - s) s t \,, \nonumber \\
c_1 &= (3 d - 16) s (s - 4 m^2) + (8 d - 44) m^2 t - 3 (d-6) s t \,, \nonumber \\ 
c_2 &= (84 - 16 d) m^2 + (7 d - 38) s - 2 (d-6) t \,, \\
c_3 &= (4 d - 22) \nonumber \, ,
\end{align}
and so, we get:
\begin{align}
\masters & =3 \, , \quad
\mathcal{P} =\{ 0 \, , -s \, ,4m^2-s \,, t \} \, .
\end{align}

\paragraph{Mixed Bases} Let us consider the decomposition of $I_{1,1,1,1,1,1,1;-1}=\langle{\phi_{2}}|\mathcal{C}]$.\\
The MIs can be chosen as:
\begin{equation}
J_{1}=I_{1,1,1,1,1,1,1;0}=\langle{e_{1}}|\mathcal{C}] \, , \quad J_{2}=I_{1,1,1,1,1,2,1;0}=\langle{e_{2}}|\mathcal{C}] \, , \quad J_{3}=I_{1,1,1,1,1,1,2;0}=\langle{e_{3}}|\mathcal{C}] \, ,  
\end{equation}
with:
\begin{align}
\hat{e}_{1} = 1 \, , \quad
\hat{e}_{2} &= \frac{5-d}{z} \, , \quad
\hat{e}_{3} = \frac{d-5}{z+s} \, .
\end{align}
Moreover we introduce the following dlog differential-stripped cocycles:
\begin{align}
\hat{\varphi}_{1} & =\frac{1}{z}-\frac{1}{z+s-4m^2} \, , \\   
\hat{\varphi}_{2} & =\frac{1}{z+s-4m^2}-\frac{1}{z+s} \, , \\
\hat{\varphi}_{3}& =\frac{1}{z+s}-\frac{1}{z-t}
\end{align}
We compute the $\mathbf{C}$ matrix:
\begin{equation}
\mathbf{C}_{ij}=\bra{e_{i}}\ket{\varphi_{j}} \, , \quad 1 \leq i,j \leq 3,
\end{equation}
with:
\begin{align}
\langle e_1 | \varphi_1 \rangle &= \frac{4 m^2 - s}{2 (d-5)} \, , & \langle e_1 | \varphi_2 \rangle &= \frac{2 m^2}{5-d} \, , & \langle e_1 | \varphi_3 \rangle &= \frac{s+t}{2 (d-5)} \, , \\
\langle e_2 | \varphi_1 \rangle &= \frac{2 (5 - d)}{d-6} \, , & \langle e_2 | \varphi_2 \rangle &= 0 \, , & \langle e_2 | \varphi_3 \rangle &= 0 \, , \\
\langle e_3 | \varphi_1 \rangle &= 0 \,, & \langle e_3 | \varphi_2 \rangle &= -2 \, , & \langle e_3 | \varphi_3 \rangle &= 2 \, .
\end{align}
Moreover we can compute the intersection numbers:
\begin{align}
\langle \phi_{2} | \varphi_1 \rangle &= \frac{(4 m^2 - s) (4 (4 d - 19) m^2 + (14 - 3 d) s - 2 (d-5) t)}{4 (d-5) (2 d-9)} \, ,  \\
\langle \phi_{2} | \varphi_2 \rangle &= \frac{m^2 ((76 - 16 d) m^2 + (7 d - 34) s + 2 (d-5) t)}{(d-5) (2 d-9)} \, , \\
\langle \phi_{2} | \varphi_3 \rangle &= \frac{(s + t) (4 m^2 + (14 - 3 d) s + 2 (d-5) t)}{4 (d-5) (2d-9)} \, .
\end{align}
Then, eq.~($\ref{eq:masterdeco:}$) yields:
\begin{equation}
I_{1,1,1,1,1,1,1;-1}=c_{1} J_{1}+c_{2} J_{2}+c_{3} J_{3} \, ,
\end{equation}
with:
\begin{align}
c_1 & = \frac{4 (4 d {-} 19) m^2 {+} (14 {-} 3 d) s {+} 2 (d{-}5) t}{4 d {-} 18} \, , \\
c_2 & = \frac{(d {-} 6) (4 m^2 {-} s) t}{2 (d{-}5) (2 d{-}9)}\, , \\ 
c_3 & = \frac{{-} 2 m^2 (s {+} t)}{2d{-}9} \, ,
\end{align}
 in agreement with \textsc{Kira}.

\section{Two-Loop Associated Higgs Production}
\label{sec:higgsproduction}

\subsection{Planar Contribution to $H{+}j$ and $HH$ Production}
\label{sec:planarhiggs}

\begin{figure}[H]
    \centering
    \includegraphics[width=0.38\textwidth]{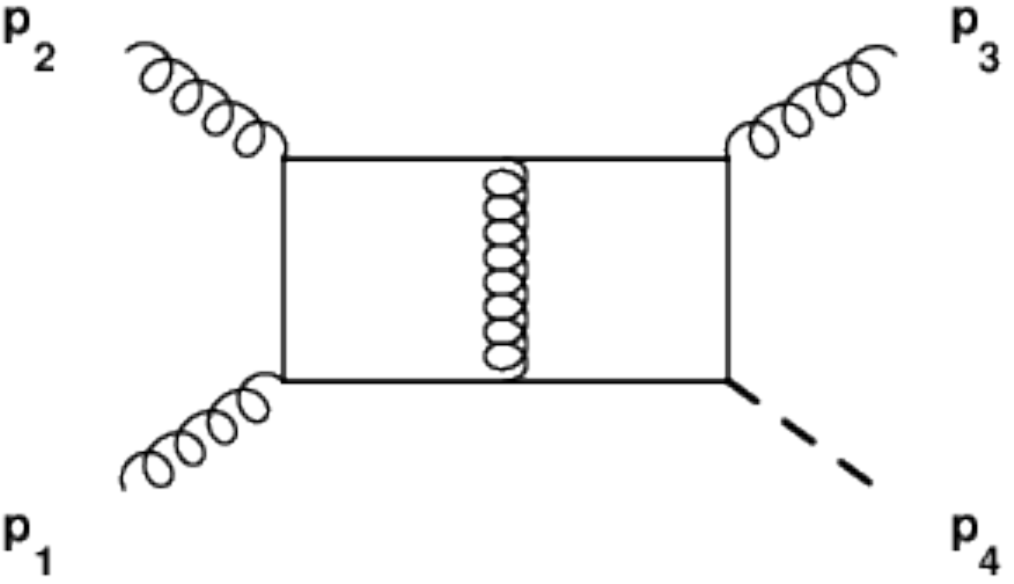}
    \hspace{1cm}
    \includegraphics[width=0.38\textwidth]{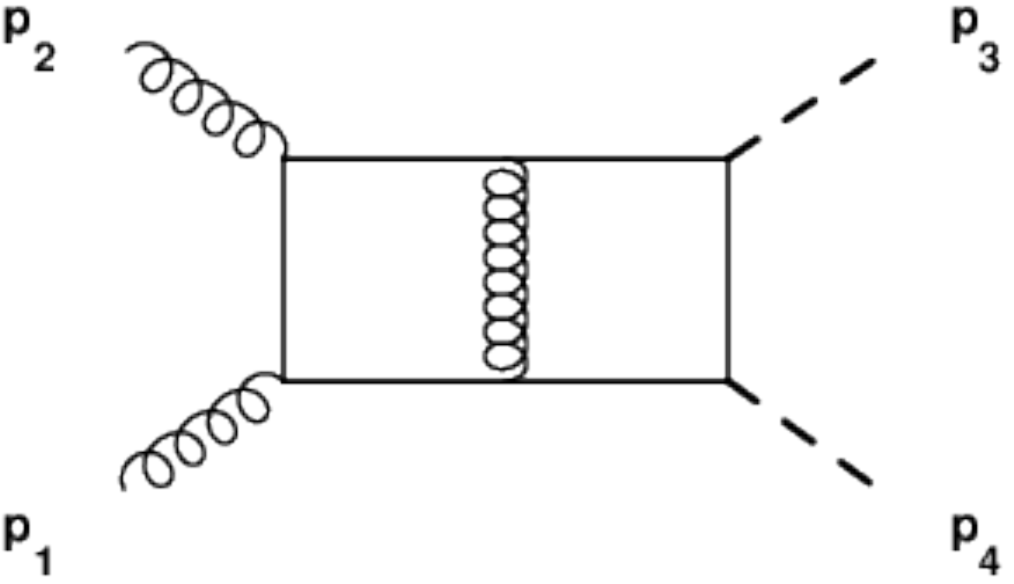}
    \caption{A planar contribution to $H{+}j$ (left) and to $HH$ (right).}
    \label{fig:higgs_plus_jet_planar}
\end{figure}

In this section we discuss planar seven-propagator integrals contributing to Higgs+jet ($H{+}j$) production~\cite{Melnikov:2016qoc, Bonciani:2016qxi, Jones:2018hbb} and double Higgs ($HH$) production~\cite{Borowka:2016ypz, Baglio:2018lrj}, portrayed in Fig.~\ref{fig:higgs_plus_jet_planar}.
We treat the two cases simultaneously as they are very similar.
The seven propagators are
\begin{align}
&D_1 = k_2^2 - m_t^2, \;\; D_2 = (k_2 {+} p_1)^2 - m_t^2, \;\; D_3 = (k_2 {+} p_1 {+} p_2)^2 - m_t^2, \;\; D_4 = (k_1 {+} p_1 {+} p_2)^2 {-} m_t^2, \nonumber \\
& D_5 = (k_1 {+} p_1 {+} p_2 {+} p_3)^2 - m_t^2, \qquad D_6 = k_1^2 - m_t^2, \qquad D_7 = (k_1 {-} k_2)^2.
\end{align}
The ISP is chosen to be: 
\begin{equation}
z = D_8 = (k_1 {+} p_1)^2 - m_t^2
\end{equation}
The kinematics is such that $p_1^2 = p_2^2 = 0$, $(p_1+p_2)^2 = s$, $(p_2+p_3)^2 = t$, and $(p_1+p_2+p_3)^2 = m_H^2$. The only difference between the cases of $H{+}j$ and $HH$ is the relation $p_3^2 = \delta m_H^2$ where $\delta=0$ for the case of $H{+}j$, while $\delta = 1$ for the case of $HH$.\\
The \lbl Baikov parametrization, on the maximal cut, gives us:
\begin{align}
u &=z^{d-5} ((z-r_1)(z-r_2))^{(4 - d)/2} ((z-r_3)(z-r_4))^{(d-5)/2} \, ,    
\end{align}
with:
\begin{align}
r_1  = &  \frac{1}{2} \left(-\sqrt{s^2-4 s m_t^2}-s\right) \, , \\
r_2  = & \frac{1}{2} \left(\sqrt{s^2-4 s m_t^2}-s\right) \, , \\
r_3  = & \Big(2 \sqrt{s \left(\delta  m_H^4-(\delta +1) t m_H^2+t (s+t)\right) \left(m_H^4 \left((\delta -1)^2 m_t^2+\delta  s\right)-2 (\delta +1) s
   m_H^2 m_t^2+s^2 m_t^2\right)}+ \nonumber\\
  & 2 \delta  s m_H^4-(\delta +1) s t m_H^2+s^2 t \Big)/ \Big((\delta -1)^2 m_H^4-2 (\delta +1) s m_H^2+s^2 \Big) \, , \\
r_{4} = & \Big(-2 \sqrt{s \left(\delta  m_H^4-(\delta +1) t m_H^2+t (s+t)\right) \left(m_H^4 \left((\delta -1)^2 m_t^2+\delta  s\right)-2 (\delta +1) s
   m_H^2 m_t^2+s^2 m_t^2\right)}+ \nonumber \\
   & 2 \delta  s m_H^4-(\delta +1) s t m_H^2+s^2 t \Big) / \Big((\delta -1)^2 m_H^4-2 (\delta +1) s m_H^2+s^2 \Big) \, .  
\end{align}
Then, we have:
\begin{align}
\omega &= \frac{q_0 + q_1 z + q_2 z^2 + q_3 z^3 + q_4 z^4}{2 z  (z-r_1) (z-r_2) (z-r_3) (z-r_4)} dz
\end{align}
where:
\begin{align}
& q_0 = 2 (d-5) r_1 r_2 r_3 r_4 \, , \\
& q_1 = - ((d-6) (r_1 + r_2) r_3 r_4 + 3 (d-5) r_1 r_2 (r_3 + r_4)) \, ,  \\
& q_2 = 4 (d-5) r_1 r_2 - 2 r_3 r_4 + (2d - 11) (r_1 + r_2) (r_3 + r_4) \, , \\
& q_3 = ((16 - 3 d) (r_1 + r_2) - (d-7) (r_3 + r_4)) \, , \\
& q_4 = 2 (d-6) \, .
\end{align}
Therefore we find:
\begin{equation}
\masters=4 \, , \quad \mathcal{P}=\{ 0, r_1 \, , r_2 \, , r_3 \, , r_4 \, \infty \} \, .    
\end{equation}

\paragraph{Mixed Bases.} Let us consider the reduction of $I_{1,1,1,1,1,1,1;-1}=\langle{\phi_{2}}|\mathcal{C}]$ in terms of $J_1=I_{1,1,1,1,1,1,1;0}= \langle{e_1}|\mathcal{C}]$, $J_2=I_{1,2,1,1,1,1,1;0}=\langle{e_2}|\mathcal{C}] $, $J_3=I_{1,1,1,1,2,1,1;0}=\langle{e_3}|\mathcal{C}]$ and $J_4=I_{1,1,1,1,1,1,2;0}=\langle{e_4}|\mathcal{C}]$ with:
\begin{align}
\hat{e}_{1} & =1 \, , \\  
\hat{e}_{2} & = \frac{5-d}{z} \, , \\
\hat{e}_{3} & = \frac{(d-5) \, s \, ( ((1 + \delta) m_H^2 - s - 2 t) z - st )}{((1-\delta)^2 m_H^4 - 2 (1 + \delta) m_H^2 s + s^2) (z - r_3) (z-r_4)} \, , \\
\hat{e}_{4} & =\frac{(d-5) s (2 m_t^2 + z)}{(4 m_t^2 - s) z^2} \, .
\end{align}
Moreover we can introduce the following differential-stripped cocycles:
\begin{align}
\hat{\varphi}_1 &= \frac{1}{z} - \frac{1}{z{-}r_1} \,, & \hat{\varphi}_2 &= \frac{1}{z{-}r_1} - \frac{1}{z{-}r_2} \,, \qquad \\ 
\hat{\varphi}_3 &= \frac{1}{z{-}r_2} - \frac{1}{z{-}r_3} \,, \!\!\!\!\! & \hat{\varphi}_4 &= \frac{1}{z{-}r_3} - \frac{1}{z{-}r_4} \,.
\end{align}
We compute the $\mathbf{C}$ matrix:
\begin{equation}
\mathbf{C}_{ij}=\bra{e_{i}}\ket{\varphi_{j}} \, , \quad 1 \leq i , j \leq 4 \, ,     
\end{equation}
with:
\begin{align}
\bra{e_{1}}\ket{\varphi_{1}} & = \frac{r_1}{d-5} \,, & \bra{e_{1}}\ket{\varphi_{2}} & = \frac{r_2-r_1}{d-5} \,, & \bra{e_{1}}\ket{\varphi_{3}} & = \frac{r_3-r_2}{d-5} \,, \\
\bra{e_{1}}\ket{\varphi_{4}} & = \frac{r_4-r_3}{d-5} \,, & \bra{e_{2}}\ket{\varphi_{1}} & = -1 \,, & \bra{e_{2}}\ket{\varphi_{2}} & = 0 \,, \\[-12mm] \nonumber
\end{align}
\begin{align}
\bra{e_{2}}\ket{\varphi_{3}} \,=\, \bra{e_{2}}\ket{\varphi_{4}} \,=\, \bra{e_{3}}\ket{\varphi_{1}} \,=\, \bra{e_{3}}\ket{\varphi_{2}} \,=\, 0 \,, \\[-12mm] \nonumber
\end{align}
\begin{align}
\bra{e_{3}}\ket{\varphi_{3}} & = \frac{2 s (s t + r_3 (s + 2 t -(1 + \delta) m_H^2 ))}{(r_3 - r_4) ((1 - \delta)^2 m_H^4 - 2 (1 + \delta) m_H^2 s + s^2)} \, , \\ 
\bra{e_{3}}\ket{\varphi_{4}} & = \frac{2 s ( (r_3 + r_4) ((1 + \delta) m_H^2 - s - 2 t) - 2 s t )}{(r_3 - r_4) ((1 - \delta)^2 m_H^4 - 2 (1 + \delta) m_H^2 s + s^2)}\,, \\
\bra{e_{4}}\ket{\varphi_{1}} & = \frac{s \, [ (d-6) r_1 r_2 r_3 r_4 + m_t^2 ((d-6) r_2 r_3 r_4 - (d-4) r_1 r_3 r_4 + (d-5) r_1 r_2 (r_3 + r_4))]}{(d-6) r_1 r_2 r_3 r_4 (4 m_t^2 - s)} \, , \\
\bra{e_{4}}\ket{\varphi_{2}} & = \frac{(r_1-r_2)}{r_1 r_2} \, \frac{2 (d-5) m_t^2 s}{(d-6) (4 m_t^2 - s)} \, , \\
\bra{e_{4}}\ket{\varphi_{3}} & = \frac{(r_2-r_3)}{r_2 r_3} \, \frac{2 (d-5) m_t^2 s}{(d-6) (4 m_t^2 - s)} \, , \\
\bra{e_{4}}\ket{\varphi_{4}} & = \frac{(r_3-r_4)}{r_3 r_4} \, \frac{2 (d-5) m_t^2 s}{(d-6) (4 m_t^2 - s)} \, .
\end{align}
We notice that only \emph{two} of the \emph{sixteen} entries of $\mathbf{C}$ have explicit dependence on $\delta$.\\
Moreover we can compute the intersection numbers:
\begin{align}
\bra{\phi_{2}}\ket{\varphi_{1}} & = \frac{r_1 ((3d-14) r_1 + (d-4) r_2 - (d-5) (r_3 + r_4))}{2 (d-5) (d-4)} \, , \\
\bra{\phi_{2}}\ket{\varphi_{2}} & = \frac{(r_2 - r_1) ((3d-14) ( r_1 + r_2) - (d-5) (r_3 + r_4))}{2 (d-5) (d-4)} \, , \\
\bra{\phi_{2}}\ket{\varphi_{3}} & = \frac{(r_3 - r_2) ((d-4) r_1 + (3d-14) r_2 + (d-5) (r_3 - r_4))}{2 (d-5) (d-4)} \, ,  \\
\bra{\phi_{2}}\ket{\varphi_{4}} & = \frac{(r_4 - r_3) ((d-4) (r_1 + r_2) + (d-5) (r_3 + r_4))}{2 (d-5) (d-4)} \, .
\end{align}
From now on, we reintroduce the explicit values of $\delta$.
\paragraph{ $\bullet$ H+j reduction: $(\delta=0)$.} The final reduction~(\ref{eq:masterdeco:}) reads:
\begin{equation}
I_{1,1,1,1,1,1,1;-1}=c_1 J_1 + c_2 J_2 +c_3 J_3+ c_4 J_4 \, ,    
\end{equation}
with:
\begin{align}
c_1 &= \frac{s \, \big( (3d - 14) ( 2 m_T^2 (m_H^2 - s) - st) + 4 (2d-9) m_t^2 t \big)}{2 (d-4) ( s t + 2 m_t^2 (m_H^2 - s - 2 t))} \, , \\
c_2 &= \frac{2 m_t^2 s ( m_t^2 (m_H^2 - s - 2 t) + st )}{(d-4) ( s t + 2 m_t^2 (m_H^2 - s - 2 t))} \, , \\
c_3 &= \frac{2 m_t^2 \big( s t (s + t - m_H^2) - m_t^2 (s + 2 t - m_H^2)^2 \big)}{(d-4) ( s t + 2 m_t^2 (m_H^2 - s - 2 t))}, \\
c_4 &= \frac{(d-6) \, s t \, (s-4 m_t^2) \, (4 m_t^2 (m_H^2 - s - t) + s t)}{2 (d-5) (d-4) (m_H^2 - s) ( s t + 2 m_t^2 (m_H^2 - s - 2 t))} \, .
\end{align}
The result is in agreement with \textsc{Kira}.
\paragraph{ $ \bullet$ HH production: $(\delta=1)$.} The final reduction~(\ref{eq:masterdeco:}) reads:
\begin{equation}
I_{1,1,1,1,1,1,1;-1}=c_1 J_1 + c_2 J_2 +c_3 J_3+ c_4 J_4 \, ,
\end{equation}
with:
\begin{align}
c_1 &= \frac{s \, \big( (4 (d{-}5) m_H^4 {-} (3d{-}14) s ( 4 m_H^2 {-} s)) t {-} 2 m_t^2 (4 m_H^2 {-} s) ((3d{-}14) (2 m_H^2 {-} s) {+} 2 (9 {-} 2 d) t) \big)}{2 (d-4) (4 m_H^2 - s) (2 m_t^2 (2 m_H^2 - s - 2 t) + s t)} \, , \\
c_2 &= \frac{2 m_t^2 s \, \big( m_t^2 (4 m_H^2 - s) (2 m_H^2 - s - 2 t) + 2 m_H^4 (m_H^2 - t) - s^2 t - m_H^2 s (m_H^2 - 4 t) \big)}{(d-4) (4 m_H^2 - s) (2 m_t^2 (2 m_H^2 - s - 2 t) + s t)} \, , \\
c_3 &= \frac{2 (m_H^4 - m_t^2 (4 m_H^2 - s)) \, (m_t^2 (s + 2 t - 2 m_H^2)^2 - s t (s + t - 2 m_H^2))}{(d-4) (4 m_H^2 - s) (2 m_t^2 (2 m_H^2 - s - 2 t) + s t)}\, , \\
c_4 &= \frac{(d-6) \, (2 m_H^2 - s) \, (4 m_t^2 - s) \, ( s t^2 - 4 m_t^2 (m_H^4 + (s + t - 2 m_H^2) t ) )}{2 (d-5) (d-4) (4 m_H^2 - s) (2 m_t^2 (2 m_H^2 - s - 2 t) + s t)} \, .
\end{align}
The result is in agreement with \textsc{Kira}.\\
\\
Let us mention that no relations such as $\delta^2=\delta$ have been imposed; this means that $\delta$ is effectively acting as an extra mass scale. Thus, taking for instance $\delta \to m^2_{Z} / m^2_{H}$ would correspond to a contribution to $H+Z$ production.

\subsection{Non-Planar Contribution to $H{+}j$ Production}
\label{sec:nonplanarhiggs}

\begin{figure}[H]
    \centering
    \includegraphics[width=0.38\textwidth]{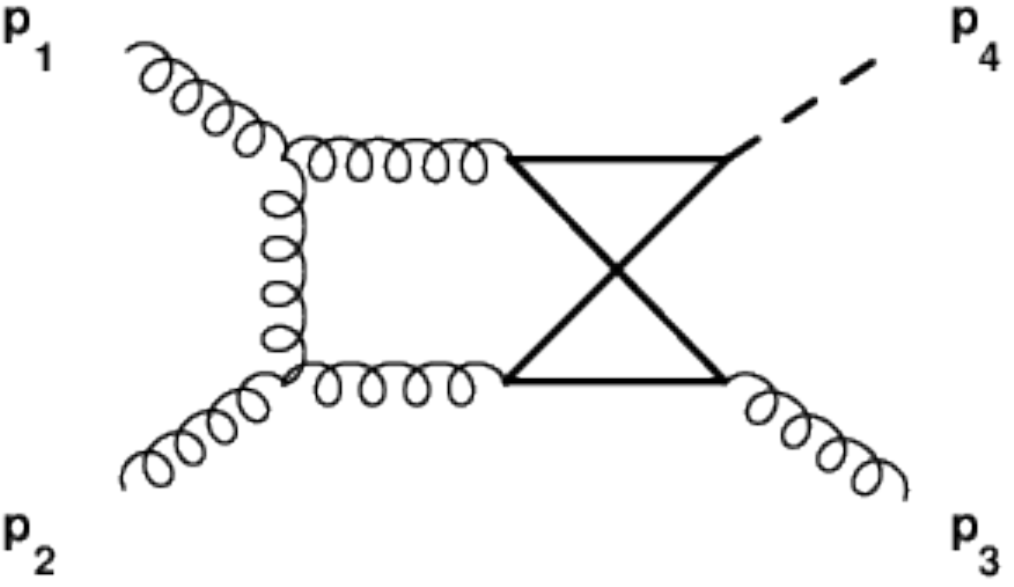}
    \caption{Non-planar $(H{+}j)$-production.}
    \label{fig:Non_planar_2L_Higgs_plus_jet}
\end{figure}
Let us consider one of the the non-planar integral families which contributes to the $H+j$ production \cite{Melnikov:2016qoc} at hadron colliders, depicted in Fig.~\ref{fig:Non_planar_2L_Higgs_plus_jet}.\\
The independent (incoming) momenta are: $\{ p_{1} \, , p_{3} \, , p_{4} \}$ with $p_1^2 = p_2^2 = p_3^2=0$ and $p_{4}^{2}=m_{H}^{2}$. We define the kinematic invariants as $(p_{3}+p_{4})^2=s$ and $(p_{1}+p_{4})^2=t$. 
The denominators are defined as:
\begin{equation}
\begin{split}
& D_{1}=k_1^2 \, , \quad D_{2}=\left(k_1+p_1\right){}^2 \, , \quad D_{3}=\left(k_1-p_3-p_4\right){}^2 \,, \\
& D_{4}=\left(k_2-p_3\right){}^2-m_t^2 \, \quad D_{5}=k_2^2-m_t^2 \, , \quad D_{6}=\left(k_1-k_2\right){}^2-m_t^2 \,, \\
& D_{7}=\left(k_1-k_2-p_4\right){}^2-m_t^2 \,.
\end{split}
\end{equation}
We choose one ISP as:
\begin{equation}
z = D_{8} = (k_1-p_3)^2.
\end{equation}
We use the \lbl form of the Baikov representation and after performing the maximal cut as defined in eq.~(\ref{eq:def:cut-integral}), we obtain
\begin{align}
u &= \frac{\left(-m_H^2+s+t+z\right){}^{d-5} \left(z \left(m_H^2-s-z\right)+4 s m_t^2\right){}^{\frac{d-5}{2}}}{\sqrt{z \left(-m_H^2+s+z\right)}} \, , \\
\omega &= \frac{q_{0}+q_{1} \, z +q_{2} \, z^{2}+ q_{3} \, z^{3} + q_{4} \, z^{4}}{2 z \left(-m_H^2+s+z\right) \left(-m_H^2+s+t+z\right) \left(z \left(-m_H^2+s+z\right)-4 s m_t^2\right)} \, dz \,,
\end{align}
where,
\begin{equation}
\begin{split}
q_{0}=& 4 s m_t^2 (m_H^2-s) (m_H^2-s-t) \, , \\
q_{1}= & 8 m_t^2 s t - (d-6) (m_H^2 - s)^2 (m_H^2 - s - t) + 4 (2d-13) m_t^2 (m_H^2 - s) s \, , \\
q_{2}=& 2 (3d - 17) (m_H^2 - s)^2 - (d-6) (8 m_t^2 s + 3 (m_H^2 - s) t) \, , \\
q_{3}=& 2 (d-6) t - (9d-50) (m_H^2 - s) \, , \\
q_{4}=& 4 d-22 \, , 
\end{split}
\end{equation}
So, we get
\begin{align}
\masters &=4 \, , \\
\mathcal{P}=& \{ 0 , \, m_H^2{-}s , \, \tfrac{1}{2} (m_H^2{-}s{-}\rho), \, \tfrac{1}{2} (m_H^2{-}s{+}\rho) , \, m_H^2{-}s{-}t , \, \infty \}  \, ,
\end{align}
where,
\begin{equation}
\rho=\sqrt{m_H^4 - 2 s m_H^2 + 16 s m_t^2+s^2} \, .
\end{equation}

\paragraph{Mixed Bases} Let us consider the decomposition of $I_{1,1,1,1,1,1,1;-1}= \langle{\phi_{2}}|\mathcal{C}]$. We define the master integrals as: $J_{1}=I_{1,1,1,1,1,1,1;0}=\langle{e_{1}}|\mathcal{C}]$, $J_{2}=I_{1,2,1,1,1,1,1:0}=\langle{e_{2}}|\mathcal{C}]$, $J_{3}=I_{1,1,1,2,1,1,1;0}=\langle{e_{3}}|\mathcal{C}]$ and $J_{4}=I_{1,1,1,1,2,1,1;0}=\langle{e_{4}}|\mathcal{C]}$, where
\begin{equation}
\begin{split}
\hat{e}_{1}=& 1 \, , \\
\hat{e}_{2}= & \frac{(d-5) \left(m_H^4 - m_H^2 (2 s+t+z)+s^2+s (t+z)+2 t z\right)}{s (-m_H^2+s+t+z)^2} \, , \\
\hat{e}_{3} = & \frac{(d-5) (s+z)}{z (m_H^2-s-z) + 4 s m_t^2} \, , \\ 
\hat{e}_{4}= & \frac{(d-5) (m_H^2-z)}{z (m_H^2-s-z)+4 s m_t^2} \, . \\
\end{split}
\end{equation}
Moreover we introduce the following dlog differential-stripped cocycles:
\begin{equation}
\begin{split}
\hat{\varphi}_{1}=& \frac{1}{z}-\frac{1}{-m_H^2+s+z} \, , \\
\hat{\varphi}_{2}= & \frac{1}{-m_H^2+s+z}-\frac{1}{\frac{1}{2} \left(-m_H^2+\rho +s\right)+z} \, , \\
\hat{\varphi}_{3}= & \frac{1}{\frac{1}{2} \left(-m_H^2+\rho +s\right)+z}-\frac{1}{\frac{1}{2} \left(-m_H^2-\rho +s\right)+z} \, , \\
\hat{\varphi}_{4}=& \frac{1}{\frac{1}{2} \left(-m_H^2-\rho +s\right)+z}-\frac{1}{-m_H^2+s+t+z} \, .
\end{split}
\end{equation}
Then we         
compute the $\mathbf{C}$ matrix:
\begin{equation}
\mathbf{C}_{ij}=\bra{e_{i}}\ket{\varphi_{j}} \, , \quad 1 \leq i,j \leq 4,
\end{equation}
with:
\allowdisplaybreaks{
\begin{align}
\bra{e_{1}}\ket{\varphi_{1}} &= \frac{m_H^2-s}{2 (d-5)} \, , \\
\bra{e_{1}}\ket{\varphi_{2}} &= \frac{m_H^2+\rho -s}{20-4 d} \, , \\
\bra{e_{1}}\ket{\varphi_{3}} &= \frac{\rho }{2 (d-5)} \, , \\
\bra{e_{1}}\ket{\varphi_{4}} &= \frac{m_H^2-\rho-s-2 t}{4 (d-5)} \, , \\
\bra{e_{2}}\ket{\varphi_{1}} &= \frac{2 (d-5) \left(m_H^2-s\right)}{(d-6) s} \, , \\ 
\bra{e_{2}}\ket{\varphi_{2}} &= \frac{2 (d-5) \left(s-m_H^2-\rho\right) \left(m_H^2-s-t\right)}{(d-6) s (m_H^2+\rho-s-2 t)} \, , \\
\bra{e_{2}}\ket{\varphi_{3}} &= \frac{2 (d-5) \rho t (m_H^2-s-t)}{(d-6) s \big( t (m_H^2-s-t)+4 s m_t^2 \big)} \,, \\
\bra{e_2}\ket{\varphi_4} &= \frac{(d-5) \big( 4 m_t^2 s (m_H^2 - s - 2 t) -(m_H^2 + \rho - s - 2 t) (m_H^2 - s - t) t \big) }{(d-6) s (4 m_t^2 s + t (m_H^2 - s - t))} \,, \\[-10mm] \nonumber
\end{align}
\begin{align}
\bra{e_3}\ket{\varphi_1} &= 0 \,, &   \bra{e_3}\ket{\varphi_2} &= \frac{\rho - m_H^2 - s}{\rho} \,, \\
\bra{e_3}\ket{\varphi_3} &= \frac{2 (m_H^2 + s)}{\rho} \,, & \bra{e_3}\ket{\varphi_4} &= -\frac{m_H^2 + \rho + s}{\rho} \,, \\
\bra{e_4}\ket{\varphi_1} &= 0 \, , & \bra{e_4}\ket{\varphi_2} &= -\frac{m_H^2 + \rho + s}{\rho} \,, \\
\bra{e_4}\ket{\varphi_3} &= \frac{2(m_H^2+s)}{\rho} \,, & \bra{e_4}\ket{\varphi_4} &= \frac{\rho - m_H^2 - s}{\rho} \,.
\end{align}
}
The other intersection numbers read
\begin{align}
\bra{\phi_{2}}\ket{\varphi_{1}}=& \frac{\left(m_H^2-s\right) \left((d-4) (m_H^2-s)+2 (d-5) t\right)}{4 (d-5) (2 d-9)} \, , \\
\bra{\phi_{2}}\ket{\varphi_{2}}= & \frac{\left(m_H^2+\rho -s\right) \left((14-3 d) (m_H^2-s)+2 (d-5) (\rho -t)\right)}{8 (d-5) (2 d-9)} \, , \\
\bra{\phi_{2}}\ket{\varphi_{3}}=& \frac{\rho  \left((d-4) (m_H^2-s) + 2 (d-5) t\right)}{4 (d-5) (2 d-9)} \, , \\
\bra{\phi_{2}}\ket{\varphi_{4}}= & \frac{\left(-m_H^2+\rho +s+2 t\right) \left((14-3 d) (m_H^2-s) + 2 (d-5) (t-\rho )\right)}{8 (d-5) (2 d-9)} \, .
\end{align}
Then, by means of eq.~(\ref{eq:masterdeco:}), we obtain the following final reduction:
\begin{equation}
I_{1,1,1,1,1,1,1;-1}=c_{1} \, J_{1} + c_{2} \, J_{2} + c_{3} \, J_{3}+ \, c_{4} \, J_{4} \, ,
\label{eq:HPlusJNP2L:red}
\end{equation}
with
\begin{align}
c_1= & \Big(4 (4d-19) m_t^2 s (s-m_H^2) t - (5d-22) (m_H^2 - s)^2 t^2 + 8 (d-5) m_t^2 s t^2 \nonumber \\
& + 2 (2d-9) (m_H^2 - s) t^3 + (d-4) (m_H^2 - s)^2 (2 m_t^2 s + (m_H^2 - s) t) \Big) \Big/ \nonumber \\
& \Big( 2 (2d-9) (m_H^2 - s - 2t) (2 m_t^2 s + t (m_H^2 - s - t)) \Big) \, , \\
c_{2}= & \frac{(d-6) s t \left(m_H^2-s-t\right) \left(4 s m_t^2+t (m_H^2-s-t)\right)}{2 (d-5) (2 d-9) \left(m_H^2-s-2 t\right) \left(2 s
   m_t^2+t (m_H^2-s-t)\right)} \, , \\
c_{3}=& \Big(2 s m_t^2 \left(m_H^4 \left(2 s m_t^2-t (3 s+5 t)\right)+t m_H^2 \left(4 s m_t^2+t (s+3 t)\right)+2 t m_H^6+ \right. \nonumber \\
& \left. s \left(t (s+t) (s+3 t)-2 m_t^2
   \left(s^2+6 s t+4 t^2\right)\right)\right) \Big) /  \Big((2 d-9) \left(m_H^2+s\right) \nonumber \\
& \left(m_H^2-s-2 t\right) ( 2 s m_t^2 + t (m_H^2-s-t)) \Big) \, ,\\
c_{4}= & \Big( 2 s m_t^2 \left(-t \left(-3 s m_H^2 \left(4 m_t^2+s\right)+m_H^6+2 s^2 \left(2 m_t^2+s\right)\right)-3 t^3 \left(m_H^2+s\right)+\right. \nonumber \\
& \left. t^2 \left(s
   m_H^2+4 m_H^4-s \left(8 m_t^2+5 s\right)\right)+2 s m_t^2 \left(s-m_H^2\right) \left(m_H^2+s\right)\right) \Big) \Big/ \Big( (2 d-9) \nonumber \\
   & \left(m_H^2+s\right) \left(m_H^2-s-2 t\right) \left(2 s m_t^2 + t (m_H^2-s-t)\right) \Big) \, .
\label{eq:reduction_H+j_nonplanat_2L_interx}
\end{align}

\paragraph{Checks.} The IBP reduction on the maximal-cut
$D_i=0 \ , i=1,\ldots,7$, and negative powers of $D_8$ and $D_9 = (k_2+p_1)^2$ , 
performed with {\sc KIRA}, 
leaves us with 6 MIs, chosen as,
\begin{align}
J_1 &= I_{1,1,1,1,1,1,1;0,0} \, , & J_2 &= I_{1,2,1,1,1,1,1;0,0} \, , & J_3 &= I_{1,1,1,2,1,1,1;0,0} \, , \nonumber \\
J_4 &= I_{1,1,1,1,2,1,1;0,0} \, , & J_5 &= I_{1,1,1,1,1,2,1;0,0} \, , & J_6 &= I_{1,1,2,1,1,1,1;0,0} \, . 
\end{align}
Adding the IBP identities obtained by reducing, on the same hepta-cut, the 8-denominator integral family built by allowing $D_8$ to appear as a propagator as well, the number of MIs is reduced to 5 - an example of an additional relation (on the maximal-cut) being:
\begin{align}
J_6 &= \frac{10-2d}{s} J_1 + \frac{(2m_t^2 - m_H^2) s + m_H^4}{m_H^2 s} J_3 + \frac{2m_t^2}{s} J_4 + \frac{s(m_H^2-2 m_t^2) + 2 m_H^2 m_t^2}{m_H^2 s} J_5 \,.
 \label{eq:HPlusJNP2L:ER}
\end{align}
Moreover, by applying to $J_5$ the self-similarity transformation, 
\begin{eqnarray}
k_{1} \rightarrow -k_{1}-p_{1}-p_{2} \, , \quad k_{2} \rightarrow -k_{2}+p_{3} \, , \quad p_{1} \leftrightarrows p_{2} \, ,
\end{eqnarray}
(mapping the set of denominators $D_i=0 \ , i=1,\ldots,7$ into itself), together with IBP identities,
we obtain a second relation
\begin{equation}
J_{5}=\frac{s}{m_H^2+s} \, J_{3}  -\frac{m_H^2}{m_H^2+s} \, J_{4}
\label{eq:sec_sym_H+j_nonplanar_2L}
\end{equation}
bringing the number of master integrals on the maximal cut down from 6 to 4, as expected from intersection theory. 
We verified that after using these 2 extra relations, the reduction of $I_{1,1,1,1,1,1,1;-1,0}$ (in terms of $J_i, i =1,\ldots, 4$) with \textsc{Kira} is in perfect agreement with the eq.~\eqref{eq:HPlusJNP2L:red}, and additionally 
we have verified $\masters=4$ with a numerical evaluation of the integrals on the maximal cut combined with the high-precision arithmetic PSLQ algorithm
\cite{Bailey09pslq:an} (80 digits accuracy).

\subsection{Non-planar contribution to $HH$ Production.}
\label{sec:nonplanardoublehiggs}
\begin{figure}[H]
    \centering
    \includegraphics[width=0.38\textwidth]{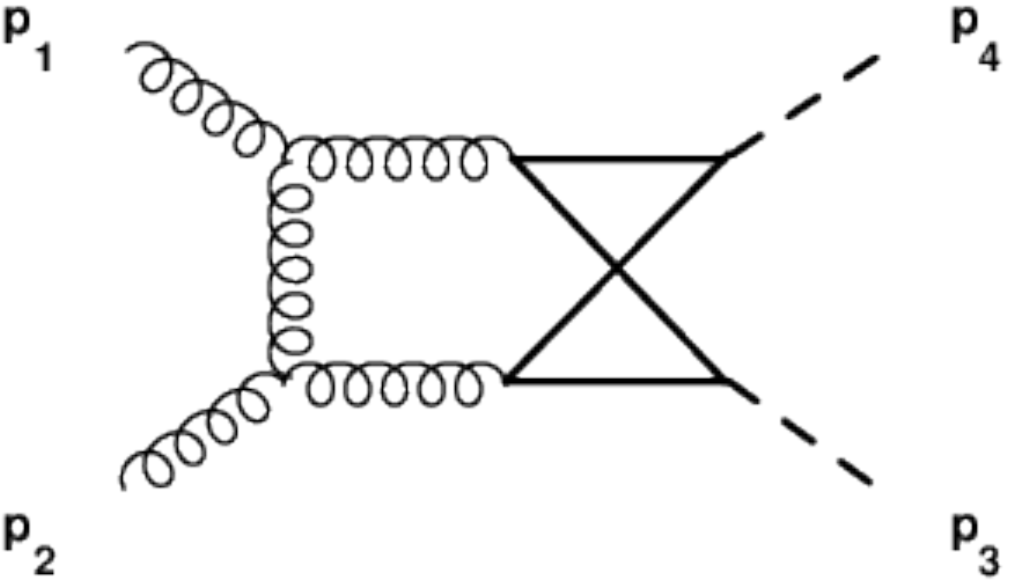}
    \caption{Non-planar $(HH)$-production.}
    \label{fig:Non_planar_2L_Higgs_Higgs}
\end{figure}
Let us consider the non-planar contribution to $HH$ production~\cite{Borowka:2016ypz, Baglio:2018lrj} in Fig.~\ref{fig:Non_planar_2L_Higgs_Higgs}.\\
The denominators are:
\begin{equation}
\begin{split}
& D_{1}=k_1^2 \, , \quad D_{2}=\left(k_1+p_1\right){}^2 \, , \quad D_{3}=\left(k_1-p_3-p_4\right){}^2 \, \\
& D_{4}=\left(k_2-p_3\right){}^2-m_t^2 \, \quad D_{5}=k_2^2-m_t^2 \, , \quad D_{6}=\left(k_1-k_2\right){}^2-m_t^2 \, , \\
& D_{7}=\left(k_1-k_2-p_4\right){}^2-m_t^2 \,
\end{split}
\end{equation}
while the ISP is:
\begin{equation}
z = D_{8} = (k_1-p_3)^2.
\end{equation}
The kinematics is such that: $p_{1}^2 = p_2^2 = 0$, $p_{3}^2=p_{4}^2=m_{H}^2$, $s=(p_{3}+p_{4})^2$, and $t=(p_{1}+p_{4})^2$.\\
The Loop-by-Loop Baikov parametrization gives us:
\begin{align}
u & =\frac{\left(\left(z-r_1\right) \left(z-r_2\right)\right){}^{\frac{d-5}{2}} \left(z-r_5\right){}^{d-5}}{\sqrt{\left(z-r_3\right)
   \left(z-r_4\right)}}, \\
\omega & = \frac{1}{2} \left(\frac{d-5}{z-r_1}+\frac{d-5}{z-r_2}+\frac{2 (d-5)}{z-r_5}+\frac{1}{r_3-z}+\frac{1}{r_4-z}\right) dz,
\end{align}
with:
\begin{align}
r_1 & \to m_H^2-\frac{s}{2} -\frac{1}{2} \sqrt{s \left(-4 m_H^2+16 m_t^2+s\right)}, \\
r_2 & \to m_H^2- \frac{s}{2} +\frac{1}{2}\sqrt{s \left(-4 m_H^2+16 m_t^2+s\right)}, \\
r_3 & \to m_H^2-\frac{s}{2} -\frac{1}{2} \sqrt{s \left(s-4 m_H^2\right)}, \\
r_4 & \to m_H^2-\frac{s}{2} +\frac{1}{2} \sqrt{s \left(s-4 m_H^2\right)},     \\
r_5 & \to 2 m_H^2-s-t.
\end{align}
Therefore we infer:
\begin{equation}
\masters = 4, \quad \mathcal{P}=\{ r_1, r_2, r_3, r_4, r_5, \infty \}.
\end{equation}
\paragraph{Mixed Bases.} We consider the decomposition of $I_{1,1,1,1,1,1,1;-1}=\langle{\phi_{2}}| \mathcal{C}]$ in terms of $J_{1}=I_{1,1,1,1,1,1,1;0}=\langle{e_{1}}|\mathcal{C}]$, $J_{2}=I_{1,2,1,1,1,1,1;0}=\langle{e_{2}}|\mathcal{C}]$, $J_{3}=I_{1,1,1,2,1,1,1;0}=\langle{e_{3}}|\mathcal{C}]$ and $J_{4}=I_{1,1,1,1,2,1,1;0}=\langle{e_{4}}|\mathcal{C}]$, where:
\begin{align}
\hat{e}_{1} & =1, \\
\hat{e}_{2} & =\frac{(d-5) \left(2 m_H^4 - 2 m_H^2 (2s+t+z) + s (s+t+z)+2 t z\right)}{s \left(z-r_5\right){}^2}, \\
\hat{e}_{3} & = \frac{(d-5) \left(m_H^2-s-z\right)}{\left(z-r_1\right) \left(z-r_2\right)}, \\
\hat{e}_{4} & =-\frac{(d-5) \left(m_H^2-z\right)}{\left(z-r_1\right) \left(z-r_2\right)}. 
\end{align}
Moreover, we introduce the following differential-stripped cocycles:
\begin{align}
\hat{\varphi}_{1} & =\frac{1}{z-r_1}-\frac{1}{z-r_2}, \quad \hat{\varphi}_{2}=\frac{1}{z-r_2}-\frac{1}{z-r_3},\\
\hat{\varphi}_{3} & =\frac{1}{z-r_3}-\frac{1}{z-r_4}, \quad \hat{\varphi}_{4}=\frac{1}{z-r_4}-\frac{1}{z-r_5}.
\end{align}
We can compute the $\mathbf{C}$ matrix:
\begin{equation}
\mathbf{C}_{ij}=\bra{e_{i}}\ket{\varphi_{j}}, \quad 1 \leq i,j \leq 4,
\end{equation}
and the intersection numbers:
\begin{equation}
\bra{\phi_{2}}\ket{\varphi_{j}} \quad 1 \leq i \leq 4.
\end{equation}
Finally eq.~ ($\ref{eq:masterdeco:}$) yields:
\begin{equation}
I_{1,1,1,1,1,1,1;-1}= c_1 J_1 +c_2 J_2 + c_3 J_3 +c_4 J_4,
\end{equation}
with:
\begin{align}
c_{1}= & -\frac{8 (d-5) s \, m_t^4 \left(-2 m_H^2+s+2 t\right)}{(2 d-9) \left(4 m_H^2-8 m_t^2-s\right) \left(m_H^4-2 t m_H^2-2 s m_t^2+t
   (s+t)\right)} \nonumber \\
   & \; -\frac{(d-5) s \left(4 m_H^2-s\right) \left(4 m_H^2-16 m_t^2-s\right)}{2 (2 d-9) \left(-2 m_H^2+s+2 t\right) \left(4 m_H^2-8
   m_t^2-s\right)}+m_H^2-\frac{s}{2} \,, \\
c_{2}=& \frac{(d-6) s \left(m_H^4-2 t m_H^2+t (s+t)\right) \left(m_H^4-2 t m_H^2-4 s m_t^2+t (s+t)\right)}{2 (d-5) (2 d-9) \left(-2 m_H^2+s+2 t\right)
   \left(m_H^4-2 t m_H^2-2 s m_t^2+t (s+t)\right)} \,, \\
c_{3}=& \left( 2 s m_t^2 \left(m_H^4 \left(8 m_t^2-s-5 t\right)-4 m_H^2 m_t^2 (s+4 t)+t m_H^2 (3 s+7 t)+m_H^6 \right. \right. \nonumber \\ 
& \left. \left.  +2 m_t^2 \left(s^2+6 s t+4 t^2\right)
 -t (s+t)
   (s+3 t)\right)\right)/ \left((2 d-9) \left(2 m_H^2-s-2 t\right) \right. \nonumber\\    & \left. \left(m_H^4-2 t m_H^2-2 s m_t^2+t (s+t)\right)\right) \,, \\
c_{4}=& \left(2 s m_t^2 \left(m_H^4 \left(8 m_t^2+2 s+13 t\right)+4 m_H^2 m_t^2 (s-4 t)-t m_H^2 (9 s+11 t)-5 m_H^6 \right. \right. \nonumber \\   
 & \left. \left. +m_t^2 \left(8 t^2+4 s t-2 s^2\right)+t
   (s+t) (2 s+3 t)\right) \right) / \left((2 d-9) \left(2 m_H^2-s-2 t\right) \right. \nonumber  \\
& \left. \left(m_H^4-2 t m_H^2-2 s m_t^2+t (s+t)\right) \right) \,.
\end{align}
The result is in agreement with \textsc{Kira}.
\paragraph{Differential Equation in Mixed bases : } We choose the variable $x = s$ with respect to which, we build the system of differential equations here. For simplicity, we choose the following arbitrary phase-space point.
\begin{align}
    t=5 \qquad m_H^2 = 3 \qquad m_t^2 =1
\end{align}
Then we compute the $\sigma (x)$ and build the $\Phi_i(x)$ as defined in Sec.~\ref{sec:DE}. They are
\begin{align}
    \sigma &= \frac{1}{2} \left(\frac{d \left(z-4 \right)}{\left(z-r_1\right) \left(z-r_2\right)}+\frac{2
   d}{z-r_5}+\frac{20}{\left(z-r_1\right) \left(z-r_2\right)}\right) \nonumber \\
   &+\frac{1}{2} \left(\frac{5 z}{\left(r_1-z\right)
   \left(z-r_2\right)}+\frac{z}{\left(r_3-z\right) \left(z-r_4\right)}+\frac{10}{r_5-z}\right)
\end{align}
\begin{align}
    \langle{\Phi_1(x)}| = &-\frac{(d-5) (4-z)}{2 \left(z-r_1\right) \left(z-r_2\right)}+\frac{d-5}{z-r_5}-\frac{z}{2
   \left(z-r_3\right) \left(z-r_4\right)},  \nonumber \\
   \langle{\Phi_2(x)}| = &\left(\frac{(d-5) \left(s^2+s (z+5)-6 (2 s+z+5)+10 z+18\right)}{s \left(z-r_5\right){}^2}\right) \nonumber \\
   & \left(-\frac{(d-5) (4-z)}{2
   \left(z-r_1\right) \left(z-r_2\right)}+\frac{d-5}{z-r_5}-\frac{z}{2 \left(z-r_3\right)
   \left(z-r_4\right)}\right) \nonumber \\
   &-\frac{
   (d-5) \left(s^3-6 \left(3 s^2+3 s(z+5)+z^2+20 z+25\right)\right)}
   {s^2\left(z-r_5\right){}^3} \nonumber \\
   &+\frac{ (d-5) \left(s^2 (z+5)+30 s z+54 (s+z+5)+10 z (z+5)-108\right)} {s^2\left(z-r_5\right){}^3}, \nonumber \\
   \langle{\Phi_3(x)} | =& \left(\frac{(d-5) (-s-z+3)}{\left(z-r_1\right) \left(z-r_2\right)}\right) \left(-\frac{(d-5) (4-z)}{2 \left(z-r_1\right)
   \left(z-r_2\right)}+\frac{d-5}{z-r_5}-\frac{z}{2 \left(z-r_3\right)
   \left(z-r_4\right)}\right) \nonumber \\
   & +\frac{(d-5)
   (3-z)}{\left(z-r_1\right){}^2 \left(z-r_2\right){}^2}, \nonumber \\
   \langle{\Phi_4(x)} | =& \frac{(d-5) (3-z) (z-4)}{\left(z-r_1\right){}^2 \left(z-r_2\right){}^2} \nonumber \\
   &-\left(\frac{(d-5) (3-z)}{\left(z-r_1\right) \left(z-r_2\right)}\right)
   \left(-\frac{(d-5) (4-z)}{2 \left(z-r_1\right) \left(z-r_2\right)}+\frac{d-5}{z-r_5}-\frac{z}{2
   \left(z-r_3\right) \left(z-r_4\right)}\right).
\end{align}
Now, by means of eq.~\eqref{eq:A:diffeq}, we compute the entries of the $\bf{A}$ matrix, which are written below.
\begin{align}
    \mathbf{A_{11}} &= \frac{1}{10} \left(\frac{5 (d-7)}{s}+\frac{8 (d-5)}{s-12}+\frac{21 (d-5)}{3 s+4}-\frac{25 (d-4)}{5 s+4}\right), \nonumber \\
    \mathbf{A_{12}} &=\frac{(d-6) (5 s-12)}{2 (d-5) (s-12) (3 s+4)}, \nonumber \\
    \mathbf{A_{13}} &= -\frac{2}{s-12}, \nonumber \\
    \mathbf{A_{14}} &= -\frac{16 (s-1)}{(s-12) (3 s+4)}, \nonumber \\
    \mathbf{A_{21}} &=\frac{2 (d-5)^2 (15 s+4)}{s (3 s+4) (5 s+4)}, \nonumber \\
    \mathbf{A_{22}} &= -\frac{2 \left(s \left(8 (d+7) s+15 s^2+160\right)+64\right)}{s (s+4) (3 s+4) (5 s+4)}, \nonumber \\
    \mathbf{A_{23}} &= -\frac{4 (d-5)}{s (s+4)}, \nonumber \\
    \mathbf{A_{24}} &= -\frac{4 (d-5) (s (29 s+96)+48)}{s (s+4) (3 s+4) (5 s+4)}, \nonumber \\
    \mathbf{A_{31}} &= \frac{6 (d-5)^2 ((s-2) s+8)}{(s-12) s (s+4) (3 s+4)}, \nonumber \\
    \mathbf{A_{32}} &= \frac{(d-6) (5 s-12)}{(s-12) (s+4) (3 s+4)}, \nonumber \\
    \mathbf{A_{33}} &= \frac{1}{4} \left(\frac{5-d}{s-12}+\frac{3 d-20}{s}+\frac{4 d-21}{s+4}-\frac{10 (d-4)}{5 s+4}\right), \nonumber \\
    \mathbf{A_{34}} &= -\frac{(29 d-148) s^2+48 (d-4)+32 s}{(s-12) s (s+4) (3 s+4)}, \nonumber \\
    \mathbf{A_{41}} &= \frac{2 (d-5)^2 (s (3 s-2)-24)}{(s-12) s (s+4) (3 s+4)}, \nonumber \\
    \mathbf{A_{42}} &= \frac{3 (d-6)}{(s-12) (3 s+4)}, \nonumber \\
    \mathbf{A_{43}} &= \frac{d (12-5 s)+26 s-72}{(s-12) s (s+4)}, \nonumber \\
    \mathbf{A_{44}} &=  \frac{1}{20} \left(\frac{5 (d-10)}{s}-\frac{11 (d-5)}{s-12}+\frac{5 (2 d-11)}{s+4}+\frac{18 (d-5)}{3 s+4}-\frac{50 (d-4)}{5 s+4}\right).
\end{align}
This is in agreement with the result obtained from \textsc{Reduze} and \textsc{Kira}.

\section{Two-Loop Pentabox}\label{sec:pentabox}

In general, the prefactor $K'$ appearing in eq.~($\ref{eq:def:cut-integral}$) can be factorized in a component proportional to the kinematic variables, and another which depends only on the dimensional parameter:
\begin{equation}\label{eq:diffeqK}
    K'=\kappa(d) \, K''(d,v_{ij}) 
\end{equation}
with $v_{ij} \equiv p_i \cdot p_j$.\\
The factor $\kappa(d)$ do not affect neither IBPs nor  differential equations, and therefore we disregard it in the following. \\
From here on we refer to $K''$ as $K$ to ease our notation.

\subsection{Planar Diagram}
\label{sec:pentabox:planar}
\begin{figure}[H]
    \centering
    \includegraphics[width=0.38\textwidth]{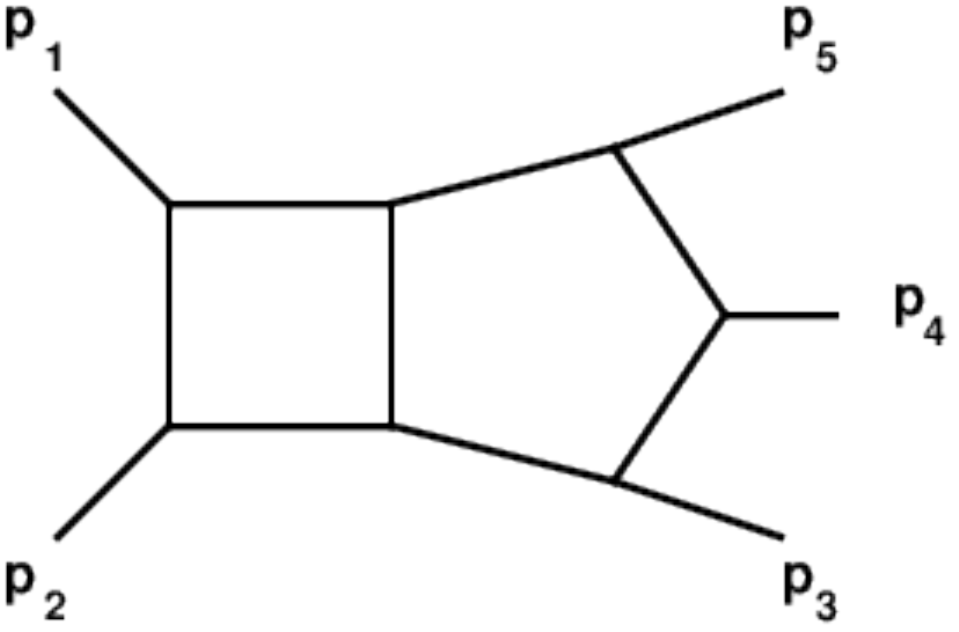}
    \caption{Massless planar pentabox.}
    \label{fig:massless_planar_pentabox}
\end{figure}

Let us consider the massless five-point planar topology at two loops~\cite{Papadopoulos:2015jft, Gehrmann:2015bfy} in Fig.~\ref{fig:massless_planar_pentabox} with the following list of denominators:
\begin{equation}
\begin{split}
& D_{1}=k_1^2 \, , \quad D_{2}=\left(k_1 + p_1\right){}^2 \, , \quad D_{3}=\left(k_1 + p_1 + p_2\right){}^2 \, , \quad D_{4}=\left(k_1 - k_2\right){}^2 \, ,\\
& D_{5}=\left(k_2 + p_1 + p_2\right){}^2 \, , \quad D_{6}=\left(k_2 + p_1 + p_2 + p_3\right){}^2 \, , \\
& D_{7}=\left(k_2 + p_1 + p_2 + p_3 + p_4\right){}^2 \, , \quad D_{8}=k_2^2 \, .
\end{split}
\label{eq:pentaboxprops}
\end{equation}
The ISP considered in the \lbl procedure is:
\begin{equation}
z = D_{9} = \left(k_2 + p_1\right){}^2 \, .
\label{eq:pentaboxISP}
\end{equation}
We find:
\begin{equation}
\label{eq:def:uLBL:pentabox}
\begin{split}
    u = & \; z^{\alpha } (a + z)^{\beta } \left(b + e z + f z^2\right)^{\alpha } \, , \\
    K = & \; v_{34}^{\frac{d-6}{2}} v_{45}^{\frac{d-6}{2}} v_{12}^{\frac{d-6}{2}} \Big(\left(v_{15}-v_{23}\right){}^2 v_{12}^2+2
   \big(\left(v_{15}-v_{23}\right) v_{23} v_{34}+\big(v_{15} \left(v_{23}-v_{15}\right)
    \\
   &  +\left(v_{15}+v_{23}\right)
   v_{34}\big) v_{45}\big) v_{12}+\left(v_{23} v_{34}+\left(v_{15}-v_{34}\right)
   v_{45}\right){}^2\Big){}^{\frac{5-d}{2}} \, ,
   \end{split}
\end{equation}
where:
\begin{align}
        \alpha & = \frac{d}{2}-3 \, ,\qquad \beta = 2-\frac{d}{2} \, ,\qquad a = 2 \, v_{12} \, ,\qquad b =4 \, v_{12} \, v_{15} \, v_{23} \, ,\\
        e & = -2 \, \left(v_{12} \, v_{15} - v_{45} \, v_{15} + v_{12} \, v_{23} - v_{23} \, v_{34} + v_{34} \, v_{45}\right) \, ,\qquad f = v_{12} {-} v_{34} {-} v_{45},
\end{align}
Thus:
\begin{align}
\omega & =\left( \frac{\beta }{a+z}+\frac{\alpha  (b+z (2 e+3 f z))}{z (b+z (e+f z))} \right) \,  dz \, , \\
\masters & =3 \, , \qquad \mathcal{P}=\{0 \, , -a \, , \frac{-\sqrt{e^2-4 b f}-e}{2 f} \, , \frac{\sqrt{e^2-4 b f}-e}{2 f} \, , \infty  \}.
\end{align}
We observe that the combination $e^2-4 b f$ is proportional to the Gram determinant $\Delta=|2 p_i \cdot p_j|$ with $1 \le i,j, \le 4$.
Similar relations hold for the cases studied in the other multileg cases (see also \cite{Mattiazzithesis,Gasparottothesis}) .
\paragraph{Monomial Basis.}
Let us consider the decomposition of $I_{1,1,1,1,1,1,1,1;-3}=\langle{\phi_{4}}| \mathcal{C}]$ in terms of $J_{1}=I_{1,1,1,1,1,1,1,1;0} = \langle \phi_1 | {\cal C} ]$, $J_{2}=I_{1,1,1,1,1,1,1,1;-1} = \langle \phi_2 | {\cal C} ]$, and $J_{3}=I_{1,1,1,1,1,1,1,1;-2} = \langle \phi_3 | {\cal C} ]$.
We can compute the $\mathbf{C}$ matrix:
\begin{equation}
{\bf C}_{ij}=\bra{\phi_{i}}\ket{\phi_{j}} \, , \quad i,j=1,2,3,
\end{equation}
and the intersection numbers:
\begin{equation}
\bra{\phi_{4}}\ket{\phi_{i}} \, , \quad i=1,2,3:    
\end{equation}
Then, eq.~($\ref{eq:masterdeco:}$) yields:
\begin{equation}
I_{1,1,1,1,1,1,1,1;-3}= c_{1} \, J_{1} + c_{2} \, J_{2} + c_{3} \, J_{3},
\end{equation}
with the coefficients:
\begin{equation}
\begin{split}
& c_{1}=-\frac{a (\alpha +1) b}{f (3 \alpha +\beta +4)} \, , \\
& c_{2}=-\frac{2 a (\alpha +1) e+b (\alpha +\beta +2)}{f (3 \alpha +\beta +4)} \, , \\
& c_{3}=-\frac{3 a (\alpha +1) f+e (2 \alpha +\beta +3)}{f (3 \alpha +\beta +4)}\, .
\end{split}
\end{equation}
In agreement with \textsc{Reduze}.
\paragraph{Differential Equations in Monomial Basis.}

We define the variable $x = v_{12}$ with respect to which we build the system of differential equations. In order to do it, one also needs
\begin{align}
\sigma(x) & = 
\partial_x \log ( u ) \\
 & = 
\frac{\left(z{-}2 v_{23}\right) \left(\left(z{-}2 v_{15}\right) ((d{-}6) z{-}4 x)+2 (d{-}4) v_{34} z\right)+2 (d{-}4) v_{45} z
   \left({-}2 v_{15}{+}2 v_{34}{+}z\right)}{2 (2 x+z) \left(\left(z-2 v_{23}\right) \left(x \left(z-2 v_{15}\right)-v_{34}
   z\right)-v_{45} z \left(-2 v_{15}+2 v_{34}+z\right)\right)}\nonumber
\end{align}
and $\langle \Phi_i(x) | = \langle (\partial_x + \sigma(x)) \phi_i |$, which are:
\begin{eqnarray}
\langle \Phi_1(x) | &=&
 \sigma dz,
   \\
\langle \Phi_2(x) | &=&
   z \, \sigma dz,
   \\
\langle \Phi_3(x) | &=&
  z^2 \, \sigma dz.
\end{eqnarray}
Then, according to the procedure described in Sec.~\ref{sec:DE}, we can compute the analytic expression of $\mathbf{A}$. \\
For readability we present the result in a single phase space point:
\begin{equation}
       v_{23} = \frac{1}{2} , \quad
       v_{34} = \frac{1}{3} , \quad
       v_{45} = \frac{1}{5} , \quad
       v_{15} = \frac{1}{7} ,
\end{equation}
The entries of $\mathbf{A}$ read:
\begin{align}
   {\bf A}_{11} & =  \frac{d-6}{x}+\frac{2 (d-4)}{10 x+3}-\frac{735 (d-4)}{134 (21 x-4)}+\frac{22556 d+375 (2263-482 d) x-96589}{67 (5 x
   (375 x-38)+243)} \, , \nonumber \\
   {\bf A}_{12} & = \frac{4 (x (45 x (11830 x{+}21893){-}268886){-}459){-}d (x (45 x (8890 x{+}19219){-}272918){+}5589)}{4 x (10
   x+3) (21 x-4) (5 x (375 x-38)+243)} \, , \nonumber\\
   {\bf A}_{13} & = \frac{7 (d-4) (15 x-8) (615 x-1103)}{2 x (10 x+3) (21 x-4) (5 x (375
   x-38)+243)}\, , \nonumber\\
   {\bf A}_{21} & = -\frac{15 (d-4) x (615 x-1103)}{(10 x+3) (21 x-4) (5 x (375 x-38)+243)} \, , \nonumber\\
   {\bf A}_{22} & = \frac{d-5}{x}-\frac{16 (d-4)}{10
   x+3}-\frac{609 (d-4)}{134 (21 x-4)}+\frac{2 (d-4) (19875 x-21104)}{67 (5 x (375 x-38)+243)} \, , \\
   {\bf A}_{23} & = -\frac{7 (d-4) (25
   x (9 x (125 x-19)+376)-864)}{x (10 x+3) (21 x-4) (5 x (375 x-38)+243)}, \nonumber\\
   {\bf A}_{31} & = \frac{30 (d-4) x (25 x (9 x (125 x-19)+376)-864)}{(10 x+3) (15 x-8) (21 x-4) (5 x (375 x-38)+243)} \, , \nonumber\\
   {\bf A}_{32} & = 
   \frac{1}{1005}\bigg(-\frac{9648 (d-4)}{10 x+3}+\frac{1740 (d-4)}{21 x-4} + 
   \nonumber\\ & \qquad +\frac{-5633500 x+9 d (137875 x-9963)+250128}{5 x (375
   x-38)+243}  -\frac{5360 (3 d-13)}{15 x-8}\bigg) \, , \nonumber\\
   {\bf A}_{33} & = \frac{d{-}6}{2 x}+\frac{15 (d{-}2)}{16{-}30 x}+\frac{14 (d{-}4)}{10
   x{+}3}+\frac{672 (d{-}4)}{67 (21 x{-}4)}+\frac{126 d (257{-}875 x){+}566625 x{-}135893}{67 (5 x (375 x-38)+243)} \, ,\nonumber
\end{align}
in agreement with \textsc{Reduze}.

\subsection{Non-Planar Diagram}\label{sec:pentaboxnpl}
\begin{figure}[H]
    \centering
    \includegraphics[width=0.38\textwidth]{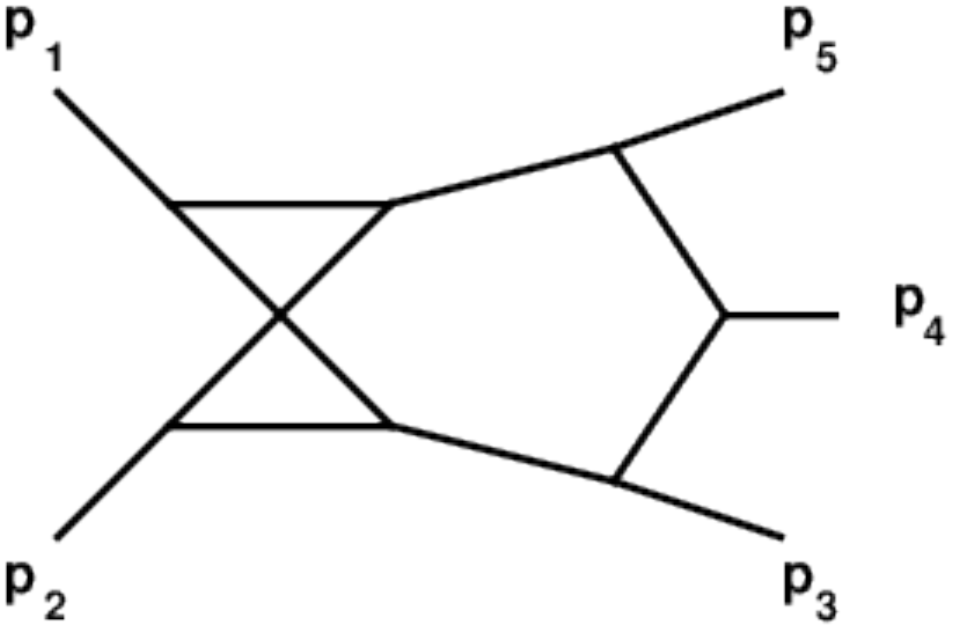}
    \caption{Massless non-planar pentabox.}
    \label{fig:massless_non_planar_pentabox}
\end{figure}
We consider the massless non-planar five-point topology at two loops~\cite{Chicherin:2018mue}, shown in Fig.~\ref{fig:massless_non_planar_pentabox}.\\
We denote the scalar products as: $v_{ij}=p_i \cdot p_j$. We consider the following list of denominators:
\begin{equation}
\begin{split}
& D_{1}=k_1^2 \, , \quad D_{2}=\left(k_1 + p_1\right){}^2 \, , \quad D_{3}=\left(k_1 -k_2 - p_2\right){}^2 \, , \quad D_{4}=\left(k_1 - k_2\right){}^2 \, ,\\
& D_{5}=\left(k_2 + p_1 + p_2\right){}^2 \, , \quad D_{6}=\left(k_2 + p_1 + p_2 + p_3\right){}^2 \, , \\
& D_{7}=\left(k_2 + p_1 + p_2 + p_3 + p_4\right){}^2 \, , \quad D_{8}=k_2^2 \, .
\end{split}
\end{equation}
The ISP considered in the \lbl approach is:
\begin{equation}
z = D_{9} = (k_2 + p_1)^2 \, .
\end{equation}
Performing a maximal cut we find:
\begin{equation}
\begin{split}
   u = &\; \left(z \, (a+z) \, \left(b+e z+f z^2\right)\right)^{\alpha } \, ,\\
   K = &\; v_{12}^{2-\frac{d}{2}} v_{34}^{\frac{d-6}{2}}
   v_{45}^{\frac{d-6}{2}} \left(v_{12}^2
   \left(v_{15}-v_{23}\right){}^2+\left(v_{23}
   v_{34}+\left(v_{15}-v_{34}\right)
   v_{45}\right){}^2+2 v_{12}
   \left(\left(v_{15}-v_{23}\right) v_{23}
   v_{34} \right. \right. \\
   & \left. \left. \; + \left(v_{15}
   \left(v_{23}-v_{15}\right)+\left(v_{15}+v_{
   23}\right) v_{34}\right)
   v_{45}\right)\right){}^{\frac{5-d}{2}} \, .
   \end{split}
\end{equation}
where
\begin{gather}\label{eq:pentabox_coefficient}
        \alpha = \frac{d}{2}-3 \, ,\qquad a = 2 \, v_{12} \, ,\qquad b  =4 \, v_{12} \, v_{15} \, v_{23} \, ,\\
        e = -2 \, \left(v_{12} \, v_{15} - v_{45} \, v_{15} + v_{12} \, v_{23} - v_{23} \, v_{34} + v_{34} \, v_{45}\right) \, ,\qquad f = v_{12} {-} v_{34} {-} v_{45} \,\nonumber
\end{gather}
Then:
\begin{equation}
\omega= \frac{\alpha  \left(2 z (a e+b)+a b+3 z^2 (a
   f+e)+4 f z^3\right)}{z (a+z) (b+z (e+f z))} \, dz \, ,
\end{equation}
and so:
\begin{equation}
\masters=3 \, , \quad \mathcal{P}=\{ 0, -a, \frac{-\sqrt{e^2-4 b f}-e}{2 f}, \frac{\sqrt{e^2-4 b f}-e}{2 f} , \infty \}. 
\end{equation}

\paragraph{Monomial Basis.}
The MIs can be chosen as: $J_{1}=I_{1,1,1,1,1,1,1,1;0}=\langle \phi_1 | {\cal C} ]$, $J_{2}=I_{1,1,1,1,1,1,1,1;-1}=\langle \phi_2 | {\cal C} ]$, and $J_{3}=I_{1,1,1,1,1,1,1,1;-2}=\langle \phi_3 | {\cal C} ]$.\\
Let us consider the decomposition of $I_{1,1,1,1,1,1,1,1;-3}=\langle \phi_4 | {\cal C} ]$ in this basis.
We can compute the ${\bf C}$ matrix,
\begin{equation}
{\bf C}_{ij}=\bra{\phi_{i}}\ket{\phi_{j}} \, , \quad i,j=1,2,3 \, ,
\end{equation}
and the additional intersection numbers:
\begin{equation}
\bra{\phi_{4}}\ket{\phi_{i}} \, , \quad i=1,2,3 \, ,
\end{equation}
and then, eq.~\eqref{eq:masterdeco:} yields:
\begin{equation}
I_{1,1,1,1,1,1,1,1,-3}= c_{1} \, J_{1} + c_{2} \, J_{2} + c_{3} \, J_{3} \, ,
\end{equation}
with:
\begin{equation}
c_{1}=-\frac{a b}{4 f} \, ,\qquad c_{2}=-\frac{a e+b}{2f} \, ,\qquad c_{3}=-\frac{3 (a f+e)}{4 f}\,
\end{equation}
in agreement with \textsc{Reduze}.

\paragraph{Differential Equations in Monomial Basis.}
Let us define the variable $x = v_{12}$ with respect to which we build the system of differential equations. Then we consider:
\begin{equation}
\begin{split}
\sigma(x) & = 
\partial_x \log ( u ) \\
 & = 
\frac{(d-6) \left(\left(z-2 v_{23}\right)
   \left(\left(4 v_{12}+z\right) \left(z-2
   v_{15}\right)-2 v_{34} z\right)-2 v_{45} z
   \left(-2 v_{15}+2 v_{34}+z\right)\right)}{2
   \left(2 v_{12}+z\right) \left(\left(z-2
   v_{23}\right) \left(v_{12} \left(z-2
   v_{15}\right)-v_{34} z\right)-v_{45} z
   \left(-2 v_{15}+2 v_{34}+z\right)\right)}
   \end{split}
\end{equation}
and $\{ \langle \Phi_i(x) | \}_{i=1,2,3}$ are given by:
\begin{eqnarray}
\langle \Phi_1(x) | &=&
 \sigma dz\, ,
   \\
\langle \Phi_2(x) | &=&
   z \, \sigma dz\, ,
   \\
\langle \Phi_3(x) | &=&
  z^2 \, \sigma dz\, .
\end{eqnarray}
Then, according to the procedure described in Sec.~\ref{sec:DE}, we can compute the analytic expression of $\mathbf{A}$. \\
For readability we present the result in a single phase space point:
\begin{equation}
    \begin{split}
       v_{23} = \frac{1}{2} , \quad
       v_{34} = \frac{1}{3} , \quad
       v_{45} = \frac{1}{5} , \quad
       v_{15} = \frac{1}{7} . \\
    \end{split}
\end{equation}
The entries of ${\bf A}$ become:
\begin{align}
   {\bf A}_{11} & =   \frac{d-6}{x}+\frac{2 (d-4)}{10
   x+3}-\frac{735 (d-4)}{134 (21
   x-4)}+\frac{22556 d+375 (2263-482 d)
   x-96589}{67 (5 x (375 x-38)+243)} \, , \nonumber \\
   {\bf A}_{12} & = \frac{42 (x (15 x (9630
   x{+}6623){-}33682){+}4167){-}d (x (45 x (26390
   x{+}20249){-}340738){+}40959)}{4 x (10 x{+}3) (21
   x{-}4) (5 x (375 x{-}38){+}243)} \, ,\nonumber \\
   {\bf A}_{13} & = \frac{7 (2
   d-9) (15 x-8) (615 x-1103)}{2 x (10 x+3)
   (21 x-4) (5 x (375 x-38)+243)} \, ,\nonumber \\
   {\bf A}_{21} & = -\frac{15 (d-4) x (615 x-1103)}{(10 x+3) (21
   x-4) (5 x (375 x-38)+243)} \, ,\nonumber \\
   {\bf A}_{22} & = \frac{4914-861
   d}{536-2814 x}+\frac{d-5}{x}+\frac{84-20
   d}{10 x+3}+\frac{-68225 d+375 (65 d-219)
   x+298917}{67 (5 x (375 x-38)+243)} \, ,\nonumber \\
   {\bf A}_{23} & = -\frac{7 (2 d-9) (25 x (9 x (125
   x-19)+376)-864)}{x (10 x+3) (21 x-4) (5 x
   (375 x-38)+243)} \, , \\
   {\bf A}_{31} & =
    \frac{30 (d-4) x (25 x (9 x (125
   x-19)+376)-864)}{(10 x+3) (15 x-8) (21 x-4)
   (5 x (375 x-38)+243)} \, ,\nonumber
   \\
   {\bf A}_{32} & = 
   \frac{450 x (3 x (45 x (500 x-227)+8107)-8504)}{67 (5 x (375
   x-38)+243)}\nonumber \\
   &\quad + \frac{-3 d (5 x (225
   x (x (2250x-1001)+763)-59368)+24192)+290304}{(10 x+3)
   (15 x-8) (21 x-4) (5 x (375 x-38)+243)} \, ,\nonumber \\
   {\bf A}_{33} & = \frac{45 (d{-}4)}{16{-}30 x}-\frac{d{-}4}{2
   x}+\frac{14 (2 d{-}9)}{10 x{+}3}+\frac{672 (2
   d{-}9)}{67 (21 x{-}4)}+\frac{d (58399{-}94875
   x){+}489750 x{-}265978}{67 (5 x (375
   x-38)+243)} \, .\nonumber
\end{align}
in 
agreement with \textsc{Reduze}.

\section{Multileg and Massive Cases}\label{sec:multileg}
\begin{figure}[H]
\centering
\subfloat[][]
{\includegraphics[width=0.3\textwidth]{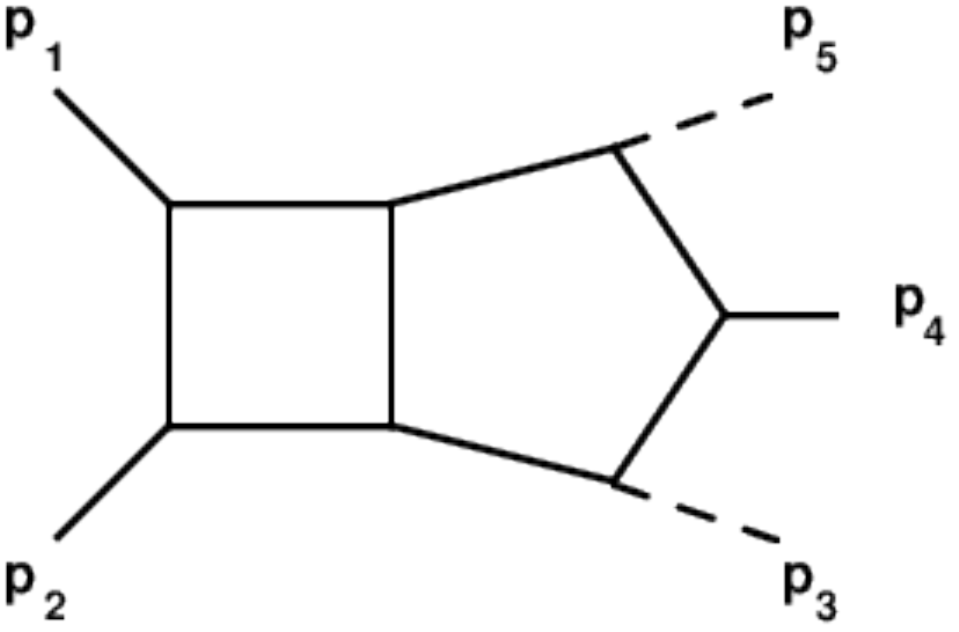}} \quad
\subfloat[][]
{\includegraphics[width=0.3\textwidth]{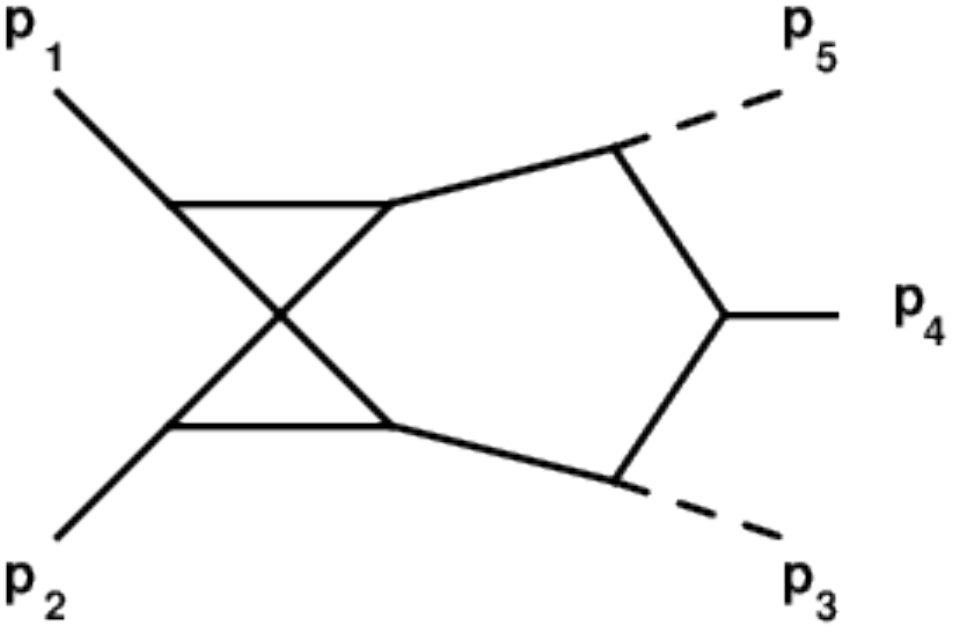}}
\\
\subfloat[][]
{\includegraphics[width=0.3\textwidth]{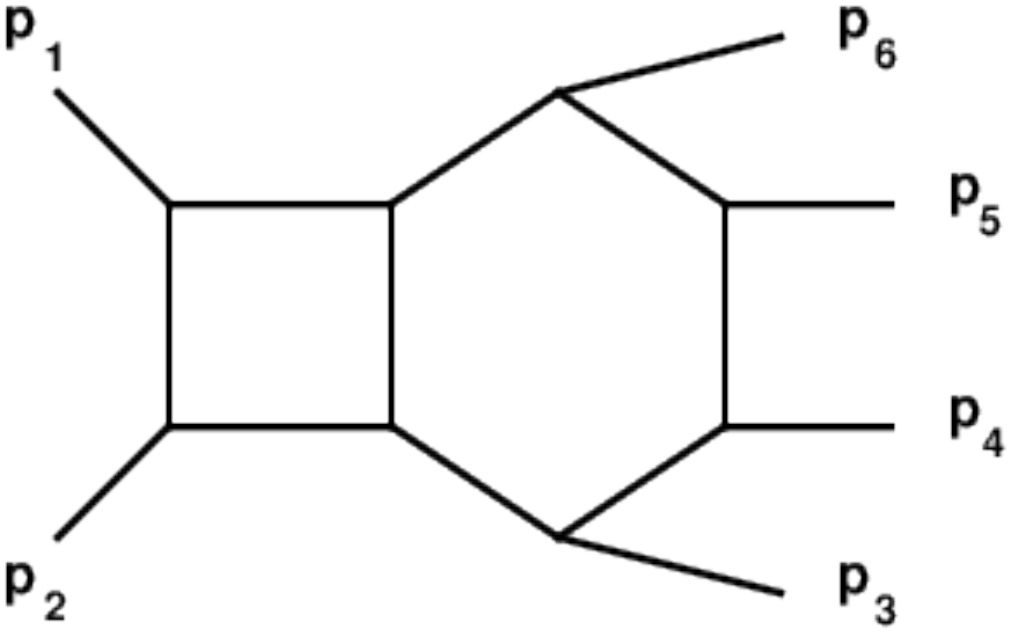}} \quad
\subfloat[][]
{\includegraphics[width=0.3\textwidth]{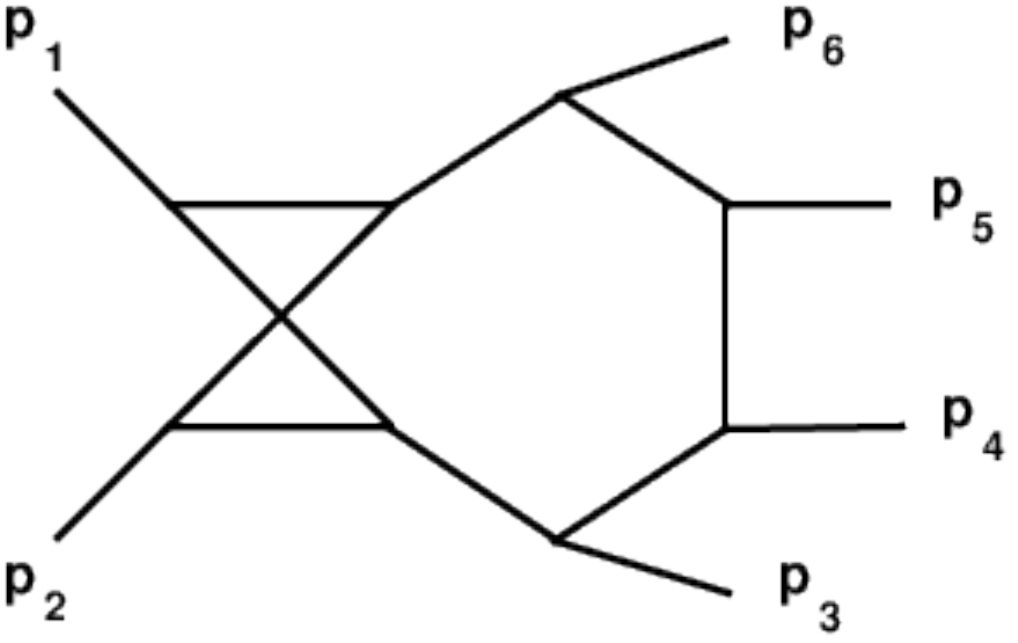}} 
\caption{Planar and non-planar pentabox with two external masses (1,2), as well as planar and non-planar hexagon-box (3,4).}
\label{fig:subfig}
\end{figure}

Let us now study how the polynomial $u$ changes when we compute Feynman integrals such as those in Sec.~\ref{sec:pentabox}, but with massive external legs, or with more massless external legs\footnote{We assume the number of space-time dimensions $d$ to not be smaller  than the number of independent external momenta.} as shown in Fig.~\ref{fig:subfig}. \\

\paragraph{$\bullet$ Case (1), planar massive pentabox:}
the external kinematic is defined by $p_3^2=p_5^2=m^2$ and $p_i^2=0$ for $i=1,2,4$, with the denominators
\begin{equation}
\begin{split}
& D_{1}=k_1^2 \, , \quad D_{2}=\left(k_1 + p_1\right){}^2 \, , \quad D_{3}=\left(k_1 + p_1 + p_2\right){}^2 \, , \quad D_{4}=\left(k_1 - k_2\right){}^2 \, ,\\
& D_{5}=\left(k_2 + p_1 + p_2\right){}^2 \, , \quad D_{6}=\left(k_2 + p_1 + p_2 + p_3\right){}^2 \, , \\
& D_{7}=\left(k_2 + p_1 + p_2 + p_3 + p_4\right){}^2 \, , \quad D_{8}=k_2^2 \, .
\end{split}
\end{equation}

\paragraph{$\bullet$ Case (2), non-planar massive pentabox:}
the external kinematic is defined by $p_3^2=p_5^2=m^2$ and $p_i^2=0$ for $i=1,2,4$, with the denominators 
\begin{equation}
\begin{split}
& D_{1}=k_1^2 \, , \quad D_{2}=\left(k_1 + p_1\right){}^2 \, , \quad D_{3}=\left(k_1 - k_2 - p_2\right){}^2 \, , \quad D_{4}=\left(k_1 - k_2\right){}^2 \, ,\\
& D_{5}=\left(k_2 + p_1 + p_2\right){}^2 \, , \quad D_{6}=\left(k_2 + p_1 + p_2 + p_3\right){}^2 \, , \\
& D_{7}=\left(k_2 + p_1 + p_2 + p_3 + p_4\right){}^2 \, , \quad D_{8}=k_2^2 \, .
\end{split}
\end{equation}

\paragraph{$\bullet$ Case (3),  planar massless hexagon-box:}
the external kinematic is defined by $p_i^2=0$ for $i=1,\dots , 5$, with the denominators 
\begin{equation}
\begin{split}
& D_{1}=k_1^2 \, , \quad D_{2}=\left(k_1 + p_1\right){}^2 \, , \quad D_{3}=\left(k_1 + p_1 + p_2\right){}^2 \, , \quad D_{4}=\left(k_1 - k_2\right){}^2 \, ,\\
& D_{5}=\left(k_2 + p_1 + p_2\right){}^2 \, , \quad D_{6}=\left(k_2 + p_1 + p_2 + p_3\right){}^2 \, , \\
& D_{7}=\left(k_2 + p_1 + p_2 + p_3 + p_4\right){}^2 \, , \quad D_{8}=k_2^2 \, , \\
& D_{9}=\left(k_2 + p_1 + p_2 + p_3 + p_4 + p_5\right){}^2\, .
\end{split}
\end{equation}

\paragraph{$\bullet$ Case (4), non-planar massless hexagon-box:}
the external kinematic is defined by $p_i^2=0$ for $i=1,\dots , 5$, with the denominators 
\begin{equation}
\begin{split}
& D_{1}=k_1^2 \, , \quad D_{2}=\left(k_1 + p_1\right){}^2 \, , \quad D_{3}=\left(k_1 - p_1 - p_2\right){}^2 \, , \quad D_{4}=\left(k_1 - k_2\right){}^2 \, ,\\
& D_{5}=\left(k_2 + p_1 + p_2\right){}^2 \, , \quad D_{6}=\left(k_2 + p_1 + p_2 + p_3\right){}^2 \, , \\
& D_{7}=\left(k_2 + p_1 + p_2 + p_3 + p_4\right){}^2 \, , \quad D_{8}=k_2^2 \, , \\
& D_{9}=\left(k_2 + p_1 + p_2 + p_3 + p_4 + p_5\right){}^2\, .
\end{split}
\end{equation}
The only ISP appearing using the \lbl procedure in these four cases is:
\begin{equation}
z = (k_2 + p_1)^2.
\end{equation}
In all the $4$ cases show in Fig.~\ref{fig:subfig}, the \lbl Baikov polynomials on the maximal cut give the common expression,
\begin{equation}
    u =z^{\alpha_i } \, (a_i + z)^{\beta_i } \, \left(b_i + e_i \, z + f_i \, z^2\right)^{\gamma_i };
\end{equation}
\paragraph{$\bullet$ Case (1), planar massive pentabox:} 
\begin{align*}
        \alpha_1 = & \gamma_1 = \frac{d}{2}-3 \, , \qquad \beta_1 =  2-\frac{d}{2} \, ,\qquad a_1 = 2 \, v_{12} \, ,\\
        b_1 = & 2 v_{12} \left(m^2 \left(v_{15}-v_{23}\right)-v_{34} \left(m^2+2
   v_{23}\right)\right) \left(m^2 \left(v_{45}-v_{23}\right)+v_{15} \left(m^2+2
   v_{45}\right)\right) \, , \\
        e_1 = \, & 2 v_{12} \left(m^2 \left(v_{15}-v_{23}\right) v_{34}+v_{45}
   \left(m^2 \left(v_{23}-v_{15}\right)+2 v_{34}
   \left(m^2+v_{15}+v_{23}\right)\right)\right) \, \\
   & -2 \left(v_{23}
   v_{34}+\left(v_{15}-v_{34}\right) v_{45}\right) \left(m^2 v_{45}+v_{34} \left(m^2+2
   v_{45}\right)\right) \, , \\
        f_1 = \, & v_{34}^2 \left(m^2+2 v_{45}\right)+2 v_{45} v_{34}
   \left(m^2-v_{12}+v_{45}\right)+m^2 v_{45}^2 \, ,
\end{align*}
\paragraph{$\bullet$ Case (2), non-planar massive pentabox:}
\begin{align*}
        \alpha_2 = & \beta_2 = \gamma_2 = \frac{d}{2}-3 \, ,\qquad a_2 =  2 \, v_{12} \, ,\\
        b_2 = &  2 v_{12} \left(m^2 \left(v_{15}-v_{23}\right)-v_{34} \left(m^2+2
   v_{23}\right)\right) \left(m^2 \left(v_{45}-v_{23}\right)+v_{15} \left(m^2+2
   v_{45}\right)\right) \, ,\\
        e_2 = & 2 v_{12} \left(m^2 \left(v_{15}-v_{23}\right) v_{34}+v_{45}
   \left(m^2 \left(v_{23}-v_{15}\right)+2 v_{34}
   \left(m^2+v_{15}+v_{23}\right)\right)\right) \\
   & -2 \left(v_{23}
   v_{34}+\left(v_{15}-v_{34}\right) v_{45}\right) \left(m^2 v_{45}+v_{34} \left(m^2+2
   v_{45}\right)\right) \, , \\
        f_2 = \, & v_{34}^2 \left(m^2+2 v_{45}\right)+2 v_{45} v_{34}
   \left(m^2-v_{12}+v_{45}\right)+m^2 v_{45}^2 \, .
   \end{align*}
In the cases concerning $6$ external legs, the number of independent kinematic variables grows a lot and the expressions for the constants become rather heavy.\\
We present them evaluated at the phase space point:
\begin{align}
   v_{12} = & 1 \, , \quad v_{13} = \frac{1}{2} \, , \quad v_{14} =  \frac{1}{3} \, , \quad v_{15} = \frac{1}{5} \, , \quad v_{23} = \frac{1}{7} \, , \nonumber \\
   \quad v_{24} = & \frac{1}{11} \, , \quad v_{25} = \frac{1}{13}, \quad v_{34} = \frac{1}{17} \, , \quad v_{35} = \frac{1}{19} \, .
\end{align}
\paragraph{$\bullet$ Case (3),  planar massless hexagon-box:}
\begin{gather}
        \alpha_3 =  \frac{d-6}{2}, \qquad \beta_3 = 2-\frac{d}{2}, \qquad \gamma_3 = \frac{d-7}{2} \, ,\qquad a_3 = 2 \, ,\nonumber\\
        b_3 =  \frac{619142135915328239231}{1450900103219383716900} \, ,\qquad e_3 = \,  -\frac{7218174020286869797}{2586274693795692900} \, ,\\
        \qquad f_3 = \,  -\frac{47636820419356249}{18440461274835600} .\nonumber
   \end{gather} 
\paragraph{$\bullet$ Case (4), non-planar massless hexagon-box:}
\begin{gather}
        \alpha_4 = \beta_4 = \frac{d}{2}-3 \, ,\qquad \gamma_4 = \frac{d-7}{2}, \qquad a_4 = 2 \, ,\nonumber\\
        b_4 =  \frac{619142135915328239231}{1450900103219383716900} \, ,\qquad e_4 = \,  -\frac{7218174020286869797}{2586274693795692900} \, , \\
        f_4 = \, \-\frac{47636820419356249}{18440461274835600}. \nonumber
   \end{gather}
We define:
\begin{equation}
\omega_i= \left(\frac{\beta_i }{a_i+z}+\frac{\gamma_i \, (e_i+2 \, f_i \, z)}{b_i+z \, (e_i+f_i \, z)}+\frac{\alpha_i }{z} \, \right) dz .
\end{equation}
So we get:
\begin{equation}
\masters=3 \, , \quad \mathcal{P}=\{ 0 \, , -a_i \, , \frac{-\sqrt{e_i^2-4 b_i f_i}-e_i}{2 f_i} \, , \frac{\sqrt{e_i^2-4 b_i f_i}-e_i}{2 f_i} \, , \infty \} \, .   
\end{equation}

\paragraph{Monomial Basis.}
The MIs are chosen to be: $J_{1}=I_{1,1,1,1,1,1,1,1;0}=\langle \phi_1 | {\cal C} ]$, $J_{2}=I_{1,1,1,1,1,1,1,1;-1}=\langle \phi_2 | {\cal C} ]$ and $
J_{3}=I_{1,1,1,1,1,1,1,1;-2}=\langle \phi_3 | {\cal C} ]$. \\
Let us consider the decomposition of $I_{1,1,1,1,1,1,1,1;-3}= \langle{\phi_{4}}|\mathcal{C}]$ in terms of this basis.\\
We compute the  the ${\bf C}$ matrix:
\begin{equation}
{\bf C}_{ij}=\bra{\phi_{i}}\ket{\phi_{j}}, \quad i,j=1,2,3,
\end{equation}
and the intersection numbers:
\begin{equation}
\bra{\phi_{4}}\ket{\phi_i}, \quad i=1,2,3,
\end{equation}
and the eq.~(\ref{eq:masterdeco:}) gives:
\begin{equation}\label{eq:multileg}
I_{1,1,1,1,1,1,1,1;-3}= c_{1} J_{1} + c_{2}J_{2} + c_{3}J_{3},
\end{equation}
with:
\begin{equation}
\begin{split}
& c_{1}=-\frac{a_i \, (\alpha_i +1) \, b_i}{f_i \, (\alpha_i +\beta_i +2 \gamma_i +4)} \, , \\
& c_{2}=-\frac{a_i \, e_i \, (\alpha_i
   +\gamma_i +2)+ b_i \, (\alpha_i +\beta_i +2)}{f_i \, (\alpha_i +\beta_i +2 \gamma_i +4)} \, , \\
& c_{3}=-\frac{a_i \, f_i
   \, (\alpha_i +2 \gamma_i +3)+e_i \, (\alpha_i +\beta_i +\gamma_i +3)}{f_i \, (\alpha_i +\beta_i +2 \gamma_i
   +4)}\, ,
\end{split}
\end{equation}
in agreement with \textsc{Reduze}, for all four cases.

\subsection{Arbitrary Number of External Legs Case}\label{sec:multileg:all}
\begin{figure}[H]
    \centering
    \includegraphics[width=0.38\textwidth]{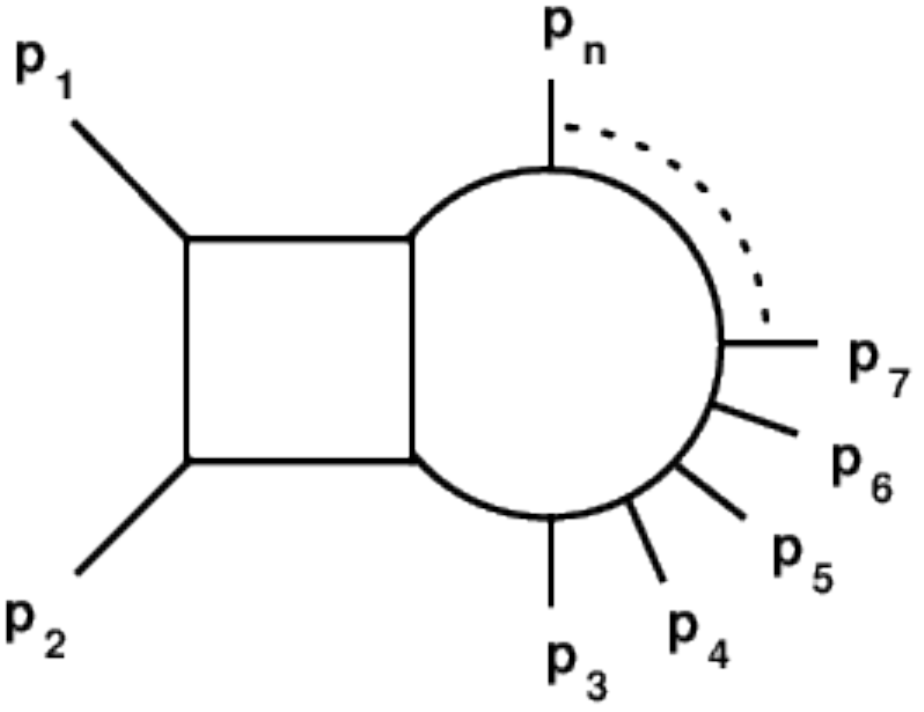}
    \caption{Multileg generalization of the topologies portrayed in Fig.~\ref{fig:subfig}.}
    \label{fig:alllegs}
\end{figure}

The direct generalization of the cases discussed above is portrayed in Fig.~\ref{fig:alllegs} . \\
Here, choosing the following list of denominators:
\begin{equation}
\begin{split}
& D_{1}=k_1^2 \, , \quad D_{2}=\left(k_1 + p_1\right){}^2 \, , \quad D_{3}=\left(k_1 + p_1 + p_2\right){}^2 \, , \quad D_{4}=\left(k_1 - k_2\right){}^2 \, ,\\
& D_{5}= \left(k_2 + p_1 + p_2 \right){}^2 \, , \quad D_{5+j}= \Big(k_2 + p_1 + p_2 + \sum_{r=1}^{j} p_{2+r} \Big)^2\\
\end{split}
\end{equation}
for a diagram with a number of external legs $E$ equal to $E=4+j$ with $j>1$, and choosing as ISP
\begin{equation}
z = D_{8+j}= (k_2 + p_1)^2.
\end{equation}
the \lbl Baikov polynomials on the maximal cut have the same structure as the previous 5 and 6 point cases, where at least one of $p_1$ and $p_2$ is massless:
\begin{equation}
    u =z^{\alpha_i } \, (a_i + z)^{\beta_i } \, \left(b_i + e_i \, z + f_i \, z^2\right)^{\gamma_i };
\end{equation}
Therefore the reduction derived in eq.~(\ref{eq:multileg}) remains valid for any number of external legs.\\
This result has been checked numerically with {\sc Reduze} up to 8 external legs.

\section{Arbitrary Loop Examples}
\label{sec:arbitraryloop}

\subsection{Planar Rocket Diagram for $H{+}j$: $(3{+}2n)$-Loop Case}\label{sec:planar3Lbeyond}
In this section we consider certain higher-loop topologies that contribute to the Higgs+jet production. As done in Sec. \ref{sec:pentabox}, we define $K$ as described in and around eq.~\eqref{eq:diffeqK}.

\begin{figure}[H]
    \centering
    \includegraphics[width=0.4\textwidth]{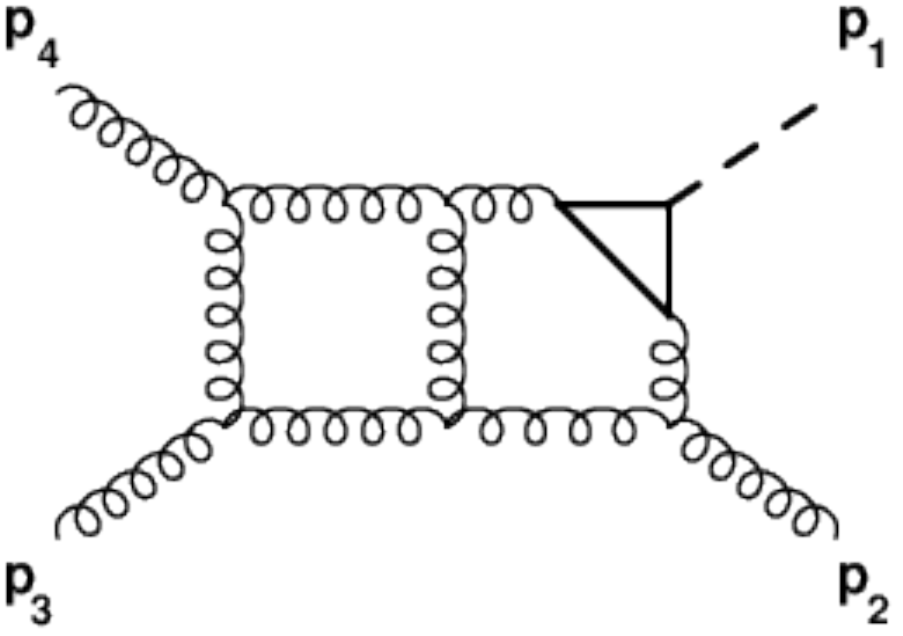}
    \caption{Planar three-loop diagram contributing to $H{+}j$ production.}
    \label{fig:3LoopHPlusJet}
\end{figure}

\begin{figure}[H]
    \centering
    \includegraphics[width=0.4\textwidth]{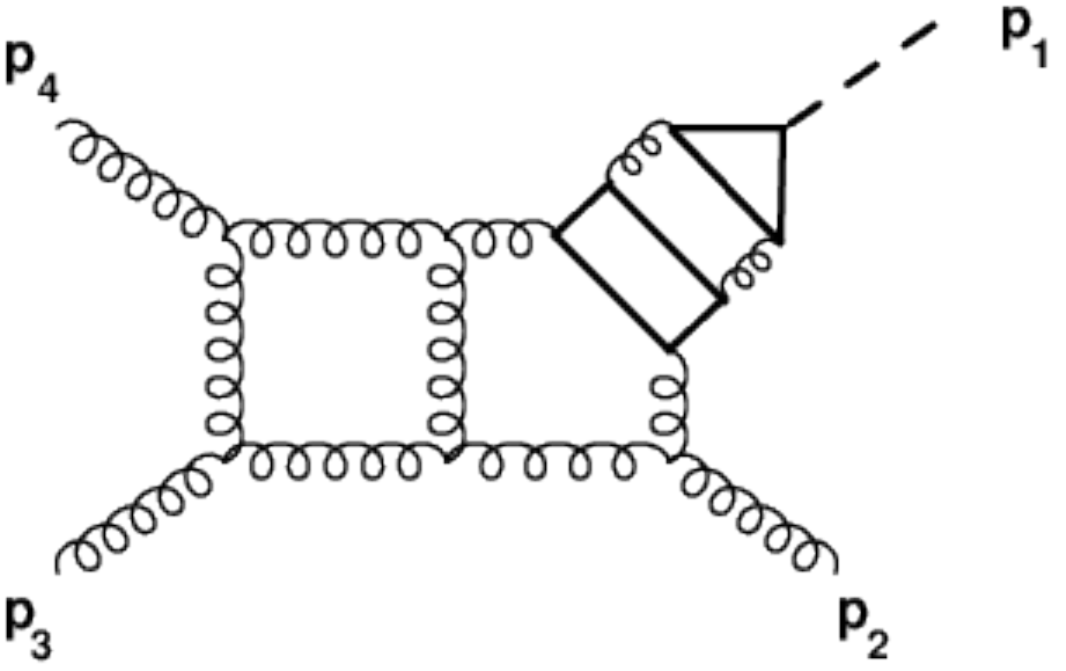}
    \caption{Planar five-loop diagram contributing to $H{+}j$ production.}
    \label{fig:5LoopHPlusJet}
\end{figure}

\begin{figure}[H]
    \centering
    \includegraphics[width=0.5\textwidth]{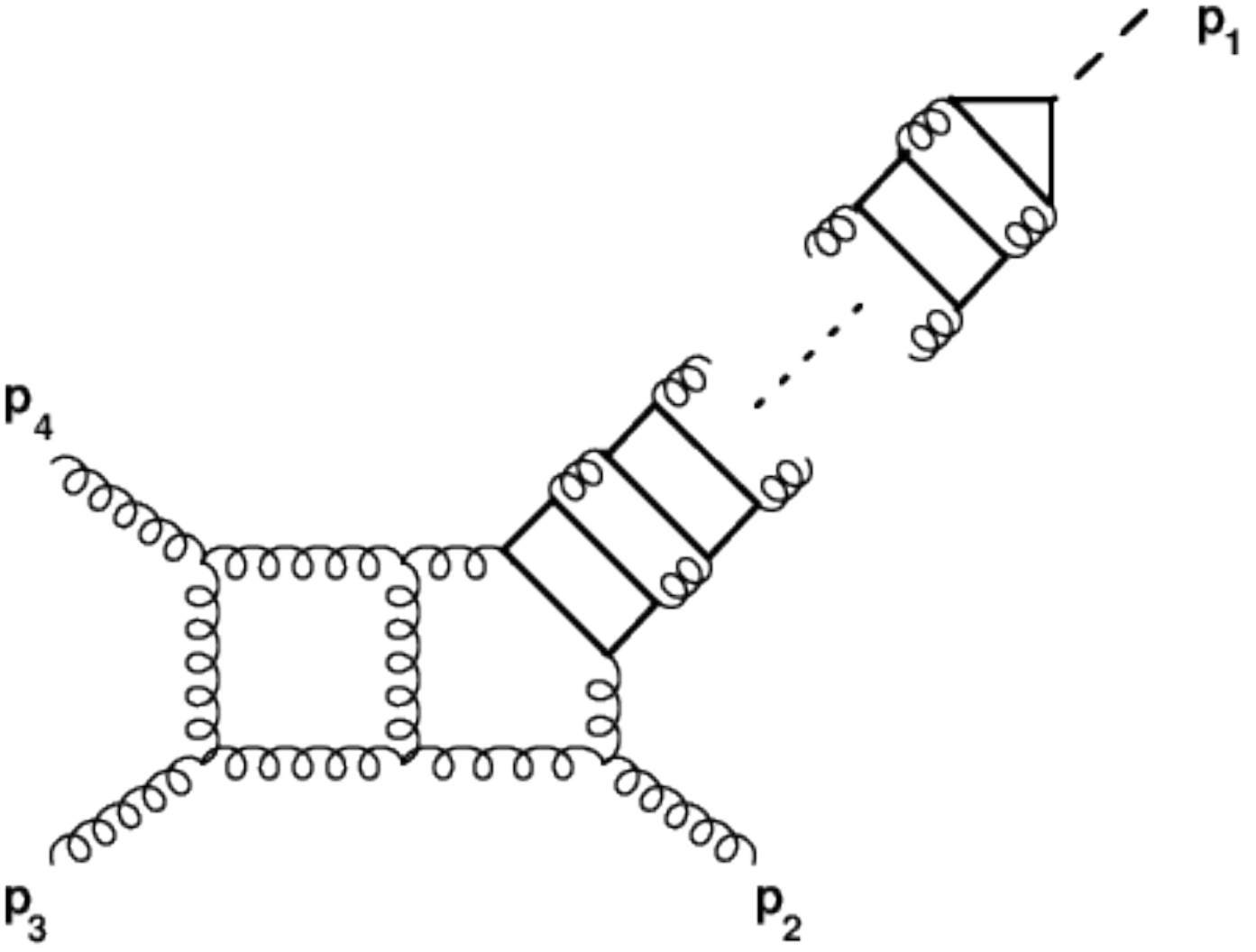}
    \caption{Planar box-rocket diagram contributing to $H{+}j$ production.}
    \label{fig:AllLoopHPlusJet}
\end{figure}

Let us consider a specific planar integral sector for Higgs+jet production from gluon fusion at three loops, depicted in Fig.~\ref{fig:3LoopHPlusJet}. The kinematics is such that: $p_1^2=m_H^2$, $p_i^2=0$ with $i=2,3$, $s=(p_1+p_2)^2$, $t=(p_2+p_3)^2$ and $(p_1+p_2+p_3)^2=0$.\\
The denominators are chosen as: 
\begin{align}
D_1&=k_1^2-m_t^2, \quad
D_2=\left(k_1-p_1\right){}^2-m_t^2, \quad
D_3=\left(k_1-k_2\right){}^2-m_t^2,\nonumber\\
D_4 &= \left(k_2-p_1\right){}^2, \quad
D_5 =\left(k_2-p_1-p_2\right)^2, \quad
D_6 = k_2^2, \quad
D_7 = \left(k_2-k_3\right){}^2, \\
D_8 &= \left(k_3-p_1-p_2\right){}^2, \quad
D_9 =  \left(k_3-p_1-p_2-p_3\right){}^2, \quad
D_{10} = k_3^2,\nonumber
\end{align}
while the ISP is:
\begin{equation}
z = D_{11} = (k_3+p_1)^2.
\end{equation}
The Loop-by-Loop Baikov representation on the maximal cut, gives:
\begin{align}
u & = \left(z-2 m_H^2\right){}^{\frac{d}{2}-3} \left(m_H^2+s-z\right){}^{2-\frac{d}{2}} \left(-2 m_H^2+t+z\right){}^{d-5}, 
\label{eq:u_for_H_plus_jet_3L_planar}
\\
K & = \frac{s^{d-6}
   t^{2-\frac{d}{2}} m_t^{d-4}
   \left(-m_H^2+s+t\right){}^{
   2-\frac{d}{2}}}{m_H^2}.
\label{eq:K_for_H_plus_jet_3L_planar}
\end{align}
On the other hand let us consider the five loop topology in Fig~\ref{fig:5LoopHPlusJet},
Given the set of denominators:
\begin{align}
D_1&=k_1^2-m_t^2 \, , \quad
D_2=\left(k_1-p_1\right){}^2-m_t^2 \, , \quad
D_3=\left(k_1-k_2\right){}^2-m_t^2 \, , \nonumber \\
D_4 &= \left(k_2-p_1\right){}^2 \, , \quad
D_5=\left(k_2-k_3\right){}^2-m_t^2 \, ,\quad
D_6=k_2^2 \, , \quad \nonumber \\
D_7 &= \left(k_3-p_1\right){}^2 -m_t^2 \, , \quad
D_8=\left(k_3-k_4\right){}^2-m_t^2 \, ,\quad
D_9=k_3^2-m_t^2 \, ,\\
D_{10} & =\left(k_4-p_1\right)^2 \, , \quad
D_{11} =\left(k_4-p_1-p_2\right)^2 \, , \quad
D_{12} = k_4^2 \, , \quad
D_{13} = \left(k_4-k_5\right){}^2 \, , \nonumber \\
D_{14} &= \left(k_5-p_1-p_2\right){}^2 \, , \quad
D_{15} =  \left(k_5-p_1-p_2-p_3\right){}^2 \, , \quad
D_{16} = k_5^2 \, . \nonumber
\end{align}
and choosing the ISP as:
\begin{equation}
z = D_{17} = (k_5+p_1)^2 \, ,
\end{equation}
the Loop-by-Loop Baikov representation on the maximal cut gives:
\begin{align}
u &= \left(z-2 m_H^2\right){}^{\frac{d}{2}-3} \left(m_H^2+s-z\right){}^{2-\frac{d}{2}} \left(-2 m_H^2+t+z\right){}^{d-5}, \\
K &= s^{d-6} t^{2-\frac{d}{2}}
   m_H^{d-9} m_t^{3 (d-4)}
   \left(4
   m_t^2-m_H^2\right){}^{\frac
   {3-d}{2}}
   \left(-m_H^2+s+t\right){}^{2-\frac{d}{2}}.
\end{align}
We notice that $u$ is exactly the same as eq. ($\ref{eq:u_for_H_plus_jet_3L_planar}$), while $K$ slightly changes from eq. ($\ref{eq:K_for_H_plus_jet_3L_planar}$). \\
\\
Iterating the Loop-by-Loop procedure to topologies with higher number of loops, we observe that the structure remains the same; thus, we can generalize that formula to the $(3{+}2n)$-loop case ($n \geq 0$) shown in~Fig.~$\ref{fig:AllLoopHPlusJet}$
In fact choosing the ISP as:
\begin{equation}
z=D_{11+6n}=\left(k_{3+2n}+p_1\right){}^2,
\end{equation}
the Loop-by-Loop Baikov representation on the maximal cut gives:
\begin{align}
u & = \left(z-2 m_H^2\right){}^{\frac{d}{2}-3} \left(m_H^2+s-z\right){}^{2-\frac{d}{2}} \left(-2 m_H^2+t+z\right){}^{d-5}, \\
K &= s^{d-6} t^{2-\frac{d}{2}}
   m_H^{(d-7) n-2} m_t^{(d-4)
   (2 n+1)} \left(4
   m_t^2-m_H^2\right){}^{-\frac{1}{2} (d-3) n}
   \left(-m_H^2+s+t\right){}^{
   2-\frac{d}{2}}.
\end{align}
And so:
\begin{align}
    \omega &= \frac{1}{2} \left(\frac{d-4}{m_H^2+s-z}+\frac{2 (d-5)}{-2 m_H^2+t+z}+\frac{d-6}{z-2 m_H^2}\right)dz, \\
\masters & = 2,\qquad \mathcal{P} = \lbrace 2 m_H^2,\;\; 2 m_H^2-t,\;\; m_H^2+s,\;\; \infty \rbrace,
\end{align}
which are valid for all the $(3{+}2n)$-loop diagrams.\\
\paragraph{Monomial Basis.}
Let us consider the decomposition of $I_{1,1,\dots,1;-3}=\langle{\phi_{4}}|\mathcal{C}]$ in terms of the MIs: $J_{1}=I_{1,1,\dots,1;0}=\langle{\phi_{1}}|\mathcal{C}]$ and $J_{2}=I_{1,1,\dots,1;-1}= \langle{\phi_2}|\mathcal{C}]$.
We compute the $\mathbf{C}$ matrix:
\begin{equation}
\mathbf{C}_{ij}=\bra{\phi_{i}}\ket{\phi_{j}} , \quad i,j=1,2 \, ,   
\end{equation}
and the intersection numbers:
\begin{equation}
\bra{\phi_{4}}\ket{\phi_{i}}, \quad i=1,2 \, ,    
\end{equation}
Then, we obtain the final decomposition by means of eq.~($\ref{eq:masterdeco:}$):
\begin{align}
I_{1,1,\dots,1;-3} = c_1 J_1 +c_2 J_2,
\end{align}
with:
\begin{align}
c_{1} = &-\frac{m_H^2 \left(9 d^2 s^2-d^2 s t-66 d s^2-14 d s t-2 d t^2+120 s^2+72 s t+16 t^2\right)}{2 (d-3) (d-2)} \nonumber\\
 & -\frac{m_H^4 \left(36 d^2 s+5 d^2 t-168
   d s-42 d t+96 s-8 t\right)}{4 (d-3) (d-2)} \\
 &-\frac{\left(5 d^2-10 d+24\right) m_H^6}{2 (d-3) (d-2)} +\frac{(d-4) s t (3 d s-10 s-4 t)}{4 (d-3)
   (d-2)}, \nonumber\\ 
c_{2} = &\frac{m_H^2 \left(9 d^2 s-d^2 t-42 d s+2 d t+24 s-16 t\right)}{2 (d-3) (d-2)} +\frac{3 \left(7 d^2-30 d+40\right) m_H^4}{4 (d-3) (d-2)} \\
&+\frac{9 d^2
   s^2+2 d^2 s t-66 d s^2-28 d s t+120 s^2+80 s t+8 t^2}{4 (d-3) (d-2)} \ ,\nonumber
\end{align}
in (numerical) agreement with \textsc{Reduze} in the \emph{three} loop case. \\

\paragraph{Differential Equation in Monomial Basis.}
We build the system of differential equations with respect to the variable s.\\
We consider:
\begin{align}
\sigma(s) = 
\partial_s \log ( u ) = -\frac{d-4}{2 \left(m_H^2+s-z\right)}.
\end{align}
Then, $\{ \langle \Phi_i | \}_{i=1,2}$ are given by:
\begin{align}
\langle \Phi_1 | &= \sigma dz,\\
\langle \Phi_2 | &= \sigma z dz.
\end{align}
Following the discussion presented in Sec.~($\ref{sec:DE}$) we determine the $\mathbf{A}$ matrix; the entries read:
\begin{align}
{\bf A}_{11} &= \frac{(d-4) (2 s+t)}{t \left(-m_H^2+s+t\right)}+\frac{2 (d-4) s}{t \left(m_H^2-s\right)}+\frac{d-6}{s}, \\
{\bf A}_{12} &= \frac{d-4}{\left(s-m_H^2\right) \left(-m_H^2+s+t\right)}, \\
{\bf A}_{21} &= \frac{(d-4) \left(s t-m_H^2 \left(2 m_H^2+6 s+t\right)\right)}{2 \left(s-m_H^2\right) \left(-m_H^2+s+t\right)}, \\
{\bf A}_{22} &= -\frac{3 (d-4)}{2 \left(-m_H^2+s+t\right)}+\frac{2 (d-4) s}{\left(s-m_H^2\right) \left(-m_H^2+s+t\right)}+\frac{d-6}{s},
\end{align}
in (numerical) agreement with \textsc{Reduze} in the \emph{three} loop case.

\subsection{Non-Planar Rocket Diagram for $H{+}j$: $(3{+}2n)$-Loop Case}\label{sec:3Lbeyond:nonplanar}
\begin{figure}[H]
    \centering
    \includegraphics[width=0.5\textwidth]{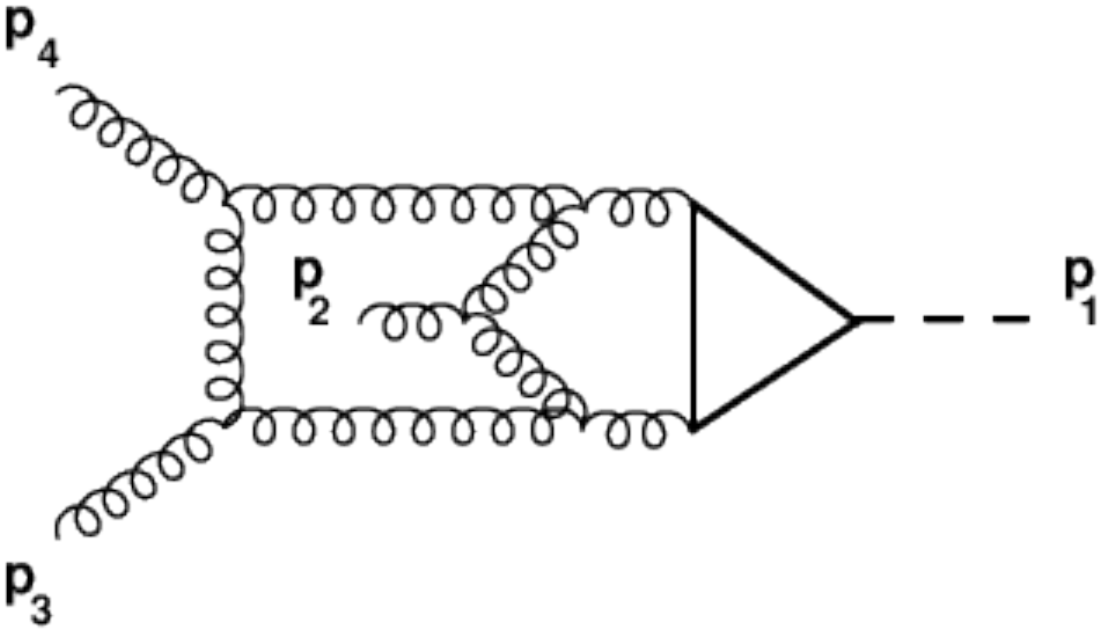}
    \caption{Non-planar three-loop diagram contributing to the $H{+}j$ production.}
    \label{fig:3LoopNPHPlusJet}
\end{figure}

Let us consider the non-planar topology for the $H+j$ production at \emph{three} loop portrayed in Fig.~$\ref{fig:3LoopNPHPlusJet}$. The kinematics is such that: $p_1^2=m_H^2$, $p_i^2=0$ with $i=2,3$, $s=(p_1+p_2)^2$, $t=(p_2+p_3)^2$ and $(p_1+p_2+p_3)^2=0$.\\
The denominators are given by:
\begin{align}
D_1&=k_1^2-m_t^2, \quad
D_2=\left(k_1-p_1\right){}^2-m_t^2, \quad
D_3=\left(k_1-k_2\right){}^2-m_t^2,\\ 
D_4 &= \left(k_2-p_1\right){}^2 \quad
D_5 =\left(k_2-k_3+p_2\right){}^2, \quad
D_6 = k_2^2, \quad
D_7 = \left(k_2-k_3\right){}^2, \\
D_8 &= \left(k_3-p_1-p_2\right){}^2, \quad
D_9 =  \left(k_3-p_1-p_2-p_3\right){}^2, \quad
D_{10} = k_3^2,
\end{align}
while the ISP is:
\begin{equation}
 z=D_{11}=\left(k_3+p_1\right){}^2.
\end{equation}
Using the \lbl Baikov representation, on the maximal cut we obtain:
\begin{align}
u & = \left(z-2 m_H^2\right){}^{\frac{d}{2}-3} \left(m_H^2+s-z\right){}^{\frac{d}{2}-3} \left(-2 m_H^2+t+z\right){}^{d-5}, \\
K &= \frac{t^{2-\frac{d}{2}}
   m_t^{d-4}
   \left(-m_H^2+s+t\right){}^{
   2-\frac{d}{2}}}{s m_H^2}.
\end{align}
As done for the planar diagram, we can infer the general structure for the corresponding $3+2n$-loop integral ($n \geq 0$) shown in Fig.~$\ref{fig:AllLoopNPHPlusJet}$:
\begin{figure}[H]
    \centering
    \includegraphics[width=0.8\textwidth]{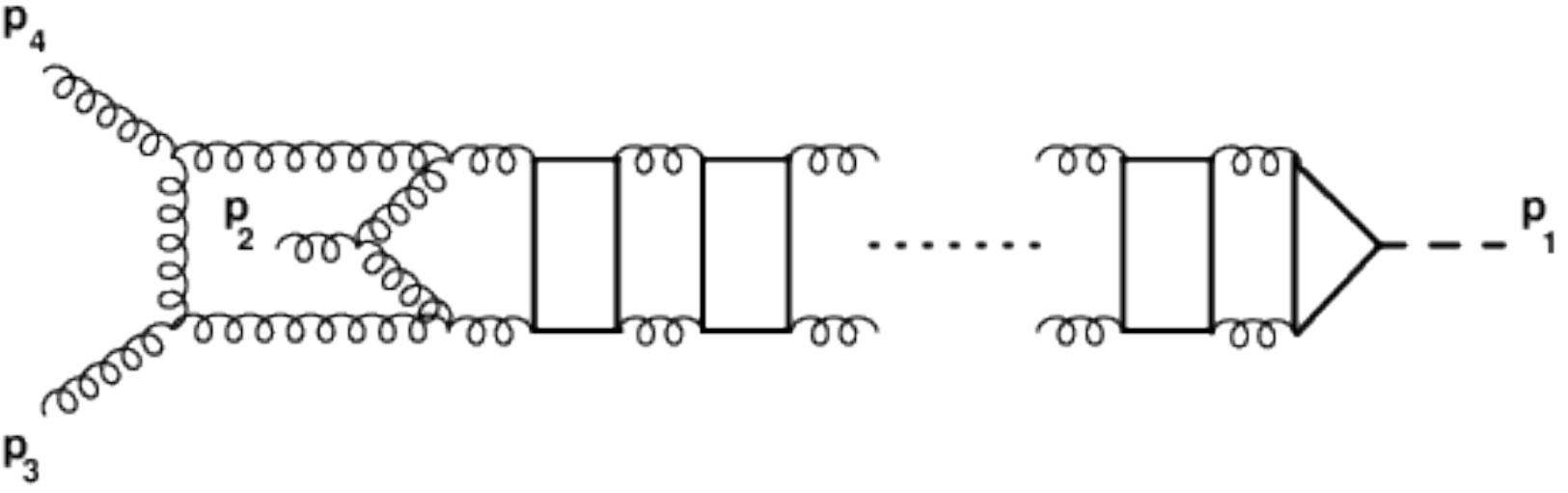}
    \caption{Non-planar box-rocket diagram contributing to $H{+}j$ production.}
    \label{fig:AllLoopNPHPlusJet}
\end{figure}
In fact choosing the ISP as:
\begin{equation}
z=D_{11+6n}=\left(k_{3+2n}+p_1\right){}^2,
\end{equation}
after the maximal cut, we find:
\begin{align}
u & = \left(z-2 m_H^2\right){}^{\frac{d}{2}-3} \left(m_H^2+s-z\right){}^{\frac{d}{2}-3} \left(-2 m_H^2+t+z\right){}^{d-5}, \\
K &= \frac{t^{2-\frac{d}{2}}
   m_H^{(d-7) n-2} m_t^{(d-4)
   (2 n+1)} \left(4
   m_t^2-m_H^2\right){}^{-\frac{1}{2} (d-3) n}
   \left(-m_H^2+s+t\right){}^{
   2-\frac{d}{2}}}{s} \, ,
\end{align}
thus:
\begin{align}
\omega &= \frac{1}{2} \left(\frac{6-d}{m_H^2+s-z}+\frac{2 (d-5)}{-2 m_H^2+t+z}+\frac{d-6}{z-2 m_H^2}\right)dz, \\
\masters & = 2,\qquad
\mathcal{P} = \lbrace 2 m_H^2,\; -t{+}2 m_H^2,\; (m_H^2{+}s),\; \infty \rbrace,
\end{align}
which are valid for all the $(3{+}2n)$-loop diagrams.\\

\paragraph{Monomial Basis.}
Let us consider the decomposition of $I_{1,1,\dots,1;-3}=\langle{\phi_4}|\mathcal{C}]$ in terms of $J_{1}=I_{1,1,\dots,1;0}=\langle{\phi_1}|\mathcal{C}]$ and $J_{2}=I_{1,1,\dots,1;-1}=\langle{\phi_2}|\mathcal{C}]$.\\
We can compute the $\mathbf{C}$ matrix:
\begin{equation}
\mathbf{C}_{ij}=\bra{\phi_i}\ket{\phi_j}, \quad i,j=1,2 \, ,
\end{equation}
and the additional intersection numbers:
\begin{equation}
\bra{\phi_4}\ket{\phi_i}, \quad i=1,2 \, ,    
\end{equation}
and finally  eq.~($\ref{eq:masterdeco:}$) gives:
\begin{equation}
I_{1,1,\dots,1;-3}=c_1 J_1 + c_2 J_2,
\end{equation}
with:
\begin{align}
c_1 = &-\frac{m_H^2 \left(9 d s^2-17 d s t+3 d t^2-30 s^2+62 s t-9 t^2\right)}{4 (2 d-7)}\nonumber \\
&-\frac{m_H^4 (108 d s-59 d t-384 s+202 t)}{8 (2 d-7)} \nonumber \\
&-\frac{(65 d-226) m_H^6}{4 (2 d-7)}+\frac{s t (3 d s-2 d t-10 s+6 t)}{8 (2 d-7)},\\
c_2 =&  \frac{m_H^2 (27 d s-20 d t-96 s+69 t)}{4 (2 d-7)}+\frac{3 (43 d-150) m_H^4}{8 (2 d-7)} \nonumber \\
&+\frac{9 d s^2-8 d s t+4 d t^2-30 s^2+30 s t-12 t^2}{8 (2
   d-7)} \ ,
\end{align}
in (numerical) agreement with \textsc{Reduze} in the \emph{three} loop case.

\paragraph{Differential Equations in Monomial Basis.}
We build the system of differential equation with respect to the variable $s$.\\
We consider:
\begin{align}\label{sigma-1}
\sigma(s) = \partial_s \log ( u ) = \frac{d-6}{2 \left(m_H^2+s-z\right)},
\end{align}
which gives $\{ \bra{\Phi_i} \}_{i=1,2}$ :
\begin{align}
\langle \Phi_1 | &= \sigma dz, \\
\langle \Phi_2 | &= \sigma z dz.
\end{align}
Then, following the discussion in Sec.~$\ref{sec:DE}$, we build the $\mathbf{A}$ matrix, with:
\begin{align}
{\bf A}_{11} &= \frac{m_H^2 ((21-4 d) s+t)+s ((d-6) t-2 s)-m_H^4}{s \left(s-m_H^2\right) \left(-m_H^2+s+t\right)},  \\
{\bf A}_{12} &= \frac{2 d-9}{\left(s-m_H^2\right) \left(-m_H^2+s+t\right)},  \\
{\bf A}_{21} &= \frac{m_H^2 ((3 d-16) t-6 (d-4) s)+(48-10 d) m_H^4+(d-4) s t}{2 \left(s-m_H^2\right) \left(-m_H^2+s+t\right)}, \\
{\bf A}_{22} &= \frac{m_H^2 ((5 d-18) s+2 t)+s ((3 d-16) s-2 t)-2 m_H^4}{2 s \left(s-m_H^2\right) \left(-m_H^2+s+t\right)}.
\end{align}
in (numerical) agreement with \textsc{Reduze} in the \emph{three} loop case.

\subsection{Planar Rocket Diagram for $H{+}j$: $(2{+}2n)$-Loop Case}\label{sec:4Lbeyond:planar}
\begin{figure}[H]
    \centering
    \includegraphics[width=0.5\textwidth]{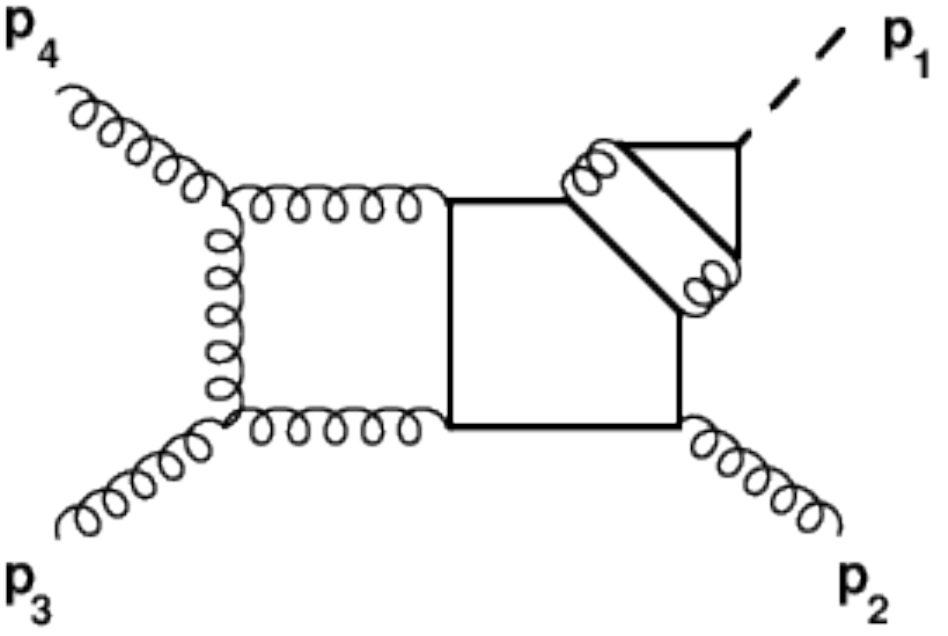}
    \caption{Planar four-loop diagram contributing to $H{+}j$ production.}
    \label{fig:4LoopHPlusJet}
\end{figure}
Let us consider the \emph{four} loop planar topology in Fig.~$\ref{fig:4LoopHPlusJet}$ which contributes to the $H{+}j$ production. The kinematics is such that: $p_1^2=m_H^2$, $p_i^2=0$ with $i=2,3$, $s=(p_1+p_2)^2$, $t=(p_2+p_3)^2$ and $(p_1+p_2+p_3)^2=0$.\\
The denominators are:
\begin{align}
& D_{1}=k_1^2-m_t^2 \, , \quad
D_{2}=\left(k_1-p_1\right){}^2-m_t^2 \, , \quad
D_{3}=\left(k_1-k_2\right){}^2-m_t^2 \, , \nonumber\\
& D_{4}=\left(k_2-p_1\right){}^2 \, , \quad
D_{5}=  \left(k_2-k_3\right){}^2-m_t^2 \, , \quad
D_{6}=  k_2^2 \, , \\
& D_{7}=  \left(k_3-p_1\right){}^2-m_t^2 \, , \quad
D_{8}=  \left(k_3-p_1-p_2\right){}^2-m_t^2 \, , \quad
D_{9}=  \left(k_3-k_4\right){}^2-m_t^2 \, , \nonumber\\
& D_{10}=  k_3^2-m_t^2 \, , \quad
D_{11}=  \left(k_4-p_1-p_2\right){}^2 \, , \quad
D_{12}=  \left(k_4-p_1-p_2-p_3\right){}^2 \,,\quad  D_{13}=k_4^2. \nonumber
\end{align}
While the ISP is:
\begin{equation}
z=D_{14}=\left(k_4+p_1\right){}^2.
\end{equation}
The \lbl Baikov representation, on the maximal cut gives:
with
\begin{align}
& u=\frac{\left(m_H^2{+}s{-}z\right){}^{2-\frac{d}{2}} \left(-2 m_H^2{+}t{+}z\right){}^{d-5} \left(2 s m_H^2{-}4 m_H^2 m_t^2{-}4 s
   m_t^2{+}z \left(4 m_t^2{-}s\right)\right){}^{\frac{d-5}{2}}}{\sqrt{z-2 m_H^2}} \, , \\
& K=s^{\frac{d-7}{2}} t^{2-\frac{d}{2}} m_H^{(d-7)
   } m_t^{2 (d-4) } \left(4
   m_t^2-m_H^2\right){}^{-\frac{1}{2} (d-3) }
   \left(-m_H^2+s+t\right){}^{2-\frac{d}{2}} \, .
\end{align}
We can generalize such a construction in order to describe the $(2{+}2n)$-loop diagram ($n \geq 0$), shown in Fig.~\ref{fig:AllLoopHPlusJetfrom4L}:
\begin{figure}[H]
    \centering
    \includegraphics[width=0.5\textwidth]{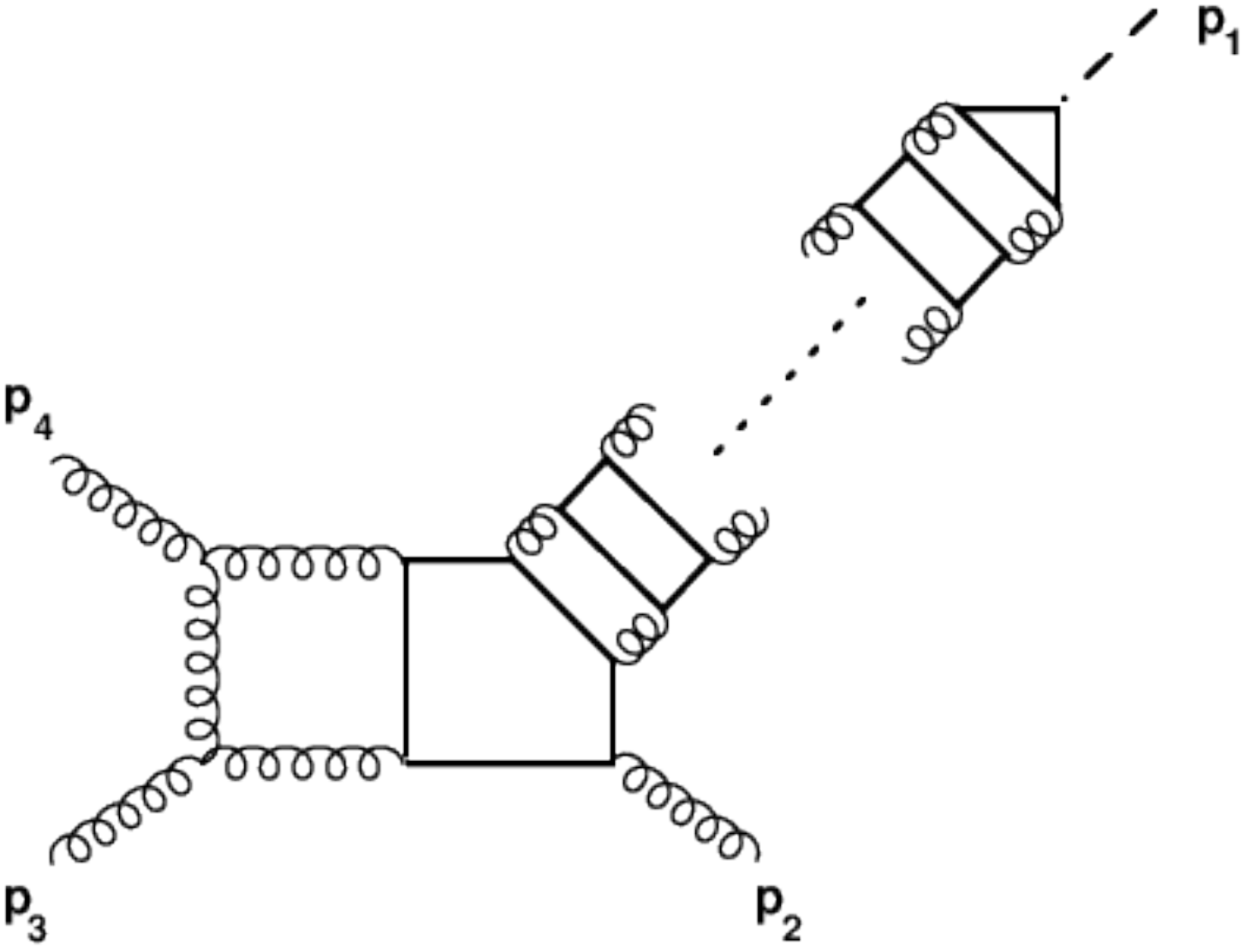}
    \caption{Planar All-loop diagram contributing to $H{+}j$ production.}
    \label{fig:AllLoopHPlusJetfrom4L}
\end{figure}
In fact choosing as ISP:
\begin{equation}
D_{8+6n}=\left(k_{2+2n}+p_1\right){}^2,
\end{equation}
we obtain:
\begin{align}
& u=\frac{\left(m_H^2{+}s{-}z\right){}^{2-\frac{d}{2}} \left(-2 m_H^2{+}t{+}z\right){}^{d-5} \left(2 s m_H^2{-}4 m_H^2 m_t^2{-}4 s
   m_t^2{+}z \left(4 m_t^2{-}s\right)\right){}^{\frac{d-5}{2}}}{\sqrt{z-2 m_H^2}} \, , \\
& K=s^{\frac{d-7}{2}} t^{2-\frac{d}{2}} m_H^{(d-7)
   n} m_t^{2 (d-4) n} \left(4
   m_t^2-m_H^2\right){}^{-\frac{1}{2} (d-3) n}
   \left(-m_H^2+s+t\right){}^{2-\frac{d}{2}} \, ,
\end{align}
from which we can evaluate:
\begin{gather}
\omega = \frac{1}{2} \left(\frac{(d-5) \left(s-4
   m_t^2\right)}{m_H^2 \left(4 m_t^2-2
   s\right)+4 m_t^2 (s-z)+s
   z}+\frac{d-4}{m_H^2+s-z}+\frac{2 (d-5)}{-2
   m_H^2+t+z}+\frac{1}{2 m_H^2-z}\right) dz     ,\nonumber\\
\masters=3 \, ,\qquad
\mathcal{P}=  \{ 2 m_H^2 \, , m_H^2{+}s \, , 2 m_H^2{-}t \, , -\frac{2 \left({-}s m_H^2{+}2 m_H^2 m_t^2{+}2 s m_t^2\right)}{s-4 m_t^2} \, , \infty \} \, .
\end{gather}
which are valid for all the $(2{+}2n)$-loop diagrams.\\

\paragraph{Monomial Basis.}
Let us consider the reduction of $I_{1,1,\dots,1;-3}=\langle{\phi_4}|\mathcal{C}]$ in terms of: $J_{1}=I_{1,1,\dots,1;0}=\langle{\phi_1}|\mathcal{C}]$ and $J_2=I_{1,1,\dots,1;-1}=\langle{\phi_2}|\mathcal{C}]$ and $J_3=I_{1,1,\dots,1;-2}=\langle{\phi_3}|\mathcal{C}]$.\\
We can compute the $\mathbf{C}$ matrix:
\begin{equation}
\mathbf{C}=\bra{\phi_i}\ket{\phi_j}, \quad i=1,2,3 \, ,    
\end{equation}
and the intersection numbers:
\begin{equation}
\bra{\phi_4}\ket{\phi_i}, \quad i=1,2,3 \, ,    
\end{equation}
thus, eq.~($\ref{eq:masterdeco:}$) leads to:
\begin{equation}
I_{1,1,\dots,1;-3}=c_1 J_1 + c_2 J_2 + c_3 J_3, 
\end{equation}
with:
\begin{align}
c_{1}= & \Big(   s m_H^2 \left(m_H^2 ((6 d-20) s+(d-10) t)+2 (d+2) m_H^4+(2-d) s t\right)+ \nonumber \\
&\quad 2 m_t^2 \left(m_H^2{+}s\right) \left(m_H^2 ((14{-}4 d) s{+}7 t)+(2{-}4 d)
   m_H^4{+}s t\right) \Big)/ \Big((d{-}2) \left(s{-}4 m_t^2\right) \Big), \,  \\
c_{2}= & \Big( 4 m_t^2 \left(m_H^2 (6 (2 d-5) s-11 t)+(10 d-11) m_H^4+s ((2 d-7) s-5 t)\right)+  \nonumber \\
&\;\; s\! \left(m_H^2 ((40{-}12 d) s{-}(d{-}18) t){+}(8{-}12 d) m_H^4{+}(d{-}2)
   s t\right) \Big) / \Big(2 (d{-}2) \left(s{-}4 m_t^2\right) \!\Big) , \\
c_{3}=& \Big(4 m_t^2 \left((13-8 d) m_H^2+(11-4 d) s+4 t\right)+ \nonumber \\
&\quad s \left((9 d-14) m_H^2+(3 d-10) s-4 t\right) \Big)/ \Big(2 (d-2) \left(s-4 m_t^2\right) \Big) \, .
\end{align}
in agreement with \textsc{Reduze} in the \emph{two} loop case.

\paragraph{Differential Equations in Monomial Basis.}
We derive:
\begin{align}
\sigma(s) = -\frac{\left(z-2 m_H^2\right) \left(-(d-5) m_H^2+(d-5) z+s\right)+4 m_t^2 \left(m_H^2+s-z\right)}{2 \left(m_H^2+s-z\right) \left(4 m_t^2\left(m_H^2+s-z\right)+s \left(z-2 m_H^2\right)\right)}.
\end{align}
The $ \{ \langle \Phi_i | \}_{i=1,2,3}$ are given by:
\begin{align}
\langle \Phi_1 | &= \sigma dz,\\
\langle \Phi_2 | &= \sigma z dz,\\
\langle \Phi_3 | &= \sigma z^2 dz.
\end{align}
Then, the $\mathbf{A}$ matrix can be computed following Sec.~($\ref{sec:DE}$); the entries are presented evaluated at the phase space point:
\begin{equation}
 m_{t}^{2}=1 \,,\qquad  m_{H}^{2}=3 \,,\qquad  t=5 \,.
\end{equation}
We find:
\begin{align}
& \mathbf{A}_{11}= \frac{d \left(13 s^2+7 s-120\right)-2 \left(s^3+19 s^2+10 s-186\right)}{s \left(s^3-9 s^2+2 s+48\right)} \, , \\
& \mathbf{A}_{12}=\frac{d \left(-2 s^2-15 s+72\right)+8 s^2+50 s-244}{(s-8) (s-3) s (s+2)} \,  ,\\
& \mathbf{A}_{13}=\frac{2 (d-3) (s-4)}{(s-8) (s-3) s (s+2)} \, , \\
& \mathbf{A}_{21}=\frac{2 \left(3 d \left(5 s^2-15 s-72\right)-59 s^2+168 s+810\right)}{(s-8) (s-3) s (s+2)} \, , \\
& \mathbf{A}_{22}=\frac{d \left(-10 s^2+28 s+240\right)-2 s^3+51 s^2-84 s-900}{s \left(s^3-9 s^2+2 s+48\right)} \, , \\
& \mathbf{A}_{23}=\frac{(d-3) \left(s^2-4 s-24\right)}{(s-8) (s-3) s (s+2)} \, , \\
& \mathbf{A}_{31}=\frac{3 d \left(5 s^4{-}35 s^3{-}132 s^2{+}648 s{+}1728\right){-}2 \left(28 s^4{-}193 s^3{-}753 s^2{+}3636 s{+}9720\right)}{(s-8) (s-4) (s-3) s (s+2)} \, , \\
& \mathbf{A}_{32}=\frac{d \left(-23 s^4{+}173 s^3{+}408 s^2{-}2088 s{-}5184\right) {+} 2 \left(41 s^4{-}307 s^3{-}712 s^2{+}3588 s{+}8784\right)}{2 (s-8) (s-4) (s-3) s (s+2)} \, , \\
& \mathbf{A}_{33}=\frac{d \left(3 s^4{-}21 s^3{-}76 s^2{+}376 s{+}384\right){-}2 \left(7 s^4{-}59 s^3{-}70 s^2{+}740 s{+}192\right)}{2 (s-8) (s-4) (s-3) s (s+2)} \, .
\end{align}
in agreement with \textsc{Reduze} in the \emph{two} loop case.

\subsection{Non-Planar Rocket Diagram for $H{+}j$: $(2{+}2n)$-Loop Case}\label{sec:4Lbeyond:nonplanar}
\begin{figure}[H]
    \centering
    \includegraphics[width=0.5\textwidth]{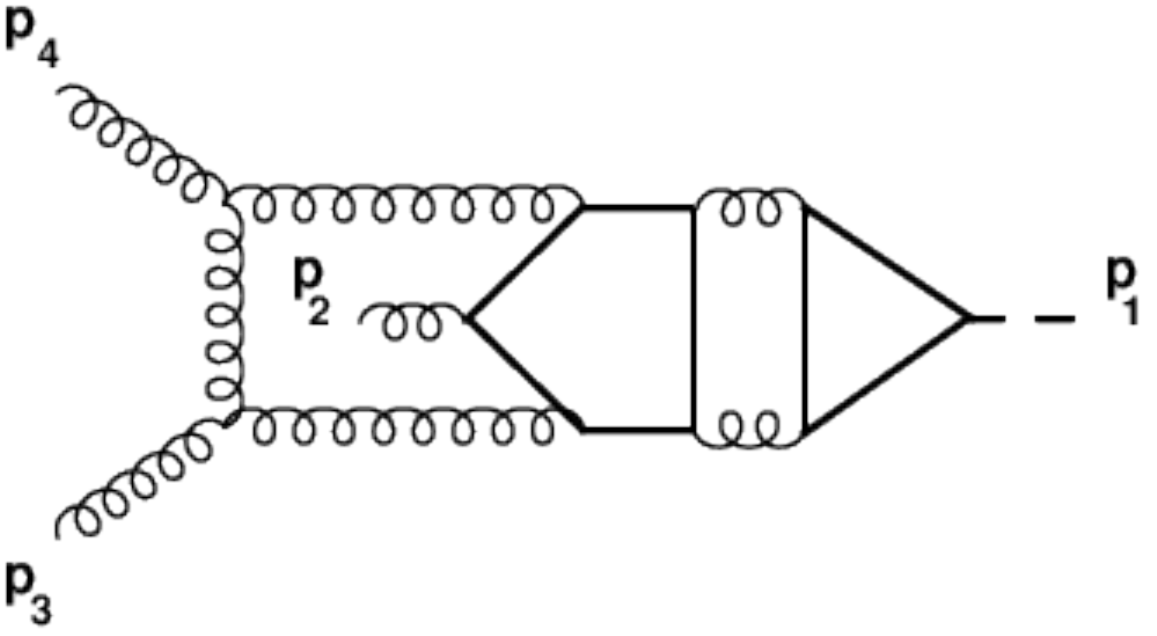}
    \caption{Non planar four loop contribution to $H{+}j$ production.}
    \label{fig:4LoopHPlusJetNP}
\end{figure}
\begin{figure}[H]
    \centering
    \includegraphics[width=0.8\textwidth]{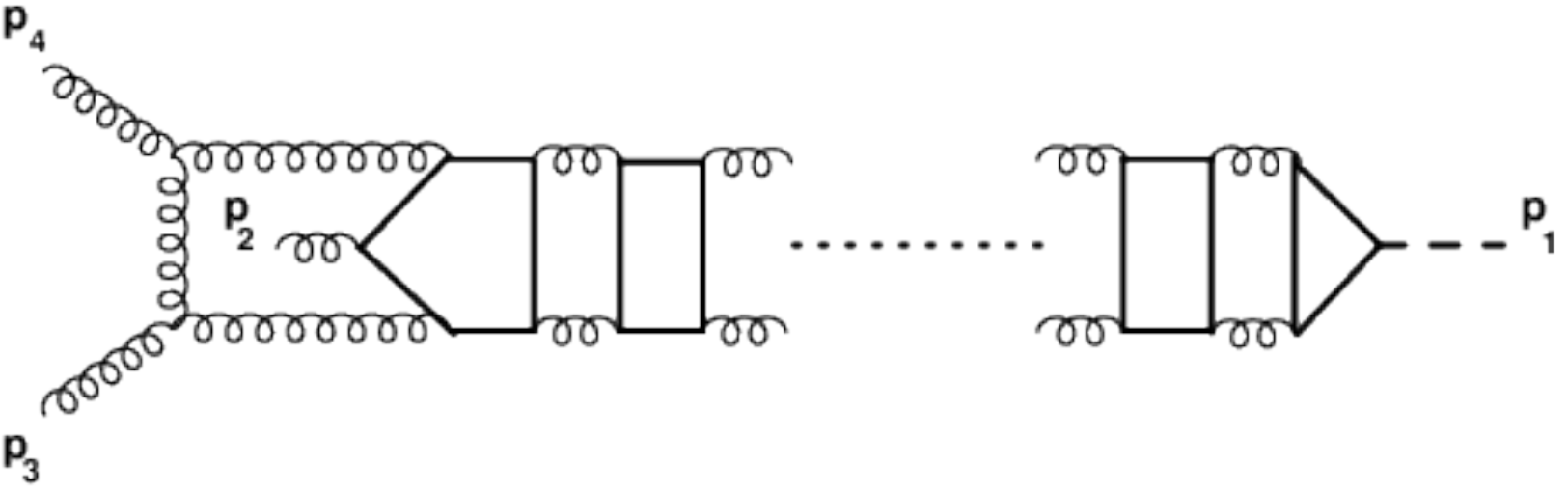}
    \caption{Non planar $(2{+}2n)$-loop contribution to $H{+}j$ production.}
    \label{fig:4LoopBeyondHPlusJetNP}
\end{figure}

Let us consider the non planar \emph{four} loop contribution to $H{+}j$ production in Fig.~$\ref{fig:4LoopHPlusJetNP}$. The kinematics is such that: $p_1^2=m_H^2$, $p_i^2=0$ with $i=2,3$, $s=(p_1+p_2)^2$, $t=(p_2+p_3)^2$ and $(p_1+p_2+p_3)^2=0$.\\
In this case the denominators are:
\begin{align}
& D_{1}=k_1^2-m_t^2 \, , \quad
D_{2}=\left(k_1-p_1\right){}^2-m_t^2 \, , \quad
D_{3}=\left(k_1-k_2\right){}^2-m_t^2 \, , \nonumber\\
& D_{4}=\left(k_2-p_1\right){}^2 \, , \quad
D_{5}=  \left(k_2-k_3\right){}^2-m_t^2 \, , \quad
D_{6}=  k_2^2 \, , \\
& D_{7}=  \left(k_3-p_1\right){}^2-m_t^2 \, , \quad
D_{8}=  \left(k_3-k_4+p_2\right){}^2-m_t^2 \, , \quad
D_{9}=  \left(k_3-k_4\right){}^2-m_t^2 \, , \nonumber\\
& D_{10}=  k_3^2-m_t^2 \, , \quad
D_{11}=  \left(k_4-p_1-p_2\right){}^2 \, , \quad
D_{12}=  \left(k_4-p_1-p_2-p_3\right){}^2 \,,\quad  D_{13}=k_4^2. \nonumber
\end{align}
We choose the ISP as:
\begin{equation}
z=D_{14}=\left(k_4+p_1\right){}^2,
\end{equation}
The \lbl Baikov representation, after the maximal cut gives:
\begin{align}
& u=\frac{\left(-2 m_H^2+t+z\right){}^{d-5} \left(\left(2 m_H^2-z\right) \left(m_H^2+s-z\right)-4
   s m_t^2\right){}^{\frac{d-5}{2}}}{\sqrt{z-2 m_H^2} \sqrt{m_H^2+s-z}} \, , \\
& K=\frac{t^{2-\frac{d}{2}} m_H^{d-7} m_t^{2 (d-4)} \left(4 m_t^2-m_H^2\right){}^{\frac{3-d}{2}}
   \left(-m_H^2+s+t\right){}^{2-\frac{d}{2}}}{s} \, .
\end{align}
As stated above, we can generalize such Baikov polynomial in order to describe the $(2{+}2n)$-loop diagram ($n \geq 0$) shown in Fig.~\ref{fig:4LoopBeyondHPlusJetNP}.
In fact choosing the ISP as:
\begin{equation}
D_{8+6n}=\left(k_{2+2n}+p_1\right){}^2,
\end{equation}
we obtain:
\begin{align}
& u=\frac{\left(-2 m_H^2+t+z\right){}^{d-5} \left(\left(2 m_H^2-z\right) \left(m_H^2+s-z\right)-4
   s m_t^2\right){}^{\frac{d-5}{2}}}{\sqrt{z-2 m_H^2} \sqrt{m_H^2+s-z}} \, , \\
& K=\frac{t^{2-\frac{d}{2}} m_H^{(d-7) n} m_t^{2 (d-4) n} \left(4
   m_t^2-m_H^2\right){}^{-\frac{1}{2} (d-3) n} \left(-m_H^2+s+t\right){}^{2-\frac{d}{2}}}{s} \, ,
\end{align}
from which we evaluate:
\begin{gather}
\omega =\frac{1}{2} \left(\frac{(d-5) \left(3 m_H^2+s-2 z\right)}{\left(z-2 m_H^2\right)
   \left(m_H^2+s-z\right)+4 s m_t^2}+\frac{2 (d-5)}{-2
   m_H^2+t+z}+\frac{1}{m_H^2+s-z}+\frac{1}{2 m_H^2-z}\right) dz,\\
\masters =4 \, ,\qquad
\mathcal{P}=  \{ 2 m_H^2 \, , m_H^2{+}s \, , 2 m_H^2{-}t \, , \rho_1 \, , \rho_2 \, , \infty \} \, .
\end{gather}
which are valid for all the $(2{+}2n)$-loop diagrams.
\paragraph{Monomial Basis.}
Let us consider the reduction of $I_{1,1,\dots,1;-4}=\langle{\phi_5}|\mathcal{C}]$ in terms of: $J_1=I_{1,1,\dots,1;0}=\langle{\phi_1}|\mathcal{C}]$, $J_2=I_{1,1,\dots,1;-1}=\langle{\phi_2}|\mathcal{C}]$, $J_3=I_{1,1,\dots,1;-2}=\langle{\phi_3}|\mathcal{C}]$ and $J_4=I_{1,1,\dots,1;-3}=\langle{\phi_4}|\mathcal{C}]$.\\
We compute the $\mathbf{C}$ matrix:
\begin{equation}
\mathbf{C}=\bra{\phi_i}\ket{\phi_j}, \quad 1 \leq i,j \leq 4,    
\end{equation}
and the intersection numbers:
\begin{equation}
\bra{\phi_5}\ket{\phi_i}, \quad 1 \leq i \leq 4,    
\end{equation}
then eq.~($\ref{eq:masterdeco:}$) gives:
\begin{equation}
I_{1,1,\dots,1;-4}=c_1 J_1 + c_2 J_2+ c_3 J_3 + c_4 J_4,
\end{equation}
with:
\begin{align}
c_{1}= & \frac{m_H^2 \left(m_H^2+s\right) \left(m_H^2 ((20-6 d) s+3 (d-2) t)+(28-10 d)
   m_H^4+(d-2) s t\right)}{2 (d-3)} \nonumber \\
  & +\frac{-2 s \, m_t^2 \left(m_H^2 ((14-4 d) s+3 t)+(10-4 d) m_H^4+s
   t\right)}{2 (d-3)} \, , \\
c_{2}= & \frac{m_H^4 (4 (15 d-46) s-13 (d-2) t)+2 s m_H^2 ((6 d-20) s-5 (d-2)
   t)}{4
   (d-3)} \nonumber \\
   &+\frac{4 s \, m_t^2 \left((17-6 d) m_H^2+(7-2 d) s+2 t\right)+8 (7 d-20) m_H^6+(2-d) s^2 t}{4
   (d-3)} \, , \\
c_{3}=& \frac{m_H^2 (4 (28-9 d) s+9 (d-2) t)+(166-57 d) m_H^4+s \left(8 (d-3) m_t^2 \right.}{4 (d-3)} \nonumber \\
&+\frac{\left. (10-3
   d) s+3 (d-2) t\right)}{4 (d-3)} \, , \\
   c_{4}=&\frac{(25 d-74) m_H^2+(7 d-22) s-2 (d-2) t}{4
   (d-3)} \, .
\end{align}
in agreement with \textsc{Reduze} in the \emph{two} loop case.

\paragraph{Differential Equations in Monomial Basis.}
In the \emph{two} loop case ($n=0$) we derive:
\begin{align}
\sigma(s)& =\frac{4 m_t^2 \left((d-5) m_H^2+(d-6) s-(d-5) z\right)+(d-6) \left(z-2 m_H^2\right)
   \left(m_H^2+s-z\right)}{2 \left(m_H^2+s-z\right) \left(\left(z-2 m_H^2\right)
   \left(m_H^2+s-z\right)+4 s m_t^2\right)} \, .
\end{align}
The $\{ \langle \Phi_i | \}_{i=1,2,3,4}$ are given by:
\begin{align}
\langle \Phi_1 | &= \sigma \, dz,\\
\langle \Phi_2 | &= \sigma \, z \, dz,\\
\langle \Phi_3 | &= \sigma \, z^2 \, dz,\\
\langle \Phi_4 | &= \sigma \, z^3 \, dz.
\end{align}
Then, the $\mathbf{A}$ matrix can be computed following Sec.~($\ref{sec:DE}$); the entries are presented evaluated at the phase space point:
\begin{equation}
 m_{t}^{2}=1 \,,\qquad  m_{H}^{2}=3 \,,\qquad  t=5 \,.
\end{equation}
We find:
\begin{align*}
 \mathbf{A}_{11}= &\frac{-2 s^5+13 s^4+92 s^3-3751 s^2-19284 s+d \left(-11 s^4-43 s^3+1063 s^2+5235
   s+4860\right)}{(s-3) s (s+2) (s+10) \left(s^2+10 s+9\right)} \\
   &+\frac{ -18144}{(s-3) s (s+2) (s+10) \left(s^2+10 s+9\right)} \, , \\
\mathbf{A}_{12}=&\frac{3 \left(-3
   s^4+2 s^3+851 s^2+4976 s+5058\right)+d \left(2 s^4-8 s^3-703 s^2-4002
   s-4077\right)}{(s-3) s (s+2) (s+10) \left(s^2+10 s+9\right)} \,  ,\\
 \mathbf{A}_{13}=&\frac{d \left(3 s^3+113
   s^2+899 s+1041\right)-2 \left(5 s^3+200 s^2+1617 s+1872\right)}{(s-3) s (s+2) (s+10)
   \left(s^2+10 s+9\right)} \, , \\
 \mathbf{A}_{14}=& -\frac{2 (2 d-7) \left(s^2+16 s+21\right)}{(s-3) s (s+2)
   (s+10) \left(s^2+10 s+9\right)} \, , \\
\mathbf{A}_{21}=&\frac{30 d \left(s^3+19 s^2+69 s+63\right)-2 \left(53 s^3+1079 s^2+3972
   s+3618\right)}{(s-3) s (s+1) (s+2) (s+10)} \, , \\
 \mathbf{A}_{22}=&\frac{-2 s^4+106 s^3+1560 s^2+6126 s-d
   \left(34 s^3+423 s^2+1628 s+1623\right)+6066}{(s-3) s (s+1) (s+2) (s+10)} \, , \\
\mathbf{A}_{23}=&\frac{d
   \left(5 s^3+80 s^2+378 s+429\right)-2 \left(9 s^3+143 s^2+677 s+768\right)}{(s-3) s
   (s+1) (s+2) (s+10)} \, , \\
\mathbf{A}_{24}=&-\frac{2 (2 d-7) \left(s^2+7 s+9\right)}{(s-3) s (s+1) (s+2)
   (s+10)} \, , \\
 \mathbf{A}_{31}=&\frac{30 d \left(s^4+19 s^3+120 s^2+297 s+243\right)-4 \left(28 s^4+535 s^3+3426 s^2+8514
   s+6939\right)}{(s-3) s (s+1) (s+2) (s+10)} \, , \\
 \mathbf{A}_{32}=&\frac{2 \left(41 s^4+813 s^3+5128 s^2+13425
   s+11853\right)-d \left(23 s^4+449 s^3+2790 s^2\right.}{(s-3) s (s+1) (s+2)
   (s+10)} \\
   &+\frac{\left. 7281 s+6453\right)}{(s-3) s (s+1) (s+2)
   (s+10)} \, , \\
\mathbf{A}_{33}=&-\frac{12 s^4+271 s^3+2061 s^2+6278 s-d \left(3 s^4+72 s^3+577 s^2+1783
   s+1779\right)+6276}{(s-3) s (s+1) (s+2) (s+10)} \, , \\
\mathbf{A}_{34}=&-\frac{(2 d-7) \left(s^3+17 s^2+70
   s+78\right)}{(s-3) s (s+1) (s+2) (s+10)} \, ,\\
 \mathbf{A}_{41}=&\frac{(s+9) \left(15 d \left(s^4+20 s^3+121 s^2+288 s+234\right)-2 \left(28 s^4+569
   s^3+3457 s^2\right.\right.}{(s-3) s (s+1) (s+2) (s+10)}\\
   &+\frac{\left. \left. 8214 s+6642\right)\right)}{(s-3) s (s+1) (s+2) (s+10)} \, , \\
 \mathbf{A}_{42}=&\frac{82 s^5+2362
   s^4+25034 s^3+119834 s^2+263028 s-d \left(23 s^5+649 s^4+6803 s^3\right.}{2 (s-3) s (s+1) (s+2) (s+10)} \\
   & \frac{\left.+32613 s^2+71874
   s+57726\right)+210492}{2 (s-3) s (s+1) (s+2) (s+10)} \, , \\
\mathbf{A}_{43}=&\frac{d \left(3 s^5+90 s^4+1135
   s^3+7030 s^2+18528 s+16578\right)-2 \left(5 s^5+162 s^4+2081 s^3\right.}{2 (s-3) s (s+1) (s+2) (s+10)}\\
   &+\frac{\left.12586 s^2+32838
   s+29376\right)}{2 (s-3) s (s+1) (s+2) (s+10)} \, , \\
 \mathbf{A}_{44}=&\frac{2 \left(s^4+62 s^3+773 s^2+2746
   s+2706\right)-d \left(s^4+40 s^3+441 s^2+1530 s+1512\right)}{2 (s-3) s (s+1) (s+2)
   (s+10)} \, .
\end{align*}
in agreement with \textsc{Reduze} in the \emph{two} loop case.

\section{Iterated One-Forms}
\label{sec:iteratedoneforms}

We consider cases of maximally cut integrals of 2-forms depending on two variables (two ISPs), and we show in a few examples, how they can be decomposed by applying the univariate intersection numbers, in one variable at a time.

In particular, we deal with integrals of the form
\begin{gather}
I_{n,m} \equiv K 
\int_{{\cal C}_1} 
\int_{{\cal C}_2} 
u \ z_1^n \, z_2^m \, dz_1 {\wedge} \, dz_2, \\
u = u(z_1,z_2) 
\ ,\\
\omega = \hat{\omega}_1 d{z_1} + \hat{\omega}_2 d{z_2} \ ,\qquad
\hat{\omega}_1  \equiv \partial_{z_1} \log u, \qquad
\hat{\omega}_2  \equiv  \partial_{z_2} \log u.
\end{gather}
As in the previous sections, the prefactor $K$ does not play any role in the decomposition formulas, and therefore it is left implicit in the following.

\noindent
\paragraph{Intersections in $z_1$.}
We rewrite $u$ as,
\begin{equation}
u = 
u_{z_1} \, , 
\end{equation}
with
\begin{eqnarray}
\hat{\omega}_{z_1} &\equiv& \partial_{z_1} \log u_{z_1} = \partial_{z_1} \log u = \hat{\omega}_1 \, .
\label{eq:generic:def:u1}
\end{eqnarray}
In this fashion,
\begin{eqnarray}
I_{n,m} 
&=& 
\int_{{\cal C}_2} J_n \, z_2^{m} \, dz_2  \, , \\
J_n &=& \int_{{\cal C}_1} u_{z_1} \, z_1^n \, dz_1
\equiv
{}_{\omega_1}\!\langle \phi_{n+1} | {\cal C}_1 ] \,.
\end{eqnarray}

For the cases at hand, we {\it assume} that the $J_n$ integral family admits $\masters_1 = 1$ master integral,
say $J_0$, defined as,
\begin{eqnarray}
J_0 &=& \int_{{\cal C}_1} u_{z_1} \, dz_1 
\equiv
{}_{\omega_1}\!\langle \phi_{1} | {\cal C}_1 ] \ ,
\end{eqnarray}
which is a function of $z_2$, {\it i.e.}, $J_0 = J_0(z_2)$.
Then, $J_n$ can be decomposed in terms of $J_0$, as 
\begin{eqnarray}
J_n = c_{n} J_0 
\quad
\Leftrightarrow 
\quad
{}_{\omega_1}\!\langle \phi_{n+1} | = 
c_{n} \ 
{}_{\omega_1}\!
\langle \phi_{1} | \ ,
\end{eqnarray}
where the coefficient $c_n$ can be obtained by intersection in $z_1$, 
using the master formula eq.~\eqref{eq:masterdeco:},
\begin{eqnarray}
c_{n}
&=&
\langle \phi_{n+1}| \phi_{1} \rangle_{\omega_1} \ 
\langle \phi_{1} | \phi_{1} \rangle_{\omega_1}^{-1} \ 
\ ,
\end{eqnarray}
and which may depend on $z_2$, {\it i.e.}, $c_n=c_n(z_2)$.

\noindent
\paragraph{Intersections in $z_2$.}
After performing all intersections in $z_1$, $I_{n,m}$ reads,
\begin{eqnarray}
I_{n,m} 
&=& 
\int_{{\cal C}_2} c_n J_0 \, z_2^{m} \, dz_2  
\ = \ 
\int_{{\cal C}_2} u_{z_2} \, \psi_{n,m} 
\ \equiv \ 
{}_{\omega_{z_2}}\!\langle \psi_{n,m} | {\cal C}_2 ] \ ,
\end{eqnarray}
where
\begin{align}
\psi_{n,m} \equiv c_n \, z_2^{m} \, dz_2 \,, 
\qquad u_{z_2} \equiv \, J_0 \, , \qquad
\hat{\omega}_{z_2} = \partial_{z_2} \log u_{z_2} \, . 
\label{eq:generic:def:u2}
\end{align}
Let us stress that $\hat{\omega}_{z_2} \ne \hat{\omega}_2$, while, by construction, $\hat{\omega}_{z_1} = \hat{\omega}_1$. \\ 
Under the assumption that $\masters_1=1$, the number $\masters_2$ of solutions of $\hat{\omega}_{z_2} = 0$ corresponds to the total number $\masters$ of MIs. 
Finally, we define a monomial basis for the $z_2$-intersection, 
 ${}_{\omega_{z_2}}\langle \phi_k | \equiv z_2^{k-1} \, d {z_2}$, and 
complete the decomposition of $I_{n,m}$, by applying the reduction by intersections 
in $z_2$ to  
${}_{\omega_{z_2}}\!\langle \psi_{n,m}|$,
\begin{eqnarray}
{}_{\omega_{z_2}}\!\langle \psi_{n,m}|
&=&
\sum_{i,j=1}^{\masters_2} 
\langle \psi_{n,m}| \phi_j \rangle_{\omega_{z_2}}
({\bf C}_{\omega_{z_2}}^{-1})_{ji} \ 
{}_{\omega_{z_2}}\!\langle \phi_i |
\ ,
\end{eqnarray}
where all intersection numbers are computed with $\omega_{z_2}$. \\ 
The above equation corresponds to the decomposition of $I_{n,m}$ in terms of $\masters_2$ master integrals $I_{0,i}$ with $i=0,1,\ldots,\masters_2{-}1$, 
\begin{eqnarray}
I_{n,m} = \sum_{i=0}^{\masters_2-1} \, c_{n,m,i} \, I_{0,i} \ ,
\end{eqnarray}
where 
\begin{eqnarray}
c_{n,m,i} &=& \langle \psi_{n,m} | \phi_j \rangle_{\omega_{z_2}}
({\bf C}_{\omega_{z_2}}^{-1})_{ji} \ .
\label{eq:itint:gendeco}
\end{eqnarray}

We apply the iterative intersections method to the two-loop sunrise and massless planar doublebox diagrams.

\subsection{Two-Loop Massless Sunrise}

In the standard Baikov approach, the sunrise type integrals considered in Sec.{\ref{sec:massless-sunrise}}, on the maximal-cut, depend on two ISPs.
The corresponding two-fold Baikov representation was studied in \cite{Bosma:2017ens}.
Accordingly, we consider the following integral family,
\begin{gather}
I_{n,m} \equiv 
\int_{{\cal C}_1} 
\int_{{\cal C}_2} 
\ u \ 
\ z_1^n \, z_2^m \ 
dz_1 \wedge dz_2 \ ,
\qquad {\cal C}_1, {\cal C}_2 = [0,\infty]
\\
u = (z_1 z_2 (1 + z_1 + z_2 ))^\gamma \ ,
\end{gather}
with
\begin{equation}
\omega = \hat{\omega}_1 d z_1 + \hat{\omega}_2 dz_2, \qquad
\hat{\omega}_1 = 
\frac{\gamma  \left(2 z_1+z_2+1\right)}{z_1 \left(z_1+z_2+1\right)}, \qquad
\hat{\omega}_2 =  
\frac{\gamma  \left(z_1+2 z_2+1\right)}{z_2 \left(z_1+z_2+1\right)} \ .
\end{equation}
We observe that by setting $\gamma=(d-4)/2$, these integrals correspond to one 
introduced in
Sec.~\ref{sec:massless-sunrise} for $s = -1$.

\subsubsection{Iterated Intersections}

We rewrite $I_{n,m}$ iteratively, as,
\begin{eqnarray}
I_{n,m} &\equiv&
\int_{{\cal C}_2} 
dz_2 
z_2^{m} \,
J_{n} \ , \\
J_{n} &\equiv& 
\int_{{\cal C}_1} 
dz_1 \ u_{z_1} \, z_1^n \ , \\
u_{z_1} &=& (z_1 z_2 (1+z_1+z_2))^\gamma \ .
\end{eqnarray}

\paragraph{Intersections in $z_1$.}
We define 
\begin{eqnarray}
\hat{\omega}_{z_1} &=& \partial_{z_1} \log\left( u_{z_1} \right) = 
\frac{\gamma  \left(2 z_1+z_2+1\right)}{z_1 \left(z_1+z_2+1\right)} = \hat{\omega}_1 \ , 
\end{eqnarray}
with $\masters_1 =1$.
The decomposition of $J_1$ in terms of $J_0$ reads
\begin{eqnarray}
J_1 &=& c_1 \ J_0 \ , 
\qquad
\Longleftrightarrow
\qquad
{}_{\omega_1}\!\langle \phi_2 | {\cal C}_1] \ = 
c_1 \ 
{}_{\omega_1}\!\langle \phi_1 | {\cal C}_1]
\\
c_1 &=& 
\langle \phi_2 | \phi_1 \rangle_{\omega_1} 
\langle \phi_1 | \phi_1 \rangle_{\omega_1}^{-1}
= - \frac{1}{2} (1+z_2)
\end{eqnarray}

\paragraph{Intersections in $z_2$.}
\begin{eqnarray}
\hat{\omega}_{z_2} &=& \partial_{z_2}\log( J_0 ) 
= 
\frac{\gamma +3 \gamma  z_2+z_2}{z_2 \left(z_2+1\right)} \ ,
\end{eqnarray}
In this case, $\masters_2 = 1$, therefore the problem has just 1 master integral,
which we chose to be ${}_{\omega_{z_2}}\!\langle \phi_1 | = dz_2 $. 
After defining
\begin{eqnarray}
\psi_{1,0} = c_1 \, dz_2 = - \frac{1}{2} (1+z_2) d z_2 \ ,
\end{eqnarray}
using eq.~(\ref{eq:itint:gendeco}),
we finally get 
\begin{eqnarray}
I_{1,0} &=& c_{1,0,0} \, I_{0,0} \ , \\
c_{1,0,0} &=& 
\langle \psi_{1,0} | \phi_1 \rangle_{\omega_1}
\langle \phi_1 | \phi_1 \rangle_{\omega_1}^{-1} = - \frac{1}{3} \ .
\end{eqnarray}
which is the expected result, in agreement with eq.~\eqref{massless-sunrise-result} for $s=-1$.

\subsection{Two-Loop Massless Double-Box}
In the standard Baikov approach, the massless double-box type integrals considered in Sec.~\ref{sec:massless-double-box}, on the maximal-cut, depend on two ISPs. The corresponding two-fold Baikov representation was studied in \cite{Bosma:2017ens}. Accordingly, we consider the following integral family,
\begin{gather}
I_{n,m} \equiv
\int_{{\cal C}_1} 
\int_{{\cal C}_2} 
\ u \ 
\ z_1^n \, z_2^m \ 
dz_1 \wedge dz_2 \ , 
\qquad {\cal C}_1, {\cal C}_2 = [0,\infty]
\\
u = (z_1 z_2 (1 + a (z_1 + z_2) + b z_1 z_2))^\gamma \ ,
\end{gather}
with
\begin{gather}
\omega = \hat{\omega}_1 d z_1 + \hat{\omega}_2 dz_2 \\
\hat{\omega}_1 = 
\frac{\gamma  \left(a \left(2 z_1+z_2\right)+2 b z_1 z_2+1\right)}{z_1 \left(a \left(z_1+z_2\right)+b z_1 z_2+1\right)}, \qquad \hat{\omega}_2 =  
\frac{\gamma  \left(2 z_2 \left(a+b z_1\right)+a z_1+1\right)}{z_2 \left(a \left(z_1+z_2\right)+b z_1 z_2+1\right)}.
\end{gather}
This family, with $\gamma=(d-6)/2$, appears in the maximal cut of the two-loop double-box introduced in Sec.~\ref{sec:massless-double-box} 
(for $a=-1/t$ and $b=-1/(s t)$).

\subsubsection{Iterated Intersections}

Rewrite $I_{n,m}$ iteratively, as,
\begin{eqnarray}
I_{n,m} &\equiv&
\int_{{\cal C}_2} 
dz_2 
z_2^{m} \,
J_{n} \ , \\
J_{n} &\equiv& 
\int_{{\cal C}_1} 
dz_1 \ u_{z_1} \, z_1^n \ , \\
u_{z_1} &=& (z_1 z_2 (1 + a (z_1 + z_2) + b z_1 z_2))^\gamma \ ,
\end{eqnarray}

We consider the decomposition of two integrals, namely 
$I_{1,0}$, and $I_{2,0}$.

\paragraph{Intersections in $z_1$.}
We define 
\begin{eqnarray}
\hat{\omega}_{z_1} &=& \partial_{z_1} \log\left( u_{z_1} \right) = 
\frac{\gamma  \left(a \left(2 z_1+z_2\right)+2 b z_1 z_2+1\right)}{z_1 \left(a \left(z_1+z_2\right)+b z_1 z_2+1\right)}\ 
= \hat{\omega}_1 \ , 
\end{eqnarray}
with $\masters_1 =1$. \\
The decomposition of $J_1$ in terms of $J_0$ reads
\begin{eqnarray}
J_1 &=& c_1 \ J_0 \ , 
\qquad
\Longleftrightarrow
\qquad
{}_{\omega_1}\!\langle \phi_2 | {\cal C}_1] \ = 
c_1 \ 
{}_{\omega_1}\!\langle \phi_1 | {\cal C}_1] \,,
\\
c_1 &=& 
\langle \phi_2 | \phi_1 \rangle_{\omega_1} 
\langle \phi_1 | \phi_1 \rangle_{\omega_1}^{-1}
= 
-\frac{a z_2+1}{2 \left(a+b z_2\right)} \ ,
\end{eqnarray}
and 
the decomposition of $J_2$ in terms of $J_0$ reads
\begin{eqnarray}
J_2 &=& c_2 \ J_0 \ , 
\qquad
\Longleftrightarrow
\qquad
{}_{\omega_1}\!\langle \phi_3 | {\cal C}_1] \ = 
c_2 \ 
{}_{\omega_1}\!\langle \phi_1 | {\cal C}_1] \,,
\\
c_2 &=& 
\langle \phi_3 | \phi_1 \rangle_{\omega_1} 
\langle \phi_1 | \phi_1 \rangle_{\omega_1}^{-1}
= 
\frac{(\gamma +2) \left(a z_2+1\right){}^2}{2 (2 \gamma +3) \left(a+b z_2\right){}^2}
\ . 
\end{eqnarray}

\paragraph{Intersections in $z_2$.}
\begin{eqnarray}
\hat{\omega}_{z_2} &=& \partial_{z_2}\log(J_0) 
= 
\frac{a^2 \left(3 \gamma  z_2+z_2\right)+a \left(2 b \gamma  z_2^2+\gamma \right)-b z_2}{z_2 \left(a z_2+1\right)
   \left(a+b z_2\right)}\ .
\end{eqnarray}
In this case, $\masters_2 = 2$, therefore the problem has 2 master integrals,
which we choose to be ${}_{\omega_{z_2}}\!\langle \phi_1 | = dz_2 $
and $ {}_{\omega_{z_2}}\!\langle \phi_2 | = z_2 \, dz_2 $. \\ 
After defining 
\begin{eqnarray}
\psi_{1,0} = c_1 \, d z_2 \ , \qquad 
\psi_{2,0} = c_2 \, d z_2 \ ,
\end{eqnarray}
we obtain the decompositions of $I_{1,0}$ and $I_{2,0}$ in terms of the master integrals 
$I_{0,0}$ and $I_{0,1}$, 
\begin{eqnarray}
I_{1,0} &=& c_{1,0,0} \, I_{0,0} + c_{1,0,1} \, I_{0,1} \ , \\
I_{2,0} &=& c_{2,0,0} \, I_{0,0} + c_{2,0,1} \, I_{0,1} \ , 
\end{eqnarray}
where the coefficients are computed using eq.~(\ref{eq:itint:gendeco}),
\begin{eqnarray}
c_{1,0,0} &=& 0 \ , \hspace*{2.5cm}
c_{1,0,1} = 1 \ , \\
c_{2,0,0} &=& -\frac{\gamma +1}{b (2 \gamma +3)} \ , \qquad 
c_{2,0,1} = -\frac{3 a^2 \gamma +3 a^2+b}{a b (2 \gamma +3)}  \ ,
\end{eqnarray}
in agreement with \textsc{Reduze}.

\paragraph{}
In App. \ref{app:LBLiterative}, we provide further applications of the Iterative Intersections approach to the decomposition of integrals belonging to the massless double-box and planar pentabox integral families, the latter being an example of three-form integral decomposition.

\section{Intersection Numbers of Two-Forms}
\label{sec:twoforms}

\allowdisplaybreaks

In this section we present an alternative algorithm for computing intersection numbers of two-forms and demonstrate how to reproduce the two-loop results of Sec.~\ref{sec:massless-sunrise} and \ref{sec:massless-double-box} from this point of view. The algorithm is an extension of Matsumoto's method \cite{matsumoto1998} to non-logarithmic differential forms. We summarize it as follows.

Let us consider an integral of the form
\be\label{two-variate-integral}
\int_{\mathcal{C}} u(x,y) \phi(x,y) \qquad\text{with}\qquad u = B_1^{\gamma_1} B_2^{\gamma_2} \cdots B_m^{\gamma_m},
\ee
where $\gamma_i$ are generic coefficients, $\phi(x,y)$ is a two-form $\phi(x,y) = \hat{\phi}\, dx{\wedge} dy$, and $\mathcal{C}$ is an integration cycle such that $u$ vanishes on its boundaries. From here we define the one-form:
\be
\omega = d\log u = \sum_{i=1}^{m} \gamma_i \left( \frac{\partial_{x} B_i}{B_i} dx + \frac{\partial_{y} B_i}{B_i} dy \right).
\ee
As before, we also define the connection $\nabla_{\omega} \equiv d + \omega\wedge$. Poles of $\omega$ form hypersurfaces $\mathcal{H}_i$. For example, associated to each factor $B_i$ in \eqref{two-variate-integral} we have:
\be
\mathcal{H}_i \equiv \{ (x,y) \; | \; B_i(x,y) = 0 \}.
\ee
It is important to remember that all the differential forms are defined on the complex projective plane $\mathbb{CP}^2$, and by choosing coordinates $(x,y) \in \mathbb{C}^2$ we committed ourselves to one particular chart on this space, which does not cover the points at infinity. In order to find all hypersurfaces, including those at infinity, it is necessary to cover the full space with other charts, e.g., $(\hat{x},\hat{y}) = (x,1/y), (1/x,y), (1/x,1/y)$. We find that they do not contribute to the cases of our interest.

The above hypersurfaces, in general, intersect at points $\mathcal{P}_{ij}$ (we assume that all intersections are transverse),
\be
\mathcal{P}_{ij} \equiv \mathcal{H}_i \cap \mathcal{H}_j \qquad\text{for}\qquad i\neq j.
\ee
It is possible that $\mathcal{P}_{ij}$ contains more than one intersection point. If more than two distinct hypersurfaces intersect at one point, i.e., $\mathcal{H}_i \cap \mathcal{H}_j \cap \mathcal{H}_k \neq \varnothing$, there is a need for a local blowup near such a point. It is not relevant to the cases we study.

\subsection{\label{two-forms-general-algorithm}General Algorithm}

The algorithm for computing the intersection number $\langle \phi_L | \phi_R \rangle_\omega$ consists of three steps.

\begin{enumerate}[leftmargin=1.3em]
	
	\item {\bf Hypersurfaces.} In the small neighbourhood of each hypersurface $\mathcal{H}_i$ construct the one-form $\psi_i$ satisfying the equation:
	\be\label{psi-i-equation}
	\nabla_{\omega}\psi_i = \phi_L\qquad\text{locally near }\mathcal{H}_i.
	\ee
	Let us show how to do it explicitly when $\mathcal{H}_i$ is a hyperplane, i.e., the corresponding $B_i(x,y)$ is a linear function in $x$ and $y$.
	
	Let us pick coordinates: normal to the hyperplane, $z_i^\perp = B_i$, and along the hyperplane $z_i^\parallel$, in such a way that they are orthonormal, i.e., $dx \wedge dy = dz_i^\perp \wedge dz_i^\parallel$ with a unit Jacobian. Treating $z_i^\parallel$ as a constant, we write an ansatz in terms of a Laurent expansion:
	\be
	\psi_i = \left(\sum_{k=\text{min}}^{\text{max}} \psi_i^{(k)} (z_i^\perp)^k + \mathcal{O}((z_i^\perp)^{k+1}) \right) dz_i^\parallel.
	\ee
	The expansion start at the order $\text{min} = \text{ord}_{z_i^\perp} (\phi_L) + 1$ and it is enough to expand until $\text{max} = -\text{ord}_{z_i^\perp} (\phi_R) - 1$. By comparing both sides of \eqref{psi-i-equation} at each order in $z_i^\perp$ we can solve for the coefficients $\psi^{(k)}_i$.
	
	\item {\bf Intersections of Hypersurfaces.} In the small neighbourhood of each point in $\mathcal{P}_{ij}$ construct the function $\psi_{ij}$ satisfying the equation:
	\be
	\nabla_{\omega} \psi_{ij} = \psi_i - \psi_j \qquad\text{locally near }\mathcal{P}_{ij}.
	\ee
	The right-hand side is known as an expansion in variables $z_i^\perp$ and $z_j^\perp$ from the previous step. Hence we change the coordinates to $(z_i^\perp, z_j^\perp)$ and write an ansatz for $\psi_{ij}$:
	\be
	\psi_{ij} = \sum_{k=\text{min}_i}^{\text{max}_i} \sum_{l=\text{min}_j}^{\text{max}_j} \psi_{ij}^{(k,l)} (z_i^\perp)^k (z_j^\perp)^l \;+\; \mathcal{O}((z_i^\perp)^{k+1}, (z_j^\perp)^{l+1}),
	\ee
	where
	\begin{align}
		\text{min}_i = \text{ord}_{z_i^\perp}(\psi_i {-} \psi_j) +1, &\qquad \text{max}_i = -\text{ord}_{z_i^\perp}(\phi_R) -1 \label{two-variate-criteria-1}\\
		\text{min}_j = \text{ord}_{z_j^\perp}(\psi_i {-} \psi_j) +1, &\qquad \text{max}_j = -\text{ord}_{z_j^\perp}(\phi_R) -1. \label{two-variate-criteria-2}
	\end{align}
	Once again, it can be solved order by order for each of the coefficients $\psi_{ij}^{(k,l)}$.
	
	\item {\bf Intersection Numbers of Two-Forms.} Finally, the intersection number $\langle \phi_L | \phi_R \rangle_\omega$ is computed as a sum over all intersection points $\mathcal{P}_{ij}$ using the double-residue formula:
	\be
	\langle \phi_L | \phi_R \rangle_\omega \equiv \sum_{\mathcal{P}_{ij}} \Res_{z_i^\perp = 0} \Res_{z_j^\perp = 0} \Big( \psi_{ij}\, \phi_R \Big).
	\ee
	In order to perform the residue computation we express $\phi_R$ as a two-form in the new coordinates $(z_i^\perp, z_j^\perp)$. Recall that upon such a change one picks up a Jacobian:
	\be
	\hat{\phi}_R(x,y)\, dx {\wedge} dy = \frac{\hat{\phi}_R(z_i^\perp, z_j^\perp)\, dz_i^\perp {\wedge} dz_j^\perp}{|\partial(z_i^\perp, z_j^\perp) / \partial (x,y)|}.
	\ee
	A given point $\mathcal{P}_{ij}$ can only contribute to this sum if $\text{min}_i \leq \text{max}_i$ and $\text{min}_j \leq \text{max}_j$ from (\ref{two-variate-criteria-1}--\ref{two-variate-criteria-2}) holds.
	
\end{enumerate}

\subsection{Two-Loop Massless Sunrise}

Let us reconsider the massless sunrise diagram from Sec.~\ref{sec:massless-sunrise}. We use propagators as in \eqref{sunrise-propagators} as well as two ISPs:
\be
x = D_4 = k_2^2, \qquad y = D_5 = (k_1 - p)^2.
\ee
The resulting maximal cut for single propagators $D_1,D_2,D_3$ reads
\be
I_{1,1,1;-n,-m} = \int_{\mathcal{C}} u(x,y)\, \phi_{-n,-m},
\ee
where
\be
u(x,y) \equiv \left(- \frac{1}{4s} x y (x+y-s)\right)^{\frac{d-4}{2}}, \qquad \phi_{-n,-m} \equiv x^n y^m\, dx {\wedge} dy.
\ee
We know from Section~\ref{sec:massless-sunrise} that this integral has $\masters=1$ master integral. Alternatively, this counting can be obtained by calculating
\be
\omega = \frac{d-4}{2} \left( \left( \frac{1}{x} + \frac{1}{x + y - s} \right) dx + \left( \frac{1}{y} + \frac{1}{x + y - s} \right) dy\right)
\ee
and finding that the critical point equation, $\omega = 0$, yields a single solution.\footnote{As emphasized in \cite{Mastrolia:2018uzb}, the number of critical points coincides with the number of master integrals $\masters$ only under certain genericity assumptions, see, e.g., \cite{aomoto2011theory}. Using the theory of hyperplane arrangements \cite{orlik2013arrangements}, $\masters$ also equals to the number of bounded chambers (connected components that do not have a boundary at infinity) of $\mathbb{R}^2 \setminus \{ u(x,y) = 0 \}$, provided that exponents $\gamma_i$ of each hyperplane $\mathcal{H}_i$ are generic. In our case there is only one given by $\{ (x,y) \, |\, x>0, y>0, x+y-s < 0\}$ for $s>0$.}

We want to decompose $\phi_{-1,0}$ in the basis of $\phi_{0,0}$. It is the most convenient to first perform the rescaling
\be
\tilde{u}(x,y) = x^3 y^3\, u(x,y), \qquad \tilde{\phi}_{-n,-m} = \frac{\phi_{-n,-m}}{x^3 y^3}
\ee
such that the integrand $u(x,y) \phi_{-n,-m} = \tilde{u}(x,y) \tilde{\phi}_{-n,-m}$ is preserved. In this way the new two-forms $\tilde{\phi}_{-1,0}$ and $\tilde{\phi}_{0,0}$ do not have poles at infinity and we can keep working in the chart $(x,y) \in \mathbb{C}^2$. Explicitly, the rescaled forms and $\omega$ are
\begin{gather}
\tilde{\phi}_{-1,0} = \frac{dx\wedge dy}{x^2 y^3}, \qquad\tilde{\phi}_{0,0} = \frac{dx\wedge dy}{x^3 y^3},\\
\tilde{\omega} = \frac{1}{2} \left(\frac{d+2}{x} + \frac{d-4}{x+y-s}\right) dx + \frac{1}{2} \left(\frac{d+2}{y} + \frac{d-4}{x+y-s}\right) dy.
\end{gather}

\subsubsection{Evaluation of Intersection Numbers}

We start by solving the equation \eqref{psi-i-equation} around each hyperplane $\mathcal{H}_i$ at finite positions associated to:
\be
B_1 = x, \qquad B_2 = y, \qquad B_3 = x+y-s.
\ee
We evaluate the necessary ingredients in the computation of $\langle \tilde{\phi}_{-1,0} |  \tilde{\phi}_{0,0} \rangle_{\tilde{\omega}}$. We ought to first compute $\psi_1, \psi_2, \psi_3, \psi_{12}, \psi_{13}, \psi_{23}$ according to the rules given in the previous subsection.

\begin{itemize}[leftmargin=1.3em]
	\item {\bf Hyperplane $\mathcal{H}_1 = \{x = 0\}$.} We choose the coordinates $(z_1^\perp, z_1^\parallel) = (x,y)$. Solving $\nabla_{\tilde{\omega}} \psi_1 = \tilde{\phi}_{-1,0}$ we find:
	\begin{align}
		\psi_1 = \bigg(&
		\frac{2}{d (z_1^\parallel)^3} \frac{1}{z_1^\perp}
		+\frac{2 (d-4)}{d (d+2) (z_1^\parallel)^3 (s-z_1^\parallel)}
		+\frac{4 (d-4) (d-1)}{d (d+2) (d+4) (z_1^\parallel)^3
			(s-z_1^\parallel)^2} z_1^\perp \nonumber\\
		&+\frac{8 (d-4) (d-1)}{(d+2) (d+4) (d+6) (z_1^\parallel)^3 (s-z_1^\parallel)^3} (z_1^\perp)^2
		+ \mathcal{O}\big((z_1^\perp)^3\big)\bigg)\, dz_1^\parallel.
	\end{align}

	\item {\bf Hyperplane $\mathcal{H}_2 = \{y = 0\}$.} We choose the coordinates $(z_2^\perp, z_2^\parallel) = (y,-x)$. Solving $\nabla_{\tilde{\omega}} \psi_2 = \tilde{\phi}_{-1,0}$ we find:
	\begin{align}
		\psi_2 = \bigg(&
		\frac{2}{(d-2) (z_2^\parallel)^2} \frac{1}{(z_2^\perp)^2}
		+\frac{2 (d-4)}{(d-2) d (z_2^\parallel)^2 (s+z_2^\parallel)} \frac{1}{z_2^\perp}
		+\frac{4 (d-4)}{d (d+2) (z_2^\parallel)^2 (s+z_2^\parallel)^2}\nonumber\\
		&+\frac{8 (d-4) (d-1)}{d (d+2) (d+4)(z_2^\parallel)^2 (s+z_2^\parallel)^3} z_2^\perp \nonumber\\
		&+\frac{16 \left(d^2-5 d+4\right)}{(d+2) (d+4) (d+6) (z_2^\parallel)^2 (s+z_2^\parallel)^4} (z_2^\perp)^2
		+ \mathcal{O}\big((z_2^\perp)^3\big)\bigg)\, dz_2^\parallel.
	\end{align}

	\item {\bf Hyperplane $\mathcal{H}_3 = \{x{+}y{-}s = 0\}$.} We choose the coordinates $(z_3^\perp, z_3^\parallel) = (x+y-s, (y-x)/2)$. Solving $\nabla_{\tilde{\omega}} \psi_3 = \tilde{\phi}_{-1,0}$ we find:
	\begin{align}
	\psi_3 = \bigg(&
	\frac{64}{(d-2) (s-2 z_3^\parallel)^2 (s+2 z_3^\parallel)^3} z_3^\perp \nonumber\\
	&-\frac{64 (7 d s-2 d z_3^\parallel -6 s+4 z_3^\parallel)}{(d-2) d (s-2
		z_3^\parallel)^3 (s+2 z_3^\parallel)^4} (z_3^\perp)^2
	+ \mathcal{O}\big((z_3^\perp)^3\big)\bigg)\, dz_3^\parallel.
	\end{align}

	\item {\bf Intersection Point $\mathcal{P}_{12} = (0,0)$.} We choose the coordinates $(z_1^\perp, z_2^\perp) = (x,y)$. Solving $\nabla_{\tilde{\omega}} \psi_{12} = \psi_1 - \psi_2$ we find:
	\begin{align}
		\psi_{12} =& 
		\frac{4 (d-4) \left(71 d^2-14 d-72\right) z_1^\perp z_2^\perp}{d (d+2)^2 (d+4)^2 s^5}
		+\frac{16 (d-4) \left(5 d^2-9 d+4\right) (z_1^\perp)^2}{(d-2) d (d+2) (d+4) (d+6) s^4 z_2^\perp} \\
		&+\frac{32 \left(d^2-5 d+4\right) (z_2^\perp)^2}{d (d{+}2) (d{+}4) (d{+}6) s^4 z_1^\perp}
		+\frac{20 \left(5 d^2-22 d+8\right) z_1^\perp}{d (d+2)^2 (d+4) s^4}
		+\frac{32 (d-4) \left(d^2-2 d+1\right) z_1^\perp}{(d{-}2) d^2 (d{+}2) (d{+}4) s^3 z_2^\perp} \nonumber\\
		&+\frac{16 (d{-}4) (d{-}1) z_2^\perp}{d^2 (d{+}2) (d{+}4) s^3 z_1^\perp}
		+\frac{16 \left(2 d^2 {-} 9 d {+} 4\right)}{d^2 (d{+}2)^2 s^3}
		+\frac{8(d-4)}{d^2 (d{+}2) s^2 z_1^\perp}
		+\frac{4 (d-4) (3 d-4)}{(d{-}2) d^2 (d{+}2) s^2 z_2^\perp} \nonumber\\
		&+\frac{4 (d-4)}{(d{-}2) d^2 s z_1^\perp z_2^\perp}
		+\frac{4\left(71 d^3 {-} 258 d^2 {-} 96 d {-} 32\right) (z_1^\perp)^2}{d (d{+}2)^2 (d{+}4) (d{+}6) s^5}
		+\frac{4 (d-4)}{(d-2) d (d+2) s (z_2^\perp)^2} \nonumber\\
		&+\frac{4 \left(643 d^4 {-} 2026 d^3 {-} 3592 d^2 {+} 5536 d {+} 384\right) (z_1^\perp)^2 (z_2^\perp)^2}{d (d+2) (d+4)^2 (d+6)^2 s^7}
		+\frac{80 (d{-}4) (d{-}1) z_2^\perp}{d (d{+}2)^2 (d{+}4) s^4} \nonumber\\
		&+\frac{4 \left(221 d^4 {-} 602 d^3 {-} 1384 d^2 {+} 672 d {+} 1408\right) (z_1^\perp)^2 z_2^\perp}{d (d+2)^2 (d+4)^2 (d+6) s^6}
		+\frac{64 \left(3 d^3 {-} 14 d^2 {+} 7 d {+} 4\right) (z_2^\perp)^2}{d(d+2)^2 (d+4) (d+6) s^5}
		\nonumber\\
		&+\frac{4 \left(191 d^4 {-} 632 d^3 {-} 844 d^2 {+} 1152 d {+} 448\right) z_1^\perp (z_2^\perp)^2}{d (d+2)^2 (d+4)^2 (d+6) s^6}
		+\frac{8 (d-4) (d-1) z_1^\perp}{(d{-}2) d (d{+}2) (d{+}4) s^2 (z_2^\perp)^2}
		\nonumber\\
		&+\frac{16 (d-4) (d-1) (z_1^\perp)^2}{(d-2) (d+2) (d+4) (d+6) s^3 (z_2^\perp)^2}
		+\frac{4}{(d-2) d z_1^\perp (z_2^\perp)^2}
		+ \mathcal{O}\big((z_1^\perp)^3, (z_2^\perp)^3\big).\nonumber
	\end{align}
	Hence the contribution to the intersection number $\langle \tilde{\phi}_{-1,0} | \tilde{\phi}_{0,0} \rangle_{\tilde{\omega}}$ from the point $\mathcal{P}_{12}$ is
	\be\label{two-form-contrib-1}
	\Res_{z_1^\perp = 0} \Res_{z_2^\perp = 0} \Big( \psi_{12}\, \tilde{\phi}_{0,0} \Big) = \frac{4 (d-4) \left(643 d^3+546 d^2-1408 d-96\right)}{d (d+2) (d+4)^2 (d+6)^2 s^7},
	\ee
	where we used that in the coordinates $(z_1^\perp, z_2^\perp)$ the two-form $\tilde{\phi}_{0,0}$ reads
	\be
	\tilde{\phi}_{0,0} = \frac{dz_1^\perp \wedge dz_2^\perp}{(z_1^\perp)^3 (z_2^\perp)^3}.
	\ee

	\item {\bf Intersection Point $\mathcal{P}_{13} = (0,s)$.} We choose the coordinates $(z_1^\perp, z_3^\perp) = (x,x+y-s)$. Solving $\nabla_{\tilde{\omega}} \psi_{13} = \psi_1 - \psi_3$ we find:
	\begin{align}
		\psi_{13} =&
		\frac{4 (d-4) (7 d-4) (z_1^\perp)^2}{d (d+2) (d+4) (d+6) s^4 z_3^\perp}
		-\frac{4 (d-6) (d-4) (z_1^\perp)^2}{d (d+2) (d+4) (d+6) s^3 (z_3^\perp)^2} \nonumber\\
		&+\frac{4 (d-4) z_1^\perp}{d (d+2) (d+4) s^3 z_3^\perp}
		+ \mathcal{O}\big((z_1^\perp)^3, (z_3^\perp)^0\big).
	\end{align}
	Hence the contribution to the intersection number $\langle \tilde{\phi}_{-1,0} | \tilde{\phi}_{0,0} \rangle_{\tilde{\omega}}$ from the point $\mathcal{P}_{13}$ is
	\be\label{two-form-contrib-2}
	\Res_{z_1^\perp = 0} \Res_{z_3^\perp = 0} \Big( \psi_{13}\, \tilde{\phi}_{0,0} \Big) = \frac{4 (d-4) (13 d-4)}{d (d+2) (d+4) (d+6) s^7},
	\ee
	where we used that in the coordinates $(z_1^\perp, z_3^\perp)$ the two-form $\tilde{\phi}_{0,0}$ reads
	\be
	\tilde{\phi}_{0,0} = \frac{dz_1^\perp \wedge dz_3^\perp}{(z_1^\perp)^3 (z_3^\perp - z_1^\perp + s)^3}.
	\ee

	\item {\bf Intersection Point $\mathcal{P}_{23} = (s,0)$.} We choose the coordinates $(z_2^\perp, z_3^\perp) = (y,x+y-s)$. Solving $\nabla_{\tilde{\omega}} \psi_{23} = \psi_2 - \psi_3$ we find:
	\begin{align}
	\psi_{23} = &
	\frac{4 (5 d-6) \left(d^2-10 d+24\right) (z_2^\perp)^2}{(d-2) d (d+2) (d+4) (d+6) s^3 (z_3^\perp)^2}
	-\frac{4 \left(19 d^3-106 d^2+128d-32\right) (z_2^\perp)^2}{(d-2) d (d+2) (d+4) (d+6) s^4 z_3^\perp}\nonumber\\
	&-\frac{4 \left(d^3-18 d^2+104 d-192\right) (z_2^\perp)^2}{(d-2) d (d+2) (d+4)
		(d+6) s^2 (z_3^\perp)^3}
	-\frac{4 (d-4) (5 d-6) z_2^\perp}{(d-2) d (d+2) (d+4) s^3 z_3^\perp}\\
	&+\frac{4 (d-6) (d-4) z_2^\perp}{(d-2) d (d+2) (d+4)
		s^2 (z_3^\perp)^2}-\frac{4 (d-4)}{(d-2) d (d+2) s^2 z_3^\perp}
	+ \mathcal{O}\big((z_2^\perp)^3, (z_3^\perp)^0\big).\nonumber
	\end{align}
	Hence the contribution to the intersection number $\langle \tilde{\phi}_{-1,0} | \tilde{\phi}_{0,0} \rangle_{\tilde{\omega}}$ from the point $\mathcal{P}_{23}$ is
	\be\label{two-form-contrib-3}
	\Res_{z_2^\perp = 0} \Res_{z_3^\perp = 0} \Big( \psi_{23}\, \tilde{\phi}_{0,0} \Big) = \frac{4 (d-4) \left(73 d^2-90 d+8\right)}{(d-2) d (d+2) (d+4) (d+6) s^7},
	\ee
	where we used that in the coordinates $(z_2^\perp, z_3^\perp)$ the two-form $\tilde{\phi}_{0,0}$ reads
	\be
	\tilde{\phi}_{0,0} = -\frac{dz_2^\perp \wedge dz_3^\perp}{(z_2^\perp)^3 (z_3^\perp - z_2^\perp + s)^3}.
	\ee

\end{itemize}

Summing up the three contributions \eqref{two-form-contrib-1}, \eqref{two-form-contrib-2}, and \eqref{two-form-contrib-3} we obtain the intersection number:
\be
\langle \tilde{\phi}_{-1,0} | \tilde{\phi}_{0,0} \rangle_{\tilde{\omega}} = \frac{36 (d-4) (3 d-4) (3 d-2) (3 d+2) (3 d+4)}{(d-2) d (d+2) (d+4)^2 (d+6)^2 s^7}.
\ee

Entirely analogous computation can be repeated for $\langle \tilde{\phi}_{0,0} | \tilde{\phi}_{0,0} \rangle_{\tilde{\omega}}$, giving
\begin{align}
	\langle \tilde{\phi}_{0,0} | \tilde{\phi}_{0,0} \rangle_{\tilde{\omega}} &= \left(\frac{12 (d-4) (3 d-4) \left(221 d^3+108 d^2-596 d-48\right)}{(d-2) d (d+2) (d+4)^2 (d+6)^2 s^8}\right) \nonumber\\
	&\quad + \left(\frac{12 (d-4) (3 d-4) (11 d-2)}{(d-2) d (d+2) (d+4) (d+6) s^8} \right) + \left(\frac{12 (d-4) (3 d-4) (11 d-2)}{(d-2) d (d+2) (d+4) (d+6) s^8} \right)\nonumber\\
	&= \frac{108 (d-4) (3 d-4) (3 d-2) (3 d+2) (3 d+4)}{(d-2) d (d+2) (d+4)^2 (d+6)^2 s^8}.
\end{align}
where in the first equality the first, second, and third terms come from the intersection points $\mathcal{P}_{12}$, $\mathcal{P}_{13}$, and $\mathcal{P}_{23}$ respectively (the last two are equal by exchange symmetry in $x$ and $y$).

\subsubsection{Basis Decomposition}

As in Sec.~\ref{sec:massless-sunrise}, we choose the master integral to be $I_{1,1,1;0,0}$ and we decompose $I_{1,1,1;-1,0}$ in this basis. Using the above intersection numbers we have:
\begin{align}
I_{1,1,1;-1,0} &= \langle \tilde{\phi}_{-1,0} | \tilde{\phi}_{0,0} \rangle_{\tilde{\omega}}\, \langle \tilde{\phi}_{0,0} | \tilde{\phi}_{0,0} \rangle_{\tilde{\omega}}^{-1}\, I_{1,1,1;0,0} \nonumber\\
&= \frac{s}{3} I_{1,1,1;0,0}
\end{align}
in agreement with \eqref{massless-sunrise-result}.

\subsection{Two-Loop Massless Double-Box}

As the next example, we reconsider the decomposition of the massless double-box integral on the heptacut from Sec.~\ref{sec:massless-double-box}. We use the propagators from eq.~\eqref{eq:double-box-propagators} as well as the two ISPs:
\be
D_8 = (k_2 - p_1)^2, \qquad D_9 = (k_1 - p_3)^2 - s - t.
\ee
As in the previous example, the two-form $\phi_{-n,-m}$ has multiple poles at infinity for $n,m \geq 0$. This time, let us introduce the coordinates
\be
x = 1/D_8, \qquad y = 1/D_9
\ee
that move all the poles to finite positions. More precisely, we have:
\be
I_{1,1,1,1,1,1,1;-n,-m} = \int_{\mathcal{C}} u(x,y)\, \phi_{-n,-m}
\ee
with
\begin{gather}
u(x,y) \equiv \left( - \frac{1}{4t(s{+}t)} x^{-2} y^{-2} \left(t y + 1\right) \left( s x + (s{+}t)y + 1\right) \right)^{\frac{d-6}{2}},\\
\phi_{-n,-m} \equiv D_8^{n} D_9^{m}\, dD_8 {\wedge} dD_9 = x^{-n-2} y^{-m-2}\, dx{\wedge}dy.\label{double-box-two-form}
\end{gather}
Hence the one-form $\omega$ is
\be
\omega = \frac{d-6}{2} \left( \left(\frac{-2}{x} + \frac{s}{sx{+}(s{+}t)y{+}1}\right) dx + \left( \frac{-2}{y} + \frac{t}{ty {+}1} + \frac{s{+}t}{sx {+} (s{+}t)y {+} 1}\right) dy \right)
\ee
and since the exponents of each hyperplane are generic and $\omega=0$ gives two solutions, we have that $N=2$ is the number of master integrals.

The four hyperplanes at finite positions are:
\be
B_1 = x, \qquad B_2 = y, \qquad B_3 = t y + 1, \qquad B_4 = s x + (s+t) y + 1.
\ee
They intersect at the five points:
\begin{gather}
\mathcal{P}_{12} = (0,0), \qquad \mathcal{P}_{13} = (0,-1/t), \qquad \mathcal{P}_{14} = (0,-1/(s{+}t)),\nonumber\\
\mathcal{P}_{24} = (-1/t, -1/s), \qquad \mathcal{P}_{34} = (1/t, -1/t).
\end{gather}
There are also multiple intersection points at infinity that do not contribute because of the form of \eqref{double-box-two-form}.

We choose the following bases of twisted cocycles and their duals:
\be
\langle \varphi_1| = | \varphi_1 \rangle = \phi_{0,0}, \qquad \langle \varphi_2| = | \varphi_2 \rangle = \phi_{-1,0}
\ee
Using the algorithm from Sec.~\ref{two-forms-general-algorithm} we arrive at the following entries of the matrix $\mathbf{C}_{ij} = \langle \varphi_i | \varphi_j \rangle_{\omega}$:
\begin{align}
\langle \varphi_1 | \varphi_1 \rangle_{\omega} &= -\frac{s^2 \left(27 d^2 s^2 {+} 48 d^2 s t {+} 20 d^2 t^2 {-} 324 d s^2 {-} 576 d s t {-} 240 d t^2 {+} 960 s^2 {+} 1704 s t {+} 708 t^2\right)}{16 (d{-}7)^2
   (d{-}5)^2}\!,\nonumber\\
\langle \varphi_1 | \varphi_2 \rangle_{\omega} &= s^2 \Big(81 d^3 s^3 {+} 162 d^3 s^2 t {+} 88 d^3 s t^2 {+} 8 d^3 t^3 {-} 1566 d^2 s^3 {-} 3114 d^2 s^2 t {-} 1672 d^2 s t^2 \nonumber\\
&\qquad\quad {-} 144 d^2 t^3 {+}10008 d s^3 {+} 19788 d s^2 t {+} 10504 d s t^2 {+} 856 d t^3 {-} 21120 s^3 {-} 41520 s^2 t \nonumber\\
&\qquad\quad {-} 21792 s t^2 {-} 1680 t^3\Big) / \Big(32 (d{-}8) (d{-}7)^2 (d{-}5)^2 \Big), \\
\langle \varphi_2 | \varphi_1 \rangle_{\omega} &= s^2 \Big(81 d^3 s^3 {+} 162 d^3 s^2 t {+} 88 d^3 s t^2 {+} 8 d^3 t^3 {-} 1350 d^2 s^3 {-} 2718 d^2 s^2 t {-} 1496 d^2 s t^2 \nonumber\\
&\qquad\quad {-} 144 d^2 t^3 {+} 7416 d
   s^3 {+} 15036 d s^2 t {+} 8392 d s t^2 {+} 856 d t^3 {-} 13440 s^3 {-} 27456 s^2 t \nonumber\\
&\qquad\quad{-} 15552 s t^2 {-} 1680 t^3 \Big)/\Big(32 (d{-}7)^2 (d{-}5)^2 (d{-}4)\Big), \\
\langle \varphi_2 | \varphi_2 \rangle_{\omega} &= -s^3 \Big(243 d^4 s^3 {+} 540 d^4 s^2 t {+} 360 d^4 s t^2 {+} 64 d^4 t^3 {-} 5832 d^3 s^3 {-} 12960 d^3 s^2 t {-} 8640 d^3 s t^2 \nonumber\\
&\qquad\qquad {-} 1536 d^3
   t^3 {+} 51948 d^2 s^3 {+} 115536 d^2 s^2 t {+} 77136 d^2 s t^2 {+} 13760 d^2 t^3 {-} 203472 d s^3 \nonumber\\
&\qquad\qquad{-} 453312 d s^2 t {-} 303552 d s t^2 {-} 54528 d
   t^3 {+} 295680 s^3 {+} 660480 s^2 t {+} 444288 s t^2\nonumber\\
   &\qquad\qquad {+} 80640 t^3\Big) / \Big(64 (d{-}8) (d{-}7)^2 (d{-}5)^2 (d{-}4) \Big).
\end{align}

Since we want to perform a reduction of the integral $I_{1,1,1,1,1,1,1;-2,0}$, we also need to compute intersection numbers:
\begin{align}
\langle \phi_{-2,0} | \varphi_1 \rangle_\omega &= -s^2 \Big(243 d^4 s^4 {+} 540 d^4 s^3 t {+} 360 d^4 s^2 t^2 {+} 64 d^4 s t^3 {-} 5022 d^3 s^4 {-} 11340 d^3 s^3 t \nonumber\\
&\qquad\qquad {-} 7788 d^3 s^2 t^2 {-} 1504 d^3 s
   t^3 {-} 16 d^3 t^4 {+} 38448 d^2 s^4 {+} 88392 d^2 s^3 t\nonumber\\
   &\qquad\qquad{+} 62724 d^2 s^2 t^2 {+} 13184 d^2 s t^3 {+} 288 d^2 t^4 {-} 129312 d s^4 {-} 303360 d s^3 t \nonumber\\
   &\qquad\qquad{-} 223128
   d s^2 t^2 {-} 51104 d s t^3 {-} 1712 d t^4 {+} 161280 s^4 {+} 387072 s^3 t {+} 296064 s^2 t^2\nonumber\\
   &\qquad\qquad{+} 73920 s t^3 {+} 3360 t^4\Big) / \Big( 64 (d{-}7)^2 (d{-}5)^2
   (d{-}4) (d{-}3) \Big),\\
\langle \phi_{-2,0} | \varphi_2 \rangle_\omega &= s^3 \Big(729 d^5 s^4{+}1782 d^5 s^3 t{+}1404 d^5 s^2 t^2{+}368 d^5 s t^3{+}16 d^5 t^4{-}20412 d^4 s^4 \nonumber\\
&\qquad\quad {-}50274 d^4 s^3 t{-}40140 d^4 s^2
   t^2{-}10848 d^4 s t^3{-}544 d^4 t^4{+}225828 d^3 s^4 \nonumber\\
   &\qquad\quad{+}561456 d^3 s^3 t{+}455592 d^3 s^2 t^2{+}127568 d^3 s t^3{+}7344 d^3 t^4{-}1233792 d^2
   s^4\nonumber\\
   &\qquad\quad{-}3102744 d^2 s^3 t{-}2566656 d^2 s^2 t^2{-}748128 d^2 s t^3{-}49184 d^2 t^4{+}3328704 d s^4 \nonumber\\
   &\qquad\quad {+}8486304 d s^3 t{+}7179648 d s^2
   t^2{+}2188160 d s t^3{+}163328 d t^4{-}3548160 s^4 \nonumber\\
   &\qquad\quad{-}9192960 s^3 t{-}7981056 s^2 t^2{-}2553600 s t^3{-}215040 t^4\Big)\nonumber\\
   &\qquad/\Big(128 (d{-}8) (d{-}7)^2
   (d{-}5)^2 (d{-}4) (d{-}3)\Big).
\end{align}
This gives us the decomposition
\begin{align}
I_{1,1,1,1,1,1,1;-2,0} &= \sum_{i,j=1}^{2} \langle \phi_{-2,0} | \varphi_j \rangle_\omega \, (\mathbf{C}^{-1})_{ji}\, \langle \varphi_i | \mathcal{C} ]\nonumber\\
&= \frac{(d-4)st}{2(d-3)} I_{1,1,1,1,1,1,1;0,0} + \frac{2t-3(d-4)s}{2(d-3)} I_{1,1,1,1,1,1,1;-1,0},
\end{align}
which agrees with the result \eqref{double-box-result}. \\ \\

Algorithms such as the one discussed in the current section can in principle be applied to the decomposition of integral families which admit a two-form representation within the \lbl Baikov approach, {\it e.g.} the non-planar two-loop pentabox and the three-loop ladder box diagrams,
\begin{eqnarray}
\raisebox{-10pt}{\includegraphics[width=0.15\textwidth]{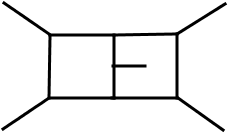}} 
\quad \ , \quad 
\raisebox{-10pt}{\includegraphics[width=0.2\textwidth]{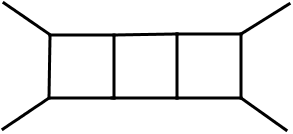}} \quad .
\nonumber
\end{eqnarray}
However, the study of those cases goes beyond the goal of the current work. \\ 

Before concluding, let us observe that
any generic Feynman integral admitting a multivariate integral representation can still be decomposed using the master decomposition formula eq.~(\ref{eq:masterdeco:}), thereby requiring the computation of {\it intersection numbers of generic rational multi-forms}. 
For maximal-cut, the integration variables correspond to irreducible scalar products, but, out of cut, they may also correspond to un-cut propagators: in the latter case, the master decomposition formula allows to determine the coefficients of the master integrals belonging to sub-sectors.
We defer these studies to future publications.

\section{Conclusions}
\label{sec:conclusions}

In this work, we gave a systematic presentation of the novel method for decomposing Feynman integrals onto a basis of master integrals by projections, which makes use of intersection numbers of differential forms~\cite{Mastrolia:2018uzb}. 
We showed advantages of this general mathematical framework by applying the decomposition-by-intersections to an extensive list of cases.

We recalled basic principles of intersection theory for hypergeometric functions, 
and established their correspondence to Feynman integrals in the Baikov representation, considering the standard formulation as well as in the more recent \lbl approach \cite{Frellesvig:2017aai}. 
We showed that within intersection theory, the integral decomposition is controlled by 
the geometric properties of the integrands, and that 
the evaluation of {\it intersection numbers} is the fundamental operation required in the  
{\it master decomposition formula}, finally yielding the direct determination of the coefficients of the integral reduction.
We elaborated on different options for the 
choice of the integral bases, and showed how the master formula can be also used to derive differential equations and dimensional recurrence relations for generic Feynman integrals.

In the first part of the work, we used the master decomposition formula to derive contiguity relations for special functions, such as the Euler $\beta$ function, the Gauss ${}_2F_1$ hypergeometric function, and the Appell $F_1$ function, belonging to the wider class of Lauricella functions.
Then, the new decomposition method was applied to  Feynman integrals. 
In particular, we focused on integrals whose maximal cuts admit 1-form integral representations, and discussed examples that have from two to an arbitrary number of loops, and/or from zero to an arbitrary number of legs, eventually corresponding to diagrams with internal and/or external massive lines. 
By limiting our analysis to 1-form integral representations, we addressed the decomposition of multi-loop integrals (on maximal cuts) which have either one irreducible scalar product (ISP), or that have multiple ISPs, but can be expressed as a one-fold integral using the \lbl approach.
In a few instructive cases, we illustrated the direct constructions of differential equations and dimensional recurrence relations for master integrals, and discussed how a different choice of the basis may impact the form of the result. Special emphasis is given to basis of {\it monomial forms} and to basis of {\it dlog forms}, 
in particular showing how the latter obeys a {\it canonical} system of differential equations.

For the cases where it was possible -- by means of public and private automatic codes, or by comparison with the literature -- 
we have verified that the decomposition formulae computed through the use of intersection numbers for 1- and 2-forms agree with the ones obtained using integration-by-parts (IBP) identities. 
In a few cases, intersection theory gave a lower number of master integrals than the one obtained by  IBP-decomposition on the maximal cut. We identified the source of the mismatch in additional, missing relations which were found at the cost of applying the IBP-reduction to integral families with a larger number of denominators.

Although the main part of this work addressed the application of intersection theory to 1-forms, 
the complete decomposition of multi-loop Feynman integrals in terms of master integrals (including also the ones corresponding to sub-diagrams) 
requires the evaluation of intersection numbers for generic $n$-forms. 
This topic does not appear to be fully covered in the differential and algebraic geometry literature. The available case of intersection numbers of dlog $n$-forms~\cite{matsumoto1998,Mizera:2017rqa} is not sufficient for Feynman integrals, which belong to the wider class of generic rational $n$-forms.
With a view towards the full extension of the formalism, in this work, we presented two novel algorithms for decomposition-by-intersections 
for cases where the maximal cuts admit a 2-form integral representation. 
They constitute important milestones for physical and mathematical research areas. Owing to the results of the research presented in this work, we are confident that the objective of the complete reduction is within reach. \\

Our results showed that, by means of intersection numbers, Feynman integrals can be decomposed in terms of master integrals directly, one-by-one, in alternative to the collective IBP-decomposition, thereby avoiding the computationally expensive system-solving strategy characterizing it. 

In spite of the rich mathematical structure behind intersection numbers, they can be computed in elementary steps using Stokes' theorem for differential forms and Cauchy's residue theorem. Both aspects played a significant role in the development of on-shell and unitarity-based methods in the modern approaches to quantum field theory amplitudes (see for instance refs. \cite{Mastrolia:2009dr,Mastrolia:2009rk}, the review \cite{Ellis:2011cr}, and the references therein). We found that they also control algebraic relations among multi-loop Feynman integrals in dimensional regularization. 

It would be interesting to study connections to the recent applications of closely-related mathematical topics to Feynman integrals, such as  
 D-module theory~\cite{Bitoun:2017nre}, 
 Hopf algebras \cite{Duhr:2012fh, Abreu:2017enx,Abreu:2017mtm}, 
 computational algebraic geometry \cite{Larsen:2015ped,Boehm:2018fpv}, 
 finite fields arithmetic \cite{Peraro:2016wsq},
 and the theory of special functions \cite{Broedel:2018qkq,Bourjaily:2018ycu}. At the same time,
applying the ideas of intersection theory to other representations of Feynman integrals than the Baikov one or to its generalizations may give us new insights on the properties of scattering amplitudes in dimensional regularization.

\appendix
\section{Critical Points and Master Integrals}
\label{app:criticalpoints}

\begin{table}[ht]
\centering
\begin{tabular}{|c|c|c|c|}
\hline $\;\;$ Integral family & Sec. & $\masters_{\rm LBL}$ & $\masters_{\rm std}$ \\ \hline 
\hline
\raisebox{-20pt}{
\includegraphics[scale=0.20]{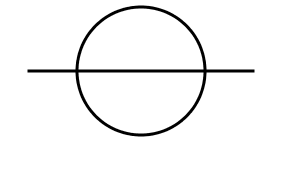}} 
&  \ref{sec:massless-sunrise}& 1 & 1 \\ \hline
\raisebox{-20pt}{
\includegraphics[scale=0.15]{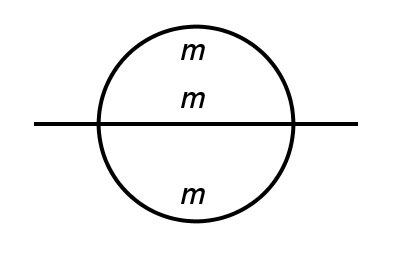}} 
&  \ref{sec:massive-sunrise}& 3 & 4 \\ \hline
\raisebox{-20pt}{\includegraphics[scale=0.28]{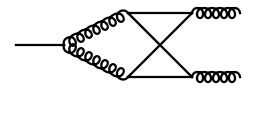}}
&  \ref{sec:nonplanartriangle}& 3 & 3 \\ \hline
\raisebox{-19pt}{\includegraphics[scale=0.28]{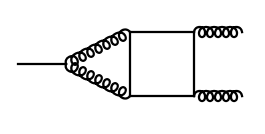}}
&  \ref{sec:planar-triangle}& 1 & 1 \\ \hline
\raisebox{-26pt}{\includegraphics[scale=0.25]{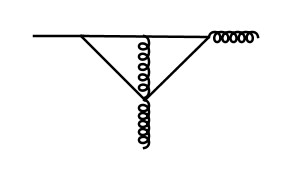}}
& \ref{sec:planardoubletriangle} & 2 & 1 \\ 
\hline
\raisebox{-25pt}{
\includegraphics[scale=0.25]{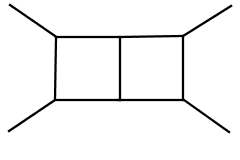}}
& \ref{sec:massless-double-box} & 2 & 2 \\ \hline
\raisebox{-27pt}{
\includegraphics[scale=0.25]{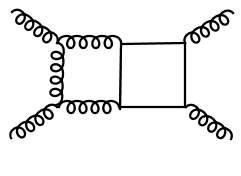}}
& \ref{sec:intmass} & 3 & 4 \\ 
\hline
\raisebox{-26pt}{
\includegraphics[scale=0.25]{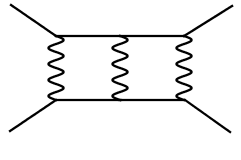}}
& \ref{sec:bhabha1} & 2 & 2 \\ 
\hline 
\raisebox{-27pt}{
\includegraphics[scale=0.25]{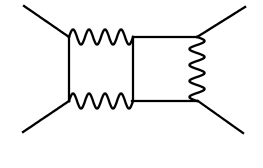}}
& \ref{sec:bhabha2} & 3 & 4 \\ \hline
\raisebox{-28pt}{
\includegraphics[scale=0.25]{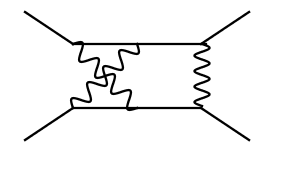}}
& \ref{sec:bhabha3} & 3 & 4 \\ 
\hline
\raisebox{-27pt}
{\includegraphics[scale=0.25]{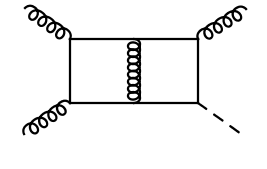}}
& \ref{sec:planarhiggs} & 4 & 4 \\
\hline
\raisebox{-28pt}
{\includegraphics[scale=0.25]{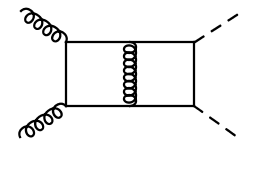}}
& \ref{sec:planarhiggs} & 4 & 4 \\ 
\hline
\raisebox{-28pt}
{\includegraphics[scale=0.25]{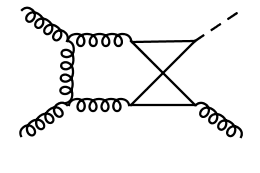}}
& \ref{sec:nonplanarhiggs} & 4 & 6 \\ 
\hline
\end{tabular}
\,
\begin{tabular}{|c|c|c|c|}
\hline $\;\;$ Integral family & Sec. & $\masters_{\rm LBL}$ & $\masters_{\rm std}$ \\ \hline 
\hline
\raisebox{-24pt}
{\includegraphics[scale=0.25]{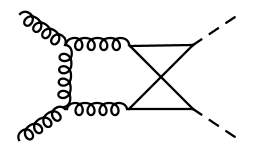}}
& \ref{sec:nonplanardoublehiggs} & 4 & 6 \\ 
\hline
\raisebox{-29pt}
{\includegraphics[scale=0.25]{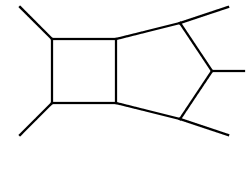}}
& \ref{sec:pentabox:planar} & 3 & 3 \\ 
\hline
\raisebox{-29pt}
{\includegraphics[scale=0.25]{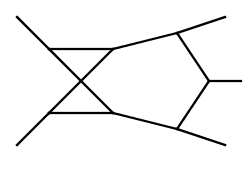}}
& \ref{sec:pentaboxnpl} & 3 & 3 \\
\hline
\raisebox{-32pt}
{\includegraphics[scale=0.25]{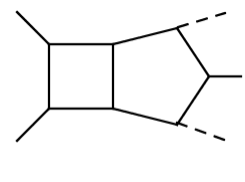}}
& \ref{sec:multileg} & 3 & 3 \\
\hline
\raisebox{-30pt}
{\includegraphics[scale=0.25]{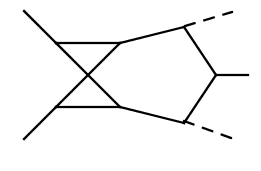}}
& \ref{sec:multileg} & 3 & 3 \\
\hline
\raisebox{-32pt}
{\includegraphics[scale=0.22]{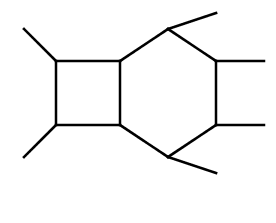}}
& \ref{sec:multileg} & 3 & 3 \\
\hline
\raisebox{-31pt}
{\includegraphics[scale=0.22]{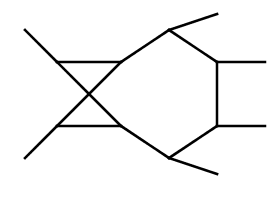}}
& \ref{sec:multileg} & 3 & 3 \\
\hline
\raisebox{-30pt}
{\includegraphics[scale=0.19]{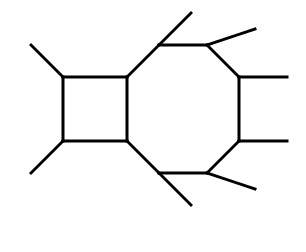}}
& \ref{sec:multileg:all} & 3 & 3 \\
\hline
\raisebox{-26pt}
{\includegraphics[scale=0.25]{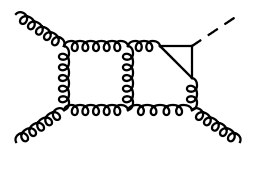}}
& \ref{sec:planar3Lbeyond} & 2 & 2 \\
\hline
\raisebox{-27pt}
{\includegraphics[scale=0.25]{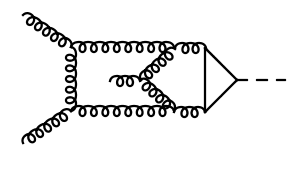}}
& \ref{sec:3Lbeyond:nonplanar} & 3 & 3 \\
\hline
\raisebox{-27pt}
{\includegraphics[scale=0.25]{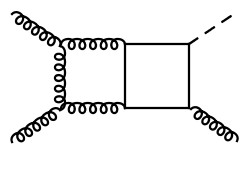}}
& \ref{sec:4Lbeyond:planar} & 3 & 4 \\
\hline
\end{tabular}

\caption{Comparisons of the number of masters obtained by the LP criterion, from \lbl ($\masters_{\rm LBL}$) and standard Baikov parametrization ($\masters_{\rm std}$).}
\label{tab:countingtable}
\end{table}

In ref. \cite{Lee:2013hzt}, it was proposed that the number of master integrals in a given sector equals the number of critical points of the Baikov polynomial on the cut corresponding to this sector: we refer to it as the Lee--Pomeransky (LP) criterion.
The LP criterion can be used to count the number of MIs either in the standard or in the \lbl version of the Baikov representation.
In this paper, we have mostly been using the latter representation to obtain integrals over $1$-forms, 
and for these examples, as shown in Tab. \ref{tab:countingtable}, we compare the number of critical points emerging within the two versions of Baikov representations. 

The relationship between the number of critical points (or the Euler characteristic) and the dimension of the integral basis hinges on several (genericity) assumptions, see, {\it e.g.}, \cite{aomoto2011theory,milnor2016morse}, a few of which are:
$(i)$ all critical points are isolated and non-degenerate;
$(ii)$ each critical point is a ``saddle point'', {\it i.e.}, the number of negative directions extending from it (the so-called \emph{Morse index}) is equal to the number of positive directions; if this is not the case then the number of critical points is only an \emph{upper bound} for the number of independent integrals;
$(iii)$ the exponents of the multi-valued function $u$ are generic enough and in particular not non-positive integers;
$(iv)$ the integral is well-defined, which means it converges in some codimension-$0$ region of the parameter space.
In some cases, we have observed a mismatch between the LP-criterion and the actual number of MIs, implying 
the violation of one of the assumptions, like the the example of Sec.~\ref{sec:planardoubletriangle}, where the number of master integrals is underestimated by the counting of critical points.%
\footnote{To our knowledge, this case was found by Roman Lee \cite{yangprivate}.} We discuss two such examples in the following subsections, and show how to overcome this issue by {\it regulating the exponents} of the Baikov polynomial, ensuring multi-valuedness of the (regulated) $u$ function around boundaries of integration.

Let us begin by considering an example where the LP criterion gives the correct number of MIs: the massless sunrise integral of 
Sec. \ref{sec:massless-sunrise}. For this integral family, after 
choosing the second ISP $y = D_5 = (k_1-p_1)^2$, the standard Baikov representation gives,
\begin{align}
u &= (z y (z+y-s))^{(d-4)/2} \ ,
\end{align}
with
\begin{align}
\omega &= d \log (u) = \frac{(d-4) (2z+y-s)}{2 z (z+y-s)} \, dz  +  \frac{(d-4) (2y+z-s)}{2 y (z+y-s)} \, dy \ .
\end{align}
The equation $\omega = 0$ has 1 solution ($z = s/3$, $y = s/3$) corresponding to 1 master integral, as it was found in Sec. \ref{sec:massless-sunrise}.

Likewise, for the double-box of Sec. \ref{sec:massless-double-box}, 
the first planar Bhabha-integral of Sec. \ref{sec:bhabha1},
and for most of the other cases (see Tab. \ref{tab:countingtable}) there is agreement between the numbers of MIs obtained from the two types of Baikov parametrization. 
Yet, for some cases, 
we find different number of master integrals in the two approaches, and in the following we discuss some of them, in detail. \\

\subsection{Planar Double-Triangle}
\label{app:doubletriangle}
The integral family of the double-triangle diagram of Sec. \ref{sec:planardoubletriangle}, within the standard Baikov representation gives
\begin{align}
u &= \Big( z \big( (s - y)^2 z - 4 m^2 s (y + z) \big) \Big)^{(d-5)/2} \ , \\
\omega &= \frac{(d-5) ((s - y)^2 z - 2 m^2 s (y + 2 z))}{z ((s - y)^2 z - 4 m^2 s (y + z))} \, dz + \frac{(d-5) ((y-s) z - 2 m^2 s)}{(s - y)^2 z - 4 m^2 s (y + z)} \, dy
\end{align}
where we defined the ISP $y = D_7 = (k_2 - p_1 - p_2)^2 - m^2$. 
In this case, $\omega = 0$ has 1 solution ($z = -s/2$, $y = s - 4 m^2$), in disagreement with the 2 MIs found in Sec.~\ref{sec:planardoubletriangle} and in the literature \cite{Aglietti:2006tp,Anastasiou:2006hc}.

As done in Sec.~\ref{sec:planardoubletriangle} for the \lbl representation, 
we introduce a regulating exponent in the powers of $z$, $u \rightarrow u z^{\rho}$, such that,
\begin{align}
u \rightarrow \big( (s - y)^2 z - 4 m^2 s (y + z) \big)^{(d-5)/2} z^{(d-5)/2+\rho} \ .
\end{align}
It is easy to verify that after building the corresponding $\omega$, 
the equation $\omega=0$ has 2 solutions in agreement with the \lbl Baikov used in the main text and with the literature.

\subsection{Internally Massive Double-Box}
\label{app:intmass}
Also for the box-type integral discussed in Sec. \ref{sec:intmass},
the two Baikov representations imply different numbers of master integrals, due to differing number of critical points between the two representations.
In particular, 
after defining the second ISP $y = D_9 = (k_1 + p_1)^2 - m^2$, 
the standard Baikov representation gives,
\begin{align}
u &= \big( m^2 s t^2 - 2 m^2 s t z - s t y z + s y^2 z + m^2 s z^2 + s y z^2 + y^2 z^2 \big)^{(d-6)/2} \,,
\end{align}
and therefore,
\begin{align}
\omega &= \frac{(d-6) \big( ( s y (y-t) + 2 y (s + y) z + 2 m^2 s (z-t) ) \, dz + z ( 2 y z + s (2 y + z - t) ) \, dy \big)}{2 \big( m^2 s (t - z)^2 + y z (y z + s (z+y-t)) \big)} \ .
\end{align}
The equation $\omega = 0$ has 4 solutions corresponding to $\masters = 4$. 
As discussed in Sec. \ref{sec:intmass} this is the number obtained by most IBP programs and also in ref. \cite{Becchetti:2017abb}, but within the \lbl approach we got $\masters = 3$, due to the extra relation given in eq. \eqref{eq:themagicrelation}. 
Also in this case, a regulating exponent, $u \rightarrow u z^{\rho}$, allows $z$ to appear in the denominator as well, so that eq. \eqref{eq:themagicrelation} can be established. Owing to this transformation, the new $\omega$ generates $\masters=3$, in agreement with what we found in Sec. \ref{sec:intmass}. \\

Also for the (second) planar Bhabha-integral of Sec. \ref{sec:bhabha2}, the non-planar Bhabha-integral of Sec. \ref{sec:bhabha3}, the non-planar $(H+j)$-integral of Sec. \ref{sec:nonplanarhiggs}, and a few other cases, the number of critical points obtained in respectively the standard and the \lbl Baikov parametrization, are not the same (see table \ref{tab:countingtable}). For the two Bhabha cases, the standard Baikov gives $\masters = 4$ as opposed to the $\masters = 3$ found in Sec. \ref{sec:bhabha}. This is also due to the existence of additional identities which are not accessible by the standard Baikov parametrization, as in the case above. 

Likewise for the non-planar $H+j$, from the standard Baikov representation one infers $\masters = 6$, as opposed to the $\masters = 4$ found in Sec. \ref{sec:nonplanarhiggs}. This is due to two missed identities: one can be generated from the IBP decomposition of an higher-sector, and one is due to an auxiliary symmetry relation, as discussed in Sec. \ref{sec:nonplanarhiggs}.

To summarize, we limit ourselves to observe that the Baikov parametrization may generate integral representations where 
the integrand is not multivalued, hence it is not fully regulated. A systematic study, beyond the scope of this work, is required, to verify whether the solutions we adopted may lead to a more general integral representation, fulfilling all the requirements indicated in \cite{aomoto2011theory,milnor2016morse}. At the same time, we have shown that within the intersection theory approach applied to the \lbl (regulated) Baikov representation, the number of critical points corresponds to the actual number of master integrals on the maximal cut, and the 
integral relations established are equivalent to the integration-by-parts identities.

\section{Loop-by-loop as Iterative intersections}
\label{app:LBLiterative}

In this appendix, we show that the Iterative Intersections approach, introduced in Sec.{\ref{sec:iteratedoneforms}}, can be used to derive the \lbl Baikov representation of the integrand. In particular we consider two examples of maximal cut diagrams with more than one ISP, which, within the \lbl parametrization, admit a univariate representation, where the integration variable is just {\it one} of all the ISPs, the examples being the two-loop massless double-box and planar pentabox. For each case, we show the decomposition of monomials built out of the products of the additional ISPs (other than the integration variable).

\subsection{Massless Double-Box}
\label{app:decompISP}
Let us begin with the reduction of the massless Double-Box integrals, discussed in Sec.~\ref{sec:massless-double-box}, when both the ISPs are present.
We start from the standard Baikov representation on the maximal cut, which depends on the \emph{two} ISPs $D_{8}=z_1$ and $D_{9}=z_2$ defined by eqs. \eqref{eq:dbisp1} and \eqref{eq:dbisp2}, and show that, within the iterative intersection method of  Sec.~\ref{sec:iteratedoneforms}, after the decomposition by intersection in one variable, say $z_2$, 
we obtain the \lbl representation of the integral. 
Subsequently, by applying the decomposition by intersection in $z_1$, we obtain the final reduction in terms of MIs, in agreement with IBPs.

Applying the standard Baikov representaion on the maximal cut, we obtain:
\begin{equation}
u_{\rm std}(z_1,z_2)= 
\left( 
z_1 z_2 
(z_1 z_2 + s (z_1 + z_2) - s t)
\right)^{\frac{d-6}{2}},
\end{equation}
\begin{equation}
\omega_{\rm std}=\hat{\omega}_{{\rm std},1} \,  d z_1 + \hat{\omega}_{{\rm std},2} \,  d z_{2}, \quad \omega_{{\rm std},1}=\partial_{z_1} \log u_{\rm std}, \quad \omega_{{\rm std},2}=\partial_{z_2} \log u_{\rm std}.
\end{equation}

The generic integral reads as,
\begin{eqnarray}
I_{1,1,1,1,1,1,1;-n,-m} &=&
\int_{  \mathcal{C}_{1} } 
dz_1
\int_{  \mathcal{C}_{2} } 
dz_2 \,
z_{1}^{n} z_{2}^{m} \,  
u_{\rm std}(z_1, z_2)  \nonumber \\
&=&
\int_{\mathcal{C}_1} z_{1}^{n}  \, d z_1 
\int_{\mathcal{C}_{2}} z_{2}^{m} \, u_{\rm std}(z_1,z_2) d z_2 .
\label{eq:factorization_Integral_STD}
\end{eqnarray}

\paragraph{Intersections in $z_2$.}
We focus on the innermost integration, namely:
\begin{equation}
\int_{\mathcal{C}_2} z_{2}^{m} \, u_{\rm std}(z_1,z_2) \, d z_2,
= \langle{\phi_{m+1}(z_2)} | \mathcal{C}_2].
\end{equation}

We observe that the equation: $\hat{\omega}_{{\rm std},2}=0$ has only \emph{one} solution, namely $\nu_{2}=1$, and the internal MIs can be chosen to be $\langle{\phi_{1}(z_2)}|\mathcal{C}_2].$\\
Therefore, using the master formula with univaraite intersections in $z_2$, we get,
\begin{equation}
\langle{\phi_{m+1}(z_2)} | \mathcal{C}_2]=  f_{m}(z_1)  \langle{\phi_{1}(z_2)}| \mathcal{C}_{2}],
\end{equation}
with
\begin{equation}
f_{m}(z_1)= \, \bra{\phi_{m+1}(z_2)}\ket{\phi_1(z_2)} \bra{\phi_1(z_2)}\ket{\phi_1(z_2)}^{-1} \ .
\end{equation}
Accordingly, the original integral reads as
\begin{eqnarray}
\int_{\mathcal{C}_1} z_{1}^{n}  \, d z_1 
\int_{\mathcal{C}_{2}} z_2^m u_{\rm std}(z_1,z_2) \, dz_2 
&=&
\int_{\mathcal{C}_1} z_{1}^{n}  \, d z_1 \,
f_m(z_1) \int_{\mathcal{C}_{2}} u_{\rm std}(z_1,z_2) \, dz_2 \nonumber \\
&=&
\mathcal{K} \, \int_{\mathcal{C}_1} z_{1}^{n}  \, d z_1 \,
f_m(z_1) \, u_{\rm LBL}(z_1) \ ,
\label{eq:key_passage}
\end{eqnarray}
where we used the direct integration in $z_2$ in the last step \cite{Frellesvig:2017aai}:
\begin{eqnarray}
\int_{\mathcal{C}_{2}} u_{\rm std}(z_1,z_2) \, dz_2 = \mathcal{K} \,  u_{{\rm LBL}}(z_1) \ , \quad 
{\rm with} \quad 
\mathcal{C}_2= \left[ 0, \frac{s \left(t-z_1\right)}{s+z_1} \right],
\end{eqnarray}
where $u_{\rm LBL}(z_1)$ was defined in eq.~\eqref{eq:u_LBL_Massless_DB}, ${\cal C}_2$ is chosen such that $u_{\rm std}(\partial {\cal C}_2)=0$ and $\mathcal{K}$ is an irrelevant overall constant.
The expression in eq.~\eqref{eq:key_passage} corresponds to the \lbl Baikov representation of the original integral $I_{1,1,1,1,1,1,1;-n,-m}$ (on the maximal-cut). 
In particular, for $D_9^m = z_2^{m}$, $m=1,2,3$, the expression of $f_m$ read,
\begin{eqnarray}
D_9 \to f_{1}(z_1) &=&  \frac{s \left(t-z_1\right)}{2 \left(s+z_1\right)} \ , \\
D_9^2 \to f_{2}(z_1) &=&  {(d-2) \over (d-3)} \left( \frac{s \left(t-z_1\right)}{2 \left(s+z_1\right)} \right)^2 \ , \\
D_9^3 \to f_{3}(z_1) &=& {d \over (d-3)} \left( \frac{s \left(t-z_1\right)}{2 \left(s+z_1\right)} \right)^3 \ .
\end{eqnarray}
Notice that $f_1(z_1)$ appeared already in eq.~(\ref{eq:massless_DB_D9_pow1}).

\paragraph{Intersection in $z_1$.} 
The reduction of $I_{1,1,1,1,1,1,1;-m,-n}$ within the \lbl approach proceeds along the same line as 
in Sec.~\ref{sec:massless-double-box}, in terms of the the two MIs $J_1=I_{1,1,1,1,1,1,1;0,0}$ and $J_2=I_{1,1,1,1,1,1,1;-1,0}$.\\
While the case $(n,m)= (1,1)$ was discussed in the abovementioned section, we hereby give the decomposition by intersections of $(n,m)=\{(1,2), (1,3) \}$.
\begin{itemize}
\item Let us consider the decomposition of 
$I_{1,1,1,1,1,1,1;-1,-2}=\langle{ f_{2}(z_1) \, z_1}|\mathcal{C}_1]$.
Using the basis $\varphi_{1,2}$ and the $\mathbf{C}$ matrix given in eqs.~(\ref{eq:right_dlog_basis_Massless_DB}, \ref{eq:C_matrix_Mixed_basis_Massless_DB}), as well as the the intersection numbers:
\begin{equation}
\begin{split}
\bra{f_2(z_1) z_1}\ket{\varphi_{1}} = &\frac{3 \left(5 d^2-41 d+82\right) s^3 t}{4 (d-5) (d-4)
   (d-3)}+\frac{3 (3 d-14) (3 d-10) s^4}{8 (d-5) (d-4) (d-3)}+\frac{s^2
   t^2}{2 (d-3)}, \\
\bra{f_2(z_1) z_1}\ket{\varphi_{2}}  = & -\frac{3 \left(13 d^2-104 d+204\right) s^3 t}{8 (d-5)
   (d-4) (d-3)}-\frac{3 (3 d-14) (3 d-10) s^4}{8 (d-5) (d-4)
   (d-3)} \\
   & -\frac{3 (d-4) s^2 t^2}{2 (d-5) (d-3)}.
\end{split}
\end{equation}
Then the master decomposition formula eq.~(\ref{eq:masterdeco:}) gives:
\begin{equation}
I_{1,1,1,1,1,1,1;-1,-2}= \frac{(10-3 d) s^2 t}{4 (d-3)} \,  J_{1} + \frac{s ((9 d-30) s+2 (d-4) t)}{4(d-3)}J_{2},
\end{equation}
in agreement with \textsc{LiteRed}.
\item Analogously, for the decomposition of 
$I_{1,1,1,1,1,1,1;-1,-3}=\langle{ f_{3}(z_1) \, z_1}|\mathcal{C}_1]$, with 
\begin{equation}
\begin{split}
\bra{f_{3}(z_1) \, z_1}\ket{\varphi_{1}}= & -\frac{3 (3 d-10) \left(6 d^2-47 d+88\right) s^4 t}{8 (d-5) (d-4)
   (d-3) (d-2)} -\frac{3 (3 d-14) (3 d-10) (3 d-8) s^5}{16 (d-5) (d-4)
   (d-3) (d-2)} \\
   & -\frac{(7 d-26) s^3 t^2}{4 (d-3) (d-2)}+\frac{s^2 t^3}{2
   (d-3) (d-2)}, \\
\bra{f_{3}(z_1) \, z_1}\ket{\varphi_{2}}= &  \frac{3 (3 d-10) \left(15 d^2-114 d+208\right) s^4 t}{16
  (d-5) (d-4) (d-3) (d-2)}+\frac{\left(29 d^2-220 d+420\right) s^3
  t^2}{8 (d-5) (d-3) (d-2)} \\
  & +\frac{3 (3 d-14) (3 d-10) (3 d-8) s^5}{16
  (d-5) (d-4) (d-3) (d-2)}+\frac{(d-6) s^2 t^3}{4 (d-3) (d-2)}.  
\end{split}
\end{equation}
Therefore the master decomposition formula eq.~(\ref{eq:masterdeco:}) gives:
\begin{equation}
\begin{split}
I_{1,1,1,1,1,1,1;-1,-3}= & \frac{s^2 t \left((9 (d-6) d+80) s+2 (d-4)^2 t\right)}{8 (d-3)
   (d-2)} J_{1}\\
   & +\frac{s \left(-3 (9 (d-6) d+80) s^2-4 (d-4) (3 d-10) s t+4
   (d-4) t^2\right)}{8 (d-3) (d-2)} J_{2},
\end{split}
\end{equation}
in agreement with \textsc{LiteRed}.
\end{itemize}

\subsection{Planar Pentabox}

We now focus on the Planar Pentabox integrals of Sec.~\ref{sec:pentabox:planar}, whose maximal-cut depends on three ISPs, say $z_1, z_2$ and $z_3$, defined as,
\begin{align}
    D_9 = z_1 \ , \quad
    D_{10} = (k_1 + p_1 + p_2 + p_3)^2 = z_2 \quad \text{and} 
    \quad D_{11} = (k_1 + p_1 + p_2 + p_3 + p_4)^2 = z_3 \ ,
\end{align}
yielding
\begin{align}
    u_{\text{std}}(z_1,z_2,z_3) &= \Big( 2 v_{12} v_{34} z_1 (2 v_{51} - z_1) (2 v_{45} - z_2) z_2 - v_{34}^2 z_1^2 (2 v_{45} - z_2)^2 \nonumber \\
    &\;\; + 2 v_{34} z_1 \big( v_{45} z_1 (2 v_{45} - z_2) - v_{12} (2 v_{45} (z_1 + 2 z_2) + z_1 z_2 - 2 v_{23} (2 v_{45} + z_2)) \big) z_3 \nonumber \\
    &\;\; - \big( v_{45} z_1 z_3 + v_{12} (z_1 z_2 + 2 v_{23} z_3 - z_1 z_3 - 2 v_{51} z_2 ) \big)^2 \Big)^{(d-7)/2} \ ,
\end{align}
Let us consider the decomposition of a generic monomial $z_1^k z_2^n z_3^m$.
After applying the Iterative Intersections, first in $z_3$ and later in $z_2$, one obtains
\begin{eqnarray}
\int_{ \mathcal{C}_{1} } 
 \!\! dz_1 \!\! 
\int_{ \mathcal{C}_{2} }  \!\! dz_2 \!\! 
\int_{ \mathcal{C}_{3} }  \!\! dz_3 \, 
z_1^k \, z_2^n \, z_3^m u_{\rm std}(z_1,z_2,z_3) &=&
\int_{ \mathcal{C}_{1} } \!\! dz_1 \ z_1^k \, f_{n,m}(z_1) 
\int_{ \mathcal{C}_{2} }  \!\! dz_2 
\int_{ \mathcal{C}_{3} }  \!\! dz_3 \, u_{\rm std}(z_1,z_2,z_3) \nonumber \\
& = & \mathcal{K} \, 
\int \!\! dz_1 \ z_1^k \, f_{n,m}(z_1) \, u_{\rm LBL}(z_1) \ ,
\end{eqnarray}
where we used the identity,
\begin{eqnarray}
\int_{{\cal C}_2} dz_2 
\int_{{\cal C}_3} dz_3 \, u_{\rm std}(z_1,z_2,z_3) &=& \mathcal{K} \,  u_{\rm LBL}(z_1) \ , 
\end{eqnarray}
with $u_{\rm LBL}$ given in eq.(\ref{eq:def:uLBL:pentabox}), and $\mathcal{K}$ an overall constant.
The two integration contours are 
\begin{align}
{\cal C}_3 = \left[ \tfrac{\alpha - 2 \sqrt{\beta}}{((v_{45} - v_{12}) z_1 + 2 v_{12} v_{23})^2}, \tfrac{\alpha + 2 \sqrt{\beta}}{((v_{45} - v_{12}) z_1 + 2 v_{12} v_{23})^2} \right] , \;\; \text{and} \;\; {\cal C}_2 = \left[0, \tfrac{2 ((v_{45} - v_{12}) z_1 + 2 v_{12} v_{23})}{2 v_{12} + z_1} \right] ,
\end{align}
with
\begin{align}
    \alpha &= v_{12} (2 v_{51} - z_1) (2 v_{12} v_{23} - v_{12} z_1 + v_{45} z_1) z_2 + v_{34} z_1 \big( v_{45} z_1 ( 2 v_{45} - z_2) \nonumber \\
    & \quad - v_{12} (2 v_{45} z_1 + 4 v_{45} z_2 + z_1 z_2 - 2 v_{23} (2 v_{45} + z_2)) \big) \ , \\
    \beta &=  v_{12} \, v_{34} \, v_{45} \, z_1 \, z_2 \, \big(z_1 (2 v_{45} - z_2) + 2 v_{12} (2 v_{23} - z_1 - z_2) \big) \, \times \nonumber \\
& \quad \big( v_{34} (2 (v_{23} - v_{45}) - z_1) z_1 + (2 v_{51} - z_1) ( (v_{45} - v_{12}) z_1 + 2 v_{12} v_{23}) \big)  \ .
\end{align}
In particular, we provide the $f_{n,m}$ for $(n,m) =\{ (1,0),\,(0,1),$ $(1,1),\,(2,0),\,(0,2)\}$:
\begin{align}
    D_{10}         &\rightarrow f_{1,0}(z_1) = \frac{(v_{34} - v_{12}) z_1 + 2 v_{12} v_{51}}{z_1 + 2 v_{12}}\,, \\
    D_{11}         &\rightarrow f_{0,1}(z_1) = \frac{(v_{45} - v_{12}) z_1 + 2 v_{12} v_{23}}{z_1 + 2 v_{12}}\,, \\
    D_{10} D_{11}  &\rightarrow f_{1,1}(z_1) = \Big( (d-4) v_{34} v_{45} z_1^2  -  4 v_{12} v_{34} v_{45} z_1 + 
 (d-2) \big( v_{12} (v_{12}{-}v_{34}{-}v_{45}) z_1^2 \nonumber \\ 
 & \qquad\qquad\qquad - 2 v_{12} (v_{12} v_{23} - v_{23} v_{34} + v_{12} v_{51} - v_{45} v_{51}) z_1 + 4 v_{12}^2 v_{23} v_{51} \big) \Big) \Big/ \nonumber \\
 & \qquad\qquad\qquad\quad \Big( (d-3) (z_1 + 2 v_{12})^2 \Big)\,, \\
   D_{10}^2        &\rightarrow f_{2,0}(z_1) = \frac{d-2}{d-3} \left( \frac{(v_{34} - v_{12}) z_1 + 2 v_{12} v_{51}}{z_1 + 2 v_{12}} \right)^2 , \\
   D_{11}^2        &\rightarrow f_{0,2}(z_1) = \frac{d-2}{d-3} \left( \frac{(v_{45} - v_{12}) z_1 + 2 v_{12} v_{23}}{z_1 + 2 v_{12}} \right)^2 .
\end{align}
The final decomposition in terms of the monomial basis used in Sec.~\ref{sec:pentabox:planar}
can be achieved, 
using univariate intersection in $z_1$, in agreement with {\sc Kira}.

\acknowledgments

We acknowledge C. Duhr, C. Papadopoulos, and E. Remiddi for interesting discussions, and G. Ossola for comments on the manuscript. We wish to thank Y. Zhang, for pointing out the example in Sec.~\ref{sec:planardoubletriangle} and for feedback on the manuscript. S.M. thanks K.~Matsumoto for useful correspondence. 
H.F., F.G., P.M., and L.M. would like to thank the organizers of {\it Amplitudes in the LHC era} workshop,
for the kind hospitality at the GGI Institute in Florence, where part of this work has been carried out.
P.M. wishes to thank Damiano M. for bringing ref.~\cite{hwa1966homology} to his attention, ahead of time. \\
CloudVeneto is acknowledged for the use of computing and storage facilities. \\
The figures were drawn with Jaxodraw~\cite{Binosi:2008ig} based on Axodraw~\cite{Axodraw:1994}.

The work of F.G., S.L, M.K.M., and P.M., is supported by the Supporting TAlent in ReSearch at Padova University (UniPD STARS Grant 2017 ``Diagrammalgebra''). 
The work of H.F. is part of the HiProLoop project funded by the European Union’s Horizon 2020 research and innovation programme under the Marie Sk{\l}odowska-Curie grant agreement 747178.
This research was supported in part by Perimeter Institute for Theoretical Physics. Research at Perimeter Institute is supported by the Government of Canada through the Department of Innovation, Science and Economic Development Canada and by the Province of Ontario through the Ministry of Research, Innovation and Science.

\bibliographystyle{JHEP}
\bibliography{references}

\end{document}